\documentclass{kapproc} % Computer Modern font calls

\input{psfig}

\let\footnote\savefootnote

\let\footnoterule\savefootnoterule 

\kluwerbib
\def\lessapprox{\setbox0=\hbox{$<$}\setbox1=\hbox{$\sim$}
     \lower0.5\ht0
     \hbox{ \vbox{\baselineskip=0pt\lineskip=0.5pt\box0\box1} }}
\def\greatapprox{\setbox0=\hbox{$>$}\setbox1=\hbox{$\sim$}
     \lower0.5\ht0
     \hbox{ \vbox{\baselineskip=0pt\lineskip=0.5pt\box0\box1} }}
\def\d3k{{\displaystyle {\rm d}{\bf k} \over \displaystyle (2\pi)^3}}

\begin{document}

\articletitle{Froth across the Universe\ \\
\medskip\\
Dynamics and Stochastic Geometry\\
of the Cosmic Foam}
\chaptitlerunninghead{Froth across the Universe}

\vskip -0.25cm
\author{Rien van de Weygaert}
\affil{Kapteyn Institute, University of Groningen, Groningen, 
the Netherlands}
\email{weygaert@astro.rug.nl}

\vskip -0.5cm
\begin{abstract}
The interior of the Universe is permeated by a tenuous space-filling frothy 
network. Welded into a distinctive foamy pattern, galaxies accumulate in walls, filaments 
and dense compact clusters surrounding large near-empty void regions.
As borne out by a large sequence of computer experiments, such 
weblike patterns in the overall cosmic matter distribution do represent 
a universal but possibly transient phase in the gravitationally propelled 
emergence and evolution of cosmic structure. We discuss the properties of this striking 
and intriguing pattern, describing its observational appearance, seeking to 
elucidate its dynamical origin and nature and attempting to frame a geometrical 
framework for a systematic evaluation of its fossil content of information on the 
cosmic structure formation process. 

An extensive discussion on the gravitational formation and 
dynamical evolution of weblike patterns attempts to put particular emphasis 
on the formative role of the generic anisotropy of the 
cosmic gravitational force fields. These tidal fields play an 
essential role in shaping the pattern of the large scale cosmic 
matter distribution. A profound investigation of their role will be a key 
element in understanding the implications of the observed cosmic foam for 
the very process of cosmic structure formation. 

The apogee of this contribution is reached in the specific attention for  
the geometric and stochastic aspects of the cosmic fabric. Its distinct geometric 
character -- linking various distinct anisotropic morphological elements 
into a global all-encompassing framework -- and the stochastic nature of 
this assembly provides the cosmic web with some unique and at first unexpected 
properties. The implications for galaxy clustering and its potential as 
discriminating of the galaxy distribution are discussed on the basis of 
its relevant branch of mathematics, stochastic geometry. Central within this 
context are Voronoi tessellations, which have been found to represent a 
surprisingly versatile model for spatial cellular distributions, whose 
flexibility and efficient exploitation warrant a central role in systematic 
assesments of the {\it cosmic foam}. 
\end{abstract}

\vskip -0.25cm
\begin{keywords}
Large-scale structure of the Universe, Methods: statistical, numerical
\end{keywords}

\begin{figure}
\vskip -.0truecm
\centering\mbox{\hskip -2.0truecm\psfig{figure=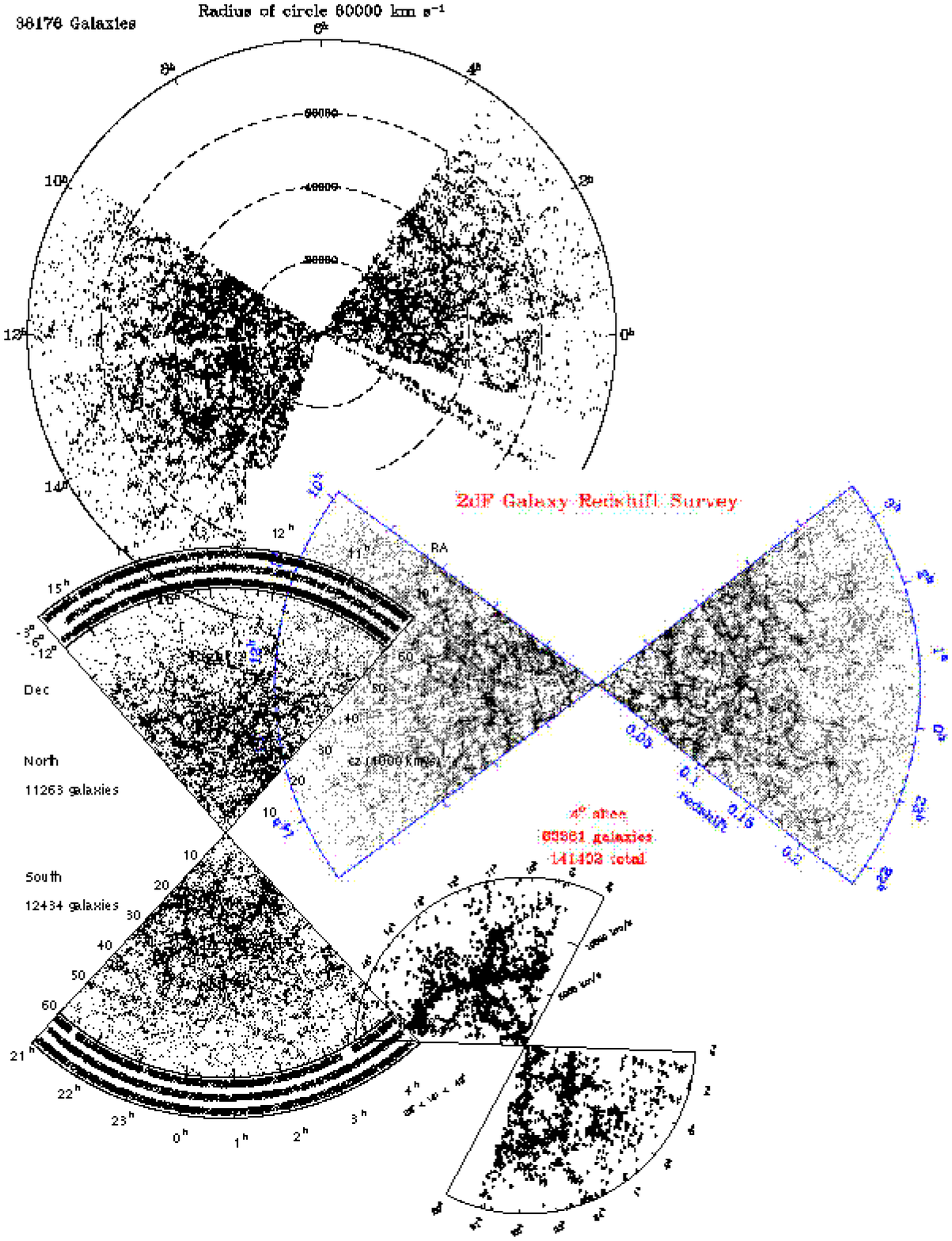,height=21.5cm}}
\end{figure}
\vfill\eject
\begin{figure}[h]
\caption{The development of our Megaparsec cosmos worldview over the past two decades. 
A compilation of the galaxy distribution charted in four major galaxy redshift survey 
campaigns. The CfA2/SSRS survey (bottom righthand figure, courtesy L. da Costa) formed 
the first stage in disclosing the existence of a complex spatial pattern in the cosmic 
galaxy distribution. The Las Campanas redshift survey (bottom lefthand figure, 
courtesy LCRS team) confirmed the ubiquity and reality of these foamy patterns 
over vast reaches of our Universe. Moreover, it also provided evidence for sizes 
of the corresponding inhomogeneities not to surpass scales of $100-200h^{-1}\hbox{Mpc}$. 
With the arrival of the major and uniformly defined galaxy redshift campaigns of the 2dF survey 
(central frame, courtesy 2dF Galaxy Redshift Survey team) and the overwhelming 
1 million galaxy redshift Sloan SDSS survey (top lefthand frame: preliminary galaxy 
redshift map, kindly provided by M. Strauss, with courtesy of the SDSS consortium) the fabric and 
the kinematics of the local Universe will get firmly established and provide a 
major resource for systematic scientific studies of all aspects of cosmic 
structure formation. Courtesy: L. da Costa; LCRS team (Shectman, S., Schechter, P., Oemler, G., 
Kirshner, B., Tucker, D., Landy, S., Hashimoto, Y. \& Lin, H.); the 2dF consortium, with special 
thanks to J. Peacock; the SDSS consortium, with special gratitude to M. Strauss.}
\end{figure}
\begin{flushright}
\footnote{Lennon \& McCartney, 1970, Let it Be (EMI records)}
\vskip -0.5truecm
\end{flushright}
\section{\rm{\Large ACROSS THE UNIVERSE ... }}
\smallskip
Over the past two decades we have witnessed a paradigm shift in our 
perception of the Megaparsec scale structure in the Universe. As increasingly 
elaborate galaxy redshift surveys charted ever larger regions 
in the nearby cosmos, an intriguingly complex and salient foamlike network 
came to unfold and establish itself as th\'e quintessential characteristic 
of the cosmic matter and galaxy distribution. 

In a great many physical systems (see Fig. 55), the spatial organization of 
matter is one of the most readily observable manifestations of the forces and processes 
forming and moulding them. Richly structured morphologies are usually the consequence 
of the complex and nonlinear collective action of basic physical processes. Their 
rich morphology is therefore a rich source of information on the   
combination of physical forces at work and the conditions from which the systems 
evolved. In many branches of science the study of geometric patterns has therefore  
developed into a major industry for exploring and uncovering the underlying physics 
(see e.g. Balbus \& Hawley 1998). 

The vast Megaparsec cosmic web is undoubtedly one of the most striking examples of 
complex geometric patterns found in nature. Revealed through the painstaking efforts 
of redshift survey campaigns, it has completely revised our view of the matter 
distribution on these cosmological scales. Figure 1 forms a telling testimony 
of the gradual unfolding of the cosmic foam patterns in the galaxy distribution 
by a sequel of ever deeper probing galaxy redshift surveys. It depicts a compilation of 
the CfA2/SSRS survey (courtesy, L. da Costa), the Las Campanas redshift survey 
(courtesy: LCRS team), the 2dF survey (courtesy 2dF Galaxy Redshift Survey team) and 
the first impression of the Sloan SDSS redshift survey (with thanks to M. Strauss). 

In its own right, the vast dimensions and intricate composition of the cosmic foam 
make it one of the most imposing and intriguing 
patterns existing in the Universe. Its wide-ranging importance stems from its 
status as a cosmic fossil. On the typical scale of tens up to a few hundred 
Megaparsecs it is still relatively straightforward to relate the configuration 
at the present cosmic epoch to that of the primordial matter distribution from which 
it has emerged. With the cosmic foam seemingly representing this phase, it 
assumes a fundamental role in the quest for understanding the origin of all structures 
in the Universe. It represents a key element in the search for a compelling theoretical 
framework that offers and self-consistent explanation and description for the breathtaking 
variety and wealth of structures and objects that populate the present-day Universe,  
making it such a fascinating world to live in. 

The emergence and formation of structure out of the almost perfectly smooth, virtually 
featureless, pristine Universe still remains one of the major unsettled issues in 
astrophysics. Ultimately, the intention is the framing of a theory that not only concerns 
the global cosmological aspects embodied in the FRW models but also includes a fully 
self-consistent explanation for the configuration and evolution of its interior mass 
distribution is. At present, the search for the necessary extension of the 
Friedmann-Robertson-Walker models towards such an all-embracing cosmological theory 
figures as of the most active branches of modern astrophysical research. 

The Friedmann-Robertson-Walker models -- based on the premise of a homogenous 
and isotropic Universe whose gravitationally driven evolution is drafted in terms of 
General Relativity theory -- have proven to provide a remarkably 
succesfull description of the structure, evolution and thermal history of 
the global Universe. The gradual accumulation of an impressive array of 
observational evidence has been so compelling that we have come to regard the 
``Hot Big Bang'' model as a central tenet of our scientific worldview. 

Yet, the FRW cosmological framework cannot be considered complete, as  
it suffers from a fundamental deficiency. It does not comprise an 
implicit explanation for what is after all one of its most visible 
characteristics, the state of its material content. The bare FRW 
cosmology is fundamentally incapable of addressing the question of 
why matter has condensed into a hierarchy of distinct objects and a 
variety of more or less coherent structures -- planets, stars and 
galaxies, as well as the vast clusters and Megaparsec superclusters. 

Indeed, while in itself one of the most fundamental issues in 
astrophysics, appealing to one of the most profound questions  
occupying mankind since the dawn of civilization -- the quest for the 
origin of our world and that of its constituents -- it is also 
a prominent issue for a variety of additional reasons. The study of  
geometry, structure, evolution and dynamics of our Universe would 
be an idle and unyielding enterprise if we would not have the full 
arsenal of astronomical objects (ranging from stars, gas clouds to 
galaxies and clusters) to function as basic probes enabling 
the measurement of the relevant physical quantities. Yet, lacking 
a precise understanding of their position and origin within the 
grander cosmological context, interpreting the measured 
information will always involve an element of uncertainty and 
arbitrariness. Moreover, perhaps the most essential of all conditions for 
advancements in answering these questions at all, the formation of structure 
paved the road for the rise of an inquiring intelligence,  ... 

This contribution revolves around the central position of the 
cosmic web in the investigations intent to solve the remaining riddles of the 
structure formation process. Particular attention is directed towards 
the stochastic and geometric properties of the cosmic web. While its complex 
cellular morphology involves one of the most outstanding and evident aspects of 
the cosmic foam, it has also remained one defying simple definitions which may 
be the cause of it having remained one of the least adressed aspects. The geometry of the 
cosmic foam may be described as a nontrivial stochastic assembly of various anisotropic 
and asymmetric elements. A major deficiency in the vast majority of studies on the 
large scale distribution of galaxies has been the lack of suitable quantitative and 
statistical characterizations of the truely fundamental aspects of the cosmic foam 
geometry, and the subsequent description in mere qualitative and not very 
decisive terms. Such patterns escape 
descriptions by appropriate simplified approximations. A statistical description in terms 
of a few conventional parameters will almost certainly fail, after all its highly 
nonlinear pattern implicates significant values for a range of higher order 
correlation functions. The limited mathematical machinery, in turn, has been 
a major obstacle in exploiting the potentially large information content of 
the cosmic web. 
\begin{figure}
\centering\mbox{\psfig{figure=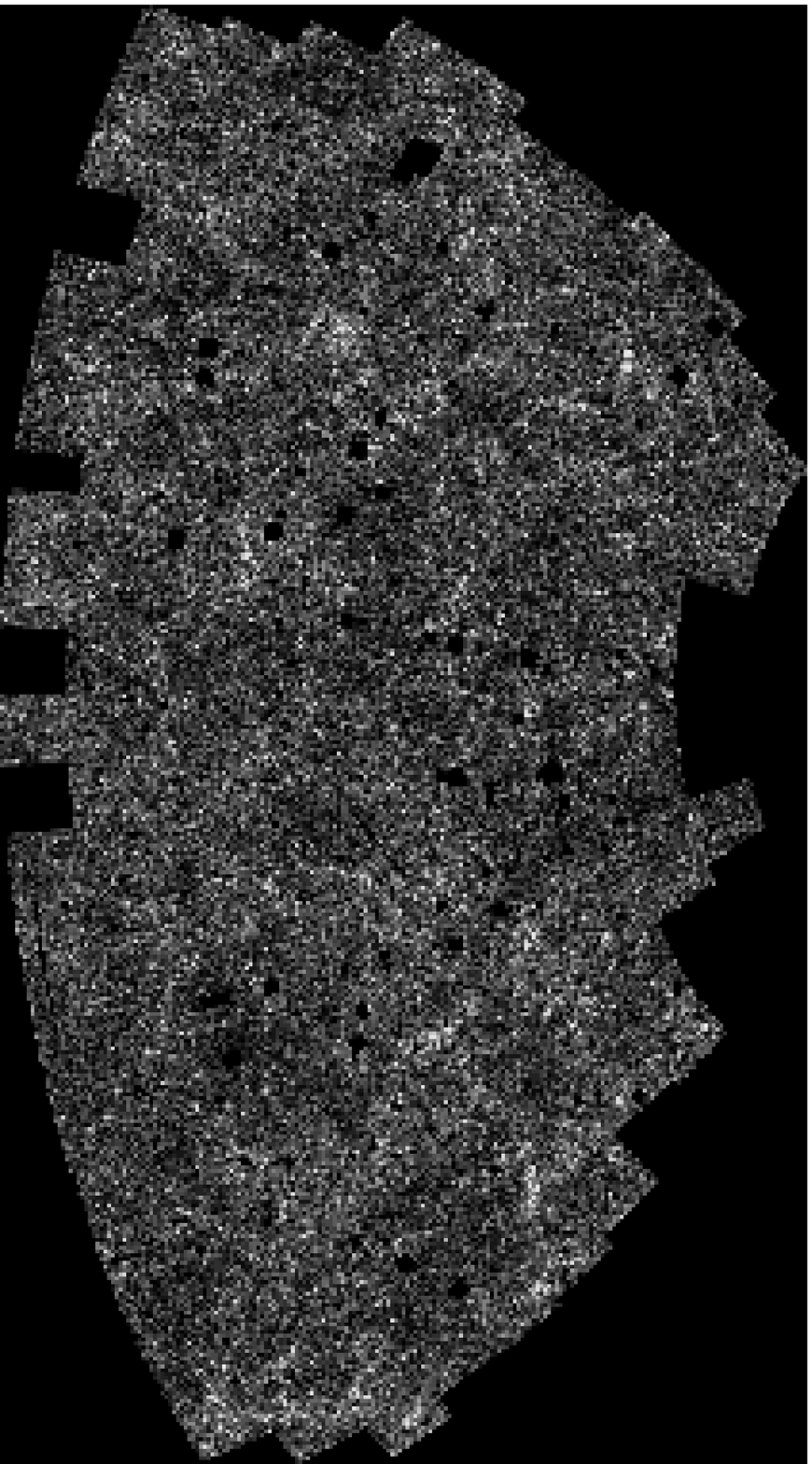,height=20.0cm}}
\end{figure}
\vfill\eject
\begin{figure}[h]
\caption{The sky distribution of galaxies in the APM survey. This 
uniformly defined galaxy map comprises $\approx$ 2 million galaxies 
with a magnitude in between $m=17$ and $m=20.5$, located within an 
angular region of 4300 square degrees on the southern sky. The survey 
is based on objective machine scans of 185 UK Schmidt plates, each 
of $6^{\circ} \times 6^{\circ}$. The resulting projected galaxy 
distribution provides ample evidence f0r the existence of 
large inhomogeneities. However, although superior for the large number 
of objects, for an overall impression of topology and morphology 
of the spatial galaxy distribution galaxy redshift surveys 
remain instrumental. See Maddox et al. 1990a,b. 
Courtesy: S. Maddox, G. Efstathiou, W. Sutherland, en D. Loveday}
\end{figure}
Hence we attempt to lay the foundations for a compelling geometrical 
framework enabling us to analyze the cosmic foam to a more 
profound and substantial extent than hitherto customary. For the 
appropriate concepts and instruments, we have delved into the mathematical field 
of stochastic geometry.  This branch of mathematics deals with a stochastic context 
of geometric objects and concepts. By implication it is also the field adressing the 
issues of spatial point clustering, a prominent point of attention in the study of the 
galaxy distribution. In particular, we emphasize the virtues of one of stochastic 
geometry's basic concepts, Voronoi tessellations. The phenomological similarity of 
Voronoi foams to the cellular morphology seen in the galaxy distribution is 
suggestive for its further exploration. Indeed, we will indicate that such 
similarity is a consequence of the tendency of gravity to shape and evolve structure 
emerging from a random distribution of tiny density deviations into a network of 
anisotropically contracting features. The application of Voronoi tessellations 
gets firmly vindicated by a thorough assessment of its spatial clustering 
properties. They provide us with a succesfull, surprisingly versatile geometrical model 
for spatial cellular distributions. Its high flexibility and applicability to a large 
variety of situations, enables us to systematically study the consequences of the 
existence of a cellular network for spatial clustering of galaxies and other cosmic objects. 
Indeed, we will show that some well-known spatial clustering properties of 
galaxies may indeed ultimately and intricately stem from the very network geometry of the 
cosmic galaxy distribution itself. It is within the context of these spatial 
statistical tests that unexpected profound `scaling' symmetries were uncovered, shedding 
new light on the intricacies of spatial clustering.

The path of delving into the secrets of cosmic structure by means of such 
cellular geometries is following a tradition almost as old as mankind has come 
to realize that the elevated realm of mathematics paves the road towards 
understanding the workings of the world of `necessity'. It was Plato who saw a world 
of geometric forms underlying the manifestations of the `Becoming' (Plato, $\approx$ 
355-350 B.C.). Most purely he was succeeded by Descartes (Fig. 33,  
Descartes 1664), combining geometric objects into a tessellating pattern in an unsuccesfull 
and not fully appreciated attempt for explaining the causal action propelling our 
solar system. We think the world at large may be the proper sphere in which to to pursue 
these momentous ventures of inquiry. 
\begin{figure}
\vskip -1.5truecm
\centering\mbox{\hskip -2.0truecm \psfig{figure=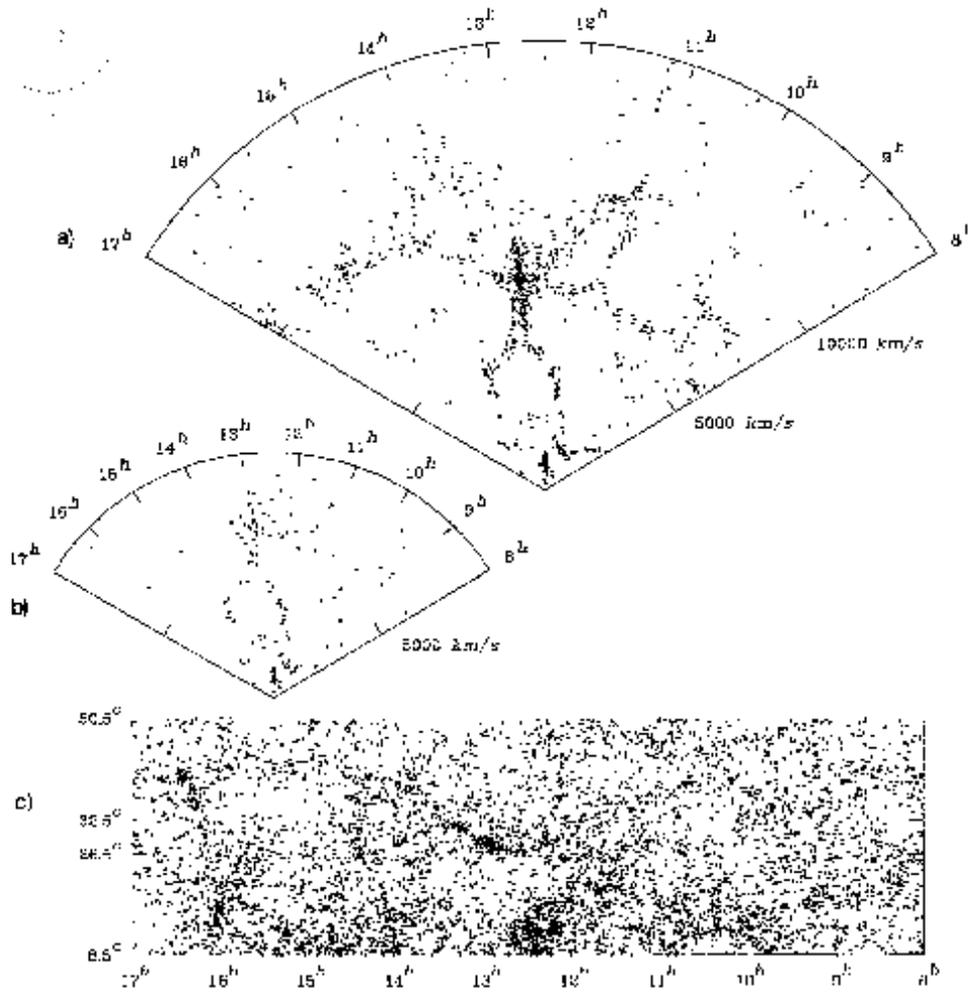,width=16.5cm}}
\vskip -7.25truecm
\caption{The revelation of the cosmic foam. The first published 
``Slice of the Universe'' from the CfA2 survey, by de Lapparent, Geller \& Huchra (1986). It comprises all galaxies 
with an apparent magnitude $m \leq 15.5$ in a narrow $6^{\circ}$ slice in a region 
towards the Coma Cluster ($(\alpha,\delta) \approx (13^h,27^{\circ}), see c))$. This stereological sampling 
of the galaxy distribution represents a highly efficient method to obtain an impression of the 
overall spatial pattern. The result was the at first surprising finding of the apparently 
idiosyncratic cellular galaxy distribution. Note that the shallower CfA1 survey (cf. b), probing to 
$m<14.5$ at hindsight contained a hint for these nontrivial patterns, be it not yet sufficiently 
convincing. Rightfully, this slice may be seen as a historic document, being a turning point for 
our view of the Universe's structure. Courtesy: V. de Lapparent, M. Geller 
\& J. Huchra. Reproduced by permission of the AAS.}
\end{figure}
\bigskip
\section{\rm{\Large WORLDWIDE WEB: ... }}
\begin{flushright}
{\rm{\Large the Foamy Distribution of Galaxies}}
\vskip -0.25truecm
\end{flushright}
\begin{figure}[b]
\vskip -1.0truecm
\centering\mbox{\hskip -3.0truecm \psfig{figure=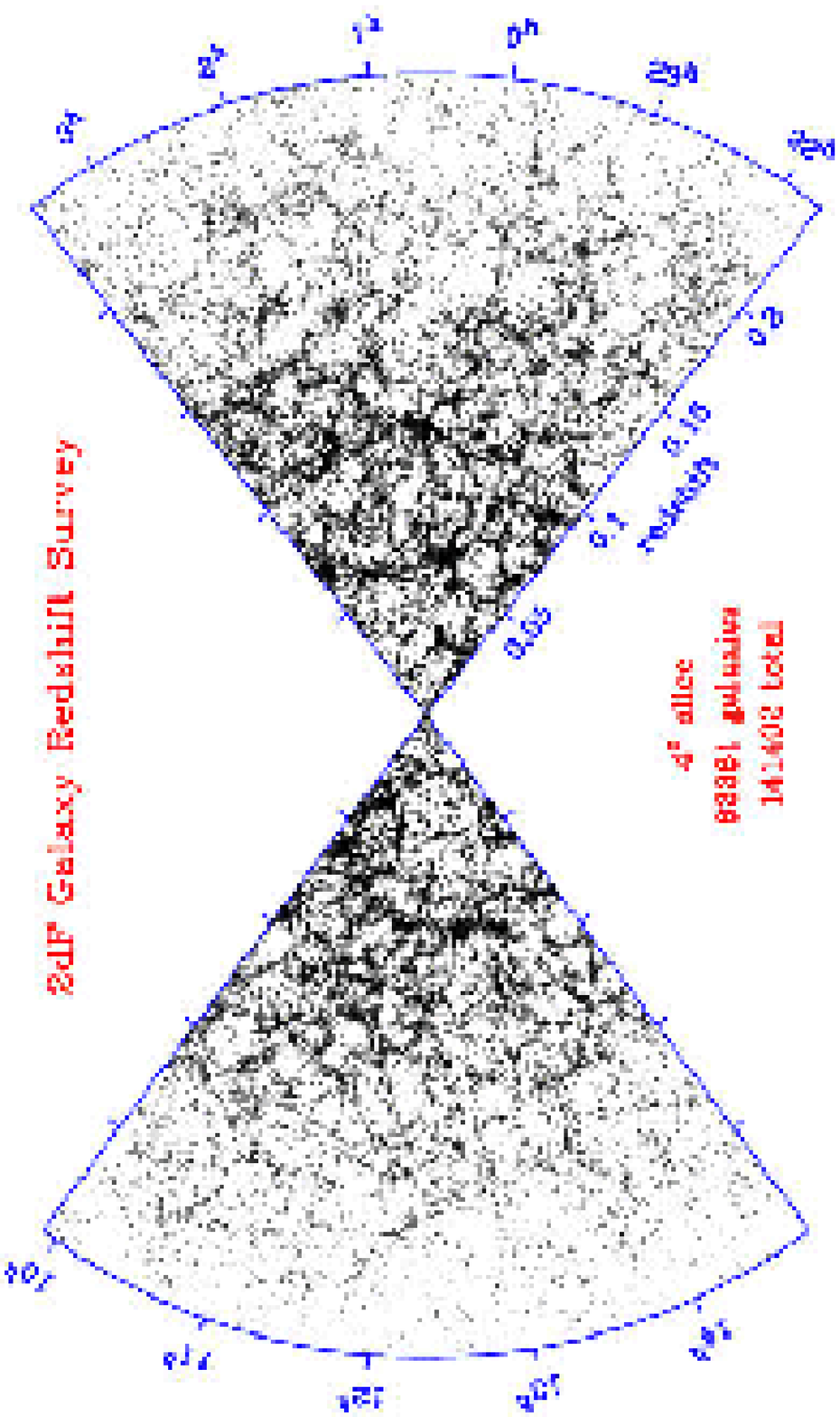,angle=295.0,height=14.5cm}}
\vskip -1.0truecm
\caption{The 2dF galaxy redshift survey. Here confined to a narrow $4^{\circ}$ slice, comprising 63361 
out of a total of 141402 galaxies, out to a redshift $z \approx 0.25$, the universality, complexity and 
intricacy of the cosmic web is strikingly displayed. Picture courtesy of the
2dF Galaxy Redshift Survey team, kindly provided by J. Peacock.}
\end{figure}
\noindent
One of the most impressive examples of a physical system displaying 
a salient geometrical morphology, and the largest in terms of sheer size, is 
the one we have encountered on Megaparsec scales in the Universe. Although at hindsight 
the projections of the spatial galaxy distribution in the form of galaxy sky maps contained 
ample hints for the existence of a complex spatial pattern (see 
the APM galaxy position sky map in Fig. 2, see Maddox et al. 1990a,b), it was the seminal publication 
of the first redshift slice by de Lapparent, Geller \& Huchra (1986, see Fig. 3) 
which offered the first direct panorama onto the cosmic tapestry. Their 
``Slice of the Universe'' may rightfully be regarded as the turning point for our 
view of the cosmic matter distribution. It changed it from an undefined amorphous, be it clumped, 
point process into one of a complex intriguing pattern. Since then, through ever more substantial 
and sophisticated observational campaigns enabled by large technological advances, the reality 
of foamlike structural arrangements has been proven to be a fundamental characteristic of the 
Universe. As we may infer from Figure 1, the past 
few decades have more than substantiated the early impression that on scales of a few up to more 
than a hundred Megaparsec, galaxies conglomerate into intriguing cellular or foamlike patterns 
pervading throughout the observable cosmos. 

A dramatic illustration of the accompanying advance in our perception of the 
``cosmic foam'' is that in Figure 4 (courtesy 2dF Galaxy Redshift Survey team), the 
recently published map of the distribution of more than 150,000 galaxies in a narrow region 
on the sky yielded by the 2dF -- two-degree field -- redshift survey.  Instead of a 
homogenous distribution, we recognize a sponge-like arrangement, with galaxies 
aggregating in filaments, walls and nodes on the periphery of giant voids. Outlined by 
galaxies populating huge {\it filamentary} and {\it wall-like} structures woven into an 
intriguing {\it foamlike} tapestry permeating the whole of the explored Universe, this 
frothy geometry of the Megaparsec universe evidently represents one of the most prominent 
aspects of the cosmic fabric (also see Fig. 14 and 15). Indeed, as we may 
infer from the preliminary map of the currently ongoing one-million galaxy redshift 
survey of the SDSS consortium (Fig. 1, topleft figure, kindly provided by Michael Strauss), as 
we probe deeper and deeper into space, the better we can appreciate the global ubiquity of the 
foamlike galaxy arrangement.

\subsection{Worldwide Web: \hfill Chains and Walls}
The closest and best studied of these massive anisotropic matter concentrations can 
be identified with known supercluster complexes, enormous structures comprising one or more 
rich clusters of galaxies and a plethora of more modestly sized clumps of galaxies. 
The sizes of the most conspicuous one regularly exceeding 100h$^{-1}$ Mpc. They are 
mostly dynamically youthful structures, not yet having reached a stage of contraction, 
and whose prominence is due to the more than average deceleration of its 
initial expansion as a consequence of its mere moderate overdensity. 
\begin{figure}[t]
\vskip -0.truecm
\centering\mbox{\hskip -1.truecm\psfig{figure=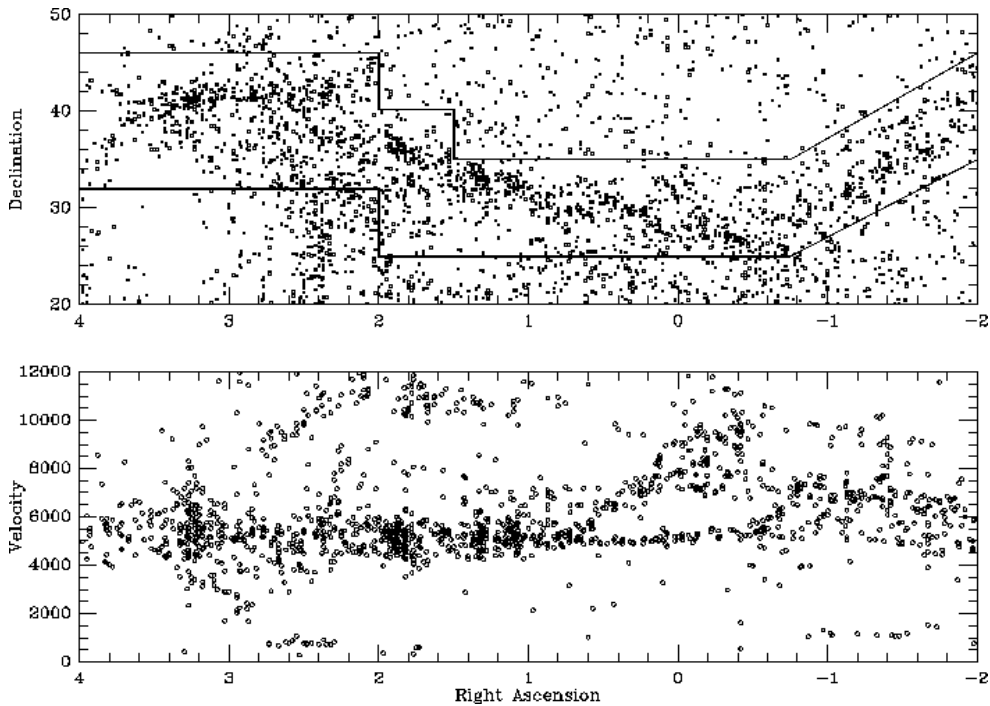,width=13.0cm,angle=270.}}
\vskip -0.2truecm
\caption{The Perseus-Pisces supercluster chain of galaxies. Separate 
two-dimensional views of the galaxy distribution in the northern 
region of the Pisces-Perseus region: sky-projected (top) and in depth (redshift, bottom). 
From Giovanelli \& Haynes 1991, 1996, kindly provided by M. Haynes.}
\vskip -0.3truecm
\end{figure}
Both our Local Group and the Virgo cluster are members of such a supercluster complex, 
the Local Supercluster, a huge flattened concentration of about fifty groups of galaxies 
in which the Virgo cluster is the dominating and central agglomeration. The Local supercluster 
is but a modest specimen of its class, counting only one rich cluster amongst its 
``subjects''. 

A far more prominent nearby representative is the Perseus-Pisces supercluster (Fig. 5). 
Its relative proximity ($\approx$ 55$h^{-1}$ Mpc), its characteristic and salient 
filamentary morphology and its favourable orientation have made it into one of the best 
mapped and meticulously studied superclusters. It is 
a huge conglomeration of galaxies that clearly stands out on the sky. The 
boundary of the supercluster on the northern side is formed by the filament 
running southwestward from the Perseus cluster. This majestic chain of 
galaxies has truely impressive proportions, a 5$h^{-1}$ wide ridge of at least 50$h^{-1}$ 
Mpc length, possibly extending out to a total length of 140$h^{-1}$ Mpc. Along this major 
ridge we see a more or less continuous arrangement of high density clusters and groups, 
of which the most notable ones are the Perseus cluster itself (Abell 462), Abell 347 and 
Abell 262. 

In addition to the presence of such huge filaments the galaxy distribution 
also contains vast planar assemblies. A striking example is the {\it Great Wall}, 
a huge planar assembly of galaxies with dimensions that are estimated to be of the 
order of $60 \times 170 \times 5$ h$^{-1}$Mpc, which has the Coma cluster 
of galaxies as its most prominent density enhancement (Geller \& Huchra 1989). 
\begin{figure}[h]
\vskip -0.truecm
\centering\mbox{\hskip -0.3truecm\psfig{figure=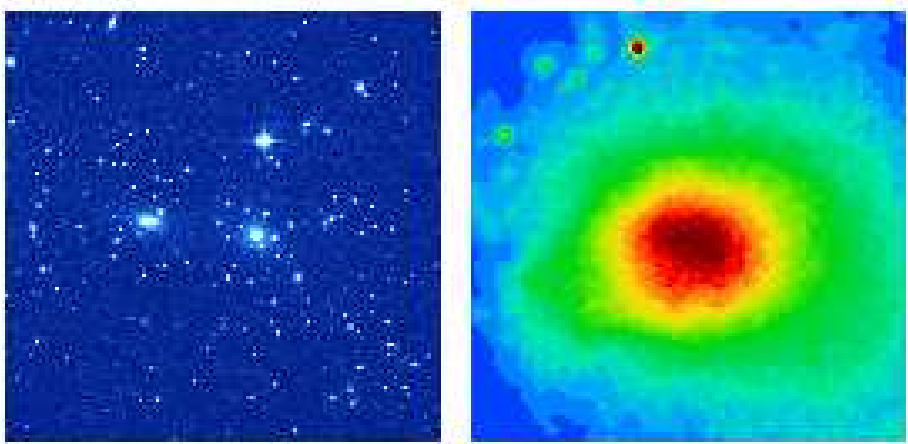,width=12.7cm}}
\vskip -0.25truecm
\caption{Comparison of optical and X-ray images of Coma cluster. Left: optical 
image; right: X-ray image (ROSAT). Courtesy: Chandra X-Ray Observatory Center.}
\end{figure}
\subsection{Worldwide Web:}
\begin{flushright}
\vskip -0.5truecm
{\bf{\large Junctions, Galaxy Clusters}}
\vskip -0.2truecm
\end{flushright}
Rather than smooth and featureless features, filaments and walls appear to be 
punctured by a variety of internal structure and density condensations. These can range
from modest groups of a few galaxies up to massive compact {\it galaxy clusters}, 
residing at the interstices of the cosmic network. The latter stand out as the most 
massive -- and likely most recent -- fully collapsed and (largely) virialized objects 
in the Universe. The richest of them contain many thousands of galaxies within a 
relatively small volume of only a few Megaparsec size. For instance, in the nearby 
Virgo and Coma clusters more than a thousand galaxies have been identified within 
a radius of a mere 1.5$h^{-1}$ Mpc around their core (see Fig. 6). The cluster galaxies are 
embedded in deep gravitational wells that have been identified as a major source of 
X-ray emission, emerging from the diffuse extremely hot 
gas trapped in these wells, possibly representing the most abundant 
state of baryonic matter in our Universe  (Fig. 6, right). The clusters may be regarded 
as a particular population of cosmic structure beacons as they typically concentrate 
near the interstices of the cosmic web, {\it nodes} forming a recognizable 
tracer of the cosmic matter distribution (e.g. Borgani \& Guzzo 2001). 

Containing many hundreds to thousands of galaxies within a compact 
region, rich clusters of galaxies are highly luminous objects at optical wavelengths 
and can be seen out to large cosmic depths. Even more conspicuous is their 
X-ray brightness, the result of the emission by the diffuse extremely hot intracluster 
gas trapped in their gravitational potential wells. The X-ray emission 
represents a particularly useful signature, an objective and clean measure of 
the potential well depth, directly related to the total mass of the 
cluster (see e.g. Reiprich \& B\"ohringer 1999). A potentially very 
promising and objective measure for the mass content of clusters is based 
on the shearing alternative measure of the cluster mass may be deduced 
from the deformation of the light path of background objects. 
However, mass determinations by gravitational lensing still 
involve very elaborate procedures, and as yet it is not feasible to 
use these as a basis for survey selections. 

\begin{figure}[t]
\vskip -0.truecm
\centering\mbox{\hskip -0.truecm\psfig{figure=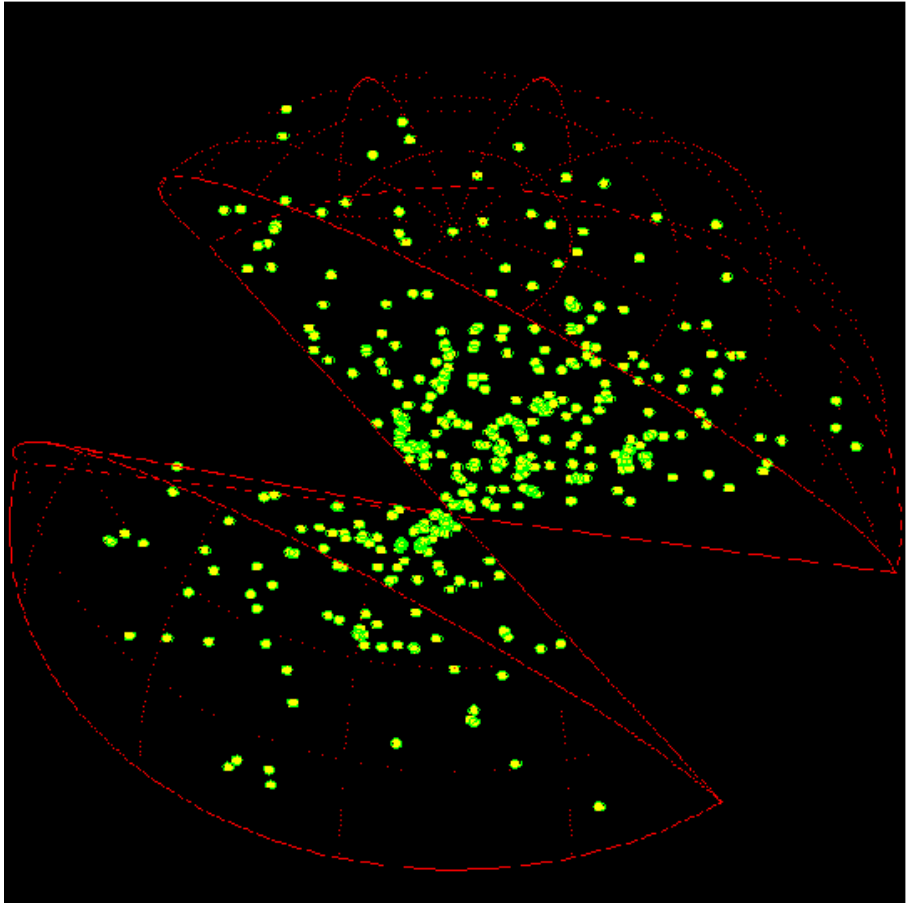,width=12.0cm}}
\vskip -0.25truecm
\caption{The spatial cluster distribution. The full volume of the X-ray REFLEX 
cluster survey within a distance of 600h$^{-1}$\hbox{Mpc}. The 
REFLEX galaxy cluster catalogue (B\"ohringer et al. 2001), 
contains all clusters brighter than an X-ray flux of $3\times 10^{-12} \hbox{erg} 
\hbox{s}^{-1} \hbox{cm}^{-2}$ over a large part of the southern sky. The missing part of the 
hemisphere delineates the region highly obscured by the Galaxy. Courtesy: Borgani 
\& Guzzo (2001). Reproduced by permission of Nature.}
\vskip -0.7truecm
\end{figure}
Through their high visibility clusters can be traced out to 
vast distances in the Universe. Hence, we can study their spatial 
distribution within large volumes of the cosmos. Thus, even though 
they represent a sparse mapping of the underlying large scale 
matter distribution, they are an ideal means of assessing its characteristics 
in very large volumes and over large scales, in particular when they 
relate to the underlying matter distribution in a direct and uncontrived 
fashion. This makes clusters into an efficient and time-saving 
probe for mapping the matter distribution over very large scales. 
A large range of observational studies, mainly based on optically 
selected samples, still display a substantial level of clumping on 
scales where clustering in the galaxy distribution has diminished below 
detectability levels. Be that due to the absence of genuine galaxy 
correlations or the fact that the clustering signal is so weak that it 
drowns in the noise, this turns out to be not so for cluster samples. 
It is in particular the Abell catalogue of optically identified galaxy 
clusters (Abell 1958; Abell, Corwin \& Olowin 1989) which has fulfilled a central 
role for the study of the large scale matter distribution on scales of 
several tens of Megaparsec (see Bahcall 1988). A wide range of observational 
studies on the basis of such optically selected samples have shown that the 
clustering of clusters is significantly more pronounced. Their two-point correlation 
function has a shape similar to  
that of galaxies, yet with a substantially higher amplitude and detectable out 
to distances of at least $\sim 50h^{-1}\hbox{Mpc}$. Mainly on instigation of 
these results, theoretical arguments were put forward motivating a simple 
linear amplified level of clumping in the cluster population on the premise 
of clusters representing the high-density peaks in a properly filtered underlying 
mass density field (Kaiser 1984). The precise values of the constant 
``linear bias'' factor with respect to that of the underlying matter field 
will depend on both cluster mass and structure formation scenario 
(Mo \& White 1996).

\begin{figure}[b]
\vskip -0.0truecm
\centering\mbox{\hskip -0.truecm\psfig{figure=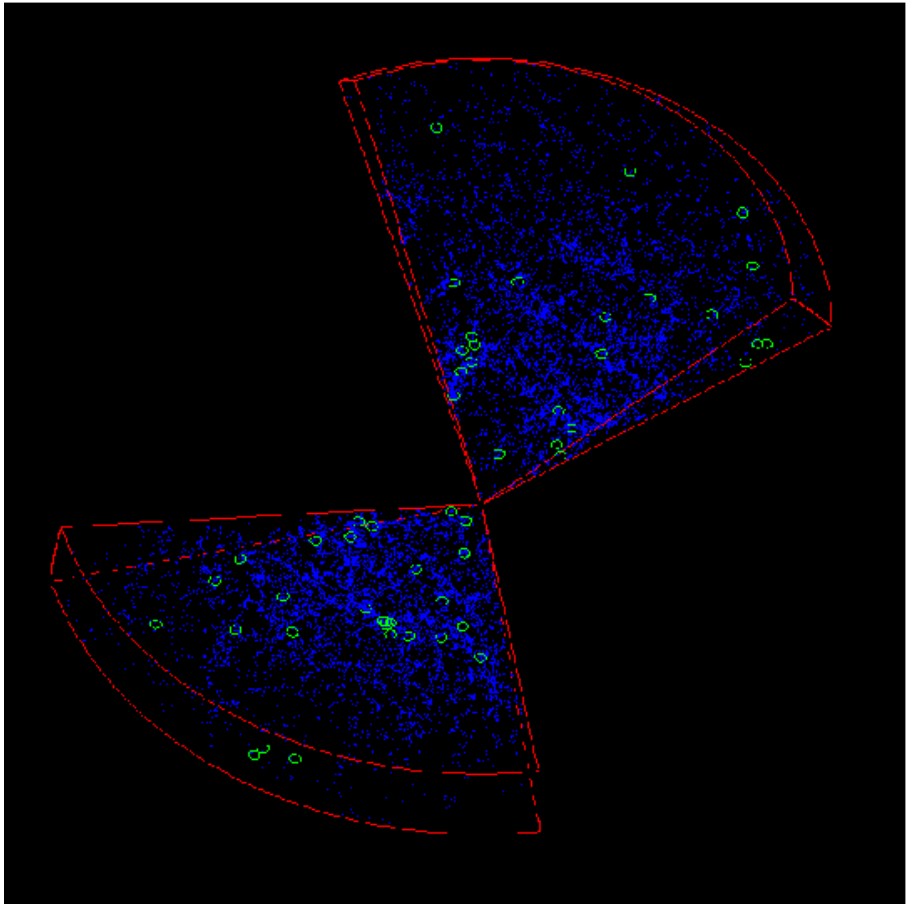,width=12.0cm}}
\vskip -0.25truecm
\caption{The spatial cluster distribution and its relation to the cosmic web. The green circles mark the positions of REFLEX X-ray clusters 
in the northern and southern slices of the Las Campanas redshift survey 
(LCRS, Shectman et al. 1996), out to a maximum distance of 600h$^{-1}$ Mpc. Underlying, 
in blue, the galaxies in the LCRS delineate a foamlike distribution of filaments, walls 
and voids. Courtesy: Borgani \& Guzzo (2001). Reproduced by permission of Nature.}
\vskip -0.7truecm
\end{figure}
A very good impression of the spatial distribution of rich clusters can be obtained 
from Fig. 7 (from Borgani \& Guzzo 2001). It depicts the spatial distribution of 
the clusters in the REFLEX galaxy cluster catalogue (B\"ohringer et al. 2001), 
containing all clusters brighter than an X-ray flux of $3\times 10^{-12} \hbox{erg} 
\hbox{s}^{-1} \hbox{cm}^{-2}$ over a large part of the southern sky. 
Maps such as these show that clusters themselves are not Poissonian distributed, 
but instead are highly clustered (see e.g. Bahcall 1988, Borgani \& Guzzo 2001). 
They aggregate to form huge supercluster complexes, coinciding with 
the filaments, walls and related features in the galaxy distribution. Indeed, in the 
case of the very highest density complexes filaments may represent such deep 
potential perturbations that also they may even light up in X-ray emission, as was 
discovered by Kull \& B\"ohringer (1999). Most galaxies in these complexes are then 
located in the diffuse regions in between 
the clusters, which according to illuminating Dutch expression should be 
seen as the ``currants in the porridge''. Note that such superclusters are mere 
moderate density enhancements on scale of tens 
of Megaparsec, typically in the order of a few times the average density. They are 
still co-expanding with the Hubble flow, be it at a slightly decelerated rate, and 
are certainly not to be compared with the collapsed, let alone virialized, and 
verily pronounced entities like clusters. Instead, it is probably most 
apt to see them as clouds of points in a stochastic spatial point process, 
clouds whose boundaries are ill-defined.

What centres in our interest is the relation between the very large scale 
cluster distribution (Fig. 7) and the underlying matter distribution,  
in particular the weblike morphology of the latter. Figure 8 (Borgani \& Guzzo 
2001) provides a catching illustration, comparing the subset of REFLEX clusters 
within the region of the Las Campanas redshift survey to that 
of the galaxy distribution in the same region ($\sim 26,000$ galaxies, 
Shectman et al. 1996). From Fig. 8 we see that the cluster distribution represents 
a mere coarse mapping of the underlying structure. Not immediately outstanding, 
a thorough spatial statistical analysis will therefore be needed to establish 
the extent and nature of the correspondence between the two distributions. 

On large scales valuable insight into the relation between the population of clusters and 
other cosmic residents has has been provided by the measurement, analysis and mutual 
comparison of large scale peculiar velocity fields of galaxies and clusters. 
By comparing the cluster population kinematics with the underlying matter and/or 
galaxy distribution, meaningful information has been obtained on the ``bias'' of clusters 
with respect to the overall matter distribution. Taking into account the vastly large 
cosmic region covered by cluster samples, a firmly established link between clusters 
and matter would provide us with the opportunity to map out the source for the peculiar 
motion of our own Local cosmic neighbourhood with respect to the cosmic background, 
as measured by the MWB dipole. For instance, Scaramella, Vettolani \& Zamorani (1991) and 
Plionis \& Valdarnini (1991) sought to establish on the basis of the cluster distribution 
within a distance of $r \approx 300h^{-1}\hbox{\rm Mpc}$ whether indeed the origin of our cosmic 
motion should be located within this volume, or whether there are indications for even larger 
cosmic structures. In a subsequent more systematic analysis, Branchini \& Plionis (1996) sought to 
relate the dipole motion implied by the cluster distribution, in an analysis similar to 
that of galaxy peculiar velocity samples, to the dipole observed in the microwave background. 
In the meantime the availability of measured galaxy peculiar velocity samples covering 
large volumes of our local cosmos have allowed to perform such analyses over relevant 
local cosmic regions, shedding substantially more light on this important issue (see 
Branchini et al. 2001). The absence of major surprises, and their rough mutual agreement, is a 
strong argument for a systematic correlation between the cluster distribution and that of 
other cosmic representatives. Related numerical studies of various structure formation 
scenarios provided a substantial theoretical foundation for such a link (Moscardini et al. 1996). 
\begin{figure}[b]
\vskip -1.truecm
\centering\mbox{\hskip -0.5truecm\psfig{figure=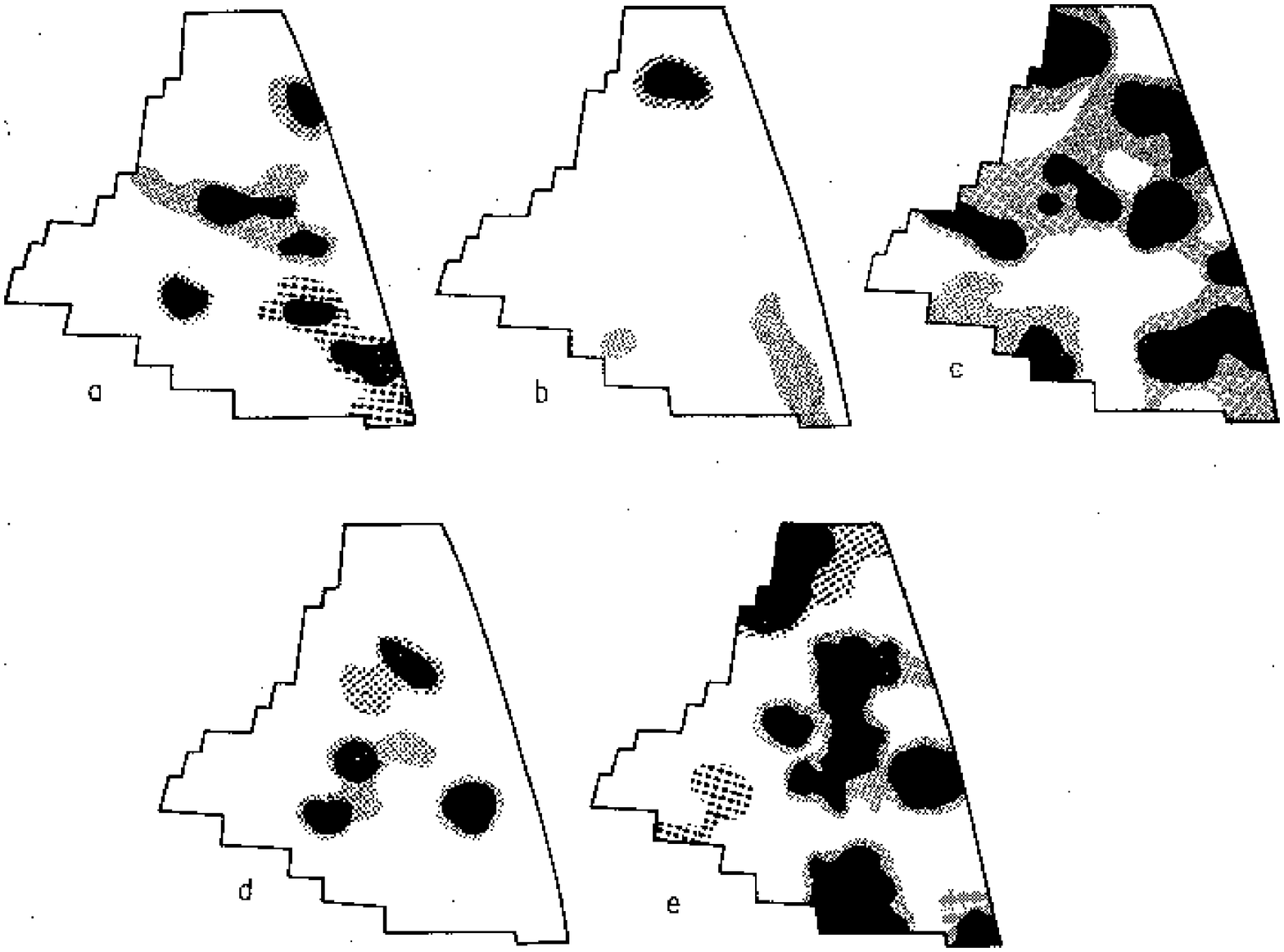,angle=270,width=15.0cm}}
\end{figure}
\begin{figure}[h]
\vskip -0.5truecm
\caption{The Bo\"otes void as revealed by the galaxy number space 
density in a sequence of five different recession velocity intervals
in the direction of the Bo\"otes constellation on the sky. The 
lowest contour represents a density equal to 0.7 of the cosmin mean, 
each higher contour represents a factor of 2 increase in density. 
Velocity ranges ($\hbox{km/s}$): (a) 7,000-12,000; (b) 12,000-17,000; 
(c) 17,000-23,000; (d) 23,000-29,000; (e) 29,000-39,000. Frame (b) 
clearly reveals a large void in the galaxy distribution, which turns 
out to be roughly spherical in outline. Figure from Kirshner et al. (1987). Reproduced 
by permission of the AAS.}
\vskip -0.3truecm
\end{figure}
These large-scale studies, involving scales $> 10h^{-1}\hbox{\rm Mpc}$, do need to be 
complemented by related small-scale studies. The detailed physical processes 
establishing a systematic relation between cluster population and the underlying matter 
distribution will ultimately be established on much smaller scales, unaivalable to 
the necessary low-resolution velocity studies. Focussing on aspects like the implied 
small-scale clustering patterns of clusters, and scrutinizing their link to the anisotropic 
geometric patterns of the cosmic foam, and systematically adressing the processes creating 
such morphological connections are absolutely necessary for the ultimate unravelling of 
this important cosmic kinship.

\subsection{Worldwide Web: \hfill the Valley of Voids}
Complementing this cosmic inventory leads to the large {\it voids}, 
one of the most intriguing and unexpected findings emanating from extensive redshift 
surveys. They revealed that the planar, linear and compact structural elements of the 
galaxy distribution appear to be located on the surface of vast underdense regions. 
These concern vast regions of space, mostly roundish in shape, practically devoid of 
any galaxy, and typical sizes in the range of $20-50h^{-1}$ Mpc. The earliest recognized 
one, the Bo\"otes void (Kirshner et al. 1981, 1987, see Fig. 9), a conspicuous almost 
completely empty spherical region (however, see Szomoru 1995) with a diameter of 
around $60h^{-1}$Mpc, is still regarded as the canonic example. The role of voids 
as key ingredients of the cosmic matter distribution has since been 
convincingly vindicated in various extensive redshift surveys, up to the recent results 
produced by Las Campanas redshift survey (Fig. 9) and the 2dF redshift survey 
(Fig. 4) and the Sloan redshift survey (see Fig. 1).
\begin{figure}[b]
\vskip -0.5truecm
\centering\mbox{\hskip -0.75truecm \psfig{figure=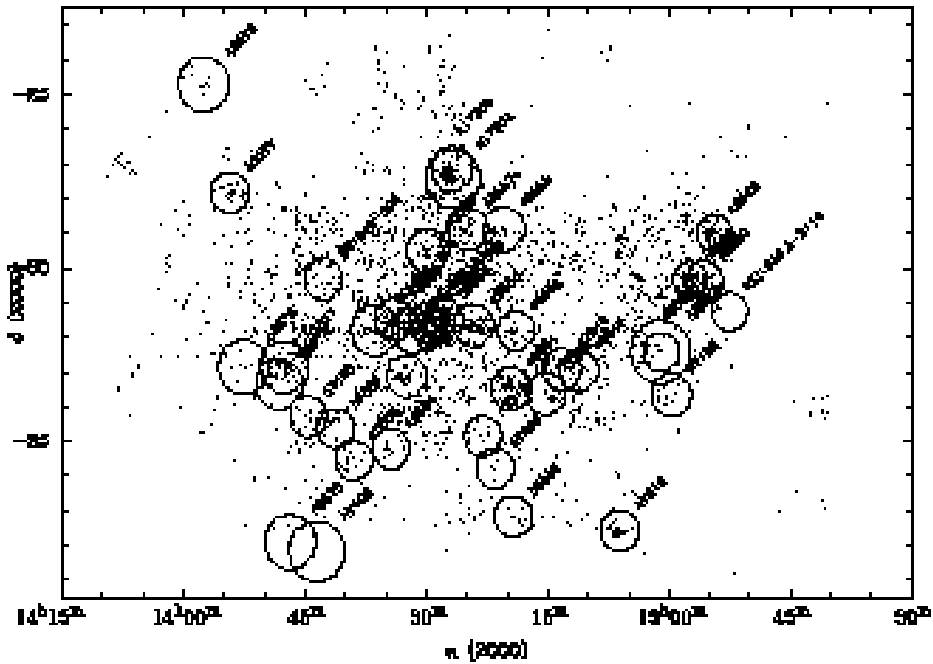,width=
12.25cm}}
\vskip -0.3truecm
\caption{The Shapley Supercluster complex. Two-dimensional sky distribution of sample 
galaxies in Shapley concentration used in study of Quintana etal. (2000). Open circles 
indicate the locations of clusters within the complex. From Quintana, Carrasco \& Reisenegger 
2000. Reproduced with permission of the AAS.}
\end{figure}
\subsection{Worldwide Web: \hfill Monster Complexes}
As our view of the spatial cosmic galaxy distribution is gradually expanding, we 
start to come across some truely awesome and rather uncharacteristic dense 
concentrations supercluster complexes. The Shapley concentration is the canonical 
example, by far the most outstanding complex in the Local Universe (see e.g. 
Raychaudhury 1989; Ettori et al. 1997; Quintana, Carrasco \& Reisenegger 2000; 
Reisenegger etal. 2000). With the first maps of the SDSS redshift survey seeing the light, a 
first qualitative assessment suggests the presence of more comparable extreme 
supercluster complex, which appears to be in line with the claim by Batuski et al. (1999) for 
the existence of other comparable structures. 

The most detailed impression of such ``monster complexes'' is offered by the 
Shapley concentration, first noted by Shapley (1930). It is an 
imposing concentration of galaxy clusters, the most massive concentration of matter 
at $z<0.1$. It is located at a distance of $\sim 140h^{-1}\hbox{Mpc}$, beyond the 
Hydra-Centaurus supercluster (at $\sim 40h^{-1}\hbox{Mpc}$). It amasses more than 
30 rich Abell galaxy clusters into a core region 
of $\sim 25h^{-1}\hbox{Mpc}$. An impression of this extraordinary concentration 
may be obtained from the depicted galaxy sample sky distribution in Fig. 10 
(from Quintana, Carrasco \& Reisenegger 2000). The central core is dominated 
by ACO clusters A3556, A3558 and A3562 and two poor clusters. Besides this 
core, one can identify an eastern part consisting of the clusters A3570, A3571, 
A3572 and A3575, and a western region formed by A3528, A3530 and A3532, while 
an elongation to the north includes A1736 and A1644. Several of these clusters 
are amongst the brightest X-ray clusters known (Ettori et al. 1997). 

Telling for its huge mass is that the Shapley concentration is probably 
responsible for about $10-20\%$ of the optical dipole observed in the motion 
of the Local Group with respect to the cosmic microwave background (e.g. 
Raychaudhury 1989). Such is also suggested by number counts 
in redshift space, which suggest that most of the supercluster has a 
density several times the cosmic average, while the two complexes 
within $\sim 5h^{-1}\hbox{Mpc}$ of clusters A3558 and A3528 have 
overdensities $\sim 50$ and $\sim 20$ times, respectively. Such 
regions are therefore far outside the ``linear'' regime of 
small density perturbations, and have indeed started contracting, although 
far from having reached the stage of collapse and virialization (Bardelli et al.  
2000). Such overdensities on scales of $\sim 25h^{-1}\hbox{Mpc}$ surely do 
stress popular theories of hierarchical structure formation by gravitational 
stability to the utmost, and consequently represent a wonderful testbed for 
the corresponding scenarios. 

Here we wish to draw special attention to these ``monster complexes'', as we will 
argue later that in fact they represent an important and natural manifestation 
of large scale cellular patterns. When clusters are concentrated near 
the junctions of the cellular network, it will induce a specific pattern 
of clustering in which a ``geometric biasing effect'' can be identified. 
The amplified level of clustering for the richest galaxy clusters may then 
be intimately linked with a concentration of such clusters in corresponding 
flattened or elongated superstructures defined by the distribution of 
the nodes of the network, the size of whom may supersede tens of Megaparsec 
(i.e. sizes comparable to the mean voidsize of $\sim 10-20h^{-1}\hbox{Mpc}$). 

\subsection{Worldwide Web: \hfill Universal Pretensions}
The first impressions of a weblike galaxy distribution suggested by the first shallow CfA slices 
(de Lapparent et al. 1986) got continuously and increasingly convincing confirmed as larger 
and more ambitious surveys expanded their reach onto greater depths of our Universe. The image 
of a vast universal cosmic foam got firmly established through the publication of the results 
of the Las Campanas redshift survey (LCRS, Shectman et al. 1996). Its chart of 26,000 galaxy 
locations in six thin strips on the sky, extending out to a redshift of $z \sim 0.1$, 
until recently represented the most representative impression of cosmic 
structure available. 

The LCRS comprised the first cosmologically representative volume of space. In the meantime, 
two ambitious enterprises, the 2dF and the Sloan digital sky survey, have embarked on 
ambitious missions to map the galaxy distribution of the Universe out to unexplored depths 
of cosmic territory, out to distances of $\sim 1000h^{-1}\rm{Mpc}$. Up to a million galaxy 
redshifts will yield an unprecedented outline of the structure of our 
cosmic environment, for the first time a fully representative and truely 
uniform sample of our Universe. Indeed, the published results of the 2dF survey (Fig. 4) can 
only be characterized as stunning in 
the detailed and refined rendering of the foamlike morphology traced out by the 
galaxy distribution. The extent and ubiquity of the foamlike patterns throughout the surveyed 
volume, extending out to a redshift $z \sim 0.2$, is truely perplexing. 
\begin{figure}[t]
\vskip -0.25truecm
\centering\mbox{\hskip -1.5truecm\psfig{figure=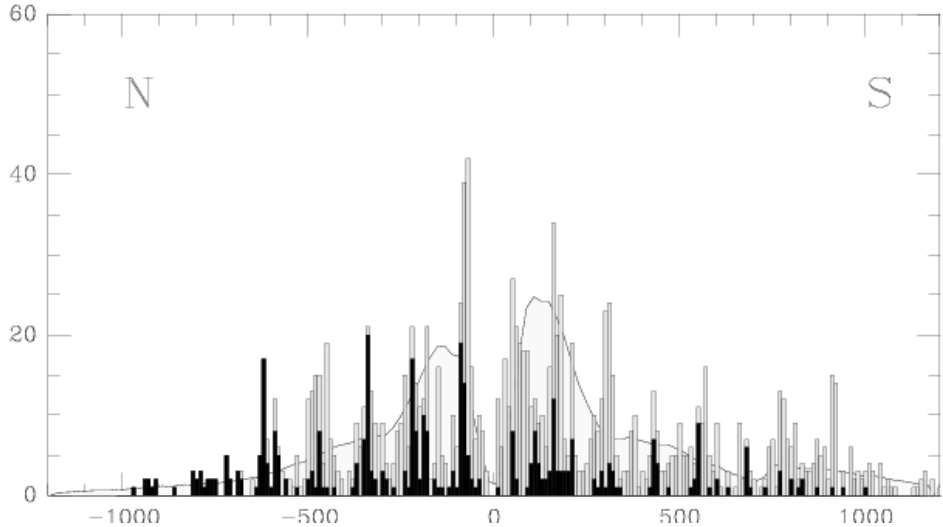,width=15.2cm,angle=270}}
\vskip -0.25truecm
\caption{A pencil beam redshift survey. The redshift distribution of galaxies 
out to a distance of of $1200h^{-1}\hbox{Mpc}$ towards the south Galactic pole (negative 
velocities) and the north Galactic pole (positive velocities). Plotted
is the number of galaxies in $10h^{-1}\hbox{Mpc}$ bins. This figure is a 
combination of several very narrow pencil beam redshift surveys, comprising fields 
of 5 to 20arcminutes. The black bars represent the number of galaxies in the
original survey of Broadhurst, Ellis, Koo \& Szalay (1990). The 
superposed dotted bars represent more recent extensions of and
additions to the original (1990) survey. The continuous curve at 
the background is the survey selection function, which combines the 
effects of the different geometries and apparent magnitude limits of 
composite survey beams. Kindly provided by Alex Szalay. }
\vskip -0.5truecm
\end{figure}
With respect to its universality, possibly most telling has been the finding that a narrow and 
very deep ``needle-shaped'' one-dimensional probe through the galaxy distribution results 
in a conspicuous pattern of sharp spikes 
separated by shallow valleys. Such pencil beam redshift surveys (Broadhurst et al. 
1990) clearly conjure up the idea of piercing through a foamlike structure of walls and 
filaments (see e.g. van de Weygaert 1991a), which suggest the cosmic foam to extend at 
least up to a redshift of $z \sim 0.5$. 

All, involve a strong confirmation of the existence of foamlike 
galaxy distributions out to large cosmic distances. It has made us realize 
that the cosmic foam is a truely universal phenomenon, extending over a vast 
realm of the observable Universe.

\subsection{Worldwide Web: \hfill Gravitational Signature}
Having observed its pronounced features reflected in the spatial distribution of 
galaxies, we should naturally wonder whether the foamlike network is indeed also 
the spatial arrangement of the (full) matter distribution. Given the fact that we 
still do not understand properly how and where galaxies did form during the 
evolution of the Universe the foamlike galaxy distribution may  
represent a biased reflection of the underflying matter distribution. In 
principal the specific foamlike features may therefore be as much be a consequence 
of the processes involved with the forming of galaxies as a result of the 
spatial matter distribution. The issue is even more pressing since we know that probably 
more than 80\% of the matter in the Universe consists of a collisionless, weakly 
interacting matter. It is certainly not self-evident that the distribution of 
baryonic matter, be it the galaxies or the diffuse intergalactic medium, does 
form a faithful representation of the spatial properties of the dominant 
species of matter. 
\begin{figure}[t]
\vskip -0.5truecm
\centering\mbox{\hskip -1.0truecm\psfig{figure=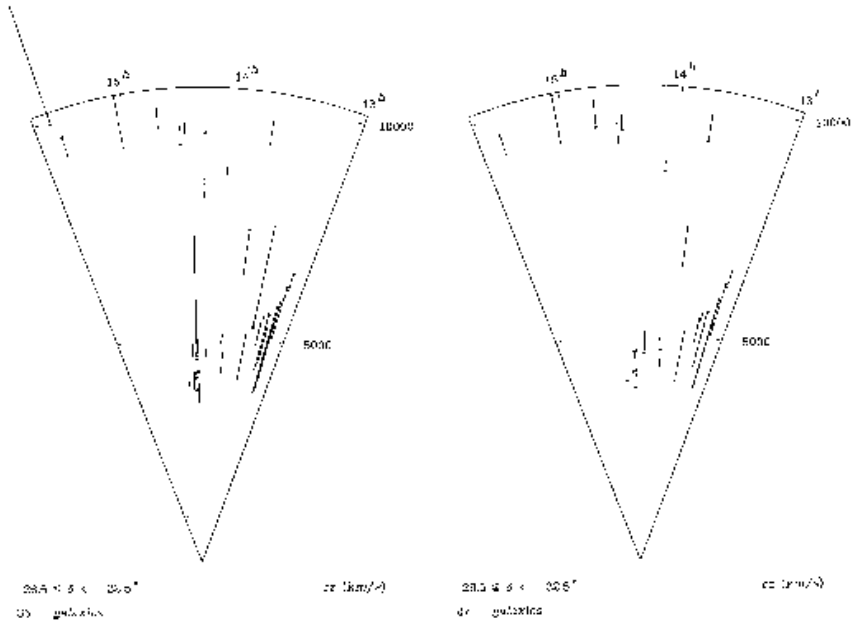,width=13.5cm}}
\vskip -8.0truecm
\caption{Velocity field for galaxies on the edge of the largest void in first CfA2 slice. 
The velocities of the galaxies were estimated on the basis of I-band Tully Fisher measurements. 
Left: all 35 galaxies with I-band TF measurements. Right: the 27 galaxies with reliable 
I-band TF distances. The tip of each galaxy velocity vector is indicated by a small 
circle. Notice the infall pattern towards the Coma cluster for the galaxies along the 
edge of the void. From Bothun etal. 1992. Reproduced by permission of the 
AAS.}
\vskip -0.5truecm
\end{figure}
Nonetheless, substantial evidence for the universal presence 
of the cosmic foam, not merely confined to the galaxy distrubition, has been 
inferred from the properties of the absorption of radiation by the intergalactic 
medium of neutral hydrogen, specifically at the Ly$\alpha$ transition. The observed 
Ly$\alpha$ forest of absorption lines in the spectra of background quasars has 
provided a rich source for exploring the spatial and thermal characteristics 
of the intergalactic medium. Interpreting them with the help of cosmohydrodynamic 
simulations of structure formation the evidence is quite compelling that the 
forest should be interpreted in terms of lines of sight piercing through 
a medium confined within a foamlike diffuse gas distribution. 

However, as far as its genuine material distribution, the most unbiased test for 
the reality of the cosmic foam is by means of its gravitational impact. Through 
meticulous work the gravitational influence of the matter distribution has 
indeed been opened to a more profound study. Two major physical effects 
provide us with a means to probe the material content of the Universe. 

Within the commonly accepted view of structure formation through gravitational 
instability (see section 3), we know that the the peculiar velocities of 
galaxies are induced by the residual gravity stemming from the inhomogenous 
matter distribution (see Dekel 1994 and Strauss \& Willick 1995 for excellent 
over- and reviews). Hence, we may use these deviations of galaxies' velocities 
from the global Hubble flow as a means to explore the underlying matter 
distribution. This has indeed developed into a major industry. A major 
complicating factor is that as yet galaxy distance estimates are still 
rather coarse, their accuracy rarely lower than $20\%$, which usually 
restricts the interpretation of the measured velocity fields to scales 
exceeding $\approx 10h^{-1}\hbox{Mpc}$. On those scales, where structure 
is still residing in the linear stage of development, the velocities 
are expected to be linearly proportional to the exerted peculiar gravitational 
force. While this has lead to succesfull reconstructions of the matter 
distribution in the Local Universe, and included the discovery of 
a nearby huge matter concentration, the Great Attractor, it also implies 
that it is very difficult to see whether we see traces of a foamlike 
matter distribution. After all, typical are its anisotropic elements, 
elongated along one or two directions, yet of a small extent in at least 
one other direction. The smoothing operations involved with these 
studies of cosmic flows therefore tend to abolish the distinctly 
anisotropic marks of a foamlike matter distribution. 

Yet, a few indications have been uncovered and appear to indicate the 
reality of a foamlike matter distribution. One strategy is to 
study in more detail the velocity field around salient filamentary 
features. The other smartly adresses the effects of voids within the 
matter distribution. Some studies have indeed been 
focussing on peculiar velocity fields near some outstanding filamentary 
features and appear to indicate faster velocities than would be expected 
on the basis of a more ``spherical'' matter distribution. Such is to be 
expected in the case of anisotropic features.  

In an early study, Bothun et al. (1992) studied the peculiar velocities of 
galaxies on the edges surrounding the largest void in the first lice 
of the CfA redshift survey (de Lapparent, Geller \& Huchra 1986). They 
found indications for significantly higher infall velocities into Coma than expected 
on the basis of merely the gravitational influence of Coma itself (see Fig. 
12). Upon close inspection of the velocity pattern in the slice (Fig. 12), we 
can clearly recognize a flow along the filament connecting onto the Coma 
cluster. On the basis of these (coarse) data one could therefore suspect the 
gravitational (pushing) influence of the void to indeed represent a significant 
contribution to the overall gravitational field. Additional analyses of the 
velocity fields in other superclusters, in particular the Perseus-Pisces 
supercluster filament, also do suggest a clear signature of a anisotropic 
infall pattern along the ridge of the complex (see e.g. Baffa et al. 1993).

The indication for the dynamical influence of voids was further substantiated by the 
far more extensive and systematic analysis of peculiar galaxy velocities 
in the Local Universe. Through the application of the POTENT procedure 
(Dekel, Bertschinger \& Faber 1990, Bertschinger et al. 1990) the 
observed radial peculiar velocities of galaxies can be used to recover the 
full three-dimensional field, smoothed on scales of $\sim 10h^{-1}\hbox{Mpc}$. 
On linear scales reconstructed 3-D velocity maps on the basis of the Mark III 
catalogue (Willick et al. 1997) can subsequently be 
applied towards reconstructing the density field in the corresponding region 
of the Local Universe. On the basis of such reconstructions, the gravitational 
impact of voids in the Local Universe can clearly be recognized (also see 
Dekel \& Rees 1994). Indeed, the POTENT reconstructions show that is necessary 
to invoke the dynamical influence of these voids to obtain a fully selfconsistent 
reconstruction of the dynamics in the Local Universe. Most interesting was 
therefore the suggestion by Dekel \& Rees (1994) to use the pushing influence 
of voids to set limits on the value of the cosmological density parameter 
$\Omega$. 
\begin{figure}[b]
\vskip -0.5truecm
\centering\mbox{\hskip -1.1truecm\psfig{figure=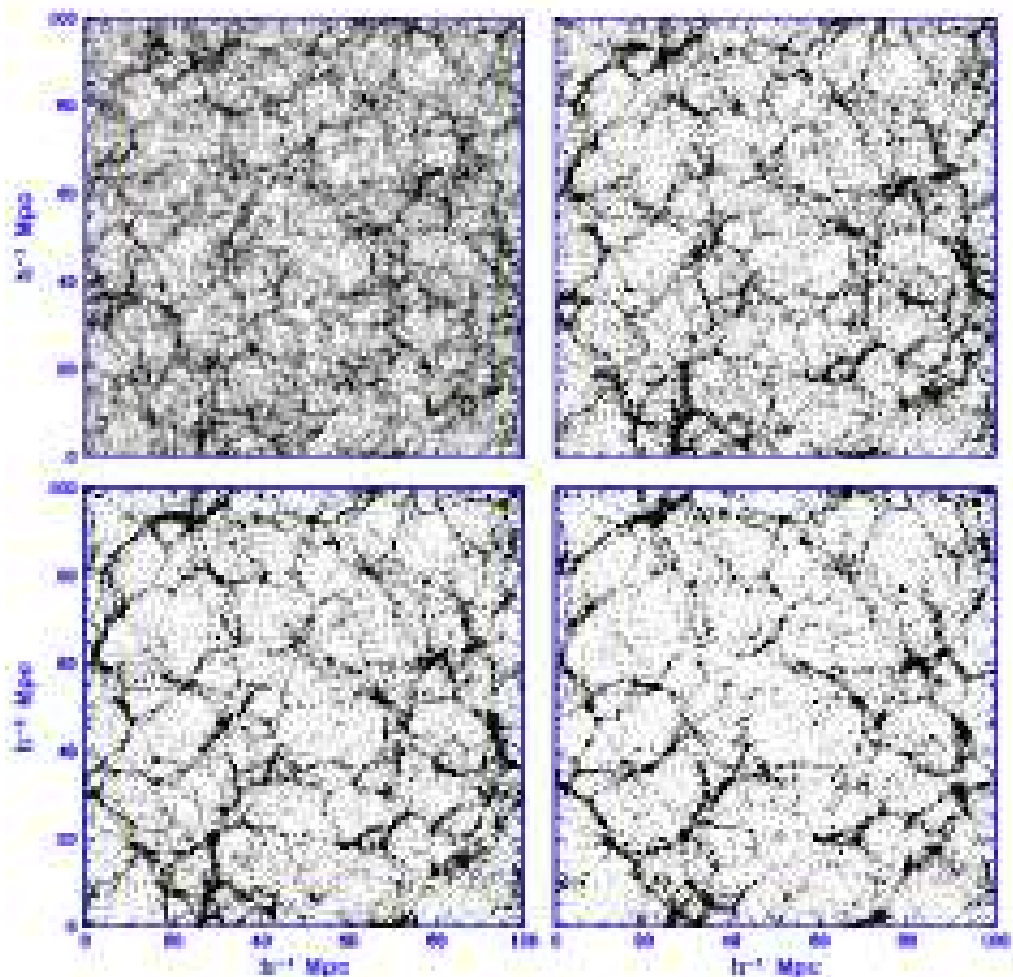,width=13.7cm}}
\vskip -0.75truecm
\caption{Evolution of structure and development of a cellular morphology in a scenario 
of structure formation through gravitational instability. Illustrated are 4 slices, 
at $a=0.2, 0.3, 0.5$ and $a=0.7$ in SCDM scenario ($\Omega_0=1.0,H_0=50\,\hbox{km/s/Mpc}$) 
from a P$^3$M N-body simulation following the clustering of 128$^3$ particles 
in a 100h$^{-1}$ Mpc box.}
\end{figure}
Probably most promising for investigating the dynamical impact of the 
cosmic web is through its influence on the trajectories of light, i.e. its 
gravitational lensing effect. While clusters of galaxies form by far the most 
outstanding sources of lensing on large Megaparsec scales, there has been 
a major effort towards detecting the signature of the more generic large 
scale structures. Theoretical evaluations by e.g. Jain, Seljak \& White (2000) have shown 
this to be a feasible and promising technique. Since the presence of a significant 
signal of cosmic shear has been inferred for by meticulous statistical analysis of wide field sky images 
(Van Waerbeke et al. 2000), we know it must indeed be possible to probe the signal of 
individual features such as those of filaments. However, despite its great promise, instrumental 
complications are still preventing the first significant reconstruction of such features 
on the basis of weak lensing measurements (there has been a claim of the detection of 
a filamentary bridge by Kaiser et al. 1998). Despite its pristine status, we may therefore 
look forward to a major amount of information on the dynamics of the cosmic web 
in the coming years. 

\subsection{Worldwide Web: \hfill the Filigree of Fantasy}
Foamlike patterns have not only been confined to the cosmos of reality. Equally important 
has been the finding that foamlike patterns do occur quite naturally in a vast 
range of structure formation scenarios within the context of the generic framework of 
gravitational instability theory. Prodded by the steep increase in computing 
power and the corresponding proliferation of ever more sophisticated and extensive 
simulation software, a large range of computer models of the structure formation process 
have produced telling images of similar foamlike morphologies (Fig. 13). 

They reveal an evolution proceeding through stages characterized by matter 
accumulation in structures with a pronounced cellular morphology, involving large 
anisotropic clustering structures such as filaments and walls. Whether or not these 
stages form a transient stage or a more permanent aspect of the matter distribution 
is not yet entirely clear, but will certainly depend on the scenario, the cosmological 
density parameter $\Omega$, and possibly various other factors. Evidently, the 
observation that numerical models seem to display the idiosyncratic tendency of 
forming foamlike patterns provides us with a firm ground for a gaining a 
more substantial insight into its formation mechanisms and conditions.
\begin{figure}[t]
\vskip -0.25truecm
\centering\mbox{\hskip -0.25truecm\psfig{figure=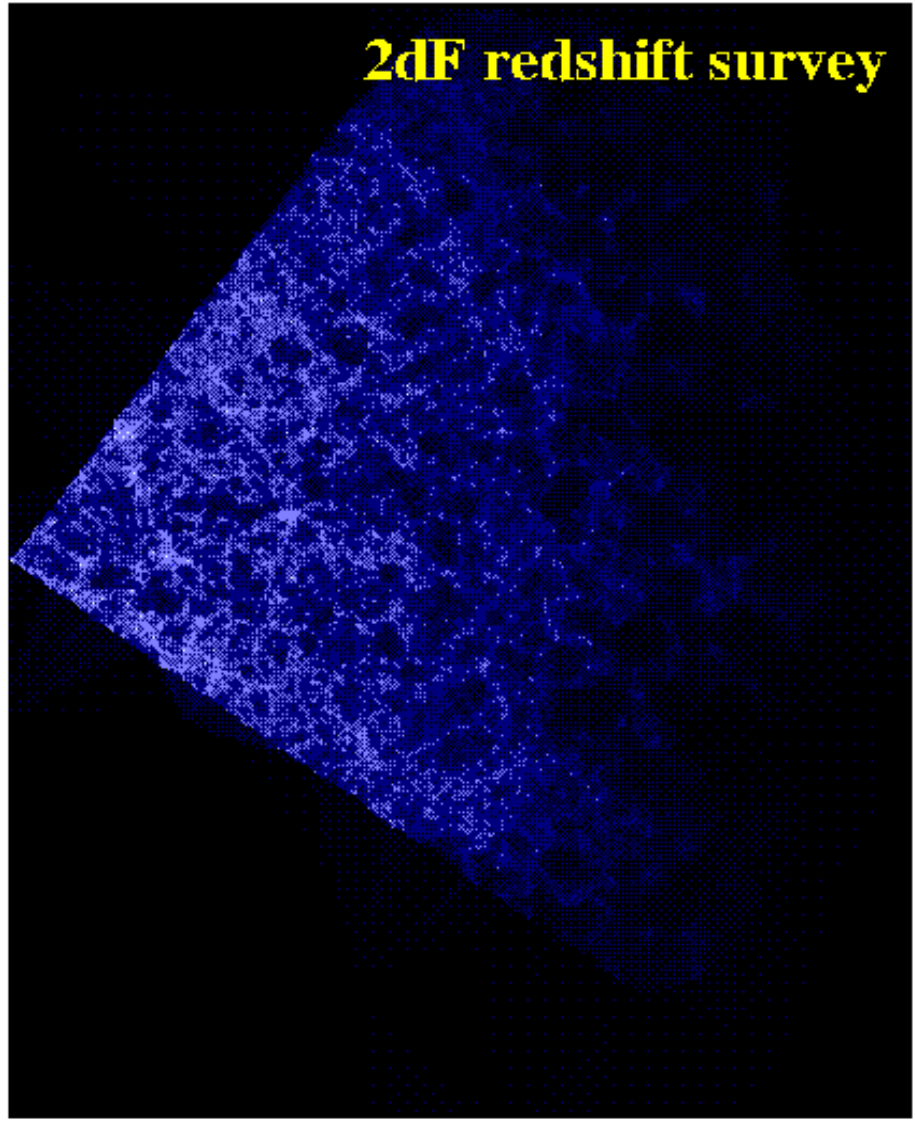,width=12.6cm}}
\vskip -0.4truecm
\caption{The Delaunay Field Estimator reconstruction of the 2dF survey field south. 
The DFE reconstruction more clearly than the galaxy distribution itself shows 
the coherence of the cosmic foam discretelt ``sampled'' by the galaxy distribution.  
Notice the detailed and refined structure which appears to be specifically 
strengthened by this fully adaptive method (from Schaap \& van de Weygaert 2002b). 
Data courtesy: the 2dF consortium.}
\vskip -0.3truecm
\end{figure}
\subsection{Worldwide Web: \hfill the Cosmic Symbiosis}
Of utmost significance for our inquiry into the issue of cosmic structure formation is 
the fact that the prominent structural components of the galaxy distribution -- clusters, 
filaments, walls and voids -- are not merely randomly and independently 
scattered features. On the contrary, we have noticed them to have arranged themselves in 
a seemingly highly organized and structured fashion, the {\it cosmic foam}. The voids 
are generically associated with surrounding density enhancements. In the galaxy 
distribution they represent both contrasting as well as complementary components, the 
vast under-populated {\it voids} being surrounded by {\it walls} and {\it filaments} 
with the most prominent and massive cosmic matter concentrations, the 
{\it clusters} of galaxies, at the intersections of the latter. 
\begin{figure}
\vskip -0.25truecm
\centering\mbox{\hskip -0.50truecm\psfig{figure=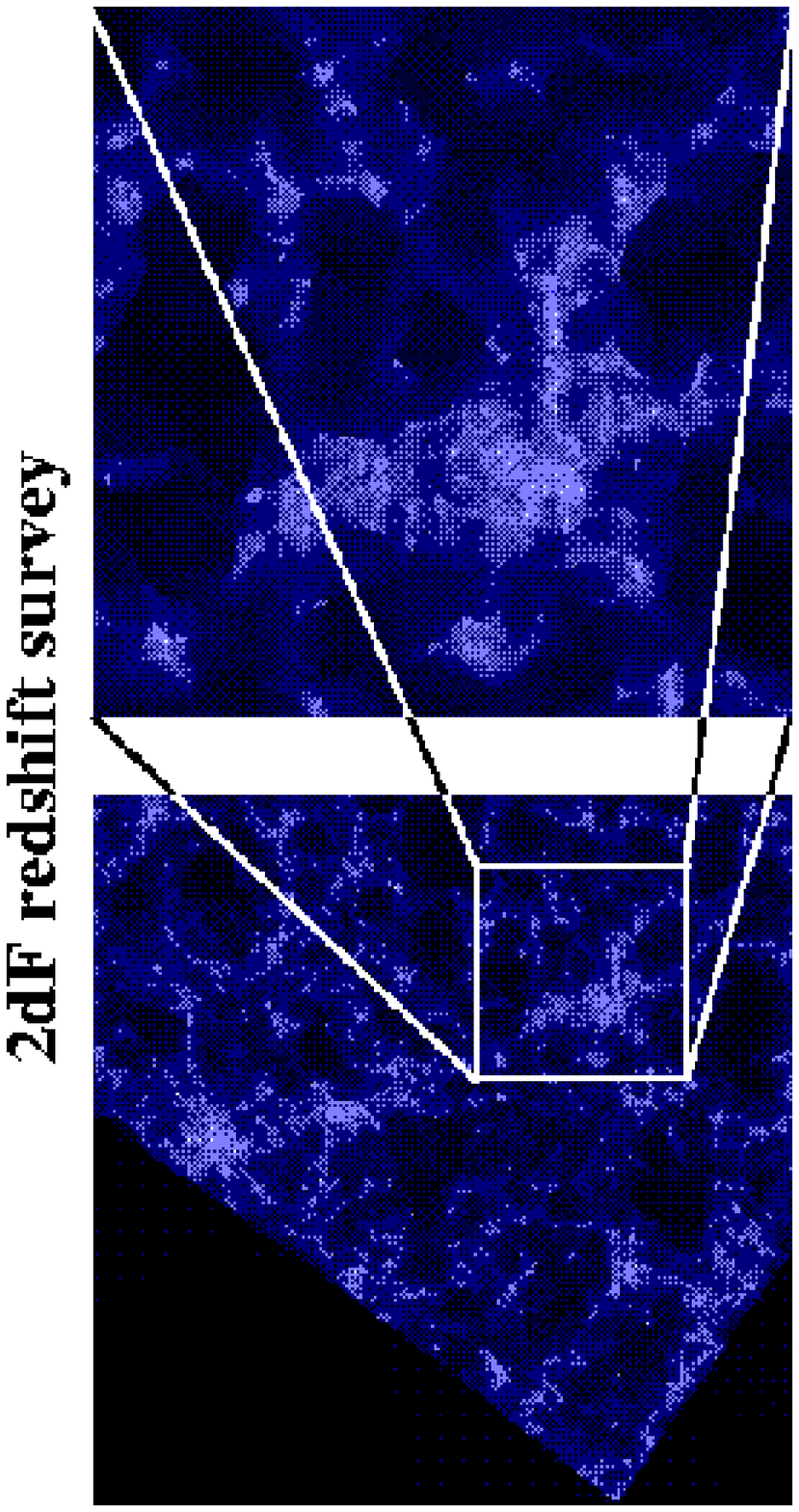,width=12.5cm}}
\vskip -0.25truecm
\end{figure}
\begin{figure}
\caption{The Delaunay Field Estimator reconstruction of the 2dF survey field south. 
In particular focussing in on a high-density junction point within the weblike structure, 
a massive matter concentration. Not only do we clearly recover the filamentary 
extensions emanating from the massive ``core'', but we can also observe the 
internal structure of these various elements (from Schaap \& van de Weygaert 2002b). 
Data courtesy: the 2dF consortium.}
\end{figure}

A major challenge will be to quantify the intricacies and cohesiveness of 
this cosmic foam geometry in a fashion befitting its rich information content. 
Such analysis should be able to yield a meaningful quantification of the 
structural content of the foamlike network, on the basis of which it will 
be possible to define distinct discriminating measures enabling a comparison 
between the various viable structure 
formation scenarios. 

The recent development of a fully adaptive method based on the Delaunay 
tessellation of the corresponding spatial point process, the {\it Delaunay Tessellation 
Field Estimator} (DTFE, see Schaap \& van de Weygaert 2000), appears to hold great promise. Based 
on the earlier work by Bernardeau \& Van de Weygaert (1996) to reconstruct 
a complete volume-covering and volume-weighted velocity field from a set of 
point-sampled velocities -- which proved to yield a significant improvement 
in reproducing the statistics of the underlying continuous velocity field -- 
it reconstructs the full and cohesive density field of which the discrete 
galaxy distribution is supposed to be a sparse sample. Without invoking any 
artificial and often structure diluting filter it is able to render both 
the {\it anisotropic} nature of the various foam elements as well as the 
{\it hierarchical} character of the distribution in full contrast (see 
Schaap \& van de Weygaert 2002a). 

The potential promise of the DTFE may be amply appreciated from its succesfull 
reconstruction of a density field from the galaxy distribution in the southern 
part of the 2dF survey (Fig. 14, cf. Fig. 4). Evidently, it manages to bring out any 
fine structural detail of the intricate and often tenuous filamentary 
structures. Notice the frequently razor-sharp rendition of thin edges 
surrounding void-like regions. Hence, it defines a volume-covering density field reconstruction 
that retains every structural detail, which will enable us to study in a much 
improved fashion the statistical and geometric properties of the foam. Indeed, 
it even appears to ``clean'' the original discrete galaxy distribution map 
by suppressing its shot noise contribution. 

To underline its capacity to dissect the internal structure of the various 
structural components, in Fig. 15 we focus in on one of the major mass 
concentrations. It nicely illustrates its location at a junction point within 
the cosmic foam. Various filamentary extensions emanate from the high-density core. 
Not only does the DTFE method elucidate the filamentary anisotropic structures and their 
mutual spatial relationship, but as well it manages to highlight automatically the 
complex internal structure of the various connected elements. 

%\bigskip
%\bigskip
\vfill\eject
\section{\rm{\Large POWER THAT BE: ...}}
\begin{flushright}
{\rm{\Large Gravity Rules the Waves}}
\end{flushright}
\smallskip\noindent
The fundamental cosmological importance of the {\it cosmic foam} is that it comprises  
features on a typical scale of tens of Megaparsec, scales at which the Universe still 
resides in a state of moderate dynamical evolution. Structures have only freshly emerged 
from the almost homogeneous pristine Universe and have not yet evolved beyond 
recognition. Therefore they still retain a direct link to the matter distribution in 
the primordial Universe, and thus still contain a wealth of direct information on 
the cosmic structure formation process. 
\medskip
\subsection{Power That Be: \hfill Gravitational Instability}
The generally accepted theoretical framework for the formation of structure is that of 
gravitational instability. The gravitational instability scenario assumes the 
early universe to have been almost perfectly smooth, with the exception of tiny density 
deviations with respect to the global cosmic background density and the accompanying tiny 
velocity perturbations from the general Hubble expansion. For a general density 
fluctuation field $\delta({\bf r}',t)=(\rho({\bf r}')-\rho^b)/\rho^b$ (Fig. 16, top 
lefthand panel), this results in a corresponding total peculiar gravitational acceleration 
${\bf g}({\bf r})$ (Fig 16, top righthand panel) field which at any cosmic position 
${\bf r}$ can be written as the integrated effect of the peculiar gravitational attraction 
exerted by all matter fluctuations throughout the Universe, 
\begin{equation}
{\bf g}({\bf r},t)= {\displaystyle 3 \Omega H^2 \over \displaystyle 8\pi}\,
\int {\rm d}{\bf r}'\,\delta({\bf r}',t)\,{\displaystyle ({\bf r}'-{\bf r})
\over \displaystyle |{\bf r}'-{\bf r}|^3}\ . 
\end{equation}
Here $\Omega$ is the cosmological density parameter quantifying the mean 
density $\rho^b$ of the Universe via the relation $4\pi G \rho^b = {3 \over 2} 
\Omega H^2$. The above relation between density field and gravitational 
field ${\bf g}$ is in essence established through the Poisson equation, 
\begin{equation}
\nabla^2 \phi = 4 \pi G {\rho^b}(t) a(t)^2 \ \delta({\bf x},t) 
= {\displaystyle 3 \over \displaystyle 2} \Omega H^2 a^2 \ \delta({\bf x},t)\ , 
\end{equation}
\noindent relating density contrast $\delta({\bf r}',t)$ and gravitational potential 
perturbation $\phi({\bf r},t)$, from which we obtain ${\bf g}=-\nabla \phi/a$. 
\begin{figure}[t]
\vskip -0.5truecm
\centering\mbox{\hskip -1.5truecm\psfig{figure=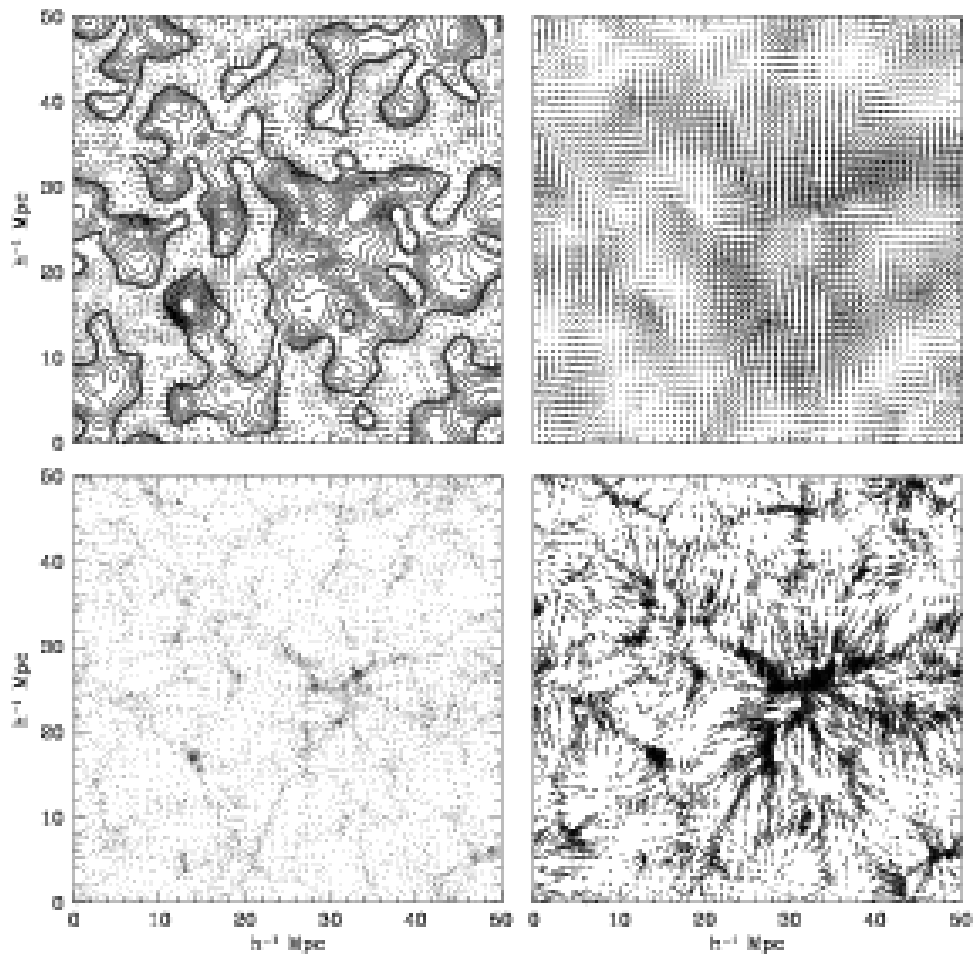,width=14.3cm}}
\vskip -0.7truecm
\caption{Gravitational Instability: schematic presentation of process. Top 
lefthand: contour map of a (Gaussian) stochastic density field. Top righthand: 
the resulting gravitational force field. Lower lefthand: resulting (nonlinear) 
particle distribution. Lower righthand: vector map corresponding velocity field}
\vskip -0.5cm
\end{figure}
The formation and moulding of structure is then ascribed to the gravitational growth of 
these primordial density- and velocity perturbations. Gravity in slightly overdense regions 
will be somewhat stronger than the global average gravitational deceleration, as will be 
the influence they exert over their immediate surroundings. In these regions the 
slow-down of the initial cosmic expansion is correspondingly stronger and, when 
the region is sufficiently overdense it may even come to a halt, turn around  
and start to contract. If or as long as pressure forces are not sufficient to counteract 
the infall, the overdensity will grow without bound, assemble more and more matter 
by accretion of matter from its surroundings, and ultimately fully collapse to form 
a gravitationally bound and virialized object. In this way the primordial overdensity 
finally emerges as an individual recognizable denizen of our Universe, their precise 
nature (galaxy, cluster, etc.) and physical conditions determined by the scale, mass 
and surroundings of the initial fluctuation. 

For a pressureless medium, the full evolution of this system of coupled 
cosmic density-, velocity- and gravity fields is encoded in three coupled 
fluid equations. Schematically, the essential aspects of this process of 
gravitational growth of structure, starting from a field primordial matter perturbations, are 
rendered in Figure 16. In addition to the Poisson equation (2), connecting the matter distribution 
to the gravitational field (left and right top panel Fig. 16), these are the {\it Euler equation} and the 
{\it continuity equation}. The Euler equation is the equation of motion describing the 
induced and corresponding matter flows (see bottom right panel Fig. 16),  
\begin{equation}
{\displaystyle \partial {\bf v} \over \displaystyle \partial t}\ +\ {\displaystyle {\dot a} \over \displaystyle a} {\bf v}
\ +\ {1 \over a}\ ({\bf v} \cdot \nabla) {\bf v}\ =\ - {1 \over a}\ \nabla \phi\ ,
\end{equation}
\noindent while the {\it continuity equation} guarantees the conservation of mass in this 
evolving system of matter migrations towards the emerging cosmic structures (see bottom left panel Fig. 16),   
\begin{equation}
{\displaystyle \partial \delta \over \displaystyle \partial t}\ +\ {1 \over a}\ \nabla \cdot (1+\delta){\bf v}\ .  
\end{equation}
\noindent 

\medskip
\subsection{Cosmic Structure: \hfill Gaussian by descent.}
Usually, the primordial density and velocity perturbation field is assumed to be 
a field of random fluctuations whose stochastic nature is that of a homogeneous and 
isotroptic spatial Gaussian process. The Gaussian nature of the random field 
$f({\bf x})$ (for which we take a zero mean, for simplicity, as in the case of the 
density excess $\delta({\bf r})$, the peculiar gravitational acceleration 
${\bf g}_{pec}$ and peculiar velocity ${\bf v}_{pec}$) implies its set of $N$-point 
joint probabilities to be given by  
\begin{equation}
{\cal P}_N\ =\ {\displaystyle \exp\left[-{\displaystyle {1 \over 2}}\,
\sum\nolimits_{i=1}^N\,\sum\nolimits_{j=1}^N\,f_i\,({\bf {\sf M}}^{-1})_{ij}
\,f_j\right] \over \displaystyle [(2\pi)^N\,(\det {\bf {\sf M}})]^{1/2}}\,
\prod_{i=1}^N\,{\rm d}f_i\,
\end{equation}
\noindent where ${\cal P}_N$ is the probability that the field $f$ has values in the 
range $f({\bf x}_j)$ to $f({\bf x}_j)+{\rm d}f({\bf x}_j)$ for each of 
the $j=1,\ldots,N$ (with $N$ an arbitrary integer and ${\bf x}_1,{\bf x}_2,
\ldots,{\bf x}_N$ arbitrary locations in the field). The matrix ${\bf {\sf M}}^{-1}$ 
is the inverse of the $N \times N$ covariance matrix ${\bf {\sf M}}$, 
\begin{equation}
M_{ij}\ \equiv \ \langle f({\bf x}_i) f({\bf x}_j) \rangle\ =\  
\xi({\bf x}_i-{\bf x}_j)\ ,
\end{equation}
\noindent in which the brackets $\langle \dots \rangle$ denote an 
ensemble average. In effect, ${\sf M}$ is the generalization of the variance 
$\sigma^2$ in a one-dimensional normal distribution. As the matrix ${\bf {\sf M}}$ 
is fully determined by the autocorrelation function $\xi(r)$, the Fourier 
transform of the power spectrum $P_f(k)$ of the fluctuations $f({\bf r})$, 
\begin{equation}
\xi({\bf r})\ =\ \xi(|{\bf r}|)\ =\ \int \d3k \ P_f(k) 
{\rm e}^{-{\rm i}{\bf k}\cdot{\bf r}}\ .
\end{equation}
\noindent This forms a statement for the full characterization of the statistical 
properties of a Gaussian random field $f$ by the power spectrum $P_f(k)$. 

Notice that the identity of $\xi({\bf r})$ and $\xi(|{\bf r})|)$ is an expression 
of the homogeneity and isotropy of the stochastic process. It means that the 
stochastic properties of the process are absolutely equivalent at every location 
and in every direction. If -- as it appears to be -- nature indeed has endowed 
us with this fortunate circumstance with respec to the matter distribution in our 
observable Universe, justified to invoke the {\it Ergodic Theorem}. On the basis of 
the latter, it is indeed a meaningful exercise to infer estimates of intrinsic 
{\it ensemble averages} of physical quantities on the basis of spatial averages 
of these quantities over merely one realization. The only provision is that the 
realization should comprise many statistically independent volumes. Given the fact 
that the one observable Universe in which we live is all we have access to, this 
is a rather welcome trait. 

\subsubsection{{\it The Power Spectrum}}
Th\'e {\it prime concept} in any (Gaussian) structure formation scenario is the 
density power spectrum $P(k)$. It embodies the relative contribution of density 
fluctuations at every relevant spatial scale $\lambda_k(=2\pi/k)$ to the full density field. 
Concretely, this is expressed through the Fourier integral over $P(k)$,  
\begin{equation}
\sigma_{\circ}^2\ =\ \int \d3k \ P(k)\ .
\end{equation}
\noindent which yields the total local density fluctuation $\sigma_{\circ}=
\langle\delta({\bf r})^2\rangle$. Having specified the density 
power spectrum $P(k)$, it is rather straightforward to set up Gaussian 
realizations of the the corresponding density field. All other (gravitional) 
physical fields and quantities are fully linked to the density field. 
The gravitational potential is related to the density field via the Poisson 
equation, and thus also the gravitational force field (see Eqn. 2). In addition, via 
the continuity equation and Euler equation we can then infer the resulting 
field of peculiar velocities (for both, see Fig. 16). While in general not 
trivial, the full velocity field can be uniquely and directly inferred in 
the linear regime (Peebles 1980). 

\medskip
\subsubsection{{\it Reality of Primordial Gaussianity}}
There are both physical and statistical arguments in favour of the  
assumption that the primordial density field in the Universe was indeed 
of a Gaussian nature. If the very early Universe went through an inflationary phase, 
quantum fluctuations would generate small-amplitude curvature fluctuations. 
The resulting density perturbation field is generally a Gaussian random
process with a nearly Harrison--Zel'dovich scale-invariant primordial power
spectrum. But, even while inflation did not occur, the density field
$\delta({\bf x})$ will be nearly Gaussian in the rather general case that its
Fourier components ${\hat \delta}({\bf k})$ are independent and have random
phases There are both physical and statistical arguments in favour of the  
assumption that the primordial density field in the Universe was indeed 
of this nature. If the very early Universe went through an inflationary phase, 
quantum fluctuations would generate small-amplitude curvature fluctuations. 
The resulting density perturbation field is generally a Gaussian random
process with a nearly Harrison--Zel'dovich scale-invariant primordial power
spectrum. But, even while inflation did not occur, the density field
$\delta({\bf x})$ will be nearly Gaussian in the rather general case that its
Fourier components ${\hat \delta}({\bf k})$ are independent and have random
phases (cf. Scherrer 1992).

Evidently, the final verdict rests on observations of the real 
world. Most fascinating has been the opening up of the window onto the 
surface of last scattering, analyzing and dissecting the cosmic microwave 
background with fantastic sensitivity and resolution. The imprint of the 
density and velocity fluctuations at the epoch of recombination on the 
last scattered photons has provided a direct impression of the primordial conditions from 
which structure in the Universe has arisen. The COBE-DMR maps of Sachs-Wolfe 
temperature fluctuations, on a relatively large angular scale of $\approx 7^{\circ}$ 
(spatial scales in the order of a Gpc), appear to be primarily of Gaussian character 
(e.g. Smoot et al. 1994).  although recently there has been a flurry of claims for 
a definite non-Gaussian signature (e.g. Ferreira, Magueijo \& G\'orski 1998). 
However, rather than being intrinsic this may be a systematic artefact (see 
Banday, Zaroubi \& G\'orski 2000). In the meantime, also skymaps of the cosmic 
microwave background at the astrophysically very interesting scales -- those whose 
angular size of a few arcminutes brings them within the realm of present-day 
recognizable features in the cosmic matter distribution -- have become available 
through the impressive work of balloon borne experiments like Boomerang and M\'axima-1. 
As yet, these high resolution maps of the CMB do not seem to indicate any significant and 
flagrant deviations from Gaussianity (Polenta et al. 2001, Wu et al. 2001). 

In spite of these strong theoretical and observational arguments in favour of 
an intrinsically Gaussian primordial Universe, the impact of even a tiny 
genuine non-Gaussian signature could be tremendous. It would involve strong  
repercussions for the subsequent development of structure. In particular so 
for the nature of the first generation of objects in the Universe. 

\medskip
\subsection{Gravitational Instability:}
\begin{flushright}
{\bf{\large Progressing Complexity}}
\vskip -0.2truecm
\end{flushright}
The early linear stages of structure formation have been succesfully and 
completely worked out within the context of the linear theory of gravitationally 
evolving cosmological density and perturbation fields (Peebles 1980). 
At every cosmologically interesting scale, it aptly and succesfully describes 
the situation in the early eons after the decoupling of radiation and matter at 
recombination. It still does so at present on those spatial scales at which 
the landscape of spatially averaged perturbations resembles a panorama of 
gently sloping hills. However, linear theoretical predictions soon fail 
after gravity surpasses its initial moderate impact and nonlinear features 
start to emerge. Soon thereafter, we start to distinguish the gradual rise of 
the complex patterns, structures and objects which have shaped our Universe into 
the fascinating world of astronomy. 
\medskip
\subsubsection{{\it Going Nonlinear}}
Once the evolution is entering a stage in which the first nonlinearities 
start to mature, it is no longer feasible to decouple the growth of 
structure on the various involved spatial and mass scales. Rapidly maturing 
small scale clumps do feel the effect of the large scale environment in which they 
are embedded. Neighbouring structures not only influence each other by 
external long-range gravitational forces but also by their impact on 
resulting matter flows. The morphology and topology of more gradually 
forming large-scale features will depend to a considerable extent on 
the characteristics of their content in smaller scale structures that 
were formed earlier on. This never-ending increasing level of complexity 
has proved to pose a daunting challenge for developing a fully consistent 
theoretical framework. No cosmogenic theory as yet has managed to integrate 
all acquired insights and observed impressions of the world around us.

\medskip
\subsubsection{{\it Stumbling upon Asymmetric Complexities}}
A major and overriding complication in the efforts to frame a complete 
theory of structure formation is the nature of the cosmic matter distribution 
itself. To a reasonably good approximation, a large proportion of astrophysically 
interesting objects possess readily exploitable intrinsic symmetries. Spherical, 
ellipsoidal and axisymmetric morphologies are amongst the most familiar 
in the astrophysical world. Often these function as the key towards unlocking 
the dynamics and kinematics of the object, its hidden internal structure and its 
evolution. Even a rather superficial evaluation of the cosmic foam (see e.g. Fig. 1) reveals 
the lack of any such useful symmetry for the cosmic matter distribution. The 
foamlike pattern itself is a supreme manifestation of an inherently 
complex system. Such complexity stems from the assumed origin of the 
cosmic matter distribution, a (linear) random and Gaussian primordial 
density and velocity fluctuation field (see former sections), subsequently 
moulded by the self-enforcing action of gravity. 

\medskip
\subsubsection{{\it From global to partial views}}
Lacking useful symmeteries, the study of structure formation has furnished an 
abundance of mutually complementary descriptions. Each focuses on one or 
more particular aspects, sometimes isolating those deemed relevant at the cost 
of neglecting others. Our ideas of the workings of the gravitational 
instability process have therefore progressed and been shaped by 
a plethora of theoretical and numerical approaches and techniques. 

In terms of methodology deveral attitudes can be discerned. One approach 
seeks to formulate approximate descriptions of the full matter distribution, valid 
during either a restricted cosmic period or in the case of a few 
particular scenarios. The well-known Zel'dovich approximation, and 
its extensions, is the best example of this strategy. Others try to follow 
the full nonlinear evolution in a few conceptually simple configurations, 
hoping to isolate the mechanisms that appear to be the most crucial, and 
henceforth seeking to trace the possible imprint in the more complex world of 
reality. Still of enormous importance, representing the foundation on which most 
of our theories and descriptions are ultimately based, are the spherical 
model and the homogeneous ellipsoidal model (see section 4.2). 

Following the Eulerian view of an evolving physical system, there have been 
impressive advances in following the nonlinear evolution of systems. This has 
produced a huge complex of advanced and complicated nonlinear perturbation 
analyses (for an extensive review, see Bernardeau et al. 2002). Although, these 
have proven to be particularly apt in uncovering and highlighting important 
statistical clues and signatures, thus defining essential tools for discriminating 
between viable formation scenarios, they are not so succesfull in guiding 
our physical intuition concerning the unfolding of the complex patterns we 
have been discussing. These can hardly be characterized by the first orders 
of a full Eulerian perturbation series. The dynamical development 
accelerates so rapidly that any serious advance, in terms of the revelant 
timescales of the physical system, is posing an almost unsurmountable amount  
of effort. 

Lagrangian approaches have proven to be the most fruitful approach in developing 
a physical intuition of the structural evolution during the more 
advanced stages. By following the matter elements on their path through 
the evolving cosmic matter field it is easier to appreciate the various 
forces and deformation tendencies acting on them. Mathematically, the 
conversion involves a transformation from an Eulerian to a Lagrangian 
time derivative,  
\begin{equation}
{\displaystyle d \over \displaystyle d t}\ =\ {\partial \over \partial t}\ + \ {\bf v}\cdot \nabla\ ,
\end{equation}
\noindent where ${\bf v}$ is the velocity of the displacing fluid element. 

The translation of self-gravitating systems into N-body computer 
simulations -- the visibly most appealing, informative and widely employed 
technique -- is the most widely known and exploited Lagrangian formulation for 
the issue of structure formation. In addition, they have undoubtedly been the 
most succesfull in capturing and shaping the imagination of a wide public, 
both of professional astronomers as well as of lay people. If we wish to appreciate 
the information and insights they can convey and, equally essential, appreciate their 
limitations it is beneficial to discuss them within a more general context. 

\medskip
\subsection{Cosmic Equipment: \hfill N-body simulations}
Arguably the most visible -- scientifically as well as PR-wise -- Lagrangian 
technique for studying cosmic structure formation is the use of computer 
models and calculations to simulate the evolution of cosmic structure through the 
force of gravity. They are unique in their ability to deal with the 
evolution of a system through the full range of linear up to highly 
nonlinear stages, in principal unconstrained and for any feasible 
configuration, independent of structural complexities and lack of 
helpful symmetries. Equally important is that the N-body approach 
is not principally restricted to purely gravitationally evolving 
systems. As evidenced by an array of computer codes developed over 
the past decade, it is rather straightforward to incorporate the 
influence and workings of a range of complicating (often dissipational) 
physical effects and processes. Sophisticated computer codes have  
been able to succesfully incorporate gravity, gasdynamical processes, atomic 
processes and a range of other complicating factors into codes 
simulating the formation and evolution of cosmic structures. 

With the exception of a few attempts towards an Eulerian implementation 
(e.g. Peebles 1987), efforts which gradually gain more ground with the inclusion 
of gas-dynamical and radiative processes, computational efforts have concerned 
N-body computer simulations. By nature, this involves a translation of the evolution 
of a system into a Lagrangian formulation, in which the computer gets instructed to 
follow the path of each particle into which the initial density field had been broken up. 
Fully nonlinear N-body computer simulations have produced the most readily 
visible, and directly appealing and accessible, descriptions of the way into 
which the gravitational instability process manages to mould the primordial 
universe into rich structural patterns. 

Truely giant technological advances over the past decades have provided us with a 
comprehensive, physically justified and badly needed visual impression of 
the way into which the mutual influence of the combined gravitational forces 
unleashed by the matter perturbations throughout the cosmos works its way 
towards the emergence of structure. It has allowed the recognition of 
some basic mechanisms during the full nonlinear evolution of self-gravitating 
systems. The spatial patterns in the resulting particle distributions, as well as 
their kinematics and dynamics, provide us with an excellent testbed for testing 
and comparing quantitatively a large range of viable structure formation 
scenarios. However, they suffer from a variety of restricting effects, and 
we should be anxious not to overinterpret and/or overestimate the results 
of their performance. 

Firstly, they can only go as far as allowed by the physics implemented into 
them. They will and cannot reveal basic new science to us. The full array of 
relevant physics, in particular when we get below scales 
of clusters of galaxies, is not confined to merely gravitational influences. 
Radiative processes, a complex interplay of various hydrodynamic processes, star 
formation processes, and the full impact of involved feedback interactions 
should make us realize the limitations of the outcome of the calculations. 
A large industry of new complex computer codes attempt to deal with at least 
a selection of these influences. They, however, also illustrate the daunting  
and possible unsurmountable challenges awaiting us in formulating a fully 
and uniquely defined scenario of structure formation, incorporating every 
necessary aspect of relevant physics. 

Even when restricting ourselves to purely gravitational systems, strictly 
only applicable to scales at which nongravitational dissipation can be 
fully discarded as irrelevant, their results should still be considered 
as merely approximate and indicative. Their dynamic range is usually 
very limited. This is true for the feasible spatial resolution, for 
the attainable mass resolution, as well as for the range of timescales 
that can be covered by them. As available computer memory will always restrict 
these, the full spatial range of fluctuations -- even influential ones -- can 
never be properly represented within a given cosmic region. A representation  
of the full primordial power spectrum of density fluctuations is beyond the 
grasp of any conceivable piece of equipment. It is therefore important that 
claims of validity of specific formation scenarios can never be fully 
justified. In addition, the limited dynamic range will also restrict the 
force resolution during the nonlinear evolution, which is particularly 
cumbersome on the smallest scales which produce the first nonlinear entities. 
In addition, we see a rapid increase in computational expense as we try to 
improve the resolution. Even though the availability of ultrafast computing 
machinery with ever growing huge memory capacity has expanded the 
achievements of cosmological N-body computer simulations to 
truely dramatic levels, the demands grow along with them, a pace not 
necessarily followed equally fast by the insights going along with them. 

An additional restriction for N-body codes is that they can strictly only be 
applied to initial value problems (see section 3.5.3). For pure theoretical 
purposes, this does not pose a restriction. For a given cosmological and 
structure formation model the initial density and velocity field are 
specified and subsequently evolved and analyzed. However, when turning 
to the observed world, we may recognize that we are usually dealing 
with analyzing situations in which most information is available for 
the present epoch. In most cases this involves nonlinear structures for 
which we are not able to extrapolate in a straightforward and direct 
fashion towards the initial conditions. Hence we are unable to 
use N-body techniques towards analyzing the implications for the 
dynamics and evolution of the observed structures. 
 
Possibly the most important limitation of computer calculations may 
be that the understanding of the dynamics and physics of the 
simulated systems is not significantly increased. {\it Simulations will not 
increase our understanding of dynamics without guidance from analytical 
approaches}, and therefore analytical approximations will be absolutely essential 
for that purpose.  Along one direction, these do provide us with a handle to 
interpret the outcome of the computer experiments. Equally important, they 
direct us in defining the best possible computer models and configurations. 

Even more important it is to be aware that full understanding involves 
the formulation of analytical descriptions, which should embody our insights into 
the systematics and regularities of a system, the true purpose of scientific 
inquiry. Analytical descriptions should therefore be the preferred endgoal, 
the computer experiments, along with the ever growing body of available 
observations, are there to guide and sharpen our insight and intuition. 
%vfill\eject
\medskip
\subsection{Cosmic Equipment:}
\begin{flushright}
{\bf{\large Analytical Lagrangian schemes}}
\vskip -0.2truecm
\end{flushright}
While N-body simulations are the most popular specimen of Lagrangian 
description of gravitationally evolving systems, we may also attempt to mould 
our physical understanding into analytical Lagrangian formulations and 
approximations. For the study of gravitational pattern formation these have 
proven to be of overriding importance. 

It is in particular the formulation of the first-order Zel'dovich approximation (Zel'dovich 1970) 
which has been of eminent importance in the study of cosmological 
structure formation. First and foremost, it did elucidate and explain qualitatively the basic 
tendencies of gravitational contraction in an evolving cosmos, in particular the tendency 
to do so anisotropically. In addition to its conceptual significance, it 
assumed extensive influence by providing computational cosmologists with the 
machinery to set up the initial conditions of cosmological N-body simulations 
for a wide variety of structure formation scenarios. 
\medskip
\subsubsection{{\it the Zel'dovich Approximation}}
By means of a Lagrangian perturbation analysis Zel'dovich (1970) proved, in a seminal 
contribution, that to first order -- typifying early evolutionary phases -- the reaction 
of cosmic patches of matter to the corresponding peculiar gravity field would be surprisingly 
simple, expressing itself in a plain ballistic linear displacement set solely by the 
initial (Lagrangian) force field. This framed the well-known {\it Zel'dovich approximation}. 

In essence, the Zel'dovich approximation is the solution of the Lagrangian equations 
for small density perturbations ($\delta^2 \ll 1$). The solution is based upon the 
first-order truncation of the Lagrangian perturbation series of the trajectories of mass 
elements, 
\begin{equation}
{\bf x}({\bf q},t)\ =\ {\bf q}\ +\ {\bf x}^{(1)}({\bf q},t)\ +\ {\bf x}^{(2)}({\bf q},t)\ +\ \ldots
\end{equation}
\noindent when considering succesive terms of $|\partial({\bf x}-{\bf q})/\partial {\bf q}|$. 
The truncation at the ${\bf x}^(1)$ term, then involves a simple linear prescription 
for the displacement of a particle from its initial (Lagrangian) comoving position 
${\bf q}$ to an Eulerian comoving position ${\bf x}$, solely determined by the 
initial gravitational potential field, 
\begin{equation}
{\bf x}({\bf q},t)\,=\,{\bf q}\,-\,D(t) \nabla \Psi({\bf q})\,.
\end{equation}
\noindent In this mapping, the time dependent function $D(t)$ is the growth rate of 
linear density perturbations, and the time-independent spatial function 
$\Psi({\bf q})$ is related to the {\it linearly extrapolated} gravitational 
potential ${\underline \phi}$\footnote{the {\it linearly extrapolated} gravitational 
potential is defined as the value the gravitational potential would assume in case the 
field would evolve according to its linear growth rate, $D(t)/a(t)$)},  
\begin{equation}
\Psi\ = \ {\displaystyle 2 \over \displaystyle 3 D a^2 \Omega H^2}\ 
{\underline \phi}\,.
\end{equation}
\noindent This immediately clarifies that what represents the major virtue of the 
Zel'dovich approximation, its ability to assess the evolution of a density field 
from the primordial density field itself, through the corresponding linearly 
extrapolated (primordial) gravitational potential. While in essence a local 
approximation, the Lagrangian description provides the starting point 
for a far-reaching analysis of the implied density field development, 
\begin{eqnarray}
{\displaystyle \rho({\bf x},t) \over \displaystyle {\rho^b}}\ =\ 
\Biggl\Vert {\displaystyle \partial {\bf x} \over \displaystyle \partial {\bf q}}
\Biggr\Vert^{-1}&\ =\ &
\Biggl\Vert \delta_{mn}-a(t)\psi_{mn}\Biggr\Vert^{-1}\nonumber\\
&\ =\ &{\displaystyle 1 \over \displaystyle [1-a(t) \lambda_1][1-a(t) \lambda_2][1-
a(t)\lambda_3]}\,,
\end{eqnarray}
\noindent where the vertical bars denote the Jacobian determinant, and
$\lambda_1$, $\lambda_2$ and $\lambda_3$ are the eigenvalues of the 
Zel'dovich deformation tensor $\psi_{mn}$,
\begin{equation}
\psi_{mn}\ = \ {\displaystyle D(t) \over \displaystyle a(t)}\,\,
{\displaystyle \partial^2 \Psi\over \displaystyle 
\partial q_m \partial q_n}
\ =\ {\displaystyle 2 \over \displaystyle 3 a^3 \Omega H^2}\,\,
{\displaystyle \partial^2 {\underline \phi} \over \displaystyle \partial q_m 
\partial q_n}\ ,
\end{equation}
\noindent which also implies the deformation matrix $\psi_{mn}$ and its eigenvalues to 
evolve as $\psi_{mn}\propto D(t)/a(t)$. On the basis of above relation, it is then 
straightforward to find the intrinsic relation between the Zel'dovich deformation 
tensor $\psi_{mn}$ and the tidal tensor $T_{mn}$,
\begin{equation}
\psi_{mn}\ = \ {\displaystyle 1 \over \displaystyle {\scriptstyle{\frac{3}{2}}}\Omega H^2 a} 
\Bigl(\ {\underline T_{mn}}\ +\ {\scriptstyle {\frac{1}{2}}}\Omega H^2\ {\underline \delta}\ \delta_{mn}\ \Bigr)\ ,
\end{equation}
\noindent where ${\underline \delta}$ and ${\underline T}_{mn}$ are the respective 
{\it linearly extrapolated} values of these quantities, i.e. ${\underline \delta}(t) \propto 
D(t)$ and ${\underline T}_{mn}\propto D/a^3$. Notice that, without loss of generality, 
we can adopt a coordinate system where the tidal tensor matrix $T_{mn}$ is diagonal, 
from which we straightforwardly find a relation between the linearly extrapolated 
${\underline \delta}$ and the deformation tensor, ${\underline\delta}(t)=a(t)\sum_m\lambda_m$.

For appreciating the nature of the involved approximation, one 
should note that the continuity equation (by definition) is always satisfied 
by the combination of the Zel'dovich approximation and the mass conservation, yet 
that they do not, in general, satisfy the Euler and the Poisson equations (Nusser 
et al. 1991). Only in the case of purely one-dimensional perturbations does the 
Zel'dovich approximation represent a full solution to all three dynamic equations. 
Indeed, as we will emphasize in section 3.5.2, the core and essential physical significance 
of the Zel'dovich approximation can be traced to this implicit assumption of the 
tidal tensor $T_{ij}$ being linearly proportional to the deformation tensor, and 
hence the velocity shear tensor. 

The central role of the Zel'dovich formalism (for a review, see Shandarin \& Zel'dovich 
1989) in structure formation studies stems from its ability to take any arbitrary 
initial random density field, not constrained by any specific restriction in terms of 
morphological symmetry or seclusion, and mould it through a simple and direct operation into a 
reasonable approximation for the matter distribution at later nonlinear epochs. It allows 
one to get a rough qualitative outline of the nonlinear matter distribution 
solely on the basis of the given initial density field. While formally the 
Zel'dovich formalism comprises a mere first order perturbative term, it turned 
out to represent a surprisingly accurate description up to considerably more 
advanced evolutionary stages, up to the point where matter flows would start to 
cross each other. 
\medskip
\subsubsection{{\it Local Lagrangian Approximations}}
The Zel'dovich approximation (1970) belongs to a class of Lagrangian approximation 
schemes in which the nonlinear dynamics of selfgravitating matter is encapsulated 
into an approximate  ``local'' formulation (Bertschinger \& Jain 1994, Hui \& 
Bertschinger 1996). In these approaches, the density, velocity gradient, and gravity 
gradient for each mass element behaves as if the element evolves independently of all 
the others once the initial conditions are specified. For instance, the evolution of 
a given mass element under the Zel'dovich approximation is completely 
determined once the initial expansion, vorticity, shear and density at this 
mass element are specified. The influence of other mass elements on the subsequent evolution of 
these quantities at this particular mass element is then assumed to be {\it fully encoded 
in the initial conditions}, and unaffected by the subsequent evolution 
of these other mass elements. Such a presumption may seem implausible in view of the 
unrestrained long-range gravitational force, yielding a noticeable influence from all 
other mass elements in the Universe. Yet, the success of the Zel'dovich 
approximation, having provided a great deal of insight into the essentials 
of nonlinear evolution of density fluctuations, demonstrates how useful such 
schemes in fact can be. 

The {\it locality} of these approximation schemes implies the evolution 
to be described by a set of {\it ordinary differential equations} for each 
mass element, with no coupling to other mass elements aside from those 
implied by the initial conditions. Note that N-body simulations are 
distinctly {\it non-local} Lagrangian descriptions. At every timestep the 
full gravitational potential set by the full cosmic matter distribution needs 
to be evaluated at the location of every mass element. 

Assessing the basic Lagrangian fluid equations readily illuminates the 
character of these local approximations and elucidates the nature of 
some basic tendencies in the evolution of the matter distribution. The deformation 
of the evolving mass element at comoving location 
${\bf x}$ is most straightforwardly encoded in the decomposition of the gradient of its 
velocity field ${\bf v}$ (the rate-of-strain tensor) into the expansion $\theta$, the 
shear $\sigma_{ij}$ and vorticity $\omega_{ij}$, 
\begin{equation}
{\displaystyle \partial v_i \over \displaystyle \partial x_j}\ = \ {1 \over 3}\ \theta \delta_{ij}\ + \ 
\sigma_{ij}\ + \ \omega_{ij}\ , 
\end{equation} 
\noindent in which the expansion $\theta$ is the trace of the velocity field gradient, 
$\sigma_{ij}=\sigma_{ji}$ the traceless symmetric part and 
$\omega_{ij}=-\omega_{ji}=\epsilon_{ijk}\omega^k$ the antisymmetric part (and $\epsilon_{ijk}$ the 
Levi-Civita symbol), 
\begin{eqnarray}
\theta\ = \ \nabla \cdot {\bf v}&\ \equiv\ &{\displaystyle \partial v_1 \over \displaystyle \partial x_1} + 
{\displaystyle \partial v_2 \over \displaystyle \partial x_2} + 
{\displaystyle \partial v_3 \over \displaystyle \partial x_3}\ ,\\
\sigma_{ij}&\ \equiv\ &{\displaystyle 1 \over \displaystyle 2} \left( 
{\displaystyle \partial v_i \over \displaystyle \partial x_j} + 
{\displaystyle \partial v_j \over \displaystyle \partial x_i}\right) - 
{\displaystyle 1 \over \displaystyle 3} (\nabla\cdot{\bf v})\,\delta_{ij}\ ,\\
\omega_{ij}&\ \equiv\ &{\displaystyle 1 \over \displaystyle 2} \left( 
{\displaystyle \partial v_i \over \displaystyle \partial x_j} -  
{\displaystyle \partial v_j \over \displaystyle \partial x_i}\right)\ ,
\end{eqnarray}
\noindent in which $2 {\vec \omega} = \nabla \times {\bf v}$. Under the assumption 
that the fluid is pressureless and irrotational (i.e. $\omega_i=0$), the evolution of a 
fluid element, whose comoving trajectory is described by ${\bf x}(t)$ is specified by 
four first order differential equations. The first is the Lagrangian 
continuity equation, concerning the evolution of the density contrast $\delta$, 
\begin{equation}
{\displaystyle d \delta \over \displaystyle d\tau}\ +\ (1+\delta)\ \theta\ =\ 0\ .
\end{equation}
\noindent in which we have been following the formulation by Bertschinger \& Jain (1994) and 
use the conformal time $\tau$ ($d\tau=dt/a$) as time coordinate, with the velocity ${\bf v}$ 
define as ${\bf v}=d{\bf x}/d\tau$. 

The second and third equation describe the 
evolution of the trace and the traceless symmetric components of the velocity 
gradient tensor. The evolution of the trace of $\nabla_i v_j$, the expansion $\theta$, 
is described by the {\it Raychaudhuri equation}, 
\begin{equation}
{\displaystyle d\theta \over d\tau}+{\displaystyle {\dot a} \over \displaystyle a}\theta + 
{\displaystyle 1 \over \displaystyle 3}\theta^2+\sigma^{ij}\sigma_{ij} = \ 
-4\pi G a^2 \rho^b \delta\ ,
\end{equation}
\noindent which follows from an evaluation of the Euler fluid equation (eqn. 3) in combination 
with the Poisson equation (eqn. 2). Similarly evaluating the symmetric  
part of the Euler equation we find that the shear $\sigma_{ij}$ evolves according to 
\begin{equation}
{\displaystyle d \sigma_{ij} \over \displaystyle d\tau}+{\displaystyle {\dot a} \over 
\displaystyle a}\sigma_{ij}+{2 \over 3} \theta \sigma_{ij}+ 
\sigma_{ik}\sigma^{k}_j-{1 \over 3}\delta_{ij}(\sigma^{kl}\sigma_{kl})\ =\ 
-T_{ij}\ ,
\end{equation}
\noindent where $T_{ij}\equiv \nabla_i \nabla_j \phi - (1/3)(\nabla^2 \phi)$ is the 
gravitational tidal field (see Eqn. 19). Within the context of general relativity $T_{ij}$ is 
the electric part of the Weyl tensor in the fluid frame. Evidently, as expected 
this equation is an expression of how the gravitational tidal field acts as a source 
term for inducing a shear component in the flow field.  

The final ``closure'' relation is that relating the gravitational field 
$\phi$ and the density contrast $\delta$, the Poisson equation (see Eqn. 2). 

An interesting observation follows from evaluating the combination of the 
continuity and Raychaudhuri equation. This leads to a second order ordinary differential 
equation describing the evolution of the density contrast $\delta$ of a fluid element: 
\begin{equation}
{\displaystyle d^2 \delta \over \displaystyle d\tau^2} + 
{\displaystyle {\dot a} \over \displaystyle a} {\displaystyle d \delta \over 
\displaystyle d t} = {\displaystyle 4 \over \displaystyle 3} (1+\delta)^{-1} 
\left({\displaystyle d \delta \over \displaystyle d \tau}\right)^2 + 
(1+\delta)\left(\sigma_{ij} \sigma_{ij}+ 4\pi G {\rho^b} \delta \right)\ .
\end{equation}
\noindent As Bertschinger \& Jain (1994) noted by means of the {\it collapse 
term}, this equation shows that in the absence of vorticity the presence of a 
nonzero shear increases the rate of growth of density fluctuations (vorticity will 
inhibit this process). In other words, the rate of growth of $\delta$ gets amplified 
in the presence of shear. This conclusion holds independently of assumptions about the 
evolution of shear or tides and is due to the simple geometrical fact that shear 
increases the rate of growth of the convergence ($\propto -\theta$) of fluid 
streamlines. With zero shear and vorticity the velocity gradient tensor is isotropic, 
corresponding to uniform spherical collapse with radial motions towards some 
centre. Equation (23) then reduces to the exact equation for the evolution of the 
mean density in the spherical model. This evidently implies the growth of uniform 
spherical perturbations to be more slowly than more generic anisotropic 
configurations with a nonzero shear term. 

Possibly the most telling illustration of this phenomenon can be found in 
the evolution of collapsing homogeneous ellipsoids. As can be found in 
section 4.2 (see Fig. 20) collapsing ellipsoids tend to slow down their 
collapse along the longest axis while they evolve far more rapidly along 
the shortest axis. Starting from a near spherical configuration, the net result 
is a rapid contraction into a highly flattened structure as the short axis has 
fully collapsed. When considering the complete (homogenous) ellipsoid as 
a ``fluid'' element, the correspondence with the reasoning above is 
straightforward. 

The one remaining complication for transforming these Lagrangian fluid equations 
into {\it local} descriptions involves the development of the driving shear term 
$\sigma_{ij}$. Evidently, its evolution depends on the behaviour of the gravitational 
tide $T_{ij}$. Obviously, its value is generically determined by the full 
matter distribution throughout the cosmic volume. From a detailed evaluation 
of the issue in its Lagrangian context, its complexity may be seen as one relating 
to a highly complicated \'and {\it non-local} expression involving $H_{ij}$, the 
Newtonian limit of the magnetic part of the Weyl tensor in the fluid frame 
(see Bertschinger \& Jain 1994).

Hui \& Bertschinger (1996) demonstrated how the Zel'dovich approximation can be 
incorporated within this context of reducing the problem to a local one by 
invoking a specific approximation. In essence it involves an implicit decision  
to discard $H_{ij}$ and the full evolution equation for $T_{ij}$ altogether
and replace it by the explicit expression for the evolution of the tidal 
tensor $T_{ij}$ in the linear regime of clustering, in which $T_{ij}$ is 
linearly proportional to the shear $\sigma_{ij}$,
\begin{equation}
T_{ij}\ =\ - {\displaystyle 3 \Omega H^2 \over \displaystyle 2 H f(\Omega)}\ 
\sigma_{ij}\ 
\end{equation}
\noindent where $f(\Omega) \approx \Omega^{0.6}$. From this consideration we 
can straightforwardly appreciate that the Zel'dovich approximation does not 
obey the Poisson equation. As a consequence, it does represent an exact approximation in 
the case of a plane-parallel density disturbance as long as the corresponding particle 
trajectories have not yet intersected others, but breaks down in the case of two- or 
three-dimensional perturbations and in general as soon as the particle tracks have 
crossed. One well-known implication is that the Zel'dovich approximation 
gives incorrect results for spherical infall. 

With the Zel'dovich approximation basically concerning the 
first order solution in Lagrangian perturbation theory, various extensions 
and elaborations were formulated attempting to extend the theory to a higher 
order, futhering its range of applicability towards more evolutionary 
more progressive situations. As can be appreciated from Eqn. (24), the Zel'dovich 
approximation essentially involves a truncation of the set of Lagrangian fluid 
equations through an implicit choice for $T_{ij}$, equating it linearly proportional to 
$\sigma_{ij}$. This implies a tinkering with the Raychaudhury 
and shear evolution equations. In the Zel'dovich approximation one therefore 
need not integrate the tidal evolution equation, the gravity field on a mass 
element given by a simple extrapolation of initial conditions. 

Pursuing along the same track, Bertschinger \& Jain (1994) and Hui \& Bertschinger 
(1996) extended this to higher-order schemes, both approximations involving the 
integration of the exact Raychaudhuri and shear evolution and shifting the 
approximation to the level of the tidal evolution equation. Bertschinger 
\& Jain (1994) tried to make the very specific assumption of setting the 
magnetic part of the Weyl tensor $H_{ij}=0$, thus characterized as the 
``nonmagnetic'' approximation. Subsequent work showed that this does not adhere 
to a well-defined or easily recognizable generic situation. Along the same lines, 
Hui \& Bertschinger (1996) defined another approximate scheme, the ``local tidal'' 
approximation. This they showed to followed more accurately the predictions of the homogeneous 
ellipsoidal model than the ``nonmagnetic'' approximation (Bertschinger \& Jain 1996) and 
the Zel'dovich approximation. Interestingly, while the ``nonmagnetic'' approximation 
implied collapse into ``spindle'' like configurations as the generic outcome of 
gravitational collapse, the more accurate ``local tidal'' approximation appears 
to agree with the Zel'dovich approximation in that ``planar'' geometries are 
more characteristic. 

A range of additional approximate schemes, mostly stemming from different considerations,
were introduced in a variety of publications. Most of them try to deal with the evolution 
of high-density regions after the particle trajectories cross and self-gravity 
of the resulting matter assemblies assumes a dominating role. Notable examples of 
such approaches are the adhesion approximation (Kofman, Pogosyan 
\& Shandarin 1990), the frozen flow approximation (Matarrese et al. 1992), the 
frozen potential approximation (Brainerd, Scherrer \& Villumsen 1993), the 
truncated Zel'dovich approximation (Coles, Melott \& Shandarin 1993), the smoothed 
potential approximation (Melott, Sathyaprakash \& Sahni 1996) and higher order 
Lagrangian perturbation theory (Melott, Buchert \& Weiss 1995). The elegant 
generalization of the Zel'dovich approximation by Giavalisco et al. (1993) will be 
described later within the context of the mixed boundary conditions problem (next 
section). An extensive and 
balanced review of the various approximation schemes may be found in Sahni \& Coles (1995).

\medskip
\subsubsection{{\it Mixed Boundary Conditions}}
For pure theoretical purposes, cosmological studies may be restricted to considerations 
involving pure {\it initial conditions} problems. For a given cosmological and 
structure formation model the initial density and velocity field are specified, and 
its evolution subsequently evolved, either by means of approximate schemes or 
by for instance by means of fully nonlinear N-body computer simulations. The outcome 
of such studies are then usually analyzed in terms of a variety of statistical 
measures. These are then compared to the outcome of the same tests in the case 
of the observed world. 

Alternatively, we may recognize that in many specific cosmological studies we 
are dealing with problems of {\it mixed boundary conditions}. Part of the large 
scale structures and velocities in the universe may have been measured at 
the present epoch, while some in the limit of very high redshift (e.g. the 
CMB). Typically, one seeks to compute the velocity field consistent with an 
observed density structure at the present epoch. Conversely, one may wish to deduce 
the density from the measured peculiar galaxy velocities. Evidently, the position 
and velocities of objects at the present epoch are intimately coupled through 
the initial conditions. 

In the linear regime, the problem of mixed boundary conditions 
is easily solved. However, in the case of the interesting features in e.g. 
the cosmic foam the density of galaxies can reach values considerably larger 
than unity, even on scales of $\sim 10h^{-1}\hbox{Mpc}$. The associated 
velocities therefore need to be computed nonlinearly. As may be obvious from the earlier 
discussions, the nonlinear computation of gravitational instabilities in the 
case of arbitrary configurations is everything but trivial. 

N-body codes are rendered obsolete in such issues involving mixed boundary 
conditions. Their application is restricted to initial value problems. A further 
complication of the nonlinear problem of mixed boundary conditions is the multivalued 
nature of the solutions. Orbit crossing makes identification of the 
correct orbits difficult, and impossible after virialization has 
erased the memory of the initial conditions. In practice, one is therefore 
often restricted to laminar flow in the quasi-linear regime. Perturbations 
may have exceeded unity but orbit crossing has not yet obstructed a 
one-to-one correspondence between the final and initial positions. 

Evidently, the Zel'dovich approximation forms a good first-order approach 
towards dealing with such issues. Indeed, it was applied to the nonlinear 
problem of mixed boundary conditions, tested and calibrated using 
N-body simulations, by Nusser et al. (1991) and Nusser \& Dekel (1992). 
\begin{figure}[t]
\vskip -0.0truecm
\centering\mbox{\hskip -0.7truecm\psfig{figure=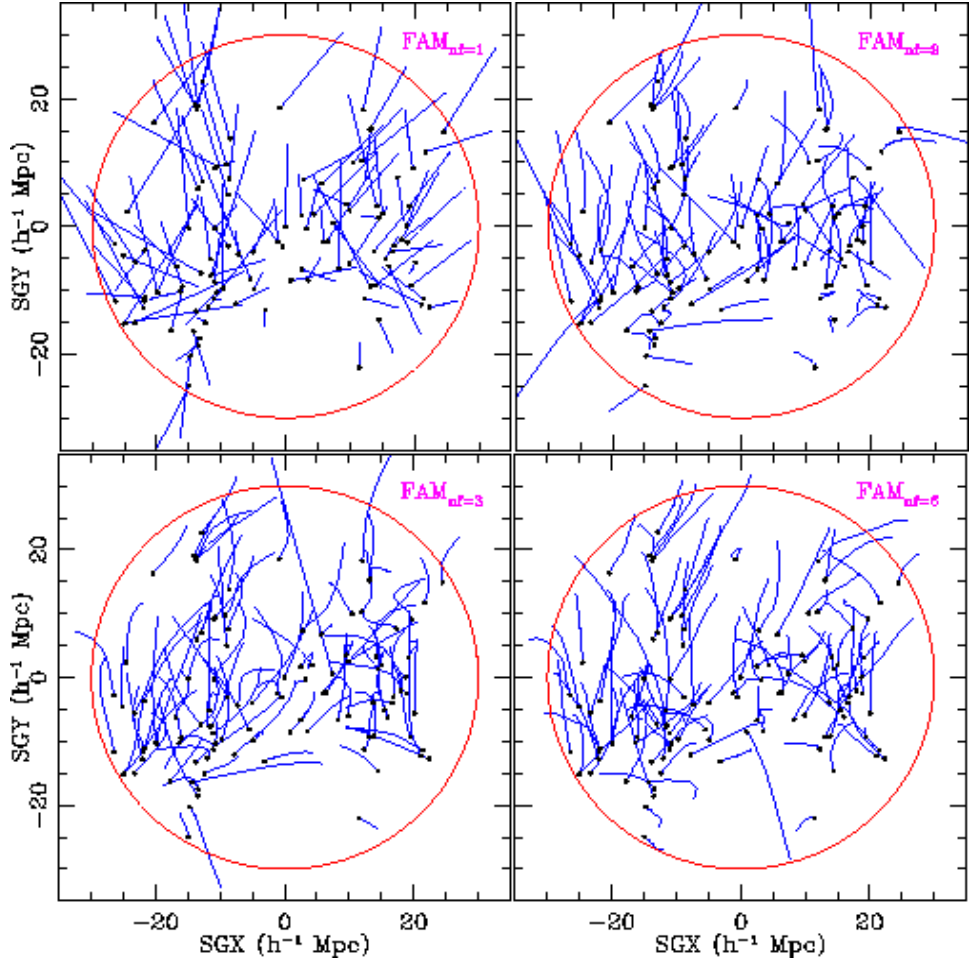,width=12.7cm}}
\vskip 0.0truecm
\caption{Projected orbits reconstructed by the Fast Action Method implementation 
of the Least Action Principle procdure. Four different levels of approximation 
are shown. The black dots represent the final (present) positions for each 
object. The blue lines indicate the trajectories followed by the objects as 
a function of time. The top-left frame shows FAM reconstructed orbits with 
$N_f=1$ (Zel'dovich approximation). Subsequently shown are $N_f=2$ (top right), 
$N_f=3$ (lower left) and $N_f=6$ (lower right).  From Romano-D\'{\i}az, Branchini 
\& van de Weygaert 2002.}
\vskip 0.0truecm
\end{figure}
In terms of the physics of the nonlinear systems, a profound suggestion was 
forwarded by Peebles (1989, 1990). He noticed that mixed boundary conditions 
naturally lend themselves to an application of Hamilton's principle. Given the 
action $S$ of a system of particles 
\begin{equation}
S\ = \ \int_0^{t_0}\ L dt \ =\ \int_0^{t_0} dt\ \sum_i\ [{1 \over 2} m_i a^2 {\dot {\bf x}}_i^2 - 
m_i \phi({\bf x}_i)\ ]\ ,  
\end{equation} 
\noindent in which $L$ is the Lagrangian for the orbits of particles with masses $m_i$ and 
comoving coordinates $x_i$. The {\it exact} equations of motion for the particles 
can be obtained from stationary variations of the action $S$. On the basis of Hamilton's 
principel one therefore seeks stationary variations of an action subject to fixed 
boundary conditions at both the initial and final time. Confining oneself to feasible 
approximate evaluation in this {\it Least Action Principle} 
approach, one describes the orbits of particles as a linear combination of suitably 
chosen universal functions of time with unknown coefficients specific 
to each particle presently located at a position ${\bf x}_{i,0}$,   
\begin{equation}
{\bf x}_i(D)\ =\ {\bf x}_{i,0}\ +\ \sum_{n=1}^{N_f}\ q_n(D)\ {\bf C}_{i,n}\ .
\end{equation}
\noindent In the above formulation we choose to use the linear growth mode 
$D(t)$ as time variable. The functions $q_n(D)$ form a set of $N_f$ 
time-dependent basis functions, while ${\bf C}_{i,n}$ are a set of free 
parameters which are determined from evaluating the stationary variations 
of the action. The basis functions $q_n(D)$ are constrained by two orbital 
constraints. To ensure that at the present time the galaxies are located 
at their observed positions ${\bf x}_i(D=1)={\bf x}_{i,0}$ we set the 
boundary constraint $q_n(1)=0$. The other boundary condition concerns the 
constraint that the peculiar motions at early epochs ($D=0$) have to vanish, 
which in turn ensures initial homogeneity, i.e. $\lim_{t \rightarrow 0} m_i a^2 
{\dot {\bf x}}_i =0$. 

The choice of base functions essentially determines the approximation scheme. Originally 
Peebles (1989, 1990) chose the base functions $q_n(D)=f_n(t)$ to be polynomials 
of the expansion factor $a(t)$. For small systems this leads to a tractable 
problem, but for larger systems it becomes exceedingly difficult to apply 
because of confusions between multivalued solutions. In an elegant contribution, 
Giavalisco et al. (1993) merged the Zel'dovich approximation into the {\it Least 
Action Principle} (LAP) scheme by expanding the formalism into one where the 
base functions are higher order polynomials of the linear growth function 
$D(t)$. In the limit of small displacements, the LAP procedure would then reproduce 
the Zel'dovich approximation. The higher order terms remove the separability 
of the temporal and spatial dependence in the Zel'dovich approximation, and allowing 
arbitrary displacements so that the orbits can be determined to any accuracy. This 
turned out to yield a rapidly converging scheme, even for highly nonlinear 
perturbations. 

A further improvement of the basic scheme of Giavalisco et al. (1993) was 
introduced by Nusser \& Branchini (2000). They based their computational 
evaluations of the action on invoking the functions 
$p_n(D)\equiv dq_n/dD$ and setting them equal to conveniently defined 
polynomials of the growth factor $D$. In addition, their {\it Fast Action Method} 
(FAM) involved a computational optimization through the evaluation of the 
gravitational potential by means of a gravitational TREECODE. This yielded 
a procedure allowing the reconstruction of the orbits of $10^4-10^5$ mass tracing 
objects back in time and reconstruct the peculiar velocities of the objects 
well beyond the linear regime. Its performance can be appreciated from 
Figure 17 (from Romano-D\'{\i}az, Branchini \& van de Weygaert 2002), in which 
4 panels show how the increasing levels of orbit expansion manage to 
probe ever deeper into the nonlinear regime. In particular, one can infer 
how the Zel'dovich approximation is naturally invoked as the 1st-order 
step (top left panel). 

\bigskip
\section{\rm{\Large THE SHAPING FORCE OF GRAVITY: ... }}
\begin{flushright}
{\rm{\Large Pattern Formation and Anisotropic Collapse}}
\end{flushright}
A characteristic aspect of the gravitational formation process is its tendency 
to develop via stages in which the cosmic matter distribution settles 
into striking anisotropic patterns. Seeking to identify the cause behind this  
shaping tendency in the evolution of cosmic structure readily leads to the generic 
anisotropic nature of the gravitational force field induced by a  
inhomogeneous and essentially stochastic matter distribution as the ultimate culprit. 

Anisotropic tidal effects are intrinsic to any cosmological scenario involving 
density deviations from the globally uniform Friedmann-Robertson-Walker 
Universe. This may be readily appreciated from considering the force field within 
a limited and bounded region of space, a ``patch'' (Bond \& Myers 1996a). 
The spatial variation of the implied force field translates into 
non-zero off-diagonal elements in the tidal force tensor, so that the gravitational 
acceleration ${\bf g}({\bf x})$ with respect to a position ${\bf x}_c$ can, 
to first order, be written as
\begin{eqnarray}
g_{i}({\bf x},t)&\ =\ &g_{i}({\bf x}_c,t)+\nonumber\\
&\ &a(t)\,\sum_{j=1}^3\,\left\{{\displaystyle 1 
\over \displaystyle 3a}(\nabla \cdot {\bf g})({\bf x}_c,t) \,\delta_{ij} - 
T_{ij}\right\} (x_j - x_{c,j})\ ,
\end{eqnarray}
\noindent where $a(t)$ is the expansion factor of the universe and $\delta_{ij}$ is the 
Kronecker delta. Hence, in addition to the zeroth-order bulk acceleration 
${\bf g}({\bf x}_c)= -\nabla \phi/a$ within a bounded cosmic region (``patch''), the 
local acceleration is modified by two terms. The divergence term 
$\nabla \cdot {\bf g}$ represents the radial and thus spherically symmetric 
infall or outflow around ${\bf x}_c$. It is the principal agent in determining the 
collapse or expansion of the regional matter distribution, yet by virtue of its 
spherically symmetric character preserves its initial shape. The latter is 
affected by the quadrupole tidal tensor $T_{ij}$, the trace-free part of 
$ -\partial g_{i} / \partial r_j=\partial^2 \phi/\partial r_i \partial 
r_j$ (${\bf r}(t)=a(t){\bf x}(t)$ being the physical position), 
\begin{eqnarray}
T_{ij}({\bf x}) &\ \equiv\ & -{\displaystyle 1 \over \displaystyle 2a} 
\left\{ {\displaystyle \partial g_{i} \over \displaystyle \partial x_i} +
{\displaystyle \partial g_{j} \over \displaystyle \partial x_j}\right\} +
{\displaystyle 1 \over \displaystyle 3a} (\nabla \cdot {\bf g}) 
\,\delta_{ij}\nonumber\\
&\ =\ & \ \ \ {\displaystyle 1 \over \displaystyle a^2} \left\{
{\displaystyle \partial^2 \phi \over \displaystyle \partial x_i \partial x_j} - 
{\displaystyle 1 \over \displaystyle 3} \nabla^2 \phi \ \delta_{ij}\right\}\ .
\end{eqnarray}
\noindent Usually higher order contributions to the tidal field are 
neglected, and attention is restricted to the quadrupolar component $T_{ij}$. The 
decisive role of the tidal field in determining the fate and final shape of an 
evolving structure has been most consistently and elaborately adressed in a seminal 
series of papers by Bond \& Meyers (1996a,b,c). They developed an elaborate ``peak-patch'' 
formalism in particular focussing on cosmic regions centered on and near peaks 
in the primordial density field. In line with their treatment, and in order 
to appreciate and exploit the impact of a force field whose configuration 
defines a nontrivial and complex spatial pattern, it proves beneficial to introduce 
a (asymptotic) distinction between ``internal'' and ``external'' gravitational fields. 
While such a bimodal partition is rather awkward and artificial in the case of a random 
density fluctuation field -- with density features embedded within a hierarchy of 
spatially larger encompassing density perturbations -- and would conform to reality 
solely in the case of clearly distinct and spatially disjunct objects, it emphasizes 
fundamentally different modes in which the gravitational impact is operating and 
effected. The ``internal'' field concerns the gravitational impact of the internal 
matter content on an emerging structure, the ``external'' that of the evolving 
external density distribution. 
\begin{figure}[t]
\vskip -0.2truecm
\centering\mbox{\hskip -0.7truecm\psfig{figure=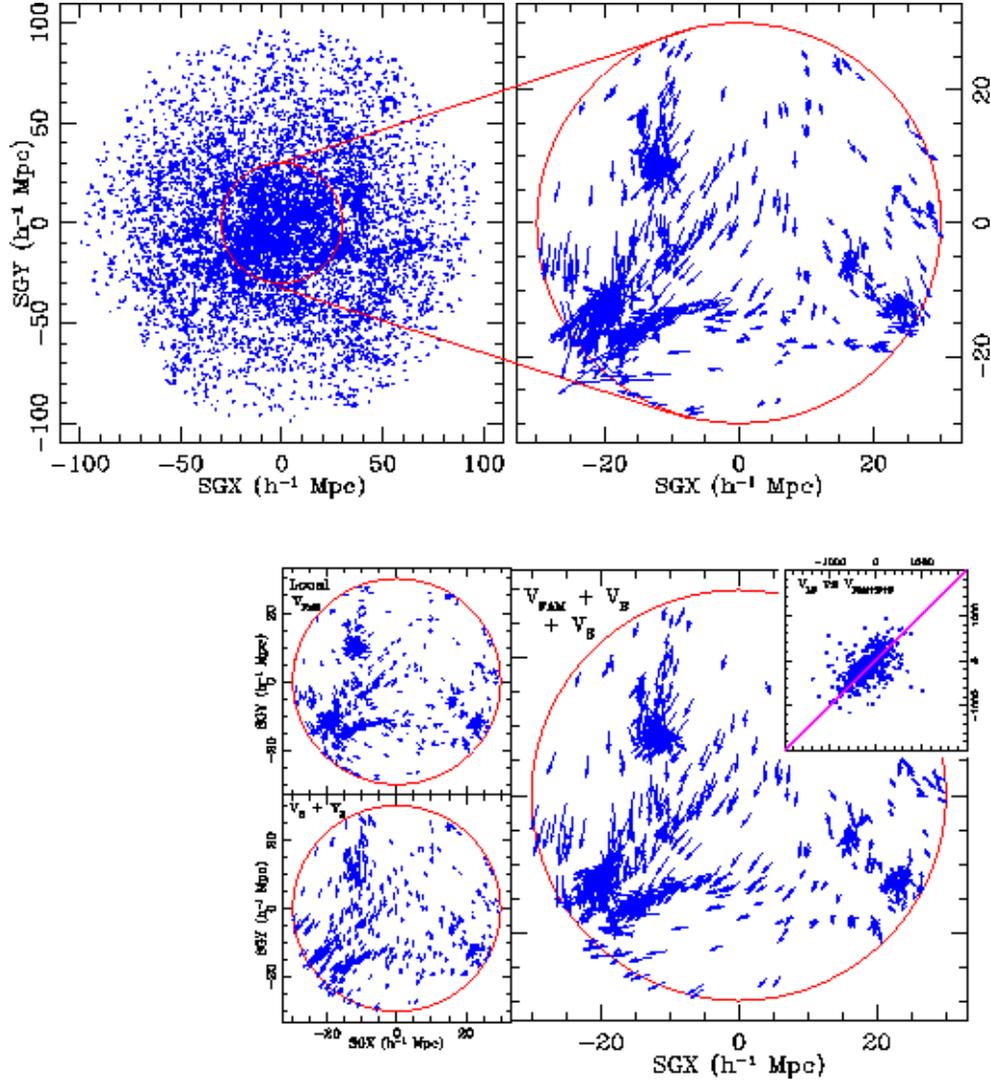,width=13.0cm}}
\vskip -0.0truecm
\caption{Left: The analyzed ``mock'' Local Supercluster + PSCz galaxy distribution, with 
the sample of ``mock'' measured Local Supercluster peculiar velocities. Right: 
FAM analysis of local velocity field:  restricted to local cosmic volume (top left). 
Including full PSCz volume: external dipole and quadrupole components (D-Q, bottom 
left). Total of locally induced velocities + D-Q external contribution (right frame): 
agreement with full ``mock'' velocities illustrated by scatter plot (insert). 
From Romano-D\'{\i}az, Branchini \& van de Weygaert 2002.}
\vskip -0.2truecm
\end{figure}

\subsection{{\it Tides in our Cosmic Backyard}}
That tidal influences are indeed at work on cosmologically relevant scales 
has been demonstrated on the basis of what may arguably be considered 
their most straightforward impact. The spatial variations in 
gravitational force within a particular cosmic region induce a 
peculiar velocity field exhibiting corresponding spatial variations, 
manifesting itself as velocity shear. On the basis of a careful analysis of 
peculiar velocities of galaxies in the Local Supercluster, Lilje, Yahil \& 
Jones (1986) estimated that at our cosmic location the velocity shear had a value 
in the order of $\sim 200\,\hbox{km/s}$ with respect to the Virgo Cluster. In a bold leap 
of imagination they argued that the source of this shear had to be a considerable mass 
concentration at a distance of $\sim 3$ times the distance to the Virgo Cluster, soon 
thereafter confirmed when Lynden-Bell et al. (1988). The latter managed to uncover a 
local velocity flow towards a ``Great Attractor'' from a painstaking analysis of a 
sample of peculiar velocities of galaxies within a radius of $\sim 60h^{-1}\hbox{Mpc}$ 
around the Local Group. Huge efforts have since been invested into improving the 
quality of both the initially spatially rather limited and coarsely sampled  
local cosmic velocity field and the ever more sophisticated tools with which one 
manages to extract cosmologically meaningful information. Recently, 
this allowed Hoffman et al. (2001) to produce an evocative reconstruction of the 
tidally induced component of the cosmic velocity field out to a distance of 
$\sim 60h^{-1}\hbox{Mpc}$. The prominent dynamical role of this local field can 
be particularly appreciated from a closer assessment of its inferred configuration, 
which appears to lead to the tentative indication of the Shapley Concentration 
marking an immense matter concentration of $\approx 1-3 \times 10^{17} \hbox{M}_{\odot}$ 
which apparently stretches out its dynamical clout out into our own cosmic realm (see 
section 2.4). Other studies involving higher quality velocity measurements, yet a less sophisticated 
dynamic model of the Local Universe, do not yet endorse such far-reaching conclusions 
(Tonry et al. 2000). However, there are strong overriding theoretical arguments for 
significant tidal influences stretching over scales in the order of 
$\approx 100h^{-1}\hbox{Mpc}$. For a set of viable structure formation scenarios, 
Romano-D\'{\i}az, Branchini \& van de Weygaert (2002) assessed in how far 
the velocities in a local region of space would be subject to significant 
and noticeable tidal forces, orginating from matter concentrations enclosed within 
a surrounding region modelled after the sampling volume of the PSCz galaxy redshift 
sample (see Fig. 18). Taking into account subtle nonlinear effects in the local velocity field 
through the application of an advanced Least Action Principle reconstruction technique 
(Peebles 1989, Nusser \& Branchini 2000) they managed to disentangle the ``internal'' 
influence in the local velocity from the that of ``external'' tidal influences 
induced by the outer mass concentrations, thus showing that the contribution 
by an external dipolar as well as a quadrupolar gravity component are necessary, 
yet sufficient, to account for the full velocity field in our local cosmic 
neighbourhood. 

\subsection{The Homogeneous Ellipsoidal Model}
As so often, understanding the essence of a physical phenomenon -- here the 
anisotropic patterns in the matter distribution, walls and filaments -- is obtained most 
readily and lucidly through the assessment an asymptotic idealization of 
more realistic and complex configurations. 

The bare essence of the driving mechanism behind the formation of 
cosmic walls and filaments is possibly best appreciated in terms of the 
dynamical evolution of homogenous ellipsoidal overdensities. In particular,   
the early work by Icke (1972, 1973) elucidated transparently the crucial 
characteristics of their development and morphology. On the basis of 
an assessment of the collapse of homogeneous ellipsoids in an expanding FRW 
background Universe -- following the formalism of Lynden-Bell (1964) and Lin, 
Mestel \& Shu (1965) -- he came to the conclusion that flattened and elongated 
geometries of large scale features in the Universe should be the norm. This description 
intrinsically involves the self-amplifying effect of a collapsing and 
progressively flattening isolated ellipsoidal overdensity. Quintessential 
was Icke's observation that gravitational instability not only involves the runaway 
gravitational collapse of any cosmic overdensity, but that it has the additional 
basic attribute of {\it inevitably amplifying any slight initial asphericity 
during the collapse}. In order to appreciate the dynamics behind the process, and 
be able to assess its action within more complex situations, it is insightful 
to focus in some detail on the evolution of homogeneous ellipsoids. 

\subsubsection{{\it Homogeneous Ellipsoidal Model: the Formalism}}
The ellipsoidal approximation involves an ellipsoidal region with a triaxially 
symmetric geometry, described in terms of its principal axes ${c}_1$, ${c}_2$ and ${c}_3$. 
The matter density in the interior of the 
ellipsoid has a constant value of $\rho^{ell}$, and the ellipsoid is embedded in a 
background with a density $\rho^b$. While the basic formalism assumes 
an isolated ellipsoid, we seek to extend this to a more generic configuration in 
which the ellipsoidal object is subjected to an external tidal field induced 
by matter fluctuations beyond its immediate neighbourhood. It should be noted 
that in principle such a configuration is a contrived one, the existence of 
an external field implying a {\it contradictio in terminis} with respect to the 
assumption of a homogeneous background. The intention therefore is to 
use this as a description reasonably approximating and illuminating relevant 
effects.  In the presence of an this external potential contribution, the 
total gravitational potential $\Phi^{(tot)}({\bf r})$ in the interior of a homogeneous 
ellipsoid is given by
\begin{equation}
\Phi^{(tot)}({\bf r})\ =\ \Phi_b({\bf r})\ + \Phi^{(int,ell)}({\bf r})\ +  
\Phi^{(ext)}({\bf r})\ ,
\end{equation}
\noindent in which $(r_1,r_2,r_3)$ figure as the coordinates in an arbitrary Cartesian 
coordinate system. In this, we have decomposed the total potential $\Phi^{(tot)}({\bf r})$ 
into three separate (quadratic) contributions,
\begin{itemize}
\item{} The potential contribution of the homogeneous background with universal 
density ${\rho^b}(t)$, 
\begin{equation}
\Phi_b({\bf r})\ =\ {2 \over 3}\pi G {\rho^b} \ (r_1^2+r_2^2+r_3^2)\ .
\end{equation}
\item{} The interior potential $\Phi^{(int,ell)}({\bf r})$ of the ellipsoidal
entity, superimposed onto the homogeneous background,
\begin{eqnarray}
\Phi^{(int,ell)}({\bf r})&\ =\ & {\displaystyle 1 \over \displaystyle 2}\,\sum_{m,n} 
\Phi^{(int,ell)}_{mn} r_m r_n \ \nonumber\\
&\ =\ &{2 \over 3}\pi G ({\rho^{ell}}-{\rho^b}) \ (r_1^2+r_2^2+r_3^2)\ + \ 
{\displaystyle 1 \over \displaystyle 2}\,\sum_{m,n} T^{(int)}_{mn} r_m r_n \ ,\nonumber\\
\ \ 
\end{eqnarray}
with $T^{(int)}_{mn}$ the elements of the traceless internal tidal shear tensor,   
\begin{equation}
T^{(int)}_{mn}\ \equiv \ {\displaystyle \partial^2 \Phi^{(int,ell)} \over 
\displaystyle \partial r_m \partial r_n}-{\displaystyle 1 \over \displaystyle
3} \nabla^2 \Phi^{(int,ell)}\ \delta_{mn}\ .
\end{equation}
\item{} The externally imposed gravitational potential $\Phi^{(ext)}$. We assume that 
the external tidal field not to vary greatly over the expanse of the ellipsoid, so that we 
can presume the tidal tensor elements to remain constant within the ellipsoidal region (cf. 
Dubinski \& Carlberg 1991). In this approximation, the elements $T_{mn}^{(ext)}$ of the 
external tidal tensor correspond to the quadrupole components of the external potential 
field, with the latter being a quadratic function of the (proper) coordinates 
${\bf r}=(r_1,r_2,r_3)$:
\begin{equation}
\Phi^{(ext)}({\bf r})\ =\ {\displaystyle 1 \over \displaystyle 2}\,\sum_{m,n} T^{(ext)}_{mn} 
r_m r_n \ .
\end{equation}
\noindent with the components $T^{(ext)}_{mn}$ of the external tidal shear tensor, 
\begin{equation}
T^{(ext)}_{mn}(t)\ \equiv\  {\displaystyle \partial^2 \Phi^{(ext)} \over 
\displaystyle \partial r_m \partial r_n}\ ,  
\end{equation}
\noindent which by default, because of its nature, is a traceless tensor. 
Note that the quadratic form of the external potential is a necessary condition for 
the treatment to remain selfconsistent in terms of the ellipsoidal formalism. 
\end{itemize}
\noindent As the ellipsoidal formalism does not include any self-consistent external potential, 
we have to impose it ourselves in an artificial way, including a specified time evolution of the 
tidal tensor components (see discussion in section on external tidal action, eqn. 49 to 51),
In essence, we impose an artificial external tidal field based on 
the assumption that the background in the immediate vicinity of the ellipsoid 
remains homogenous and that the external structures engendering the tidal field are located 
out at distances sufficiently far away from the ellipsoidal entity. This warrants 
the validity of the approximation by the quadratic equations, and assures that these 
external entities themselves are untouched themselves by the ensuing evolution of the 
object. 

The quadratic expression for the internal ellipsoidal potential contribution 
$\Phi^{(int,ell)}$ can be cast into a simplified form by choosing 
a convenient coordinate system in which the coordinate axes coincide with 
the principal axes of the ellipsoid. In this case, the expression for the 
potential of an ellipsoid with effective density $(\rho^{ell}-{\rho^b})$ 
and axes ${c}_1$, ${c}_2$ and ${c}_3$ reduces to (Lyttleton 1953) 
\begin{equation}
\Phi^{(int,ell)}({\bf r})=\pi G \ (\rho^{ell}-{\rho^b}) \,\sum_m \alpha_m
r_m^2\ ,
\end{equation}
\noindent where the coefficients $\alpha_m$ are determined by the shape of the
ellipsoid, 
\begin{equation}
\alpha_m = {c}_1 {c}_2 {c}_3 \int^\infty_0 ({c}_m^2 + \lambda )^{-1}
\prod_{n=1}^3 {1 \over \sqrt{ {c}_n^2 + \lambda} } {\rm d} \lambda\ .
\end{equation}
\noindent As a consequence of the Poisson equation the $\alpha_m$'s obey
the constraint: $\sum_{m=1}^3 \alpha_m=2$. Also, we see that the 
components of the internal tidal shear tensor are given by, 
\begin{equation}
T^{(int)}_{mn}\ =\ \ 2\pi G (\rho^{ell}-{\rho^b})\,\left(\alpha_m-{2 \over 3}\right)\ 
\delta_{mn}\ .
\end{equation}
\noindent It is easy to appreciate that in the case of a spherical perturbation all 
three $\alpha_m$'s are equal to ${2/3}$, reproducing the well-known fact that in such case 
the internal tidal tensor contributions need to vanish. 

From the quadratic nature of the total potential, the acceleration of a mass at 
a position ${\bf r}$ inside the ellipsoid, $-\nabla \Phi({\bf r})$, is a linear 
function of ${\bf r}$, 
\begin{equation}
{\displaystyle d^2 r_m \over \displaystyle d t^2} \ =\ 
-{\displaystyle 4\pi \over \displaystyle 3} G {\rho^b} r_m(t)\ -\ 
\sum_n \Phi^{(int,ell)}_{mn} r_n(t) \ -\ \sum_n T^{(ext)}_{mn} r_n(t)\ . 
\end{equation}
\noindent This leads to a linear relation between the location ${\bf r}(t)=(r_1(t),$ 
$r_2(t),r_3(t))$ of a mass element at time $t$ and 
its initial (proper) position ${\bf r}_i=(r_{1,i},r_{2,i},r_{3,i})$, 
\begin{equation}
r_m(t)\ =\ \sum_k R_{mk}(t) r_{k,i}
\end{equation}
\noindent in which the matrix $R_{mn}(t)$ is a spatially uniform matrix, solely 
dependent on time. Notice that the initial matrix $R_{mn}(t)$ is a diagonal matrix,
$R_{mn}(t_i)=R_m(t_i) \delta_{mn}$. By combining Eqn. (38) and 
Eqn. (39), the evolution of the matrix elements $R_{mk}$ is found to be described by
\begin{equation}
{\displaystyle d^2 R_{mk} \over \displaystyle d t^2} \ =\  
- {4 \pi \over 3} \pi G R_{mk}\ -\ \sum_n \Phi^{(int,ell)}_{mn} R_{nk}\ -\ 
\sum_n T_{mn} R_{nk}\ 
\end{equation}
\noindent This also implies that similar points in concentric ellipsoidal shells
behave in the same way, while the ellipsoid will remain homogeneous.

To appreciate the ramifications for the shape of an evolving 
ellipsoidal object, we exclude the torqueing and angular momentum inducing effects 
of the external tidal field. To this end, we make the simplifying assumption of the 
principal axes of the external tidal tensor to be aligned along the principal axes 
of the inertia tensor $I_{ij}$ of the ellipsoid. Of course, in realistic situations 
we would not expect to find such a perfect alignment between the tidal and the 
inertia tensors. Yet, there is a significant correlation between orientation of tidal 
field and orientation of the principal axes of a peak/dip in a Gaussian random field. 
In particular, it implies a tendency to align the strongest tidal field component along the 
smallest axis (Van de Weygaert \& Bertschinger 1996). 

The principal axis alignment implies -- when defining the coordinate system by the 
principal axes of the ellipsoid -- that all off-diagonal (external) tidal tensor elements 
vanish, $T^{(ext)}_{mn}=T^{(ext)}_{mm}\delta_{mn}$. This implies $\sum_n T^{(ext)}_{mn} R_{nk} = 
T^{(ext)}_{mm} R_{mk}$. The resulting equation for the evolution of the matrix elements
$R_{mk}$ is then described by the second order diffential equation, 
\begin{equation}
{\displaystyle d^2 R_{mk} \over \displaystyle d t^2} \ =\  
- 2 \pi G \, \left[\alpha_m \rho^{ell} + ({2 \over 3}-\alpha_m) \ {\rho^b} \ \right] 
R_{mk}\ -\ T^{(ext)}_{mm} R_{mk}\ .
\end{equation}
\noindent Evidently, the coupling between the different components $R_{mk}$ vanishes. 
As they are initially equal to zero this implies the non-diagonal
elements $R_{mk}(t)$ to remain so, $R_{mk}(t)=R_m(t) \delta_{mn}$. At time t, a
mass element initially at (proper) position ${\bf r}_i=(r_{1,i},r_{2,i},r_{3,i})$
has therefore moved to position ${\bf r}(t)=(r_1(t),r_2(t),r_3(t))=
(R_1(t)r_{1,i},R_2(t)r_{2,i},R_3(t)r_{3,i})$, in which the 
functions $R_m(t)$ evolve according to 
\begin{equation}
{\displaystyle d^2 R_{m} \over \displaystyle d t^2} \ =\  
- 2 \pi G \, \left[\alpha_m \rho^{ell} + ({2 \over 3}-\alpha_m)\ {\rho^b}\,\right] R_m\ -\ T^{(ext)}_{mm} R_{m}\ .
\end{equation}
\noindent The evolution of the ellipsoid is thus fully encapsulated in terms 
of the the functions $R_1(t)$, $R_2(t)$ and $R_3(t)$, which in essence 
should be seen as scale factors of the principal axes of the ellipsoid, 
\begin{equation}
c_m(t)\ =\ R_m(t) c_{m,i}\ .
\end{equation}
\noindent For any initial configuration of a homogeneous ellipsoid embedded 
in a FRW background Universe with current cosmic density parameter $\Omega_{\circ}$ and 
Hubble parameter $H_{\circ}$, specified by its initial characteristics at cosmic expansion 
factor $a_i$, 
\begin{itemize}
\item{} The initial principal axes, $(c_1, c_2,c_3)$.
\item{} The initial ellipsoidal density $\rho^{ell}_i$, in terms of  
density contrast $\delta_i$  
\begin{equation}
\delta_i \ \equiv \ {\displaystyle \rho^{ell}_i-{\rho^b}_i \over \displaystyle {\rho^b}_i}\ ,
\end{equation}
\noindent with respect to the background Universe with initial cosmic density 
\begin{equation}
{\rho^b}_i\ = \ {\displaystyle 3 \over \displaystyle 2}\ \Omega_{\circ} H_{\circ} 
\ \left(\displaystyle a_{\circ} \over \displaystyle a_i\right)^3\
\end{equation}
\end{itemize}
\noindent the evolution of the scale factors $R_m(t)$ can be fully recovered once  
the boundary conditions have been set, i.e. 
\begin{itemize}
\item{} The initial scale factors, 
\begin{equation}
R_m(t_i)\ = \ 1 \ 
\end{equation}
\noindent for $(m=1,2,3)$.
\item{} The initial velocity perturbation,
\begin{eqnarray}
v_m(t_i)\ =\ (dR_m/dt) \ r_{m,i}&\ =\ &v_{Hubble,m}(t_i)\ +\ v_{pec,m}(t_i)\nonumber\\
&\ =\ &H_i \ R_m r_{m,i}\ +\ v_{pec,m}(t_i)\ ,\nonumber\\
\end{eqnarray} 
\noindent for $(m=1,2,3)$, which when choosing to follow the growing mode solution of 
linear perturbation theory (Peebles 1980) is given by 
\begin{eqnarray}
v_{pec,m}(t_i)&\ =\ &{\displaystyle 2  f(\Omega_i) \over \displaystyle 3 H_i \Omega_i}
g_{pec,m}(t_i)\nonumber\\
&\ =\ &- {1 \over 2} H_i f(\Omega_i)\ \left[ \alpha_{m,i} \delta_i + {\displaystyle  
4 T^{(ext)}_{mm,\circ} \over \displaystyle 3 \Omega_0 H_0^2}\ D_i\right]\ r_{m,i}\ .
\nonumber\\
\end{eqnarray}
\end{itemize}
\noindent In the latter expression, $f(\Omega_i)$ is the corresponding linear velocity growth 
factor (Peebles 1980). The $\alpha_{m,i}$ are the values of the ellipsoidal shape factors 
$\alpha_m$ at expansion factor, which are yielded after evaluating the integral expression 
(36) for the specified initial axis ratios. For this expression, we specified the (artificially 
imposed) time development of the external tidal tensor through a growth factor 
$D(t)$, $E_{mm}/ \Omega H^2 \propto D(t)$, relating the current external tidal 
field $T^{(ext)}_{mm,\circ}$ to its initial value.  
\subsubsection{{\it Homogeneous Ellipsoidal Model:  the Approximation}}
Evidently, we have to be aware of the serious limitations of the ellipsoidal model. 
It grossly oversimplifies in disregarding important aspects like the presence of 
substructure in and the immediate vicinity of peaks and dips in the primordial 
density field, the sites it typically deals with. Even more serious is its neglect of 
any external influence, whether the secondary infall or ``collision'' with surrounding 
matter or the role of nonlocal tidal field engendered by the external mass distribution. 
\begin{figure}[b]
\centering\mbox{\hskip -0.0truecm\psfig{figure=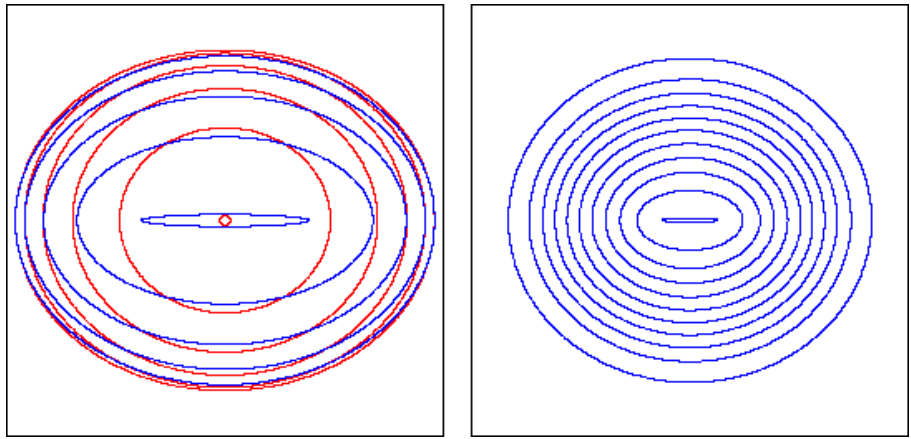,width=12.cm}}
\vskip -0.0cm
\caption{The evolution of an overdense homogeneous ellipsoid, with initial axis 
ratio $a_1:a_2:a_3=1.0:0.9:0.9$, embedded in an Einstein-de-Sitter background Universe. 
The two frames show a time sequel of the ellipsoidal configurations attained by the object, 
starting from a near-spherical shape, initially trailing the global cosmic expansion, 
and after reaching a maximum expansion turning around and proceeding inexorably towards 
ultimate collapse as a highly elongated ellipsoid. Left: the evolution depicted in 
physical coordinates. Red contours represent the stages of expansion, blue those of 
the subsequent collapse after turn-around. Right: the evolution of the same object in 
comoving coordinates, a monologous procession through ever more compact and more 
elongated configurations.} 
\end{figure}
For overdensities it represents a reasonable approximation for moderately evolved features 
like a Megaparsec (proto)supercluster, but it will be seriously flawed in the 
case of highly collapsed objects like galaxies and even clusters of galaxies. 
Nonethelss, the concept of homogeneous ellipsoids has proven to be 
particularly useful when seeking to develop approximate yet advanced descriptions 
of the distribution of virialized cosmological objects (Bond \& Myers 1996a,b,c, 
and Sheth, Mo \& Tormen 2001). 

Interestingly, in many respects the homogeneous model is a better approximation 
for underdense regions than it is for overdense ones. Overdense regions contract into 
more compact and hence steeper density peaks, so that the area in which the 
ellipsoidal model represents a reasonable approximation will continuously shrink. 
On the other hand, while voids expand and get drained, the density fields in 
the central region of the (proto)void will flatten out, so that the voids 
develop into regions of a nearly uniform density and the region of validity of 
the approximation grows accordingly. This can be readily appreciated from 
the conceptually simpler spherical model approximation. More significant,   
it was also demonstrated in the complex circumstances of voids embedded in a general 
cosmic density field for a set of N-body structure formation simulations 
(Van de Weygaert \& van Kampen 1993, see Fig. 31).  A direct repercussion 
of the flattening out of the voids is the velocity field inside them. With the exception 
of the ridges surrounding them, the quadratic potential approximation will be valid 
for most of the void's interior and therefore be characterized by a velocity field of 
excess Hubble expansion. The systematic study by Van de Weygaert \& Van Kampen (1993) 
indicated how the void velocity fields in general will evolve towards a state in which 
they become genuine ``Superhubble Bubbles'' (see section 4.7). 
Evidently, the ellipsoidal  approximation will only be useful for the interior of voids. 
At the outer fringes  
the neglect of the role of surrounding material will be rendered invalid, as 
the sweeping up of matter, the formation and subsequent induced `self-interaction' 
and the encounter with surrounding structures will become domineering effects. 
\begin{figure}[t]
\centering\mbox{\hskip -0.3truecm\psfig{figure=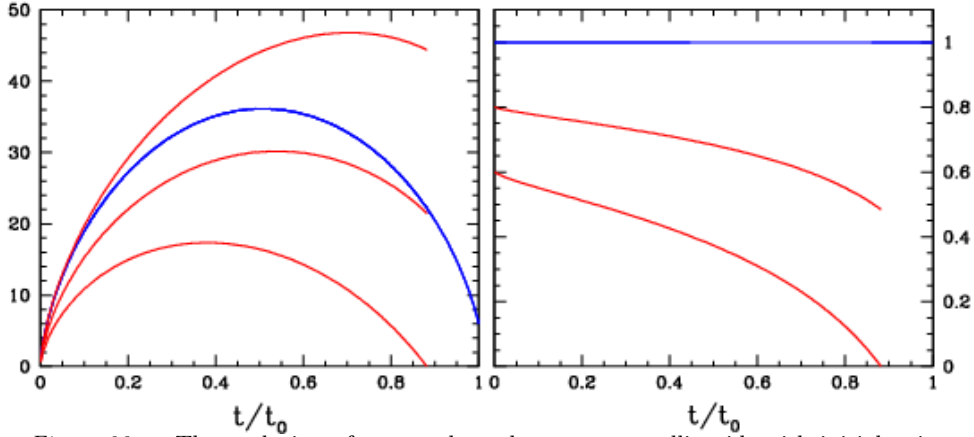,width=12.8cm}}
\vskip -0.5cm
\caption{The evolution of an overdense homogeneous ellipsoid, with initial axis 
ratio $a_1:a_2:a_3=1.0:0.8:0.6$, in an Einstein-de-Sitter background Universe. Left: 
expansion factors for each individual axis; right: axis ratios $a_2/a_1$ and 
$a_3/a_1$. The ellipsoid axes are depicted as red curves. For comparison, in blue, 
the evolution of an equivalent homogenous spherical overdensity.} 
\vskip -0.2cm
\end{figure}
\subsubsection{{\it Anisotropic Collapse:  Shaping Force of Internal Tides}}
The internal gravity field is evolving along with the evolution of the object 
itself, which renders it rather straightforward to account for it in a fully 
selfconsistent fashion. When focussing specifically on the issue of shape 
evolution, the obvious archetype for the intrinsic primordial flattening of an 
overdensity through the action of the associated ``internal tidal'' field is an 
evolving homogeneous ellipsoidal overdensity (Fig. 19). It represents the situation in 
which the geometry of the collapsing object is fully coupled to the anisotropy of the 
force field, its runaway dynamic evolution resulting from the feedback interaction 
between shape and force field of an object. 

The basic agent behind the continuing amplification of initial asphericities of a 
collapsing objects is the anisotropy in the corresponding gravitational 
force field. The gravitational force along the shortest axis of an 
ellipsoid will be stronger than along the longest axis. Along the short 
axis it will ``feel'' nothing but the increased matter concentration 
within the enclosing sphere, while along the longest axis the effective 
force will be diluted by the lower density outside the object. It is 
self-evident that this effect becomes more pronounced as the object 
develops an increasingly flattened or elongated geometry. 

A telling illustration of this behaviour can be observed in figures 10. It shows the 
evolution of a slightly overdense ellipsoid, initially almost spherical with 
(axisymmetric) axis ratios $a_1:a_2:a_3=1:0.0.9:0.9$, embedded in a background 
Einstein-de Sitter Universe. Depicted is a time sequel of 
attained geometric configurations (in the $x_1-x_2$ plane) during its development from 
the initial small near-spherical state through its turn-around phase, after which 
physical collapse sets in, towards the inexorable fate of collapse as a strongly 
elongated spindle-like object. The evolution of the ellipsoid is computed by numerical 
evaluation of the corresponding second-order differential equations for homogeneous ellipsoids 
(Icke 1973). It is interesting to observe the contrast between its evolution in physical 
coordinates (lefthand frame) and that in comoving space (righthand frame). In physical 
space we first note the ellipsoid expanding, 
trailing the global expanding Universe (red contour configurations). While the 
initial expansion slows down more and more, the ellipsoid finally reaches a maximum 
volume at its ``turnaround'' time, after which it sets in an inexorable collapse 
(blue configurations). Along with this process, during both its physical expansion 
as well as contraction, we observe the ellipsoid to assume a continuously stronger 
elongated geometry, ultimately collapsing as a spindle-like object. For contrast, the 
same process is depicted in comoving space in the righthand frame. By removing the 
effect of the expanding background Universe we obtain a better appreciation of the 
shape development of the object. Evidently, in comoving space it experiences a monologous 
contraction. Starting from its large Lagrangian volume it finally develops into 
the compact and fully collapsed object at the centre of the graph. It may be superfluous 
to note that by that stage, the simple formalism used to describe the monologous 
evolution of a homogeneous ellipsoid is rendered invalid as shell interactions and 
crossings can no longer be neglected. The latter find a partial expression in an exchange of 
energy as the process of virialization sets in. 

Quantitatively, the systematic elongating tendency of such overdense ellipsoidal objects 
may be clearly appreciated from an examination of the the expansion and subsequent 
contraction of each of the three axes of a slightly overdense triaxial ellipsoid. Figure 
20 shows a representative example, depicting the axis evolution for an ellipsoid with 
initial axis ratio $a_1:a_2:a_3=1:0.8:0.6$ (red curves). As in figure 19, the ellipsoid 
is embedded in an Einstein-de-Sitter background Universe. For contrast, the evolution 
of an initially equivalent homogeneous spherical overdensity has been added in the same 
figure (blue curve). After an initial phase of moderate expansion along all three axes we 
observe the successive turn-around along all 3 directions (lefthand frame), with the 
shortest axis being the first to turn around and with the longest one only following 
as last one after first having pursued its initial expansion. From the righthand frame, we 
can clearly see that its development is accompanied by a drastic and continuous decrease 
of both axis ratios. Even while the ellipsoid will collapse along all three axes,  it 
will ultimately do so as a highly flattened object !

It may be evident, that this secular increase of aspherical perturbations provides an 
explanation for the {\it pancake-like}, and later {\it filamentary}, appearance of 
large scale structures. 

\subsection{Anisotropic Collapse:}
\begin{flushright}
{\bf{\large External Tidal Action}}
\vskip -0.2truecm
\end{flushright}
The internal flattening is augmented, and regularly dominated, by the anisotropy of 
the gravitational force field induced by the surrounding external inhomogenous matter 
distribution, the ``external tidal '' forces. The role of the internal tidal field, the 
expression of internally anisotropic shape and/or inhomogeneous matter distribution, is 
primarily that of changing -- in situ -- the configuration and appearance of an object. 
The external tidal influence proceeds almost completely independent of the emerging 
object itself, with the possible exception of a mostly minor backreaction on the 
surrounding matter configurations. The impact of these external forces is considerably 
more versatile than, and regularly dominating over, the internal anisotropic force field.  
They play an exclusive role in spinning up a collapsing clump, making them the prime agent 
for the rotation of galaxies. In addition, external tides have a major influence on, and may 
even determine, the very outcome and fate of the gravitational collapse itself. In 
particular, it may hold a decisive influence on aspects such as collapse timescale and 
final object shape. A fully selfconsistent treatment of these effects can in principle 
only be achieved by following the overall matter distribution throughout the surrounding 
realms of the Universe. Often though, its major impact is restricted to that of the lowest 
two order terms of the force field, the dipolar and quadrupolar components.

\subsubsection{{\it External Tidal Action:  Composition and Development}}
It has proven most difficult to get a full understanding and appreciation of the extent 
of the role of external tidal forces. Partly this can be ascribed to the absence of a 
fully selfconsistent treatment, in particular with respect to the complex nature of 
their temporal evolution. The external tidal field at any one location is a mixture of 
tidal force contributions from a diversity of inhomogeneities, spanning a wide range of 
spatial scales and each evolving at their own rate. Remote and coherent large-scale 
structures may still evolve almost linearly, while neighbouring small-scale matter 
concentrations may already have reached highly nonlinear 
stages of collapse. Moreover, the composition of the different contributions will be a 
function of cosmic location. Objects embedded in highly dense regions will be 
more affected by the short-range nonlinear tidal contributions than the ones residing 
in more diluted regions, which will therefore find themselves more reacting to a 
moderately linearly evolving force field. The external influence originating from large 
scale density perturbation, still evolving linearly, will therefore also grow linearly 
according to 
\begin{eqnarray}
T^{(ext)}_{mn}(t)\ \propto\ D(t) \ \Omega(t) H^2\ \propto \ D(t)\ a(t)^{-3} &\ \propto\ & t^{-4/3} {\hskip 0.3cm} (\Omega_{\circ}=1)
\nonumber\\ 
&\ \propto\ & t^{-3} {\hskip 0.55cm} (\Omega_{\circ}\ll1)
\nonumber\\
\end{eqnarray}
\noindent with $D(t)$ the linear growth factor. The first line follows immediately from 
the well-known fact that in an $\Omega_{\circ}=1$ universe $D(t)=a(t)$, implying a tidal 
growth of $T^{(ext)}_{mn}(t) \ \propto\ a(t) H^2\ \propto\ 1/a(t)^2 \ \propto\ t^{-4/3}$. 

On the other hand, in a low-density Universe the growth of structure grinds to a halt 
at a redshift of $z \ \approx \ \left(1/\Omega_{\circ}-1\right)$. This implies a decline in 
tidal strength of $T^{(ext)}_{mn}(t) \ \propto 1/a(t)^3 \ \approx \ t^{-3}$. In essence, this is an 
expression of the structure being frozen in with the Hubble expansion, so that 
the accompanying gravitational forces -- induced by a non-evolving mass content -- 
diminish with the accompanying expansion of physical distances and the corresponding 
cosmic volume. As a telling example, we may refer to Fig. 27, where the two last rows 
of panels depict the practically unchanging cellular pattern in an $\Omega_{\circ}=0.3$ 
Universe. 

The time dependence of the tidal field will change once the inhomogeneities enter the
non-linear phase of their evolution, when the neighbouring fluctuations responsible for 
the major share of the tidal interactions have collapsed and virialized. It may be, in a 
fashion similar to that discussed for the ultimately frozen structures in a low-density 
Universes, that the cosmic entities recede from each other along with the general Hubble 
expansion of the Universe (Peebles 1969, Dubinski 1992). This gradual dilution of cosmic 
objects would lead to a more rapidly declining external tidal field, 
\begin{equation}
T^{(ext)}_{mn}(t) \ \propto\ 1/a^3 \ \propto\ t^{-2}.
\end{equation}
On the other hand, when the major share of the external tidal force is imparted by nearby 
small-scale highly nonlinear clumps, we may encounter a distinctly different situation. 
On these scales the hierarchical clustering process will usually still be in progress, 
so that the clumps will not move away from each other, but instead continue their 
congregation into ever more massive entities. One particular asymptotic example 
which may regarded as a reasonable approximation of the genuine situation is that of 
{\it ``stable clustering''} (see e.g. Jain 1997). In essence this analytically 
tractable situation presumes a conglomerate of nonlinear clumps to retain the same clustering 
configuration in physical coordinates as the Universe is expanding along. Hence, we see 
a continuing contraction in comoving space, leading to a comoving density that rises along 
with the global Hubble expansion according to $\propto a^3$. Indeed, several N-body 
experiments (see e.g. Efstathiou et al. 1988, Jing 2001) have verified that this indeed appears 
to be a proper representation of the situation pertaining at small highly nonlinear 
scales. The contribution to the total tidal field originating from such tightly   
clustered assemblies will consequently retain its strength, i.e.  
\begin{equation}
T^{(ext)}_{mn}(t)\ = \ \hbox{\rm constant}\ .
\end{equation}
Given their considerably different time dependences, the tidal influences from the 
various contributing inhomongeneities will represent continuously changing fractions  
of the full tidal force at any one location. Their relative contributions will be 
a continuously changing function of time. We may expect the large-scale linear 
perturbations to dominate in the early stages of evolution. This will change as 
nearby small-scale clumps have reached full fruition and the process of continuing 
hierarchical evolution has established itself as a prominent process on relevant scales.  
The linearly evolving contributions will gradually become superseded by the growing 
weight of the nonlinear entities. 

Moreover, the role of tidal fields is not solely a matter of the intrinsic 
evolution of the external tidal forces. Their impact as much depends on the 
configuration of the object over which they are exerted and the precise timing 
of the external forces. In early stages the external forces may still represent the 
dominant dynamical influence. However, as the object itself collapses and reaches 
a highly nonlinear phase of evolution, it largely decouples from the background and 
pursues its own lifetrack. Most external effects should then have made a detectable 
imprint. Thus, being the dominant contribution during the pristine early stages, the linear 
large-scale contributions may provide the only relevant external influence during 
an objects emergence out of the primordial density field. In the case of angular 
momentum generation, for instance, most studies indeed seem to indicate that most of the 
angular momentum should have been imparted in the early linear phase, before 
the object turns around and starts to contract. On the other hand, new results indicate 
that the nonlinear ellipsoidal collapse in the later phases may still yield a 
noticeable addition to the final angular momentum. 

In all, it will be a daunting task, possibly not even feasible, to find globally 
viable prescriptions and descriptions for the influence of external tidal onto the 
final outcome, configuration and properties of emerging structures in the 
Universe. Whether the ``internal'' or ``external'' influence is dominating has not 
been systematically answered yet, and will most likely depend to a considerable 
extent on cosmological scenario and the related coherence scale in the cosmic 
density field. Nonetheless, on the basis of currently available work we may conclude 
that to no extent it is possible to neglect the impact of external tidal fields. This was 
clearly stated by Bond \& Myers (1996a,b,c), who deemed the external tidal field of 
decisive importance for the final fate and morphology of collapsing patches in the 
cosmos, an observation on which they based their ``peak-patch'' formalism. Their arguments 
were confirmed by the succesfull improvement in predicted locations and properties of 
clusters of galaxies in a large cosmic volume. Additional support for this view 
got provided by the thorough study by Eisenstein \& Loeb (1995), who systematically adressed 
the dynamical evolution of a large number of different homogeneous ellipsoid objects   
immersed in a user-specified external tidal field. Their conclusion, possibly not unequivocally 
relevant for and applicable to a more contrived generic world, is that on average the 
external tidal field would not only set the rotation of emerging objects, but would also 
dominate over the internal force field in determining the final shape of these 
homogeneous ellipsoidal overdensities. 

\subsubsection{{\it External Tidal Action:  Tidal Torques and Spinning Galaxies}}
That external gravitational tidal forces, effected by the inhomogenous matter 
distribution in the external surroundings of an evolving cosmic structure, wield a decisive role 
in the cosmogony of galaxies was recognized since decades for one of the most crucial 
physical properties of galaxies. The tidal torques which external inhomogeneities exert 
onto collapsing objects are an exclusive instrument for imparting angular momentum. It is 
the mechanism to set dark matter haloes into a spinning mode. The resulting rotation of 
the emerging galaxy is amongst the most distinctive and discriminative properties 
of galaxies, tightly coupled to their overall morphology and structure. This is the case 
for the conspicuously dominant rotational motion of the stellar and gaseous 
components of spiral galaxies, as well as for the more mixed kinematics encountered in the 
elliptical galaxies which seem to be supported by random motions. 

Collapsing clumps of matter embedded in a global field of density fluctuations 
naturally acquire angular momentum through the winding action of the 
resulting tidal torques. Evidently, the internal gravity field is completely inept in 
providing any viable contribution to this important physical property, leaving the 
external tides as principal and exclusive agents. The torque ${\bf \tau}$ imparted by 
an external tidal field onto the material within a Lagrangian volume $V_L$ is
\begin{equation}
{\bf \tau} = \int_{V_L}\ \rho {\bf r} \ {\bf \times}\ {\bf \nabla} \Phi\ dV\,
\end{equation}
\noindent where $\Phi$ is the potential due to the external tidal field. To first order, 
$\Phi_,i \simeq T^{(ext)}_{ij} x_j$, so that the torque ${\bf \tau}$ can be seen to result from 
the coupling between tidal field $T_{ij}$ and inertial tensor $I_{kl}$ of the collapsing object, 
\begin{equation}
\tau_i\ =\ \epsilon_{ijk} T^{(ext)}_{ji} I_{kl}\ ,
\end{equation}
\noindent where $\epsilon_{ijk}$ is the Levi-Civita tensor. 

An important aspect of the tidal torque scenario is that nearly all angular momentum 
of an evolving overdensity is imparted during the early, linear phase of formation (see e.g 
Peebles 1969, Dubinski 1992). We may appreciate this to some extent by observing from the 
previous expression that -- for the case of an Einstein-de Sitter Universe -- as $I_{ij}$ 
grows as $t^{4/3}$ for a Lagrangian volume, and $T^{(ext)}_{kl}$ declines 
as $t^{-4/3}$ (see Eqn. 5), the tidal torque in the linear regime remains 
constant. Therefore, the acquired angular momentum will grow linearly in time, 
${\bf L}(t) \propto t$ (White 1984), which then would correspond to the epoch in which most 
angular momentum is generated.

Once an object has started to contract, self-gravity and noisy small-scale torques start 
to dominate over the initial coherent large-scale tidal torque. The final rotational state is 
therefore mostly in place by the time the object starts to collapse and finally 
virializes into a genuine galaxy. Even though various issues of contention remain, whether 
it concerns the fine-tuning issue of imparting a sufficient yet not overriding amount 
of angular momentum onto dissipatively collapsing baryonic galaxy disks embedded 
within dark matter haloes or the distorting effect of unceasing small-scale nonlinear 
effects such as the intermittent infall of a variety of matter clumps, the tidal 
torque mechanism has established itself as an essential aspect -- even tenet -- of 
the overall process of structure formation through gravitational instability. 

Tidal torqueing was suggested as an explanation for the origin of galactic rotation 
by Hoyle (1949). The idea was more thoroughly investigated by Peebles (1969) and 
Doroshkevich (1970) in a linear analysis of the problem, although confusion concerning 
the efficiency of the mechanism remained widespread until the analytical and numerical 
study by White (1984). Since the early studies of this mechanism, a flurry of
analytical and numerical studies (e.g. Heavens \& Peacock 1988; Ryden
1988; Hoffman 1986, 1988; Quinn \& Binney 1992; Dunn \& Laflamme 1993;
Catelan \& Theuns 1996a,b) as well as extensive and more detailed N-body 
simulations (e.g. Efstathiou \& Jones 1979; White 1984; Barnes \& Efstathiou 1987; 
Dubinski 1992, Warren et al. 1992; Eisenstein \& Loeb 1995; Sugerman, Summers \&
Kamionkowski 2000, Porciani, Hoffman \& Dekel 2001a,b) have demonstrated its viability
in a realistic cosmological setting and gradually a general consensus has been emerging 
on tidal torqueing as the basic mechanism for the origin of angular momentum of 
galaxies for structure formation scenarios based on gravitational instability. 

Intriguing would be the possibility to revert the history, and exploit observed galaxy 
rotations in an attempt to reconstruct the very source of the galaxy rotations. Tentatively, the 
imprint of a large-scale generating tidal field should be reflected in the alignment of 
induced galaxy spins. These may constitute a significant relic the cosmic structure formation 
process in case the major share of the torqueing action has been assumed by large-scale 
coherent structures like filaments. The search for such a fossil imprint has recently been the 
subject of various studies.  Statistically significant alignments of spin vectors of galaxies 
may be expected if the main source of the large-scale tidal torque would be large spatially 
coherent matter concentrations,   
filaments and  walls being the most straightforward examples, since it would lead 
to a roughly similar spin axis for neighbouring collapsing clumps. Assessing the  
prominence of such alignments would be of high interest. For one, it could potentially 
complicate the correct interpretation of the measurement of weak gravitational lensing 
effects. Particularly interesting is the suggestion by Lee \& Pen (2000) that the correlated 
galaxy spin orientations in a particular cosmic region offer a unique and alternative 
way of reconstructing the cosmic gravitational shear and potential field, and hence 
the large-scale density field. 

However, in reality the effect may be substantially clouded. Non-linear contributions to 
the spin direction may totally erase the linearly predicted alignments, as the study by 
Porciani, Dekel \& Hoffman (2001a) has indicated. Even if some level of alignment survives, 
any significant alignment will only be evident as a residual effect amidst a mostly 
random distribution of galaxy spin axes. The direction of the principal axes of 
protogalaxies -- the peaks in the primordial density field -- and that of the axes of the 
tidal shear at its location are each determined by random stochastic processes 
(see e.g. Bardeen et al. 1986) and are mutually misaligned with respect to each other.
Their stochastic distributions, however, are expected to be significantly
correlated (see e.g. van~de~Weygaert \& Bertschinger 1996, Lee \& Pen 2000).

Small or ill-defined samples of galaxy spin measurements may therefore mask 
any positive effect by statistical noise. This may be the reason for the current 
absence of a convincing detection of significant galaxy spin alignments, and 
the contradictory conclusions reached by a variety of observational attempts to 
address the issue. While some claim alignments do indeed exist (e.g. Yuan 
et al. 1997), be it a weak one, other studies do not find any significant 
evidence (Han, Gould \& Sackett 1995, Cabanela \& Dickey 1999). Given that most 
previous studies concentrated on areas where spin alignment tends to be weak, 
a substantial improvement may be accomplished by concentrating on those regions where 
we expect the strongest spin alignments, in the filamentary outliers branching 
out of clusters. If indeed detected, it would provide a tentalizing glimpse 
of the very dynamical processes underlying the shaping of the overall 
cosmic matter distribution into its salient foamlike pattern. 

\vfill\eject
\subsection{Anisotropic Collapse:}
\begin{flushright}
{\bf{\large the Zel'dovich Approximation}}
\end{flushright}
Evidently, in the world of reality there is no such thing as an isolated and 
homogeneous object, and this implies a considerably more complex and intricate 
evolution of cosmic structure. A full and self-consistent appreciation of the resulting 
emergence and morphology of cosmic structures has not yet properly settled. Often therefore 
we are confronted with mere qualitative descriptions of the intricate filamentary and 
foamlike patterns emerging in the cosmic matter distribution. 

Yet, a large share of the outstanding and inherent tendencies of the evolving patterns 
in the cosmic matter distribution may indeed be readily identified, even for generic 
circumstances where the spatial density distribution is a stochastic fluctuation field.
The Zel'dovich formalism, in essence a mere first-order Lagrangian approximation, 
which has played a major role in elucidating the qualititave aspects of the generic 
evolution of a random matter density field. Indeed, it may be argued that central role 
of the Zel'dovich formalism in structure formation studies stems from its ability 
to take any arbitrary initial random density field and mould it through a simple 
and direct operation into a reasonable approximation for the matter distribution 
at later nonlinear epochs. 

As it takes into account the complete force field, comprising that induced by 
density perturbations inside and outside of a specific region (``patch''), the 
Zel'dovich formalism is basically suited for any complex spatial configuration, 
dealing fully and self-consistently with both internal and external influences. 
Of overriding importance is the one-to-one linear nature of the relation between 
displacement and primordial local gravitational force. This allows the Zel'dovich 
approximation to identify the location of emerging structures, the sites where the 
induced migration flows accumulate the displaced matter patches. As it takes into 
account the full vector force field, it leads to an insightful relationship between 
the anisotropy of the primordial peculiar force field and morphology of the emerging matter 
distribution. Its application within a variety of cosmological scenarios demonstrated 
that flattened and elongated features are generic features of the matter distribution 
at moderate quasi-linear phases of evolution. Moreover, the coherent flattened 
structures appear to arrange themselves into a global cellular skeleton permeating 
the cosmos. 

The generic features of anisotropic collapse within generic media of cold matter 
distributions can be inferred by assessing the relation between a fluid element's 
density evolution  and the eigenvalues of its deformation tensor, and hence the tidal force tensor 
(Eqn. 13, section 3.5.1). Three instrumental observations concerning the collapse 
process can be readily made: 
\begin{itemize}
\item{} Ultimate {\it fate}: \\
Having one or more positive eigenvalues $\lambda_i$ (i=1,2,3) is sufficient 
to guarantee collapse.  Unless each individual $\lambda_m < 0$ (i=1,2,3), the fluid 
element, even if initially underdense, will undergo collapse along at 
least one of its axes at some stage during its evolution.
\item{} {\it Collapse Configurations}: \\
For fluid elements with at least one positive deformation eigenvalue, the ultimate 
collapse geometry of a fluid element can be one of 3 distinct configurations. Assuming 
that $\lambda_3>\lambda_2>\lambda_1$, a flattened {\it pancake} shape will be 
attained if only $\lambda_3>0$ and $\lambda_2<0$. While collapse has occurred 
along the corresponding axis, the element will expand along the 2 others. Likewise, 
if $\lambda_3>\lambda_2>0$ and $\lambda_1<0$, the mass element will evolve 
into an elongated {\it spindle}. A state of full collapse along all three axes 
will only be reached if all $\lambda_i>0$ (i=1,2,3). 
\item{} {\it Collapse Time Sequence}: \\
The collapse will be anisotropic and, dependent on the sign of the deformation 
eigenvalues, proceeding along a sequence of one, two or three distinct stages. Firstly, 
collapse will happen along the direction of the largest positive eigenvalue at 
$a_{1c}=1/\lambda_1$. It occurs upon the fluid element entering a flattened 
{\it pancake}. 
Subsequently, it will start contracting along a second direction, leading to collapse 
along a second axis at $a_{2c}=1/\lambda_2$, corresponding to the presence in a 
{\it filament}. Ultimately, if also $\lambda_3>0$, the fluid element will undergo 
full collapse upon a singular point at $a_{3c}=1/\lambda_3$. 
\end{itemize}
\noindent Following these considerations, it is straightforward to 
sketch a qualitative picture of the gravitational evolution of the overall 
matter distribution. Starting from a field of minor (Gaussian) fluctuations, the 
first structures to undergo collapse are {\it pancakes}, composed of fluid elements that 
collapse along the first, largest eigenvalue, axis. Subsequently, we'll see 
the formation of elongated {\it filaments} once corresponding mass elements 
start contracting along the second direction. Finally, the emergence of dense 
compact clumps will be the result of the full collapse of fluid elements 
along all 3 collapse directions. The physical nature of the emerging 
object will depend on the scale of the corresponding deformation field, 
so that a {\it galaxy halo} will emerge through full collapse on a 
scale $< 1h^{-1}\hbox{Mpc}$, and a {\it galaxy cluster} from 
a similar collapse on a scale $\sim 4h^{-1}\hbox{Mpc}$. 

\begin{figure}[t]
\centering\mbox{\hskip 0.0truecm\psfig{figure=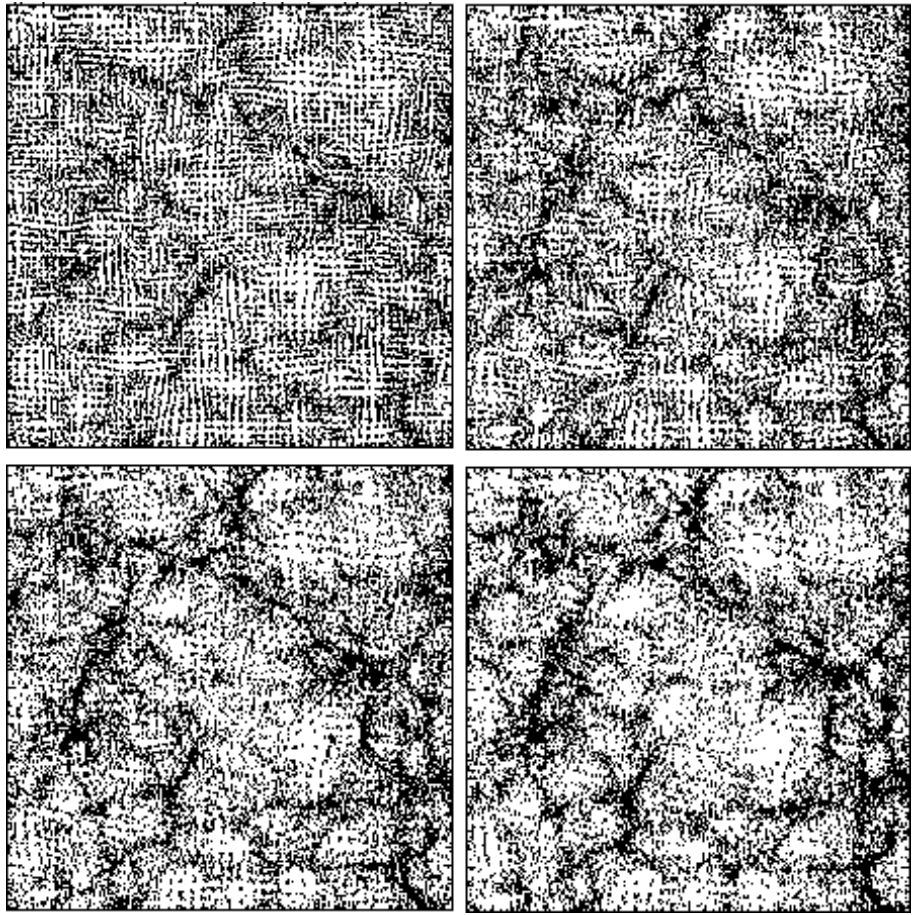,width=12.cm}}
\vskip -0.3truecm
\caption{Zel'dovich displaced particle distributions inferred from a unconstrained 
random realization of a primordial matter distribution for a SCDM cosmological 
scenario in a $50h^{-1}\hbox{Mpc}$. Time sequence from top left to bottom right, 
frames corresponding to cosmic epochs $a=0.10, 0.15, 0.20$ and $0.25$.} 
\vskip -0.5truecm
\end{figure}
The time sequence of four frames in Fig. 21 elucidates the success of the 
Zel'dovich scheme, as it does its obvious shortcomings. Portraying the Zel'dovich 
predictions for the cosmic matter distribution in a $50h^{-1}\hbox{Mpc}$ box for a SCDM 
scenario, at cosmic epochs $a=0.10, 0.15, 0.20$ and $0.25$, the gradual 
morphological procession along ``pancake'' and ``filamentary'' stages 
can be readily observed. A comparison with the results of full-scale 
N-body simulations shows that in particular at early structure formation epochs 
the predicted Zel'dovich configurations are accurately rendering the full nonlinear 
matter distributions. On the basis of the four frames in Fig. 21 we can note 
that:
\begin{itemize}
\item{} A comparison of the last 3 frames with the very first stage 
indicates that the pattern of the resulting large scale matter distribution 
is already, be it faintly, imprinted in the primordial matter configuration. 
\item{} The essence of the Zel'dovich approximation is in the deformation 
tensor $\psi_{mn}$, directly related to the tidal field tensor $T_{mn}$ 
(eqn. 14). It indicates the existence of the profound relation between the anisotropic 
force field as shaping agent and the resulting anisotropic matter concentrations, 
and the cosmic web of which these are the elementary constituents. 
\item{} The 1-1 Lagrangian-Eulerian mapping of the Zel'dovich approximation 
is only possible as long as a fluid element has not yet crossed the orbit of 
another fluid element. 
\end{itemize}
Pursuing the latter, we see the approximation starts to fail after the 
occurrence of orbit crossing, which is clearly observed through the unrealistic 
``diffusion'' of clumps, filaments and walls in the 4$^{th}$ frame. Following 
orbit crossing, we will start to see thorough gravitationally induced orbit 
mixing and energy exchange proceeding towards a ``quasi-equilibrium'' state of 
complete virialization. The Zel'dovich approximation is fundamentally 
inadequate to describe these advanced nonlinear stages. In fact, its major 
limitation stems from its virtue of restriciting itself to the initial 
fluctuation field, so that it breaks down as soon as the selfgravity of the 
emerging structures becomes so strong that the initial ``ballistic'' 
motion of the fluid element gets seriously altered, redirected, slowed down, 
and possibly even brought to a halt. Full-scale gravitational N-body simulations, 
and/or more sophisticated approximations, are necessary to deal self consistently 
with these more advanced nonlinear stages. 

From the Zel'dovich formalism we can readily infer that the overall morphology 
and spatial pattern of a cosmic density field at its ``quasilinear'' 
development stage -- i.e. the prominence of flattened structures, denser 
elongated filaments and dense compact clumps as well as 
their interconnectedness -- is sensitively dependent on the statistical 
distribution of the values of the eigenvalues $\lambda_i$. For the case of a Gaussian 
random density fluctuation field, Doroshkevich (1970) computed the probability 
distribution for the eigenvalues $\lambda_1$, $\lambda_2$, and $\lambda_3$, 
\begin{eqnarray}
P(\lambda_1,\lambda_2,\lambda_3) &\ \sim\ &
(\lambda_1 - \lambda_2)(\lambda_1-\lambda_3)(\lambda_2-\lambda_3)\nonumber\\
&\ \times\ &\exp\left\{-{\displaystyle 15 \over \displaystyle 2 \sigma^2}  
\left[\displaystyle \lambda_1^2 + \lambda_2^2 + \lambda_3^2 
-{1 \over 2}(\lambda_1 \lambda_2 + \lambda_1 \lambda_3 + 
\lambda_2 \lambda_3)\right]\right\}\nonumber\\ 
\end{eqnarray} 
This yields a probability of $8\%$ that all of the eigenvalues are 
negative, $\lambda_3<\lambda_2<\lambda_1<0$, while $92\%$ of the 
matter has one or more positive eigenvalues. From a detailed assessment of the 
deformation eigenvalue distribution (eqn. 15) for scenarios with relatively 
strong perturbations on large scales, a distinct wall- and filament-dominated 
weblike configuration during the moderate quasi-linear evolution phase is 
the natural outcome. Vast coherent wall-like and filamentary features 
characterize the matter distribution. On the other hand, when small-scale 
perturbations are so prominent that full collapse has taken place on 
small scales even before any noticeable anisotropic contraction on larger 
scales has occurred, orbit crossing and virialization have been so ubiquitous 
that the Zel'dovich description will be seriously challenged, rendering the 
validity of its predictions more than dubious.

The essential role of the anisotropy in the force field -- i.e. the tidal field -- in 
shaping the cosmic matter distribution is evidently an essential element of 
the Zel'dovich approximation, which so emphatically invokes the deformation tensor, and 
by implication the initial tidal field configuration. Yet, the Zel'dovich 
approximation restricts itself to the linearly extrapolated tidal field 
configuration of the initial density field, and fails to react properly to the rapidly 
rising gravitational strength in and around emerging matter concentrations. While it 
indicates the proper context for understanding the dynamical mechanisms behind the 
formation of the cosmic foam, we need to extend our understanding of the evidently 
tight connection between the workings of the tidal force field and the moulding of 
a cosmic foam structure. Only by elucidating this relationship up to the fully 
nonlinear regime we may hope to get a full understanding of the dynamical evolution 
of the pervasive and enduring nature of the seemingly tenuous and fragile foamlike 
network permeating our observable Universe. 

\subsection{Foam Assembly and Tidal Dynamics: \hfill}
\begin{flushright}
{\bf{\large Tidal ``Casting'' of the Cosmic Web}}
\vskip -0.2truecm
\end{flushright}
First recognized in idealized approximations such as the Zel'dovich scheme -- 
and its essence explained by the anisotropic collapse of homogeneous ellipsoids -- 
the tendency of structures to evolve through a flattened and/or elongated phase 
has been found to be a universal phenomenon. It has proven to be a characteristic 
phenomenon for a gravitationally driven formation of structure, and has been 
recovered in a vast range of viable scenarios of gravitational cosmic structure 
formation. 

A few recent studies (e.g. Bond, Kofman \& Pogosyan 1996) have drawn attention to the close and 
causal link between the generically anisotropic tidal force fields generated by typical cosmic 
matter distributions, their impact on the shape of an emerging and evolving structure, and 
the resulting cosmic foamlike structure. Indeed, if anything, the applicability 
of the Zel'dovich approximation far into the quasi-linear regime provides us with a 
rudimentary indication that there is a significant and possibly instrumental dynamical 
connection. Its remarkably long-lasting validity, by far surpassing the nominally valid 
linear regime, in combination with its basic affiliation to the primordial anisotropic force 
field (see Eqn. 14) provides a compelling indication for such a kinship. It warrants a 
considerably closer investigation of what may conceivably be an intimate and causal relationship 
between foam morphology of the cosmic matter distribution and the primordial cosmic 
force field. Such a study should involve the full range of evolutionary phases, including the 
late nonlinear stages at which the web features should actually condense out as (partially) 
virialized structures. With respect to the latter, it is important to realize that quasi-linear 
approaches like the Zel'dovich approximation cannot address the issue of whether filaments 
or walls will truely settle as genuine objects instead of dissolving, rendering them mere 
transient features. 

\subsubsection{{\it Cosmic Foam: Tidal Constraints and Connections}}
Bond, Kofman \& Pogosyan (1996) coined the word `cosmic web' in their study 
of the physical 
content of the web, in which they drew attention to their finding that knowledge of the 
value of the tidal field at a few well-chosen 
cosmic locations in some region would determine the overall outline of the 
weblike pattern in that region. This relation may be traced back to a simple 
configuration, that of a ``global'' quadrupolar matter distribution and the resulting 
``local'' tidal shear at its central site. Such a quadrupolar primordial matter 
distribution will almost by default evolve into the canonical cluster-filament-cluster 
configuration which appears so prominently in the cosmic foam. Indeed, this close 
connection between local force field and global matter distribution had been 
elucidated by means of a constrained field study by Van de Weygaert \& Bertschinger 
(1996). They, amongst others, discussed the repercussion of a specified constraint 
on the value of the tidal shear at some specific location. From the expression 
of the tidal tensor in terms of the generating density distribution, 
\begin{eqnarray}
T_{ij}({\bf r},t)\ = \ {\displaystyle 3 \Omega H^2 \over \displaystyle 8\pi}\,
\int {\rm d}{\bf r}'\,\delta({\bf r}',t)\ \left\{{\displaystyle 3 (r_i'-r_i)(r_j'-r_j)-
|{\bf r}'-{\bf r}|^2\ \delta_{ij} \over \displaystyle |{\bf r}'-{\bf r}|^5}\right\}\ - \ \nonumber\\
\nonumber\\
\ - \ {\frac{1}{2}}\Omega H^2\ \delta({\bf r},t)\ \delta_{ij}\ \ \ \ \ \ \ \ \ \ \ \ \ \ \ \ \ \ \ \ \ \ \ \ \  
\end{eqnarray}
\noindent we can immediately observe that any {\it local} value of $T_{ij}$ has {\it global} 
repercussions for the generating density field. Such {\it global} constraints are 
in marked contrast to {\it local} constraints as the value of the density contrast $\delta$ 
itself, or the shape of the local matter distribution.  
\begin{figure}[t]
\vskip -0.2truecm
\centering\mbox{\hskip -0.4truecm\psfig{figure=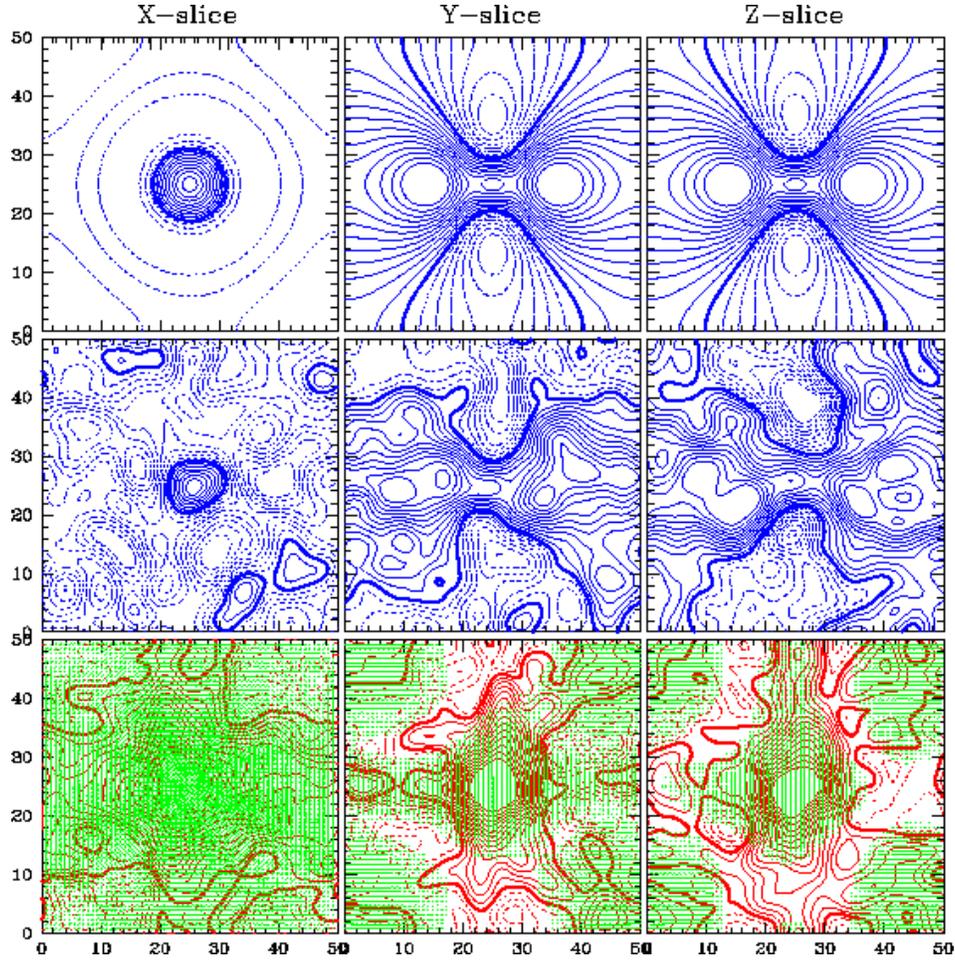,width=12.5cm}}
\vskip -0.2truecm
\caption{Constrained field construction of initial quadrupolar density pattern in a SCDM 
cosmological scenario, following the Hoffman-Ribak (1991) procedure in its field dynamical 
implementation by Van de Weygaert \& Bertschinger (1996). The tidal shear constraint is 
specified at the box centre location, issued on a Gaussian scale of $R_G=2h^{-1}\hbox{\rm Mpc}$ 
and includes a stretching tidal component along the $x$- and $y$-axis acting on a small density 
peak at the centre. Its ramifications are illustrated by means of three mutually perpendicular 
slices through the centre. Top row: the ``mean'' field density pattern, the pure signal 
implied by the specified constraint. Notice the clear quadrupolar pattern in the $y$- and 
$z$-slice,directed along the $x$- and $y$-axis, and the corresponding compact circular density 
contours 
in the $x$-slice: the precursor of a filament. Central row: the full constrained field realization, including a realization of appropriately added SCDM density perturbations. Bottom row: the 
corresponding tidal field pattern in the same three slices. The (red) contours depict the run 
of the tidal field strenght $|T|$, while the (green) tidal bars represent direction and magnitude 
of the ``compressional'' tidal component in each slice (scale: $R_G=2h^{-1}\hbox{\rm Mpc}$).} 
\vskip -0.2truecm
\end{figure}
\noindent One of the major virtues of their {\it constrained random field} construction 
technique (Bertschinger 1987, Hoffman \& Ribak 1991) is that it offers the instrument for  
translating locally specified quantities into the corresponding implied global matter 
distributions for a given structure formation scenario. In principle, the choice of possible 
implied matter distribution configurations is limitless, yet it gets substantially 
curtailed by the statistical nature of its density fluctuations, the coherence 
scale of the matter distribution and hence of the generated force field as well as the 
noise characteristics over the various spatial scales, both set by the power 
spectrum of fluctuations. 

The most straighforward example of a {\it global} constraint involves the 
local gravitational acceleration at one particular location. Such an acceleration has 
to be induced by a dipolar asymmetry in the matter distribution centered on that 
particular location. Such may be readily appreciated from the integral expression 
of ${\bf g}$ in Eqn. 1 (Van de Weygaert \& Bertschinger 1996, Fig. 3). 
It is in a similar fashion that Eqn. 16 implies a quadrupolar pattern in the 
density field distribution, its particular realization determined by the orientation 
and strength of the local tidal shear. This can be readily observed from Fig. 22, 
which is an elaboration on a similar discussion in Van de Weygaert \& Bertschinger 
(1996, Figs. 3 and 5). It provides a 3-D impression of the structure in the region 
immediately surrounding the location of the specified shear. Along each row 
we show maps in three mutually perpendicular cross-sections centered on the centre of the box.  

In the purest, noise-free, implication of the specified constraints the {\it mean field} 
in the three top panels represents the clearest depiction of the average density field 
configuration inducing the specified tidal tensor. It is very clear that the constraint 
works out into a perfect global quadrupolar field (slightly alineated along the 
$x$-axis due to the extra specification of a central elongated small-scale peak). 
Superposing {\it residual} noise fluctuations, whose amplitude is 
modified by the local correlation with the specified constraints, results 
into a representative individual realization of a matter density distribution 
that would induce the specified constraint. The outcome is depicted in the 
second row of 3 panels. The close affiliation with a strong anisotropic 
force field, within the surrounding region, can then be directly observed 
from the lower row of corresponding $x$-, $y$- and $z$-slices. The contour maps (red) reveal the 
spatial configuration of full tidal field strength ($T=\sqrt(\sum_{k=1}^3 T_k^2)$, 
with $T_k$ a tidal tensor eigenvalue) in this central region. Note that we e have crudely 
included the concept of ``external'' by (spherically) filtering the field on a 
(rather arbitrary) scale of $2h^{-1}\hbox{\rm Mpc}$. It amply illustrates 
the fact that the specified constraints work out into a maximum in the tidal 
field strength at the box centre. As is clearly borne out by the Y- and Z-slices 
(central and righthand frame), the tidal field strength contours are noticeably elongated 
along the Z-axis, culminating in a maximum at the constraint centre. Notice that this 
elongation is oriented roughly perpendicular to the stretching direction of the tidal 
shear, and along the axes along which a compressing force is acting (see footnote). A 
totally different configuration is seen in the $x$-slice. Around 
the centre no distinct orientation can be identified. A much more strongly peaked 
pattern is shown in this plane, concentrated on the box centre. 

\begin{figure}[t]
\vskip 0.1truecm
\centering\mbox{\hskip -1.truecm\psfig{figure=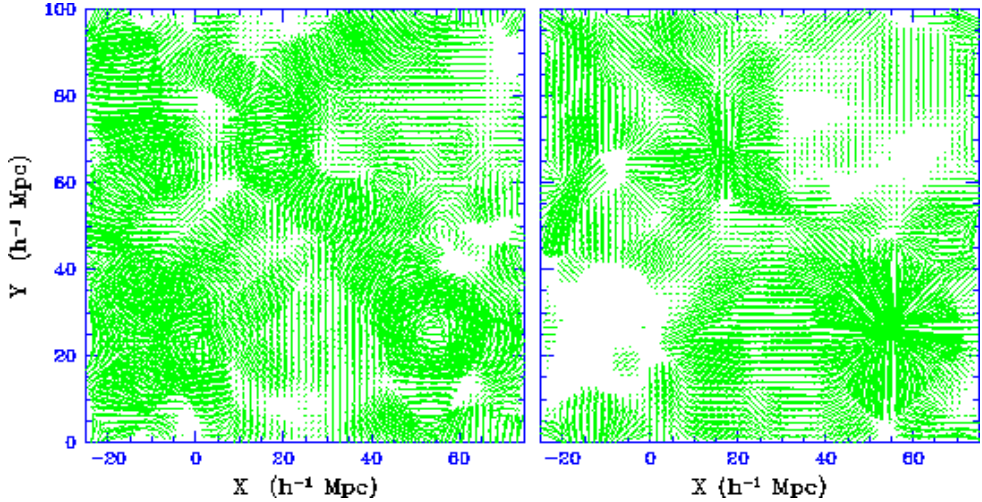,width=12.9cm}}
\vskip -0.3cm
\caption{Decomposition of the tidal field components within a slice through a 
simulation of structure formation in an $\Omega_{\circ}=0.3$ CDM Universe ($z\approx 1.5$, 
cf. fig. 18). Left: the ``compressional'' tidal component. Right: the ``dilational'' (stretching) 
tidal component). The scale of the tidal field is $R_G=2h^{-1}\hbox{\rm Mpc}$ (roughly 
rendering ``external'' tidal contributions). See footnote for further explanation.} 
\vskip -0.2truecm
\end{figure}
Further elucidating the patterns in the suggestive tidal strength contour maps are the 
superimposed ``compressional'' tidal bars (green), indicating orientation and magnitude 
of the compressional tidal components within each of the 3 mutually perpendicular planes 
slicing through the centre of the box\footnote{on the basis of the effect of a tidal 
field, we may distinguish at any one location between ``compressional'' and ``dilational'' 
components. Along the direction of a ``compressional'' tidal component $T_{c}$ (for which  
$T_c<0.0$) the resulting force field will lead to contraction, pulling together the matter 
currents. The ``dilational'' (or ``stretching'') tidal component $T_d$, on the other hand, 
represents the direction 
along which matter currents tend to get stretched as $T_d>0$. Note that within a plane, 
cutting through the 3-D tidal ``ellipsoid'', the tidal field can consist of two compressional 
components, two dilational ones or -- the most frequently encountered situation -- of one 
dilational and one compressional component. Also see fig. 14}. It shows that the central 
region wherein 
the tidal field strength assumes its maximum value, is showing up conspicuously when 
dissecting the field into its physically active components. The central region is 
clearly the one where the tidal bars have their largest size. Even more interesting is 
their corresponding orientation, coherently directed perpendicular to the $x$- and 
$y$-axis and along the $z$-axis. Thus, in the central and righthand frame we see a pattern 
of bars directed in parallel to the Z-axis, while the lefthand $x$-slice frame reveals a 
pattern of radially directed bars. 

The depicted tidal configurations seem to suggest that from the onset of the 
structure formation process on, the force field comprises a pattern of force 
anisotropies that will ultimately to strongly moulded, and folded, patterns in the matter 
distribution. 
\begin{figure}[t]
\centering\mbox{\hskip -0.4truecm\psfig{figure=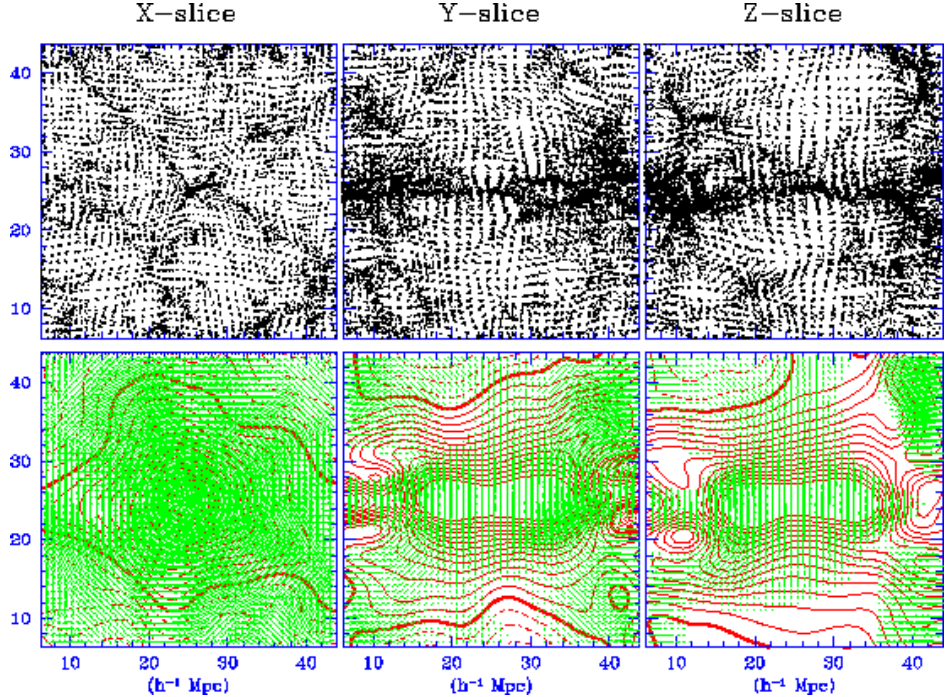,width=12.4cm}}
\vskip -0.3truecm
\caption{Illustration of connection between mass distribution in and around a filament and 
the corresponding tidal field (scale: $R_G=2h^{-1}\hbox{\rm Mpc}$). The filament has 
formed in a N-body simulation of structure formation in a SCDM cosmological scenario. Top row: the 
particle distribution in the mutually perpendicular $x$-, $y$- and $z$-slices through the centre 
of the box. Bottom row: the tidal field configuration, both tidal strength contours (red) and 
compressional tidal field bars (green) in the same slices. The similar patterns along 
in the $y$- and $z$-slices clearly reflect the striking tidal pattern along the spine of 
the filament. They are in marked contrast with the almost circularly symmetric and 
highly concentrated pattern in the $x$-slice, illustrative of the dynamical impact of the 
filament perpendicular to its spine.} 
\vskip -1.0truecm
\end{figure}

\subsubsection{{\it Tidal Connections: filaments}}
A telling illustration of the correlation between the anisotropy in the cosmic force 
field and the presence of strongly anisotropic features is provided by the multi-faceted 
impression in Fig. 24 of a conspicuous filamentary structure which emerged in an 
(128$^3$ particle) N-body simulation of structure formation in a SCDM scenario 
($\Omega_{\circ}=1.0$, $H_{\circ}=50\ \hbox{km/s/Mpc}$). The depicted particle 
distribution corresponds to a cosmic epoch at which $\sigma_8 \approx 0.7$ for the 
matter distribution, roughly corresponding to the present, and is shown in  
three mutually perpendicular planes passing through the box centre. The filament, elongated along 
the $x$-direction and oriented perpendicular to the $y-z$, appears as a dominant feature in 
the particle distribution (top row frames). Note the two massive cluster  
concentrations on either side of the filament, to the left and right end of the $x$-axis. 
These matter assemblies, in conjunction with the correspondingly large underdense volumes 
surrounding the filament perpendicular to its spinal axis, define a roughly quadrupolar 
density field. Naturally, this translates directly into a tidal force field 
with a strong compressional component perpendicular to the filament. 
\begin{figure}[t]
\vskip -0.0truecm
\centering\mbox{\hskip -1.0truecm\psfig{figure=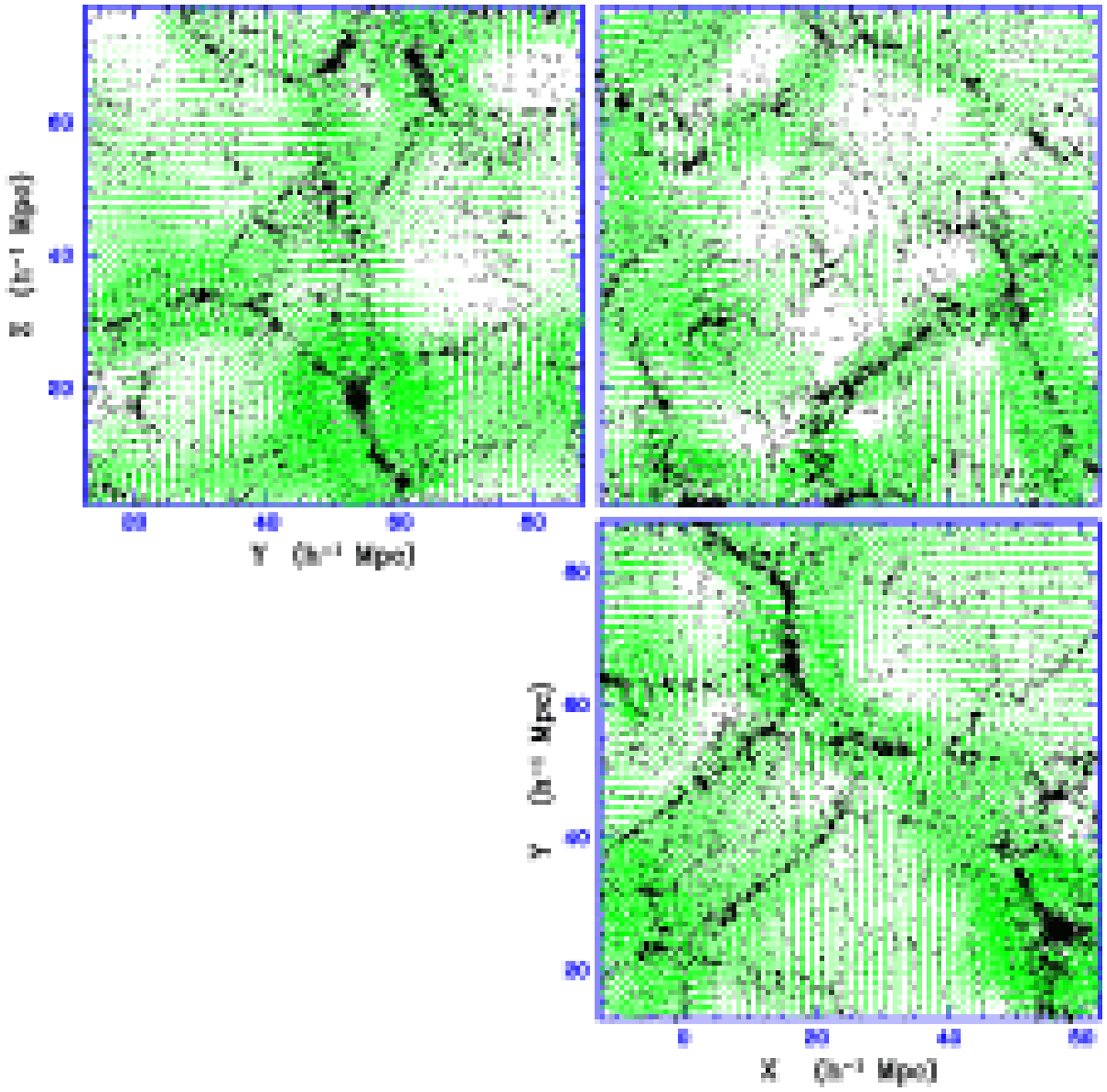,width=13.4cm}}
\vskip -0.3truecm
\caption{The compressional tidal field pattern in three mutually perpendicular slices 
through a simulation box containing a realization of cosmic structure formed in an open,  
$\Omega_{\circ}=0.3$, Universe for a CDM structure formation scenario (scale: 
$R_G=2h^{-1}\hbox{\rm Mpc}$). The matter distribution 
corresponds to the present cosmic epoch. Each frame contains the corresponding particle 
distribution within a $5h^{-1}\hbox{\rm Mpc}$ thin region centering on the slice, on which 
the related tidal bar configuration (green) is superimposed. The matter distribution, displaying 
a pronounced weblike geometry, is clearly intimately linked with a characteristic coherent 
and correlated compressional tidal bar pattern. Note the specific arrangement of the 
planar slices. Along the top row the $x$- and $y$-slice share the same vertical $z$-axis, 
while in the righthand column the $z$- and $y$-slice share the same horizontal $x$-axis.} 
\vskip -0.5truecm
\end{figure}
A striking demonstration of this intimate relationship between tidal field and the presence 
of a salient anisotropic structure like the depicted filament is presented in 
the corresponding lower row of frames in Fig. 24. The contour maps reveal how 
the maximum in tidal strength is reached in and immediately around the filament, and 
forms a plateau that appears to follow its spine along the full elongated extent, 
cut off near the position of the cluster complexes located on either side. The contours 
in the plane perpendicular to the filament complements this impression superbly, 
revealing the strongly centrally peaked and compact force strength field centered on its 
spinal column. 

Even more evocative are the superimposed ``compressional'' tidal bars (green). It shows that 
the filaments not only delineate regions of considerable tidal strength, as we can infer 
from the plateau in the map of total tidal strength and from the large size of the 
compressional bars in and immediately around the filament, but also appear to stand out 
when directing attention to the coherence of the field configuration. The filament is delineated 
by a region with an impressive coherence in both strength and orientation of the tidal bars 
along the full extent of the filament. 

\subsubsection{{\it Tidal connections: the unified web}}
Having established the strong correlation between a filament and its surrounding 
tidal field, the issue of how to interpret this within the generic context of 
a pattern of interconnected filaments -- and walls -- will further clarify the 
special correspondence and possible causal connection with the tidal field 
configurations. 

To that end, we adressed the web pattern in an N-body simulation of a CDM 
structure formation scenario in an open Universe, $\Omega_{\circ}=0.3$. In an open 
Universe a pronounced weblike pattern gets established relatively fast, after which 
further growth and development stops as soon as the expansion 
of the Universe starts to dominate over the gravitational influence of the cosmic 
matter distribution. It leads to the final fruition of cosmic structure at a 
redshift of around $z \ \approx \ \left(1/\Omega_{\circ}-1\right)$ (also see Fig. 27). 
One well-known consequence is the implication of a virtually unchanging population of 
rich clusters of galaxies over a large redshift range from $z=0$, and thus the 
presence of clusters at high redshift (see e.g. Eke, Cole \& Frenk 1996; 
Bahcall, Fan \& Cen 1997). As clusters reflect the underlying cosmic foam and 
form a particular aspect of its geometrical structure, an unchanging cluster 
population implies a virtually unchanging pattern in the cosmic matter distribution. 

Figure 25 contains an illustration of the resulting cosmic web formed at a 
redshift $z \approx 2$. A set of the three mutually perpendicular ($x$-, $y$- and 
$z$-) plane slices centered on a specified position in the box, each with a size of 
$80h^{-1}\hbox{\rm Mpc}$, evoke an impression of the three-dimensional foam pattern 
in the particle distribution (dots). Superimposed (green) are the compressional tidal 
bars. A rich amount of information is yielded by this cosmic snapshot. Notice the 
special grouping of the three frames, such that the top two ($x$- and $y$-) planes share the 
$z$-axis as the vertical axis while, with $y$-slice in its top righthand position, the 
two righthand ($z$- and $y$-) planes share the $x$-axis as horizontal axis. 

The most striking aspect of the three frames is the surprisingly {\it strong} correlation between 
the particle distribution in conspicuous -- massive yet compact and thin -- filaments and the 
spatial configurations outlined by the corresponding compressional tidal field bars (the 
correspondence with the dilational tidal component is considerably less compelling, see Fig. 23). 
All through the cosmic foam we can delineate the filaments as the sites where the compressional 
tidal forces are both strong (large tidal bars) as well as particularly strong, coherently 
and mutually parallel oriented, perpendicular to the spine of the filaments. 

Another outstanding feature of the compressional tidal field configuration concerns the 
location of the massive rich clusters in the matter distribution. Literally, these clumps in 
the particle distribution stand out as genuine ``nodes'' within the global compressional tidal 
bar pattern. It is no surprise to see them marking the maxima in the strengh of the tidal 
field, as can be straightforwardly discerned from the size of the tidal bars in their 
immediate environment. In this respect, it is most instructive to focus on the cluster in the 
lower lefthand corner of the $z$-slice. Around that cluster the field strength is so strong 
that the bars seem to weave a genuine ``bird's nest'' of overlapping and intersecting bars. 
To our idea, however, it is the contrasting topology of the bar pattern in their immediate 
environment which forms the true distinctive mark of the rich clusters within the overall 
cosmic tidal pattern. With a coherent lining up of the compressional field perpendicular 
to their spine is characteristic for filaments, clusters appear to direct the 
compressional tidal force into a closed loop of laterally oriented field bars. 
\begin{figure}[t]
\centering\mbox{\hskip -1.0truecm\psfig{figure=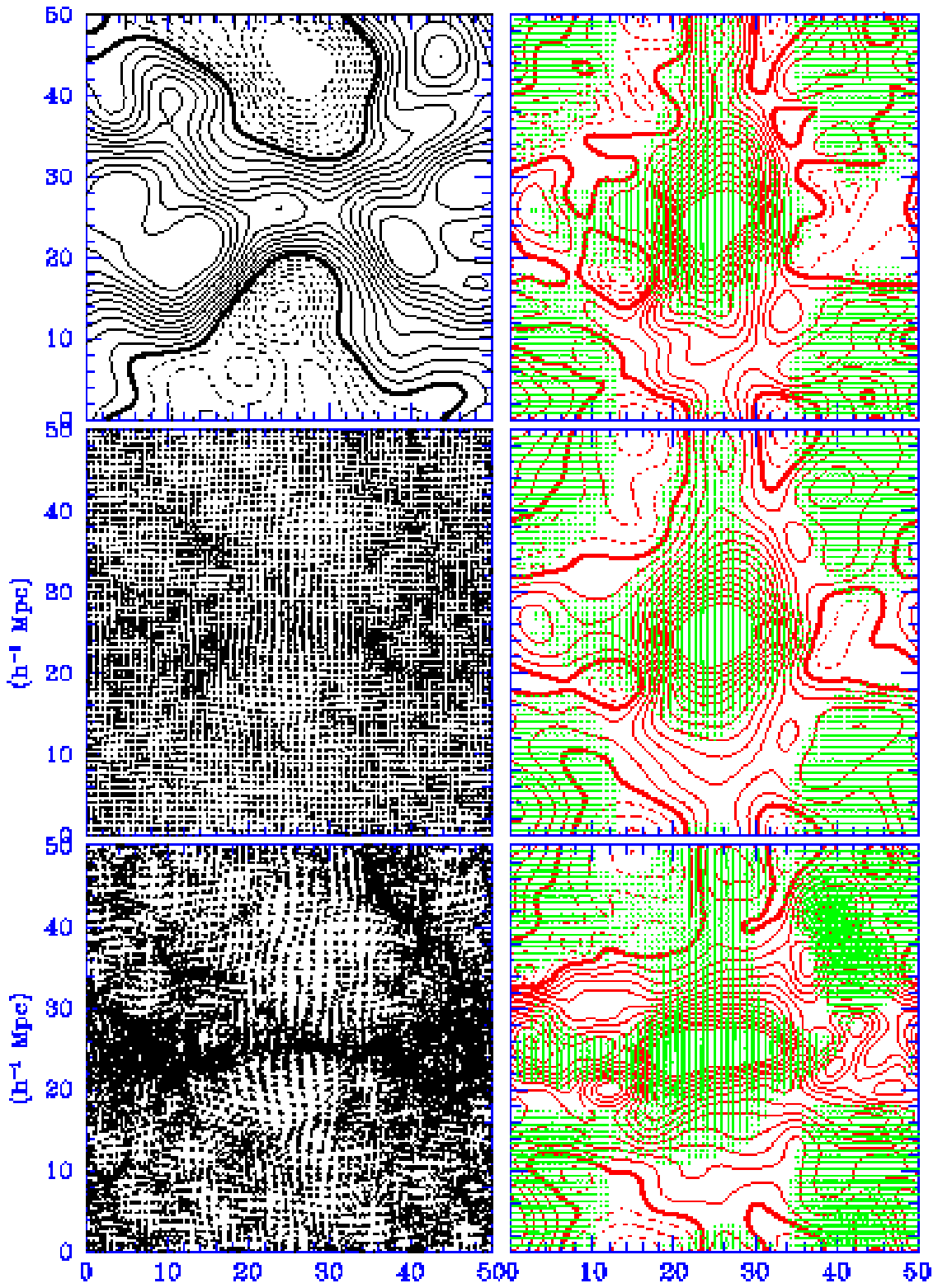,width=13.0cm}}
\end{figure}
\begin{figure}[h]
\vskip -0.5truecm
\caption{The emergence of a filament in an SCDM structure formation scenario. Based on the 
basis of a $N-body$ simulation of a particular (constrained) realization. Lefthand column: 
density/particle distribution in $z$-slice through the centre of the simulation box. Righthand 
column: the corresponding tidal field configurations, represented through the full tidal 
field strength $|T$ contour maps (red), as well as the corresponding compressional tidal 
bars (scale: $R_G=2h^{-1}\hbox{\rm Mpc}$). Top row: primordial cosmic conditions 
(primordial density field and tidal field). Centre row: $a=0.2$, the first onset of an 
emerging filament. Bottom row: the filament at the present epoch (for this realization: 
$a=0.8$. Note the formation of the filament at the site where the tidal forces peaked 
in strength from the onset onward, with a tidal pattern whose topology remains 
roughly similar, although the flattening of the tidal contours along the spine od the 
ultimately contracted filament is readily apparent.} 
\vskip -0.3truecm
\end{figure}
As a final note we wish to point out the readily apparent close relation between the 
tidal field configuration shown in Fig. 25 and the pattern of inflicted distortions 
of background galaxy images through weak lensing by the large scale matter distribution, 
and in particular by massive rich clusters (see e.g. Blandford et al. 1991, Jain, Seljak \& 
White 2000). While the maps of weak lensing distortion numerically comprise a weighted 
projection over the full three-dimensional tidal shear pattern, it is in essence a slice 
through the latter which is shown in the figures illustrated in this contribution. In fact, based 
on the full 3-D tidal field, Couchman, Barber \& Thomas (1999), devised a computational 
scheme to predict the resulting weak lensing distortion pattern. Immediately evident from 
the tidal bar maps in Fig. 25 is the reason why clusters have been the main focus of studies 
employing the observed weak lensing image deformation configurations to infer the mass of the 
enclosed cluster mass (based in particular on work of Tyson, Wenk \& Valdes, 1990 and Kaiser 
\& Squires 1993). Only in their immediate surroundings the tidal field strength is substantial 
enough, and the pattern appropriately well-behaved, that a reconstruction from the observed 
galaxy images yields a gravitational matter distribution with a sufficiently high signal/noise 
level. The more moderate to weak field values pertaining throughout the general cosmic field, and 
as yet even around filaments, have not yet produced reconstructions that would represent 
unequivocal significant mass distributions. On the other hand, comparing the relative tidal 
strengths, we may expect that cluster mass estimates on the basis of the observed shear 
pattern will to some extent be influenced by the surrounding structures (for a thorough 
study see Hoekstra 2001). Promising is that in recent years advances in detector sensitivity 
have surged to such extent that the presence of significant image distortions by the general 
field of large scale structures -- the detection of {\it ``cosmic shear''} -- has been 
demonstrated beyond doubt on the basis of intricate statistical analysis (Van Waerbeke et 
al. 2000). However, maps of the cosmic matter distribution in the general field still seem 
to be just beyond grasp of current technology. 

\subsubsection{{\it Tidal Moulding:  Shaping a Filament, a History}}
The strong correlation between the compressional components of the tidal 
field and the presence of a dense filamentary feature (or, similarly yet less pronounced, 
wall-like patterns) conjures up the question of the nature of a possible causal link between 
these two physical aspects. For investigating 
such a causal link, we in particular wish to assess whether the presence of a strong tidal 
field presages the in situ ``condensation'' of a wall, filament or related structure. The 
emergence of the cosmic web pattern could then be described as a process of casting 
the affiliated cosmic matter flows into  a primordially shaped mould, which would have 
been outlined at the onset by the tidal field induced by the tiny matter density 
fluctuations in the pristine Universe. 

Indeed, preliminary systematic investigations of such a causal connection do indicate 
the formation of the web pattern following the directions outlined by the primordial 
tidal field, specifically of its compressional components. Figure 26 is displaying the 
development of the filamentary structure of Fig. 24. In a sequel of three timesteps, from 
top to bottom rows, it displays the corresponding primordial density field ($a(t)=0.0$, top row), 
and the gradual emergence through a moderate quasi-linear phase ($a=0.5$, central row) to the final 
assembly into the salient nonlinear filament ($a_{\circ}=1$, the present epoch). While we 
see a drastic evolution in the configurations of the matter 
distribution (the left column frames depict the matter distribution in the central 
$x-z$ plane for the 3 timesteps), the configuration in the corresponding tidal field contour 
maps (frames in the righthand column, contours in red) display a considerably more moderate 
development. Qualitatively, they appear to retain their primordial signature quite 
well, be it that the later more anisotropic stages naturally involve a correspondingly 
elongated tidal field. Yet, we also clearly see the formation of the filament 
precisely there where the primordial compressional field is very strong and 
coherent. We therefore argue that a {\it mapping of the compressional tidal component 
represents a prediction for the locus of the main cosmic web features}. 

\begin{figure}[t]
\centering\mbox{\hskip -1.00truecm\psfig{figure=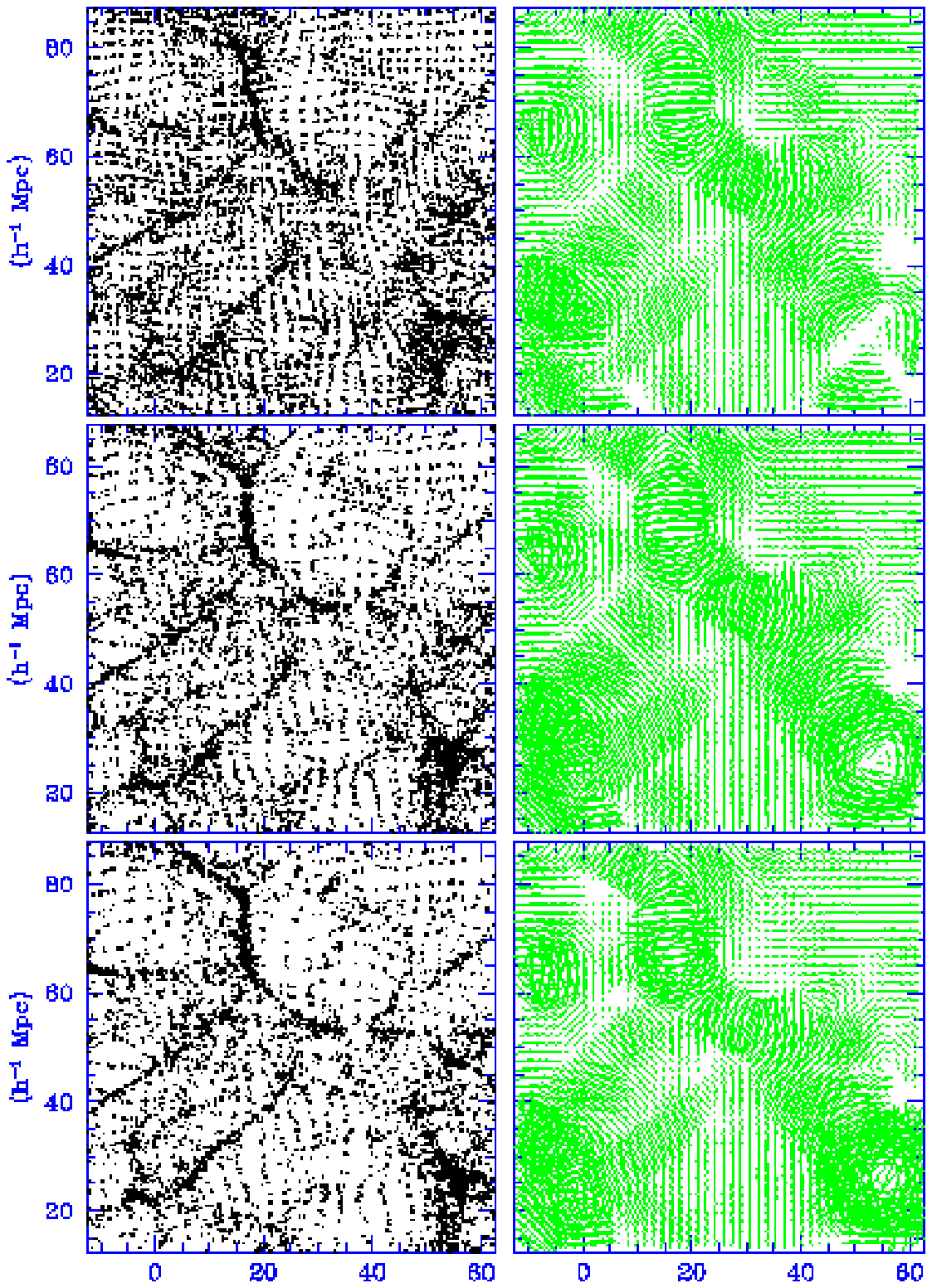,width=13.0cm}}
\end{figure}
\begin{figure}[h]
\vskip -0.5truecm
\caption{Evolution of a cosmic foam pattern in an open $\Omega_{\circ}$ CDM Universe. For three 
consecutive timesteps with expansion factors $a=0.2$, $a=0.4$ and $a=0.6$, each row shows 
the particle distribution in the central $z$-slice through a corresponding N-body simulation 
realization (lefthand frame) in conjunction with the compressional tidal bar configuration in 
the same slice (righthand frame). The geometry of the weblike matter distribution does not 
involve substantial qualitative changes. The process of structure formation appears to 
consist of a gradual inflow of matter into the web channels. These channels, sites where 
filaments appear to emerge, can be traced back to the primordial tidal field pattern. 
} 
\vskip -0.3truecm
\end{figure}
\subsubsection{{\it Tidal Moulding:  Weaving the Cosmic Tapestry, a History}}
With the gradual emergence of one particular filament seemingly predestinated by the 
tidal field configuration, it is rather logical to expect this to reflect a global process in which 
the complete cosmic foam is being cast by the overall tidal field pattern. This indeed is what 
the evolution of the cosmic web pattern in an open $\Omega_{\circ}=0.3$ FRW Universe dominated 
by CDM evidently shows in Figure 27. The particle distribution at three consecutive cosmic 
epochs ($a=0.2$, $a=0.4$ and $a=0.6$) shows the emerging cosmic foam, whose outlines have 
been marked at a relatively early time. The development of the cosmic foam essentially 
consists of a rapid increase of its density contrast as matter flows into its skeleton, 
a growth which ceases at around $a=0.3-0.4$. Evidently, the evolution between top, $a=0.2$, and 
centre panel, $a=0.4$, is substantial. This is hardly so between centre, $a=0.4$, and lower 
panel ($a=0.6$). 

The evolution and structure in the particle distribution is clearly reflected in the 
(compressional) tidal field pattern (panels right column). The tidal bars in the 
top panel already appear to outline most of the web from the onset onward. Note for intance 
the beautiful correlation between the particle content and tidal bar configuration of the 
two filaments running diagonally upward and roughly parallel from the lefthand lower corner 
of each panel. Another prominent example is the large assembly of filaments -- with clusters 
acting as nodal joints -- that runs from roughly halfway the top of the panel towards its 
righthand lower corner. It 
stands out clearly at all three consecutive timesteps. Interestingly, the filaments 
do indeed appear to be outlined from very early cosmic epochs onward while the clusters 
do not. They only show up as dominating ``tidal loops'' at later epochs, once the web 
is reaching fruition. This appears to suggest cluster locations to be determined 
by the cosmic foam or, rather, its tidal cast. In conclusion, this teaches us how 
intimately the cosmic web and the population of clusters are linked as accessories of 
the full cosmic foam machinery.
\medskip
\subsection{Hierarchical Assembly: Web Granularity}
Having identified the principal mechanism behind the foamlike geometry of 
the large scale matter distribution, it is important to point out that 
the full assembly of cosmic structure is considerably more complicated than 
a single monologous collapse into a global superstructure. 

\begin{figure}[t]
\vskip -0.3truecm
\centering\mbox{\hskip -1.0truecm\psfig{figure=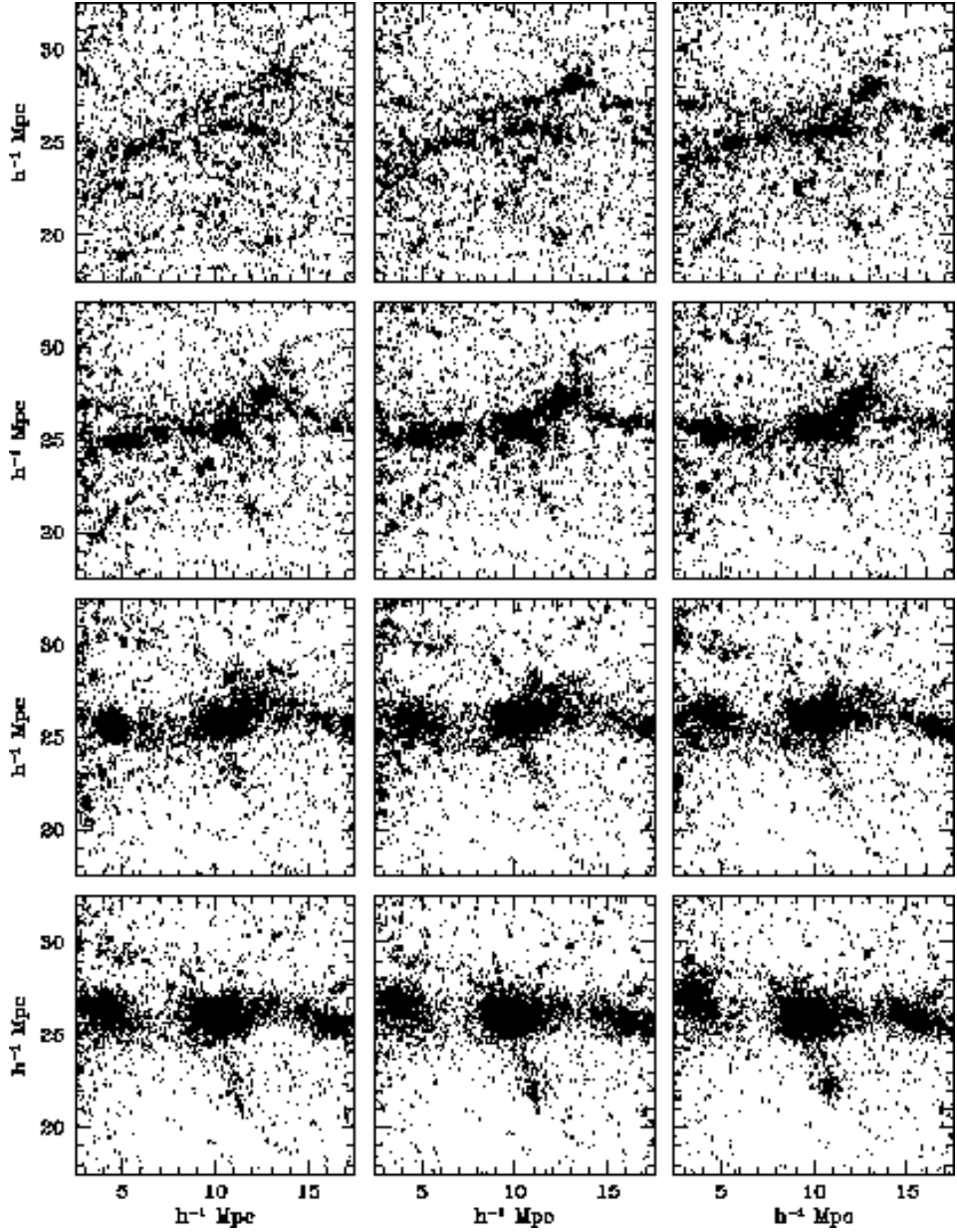,width=12.6cm}}
\vskip -0.3truecm
\caption{Gravitational Instability: hierarchical assembly of a filamentary 
structure. Focussing in on a small $15 \times 15\times 15h^{-1}\hbox{Mpc}$ 
region of a P$^3$M N-body simulation of 
structure formation in a SCDM scenario ($\Omega_0=1.0,H_0=50\,\hbox{km/s/Mpc}$), 
from the $100h^{-1}\hbox{Mpc}$ box 128$^3$ simulation shown in Fig. 12. Time is 
running from top left to bottom right, from $a=0.1$ to $a=1.0$. Notice how 
small-scale clumps aggregate into ever larger haloes, all arranged along a 
roughly filamentary configuration (oriented horizontally).} 
\vskip -0.7truecm
\end{figure}
On the contrary, the observed cosmic matter distribution is a stochastic 
superposition of a great many different fluctuations, covering a wide range of 
scales and dynamical states. Although all perturbations evolve simultaneously 
through their mutually effected cosmic gravity field, they do so by their 
own individual rate, chiefly determined by scale and physical configuration. 
The decisive quantity is their primordial amplitude, setting the dynamical 
timescale and evolution rate. Hence, the key significance of the fluctuation power 
spectrum, the function specifying the expected perturbation amplitudes 
at each spatial scale. 

What emerges is a picture of cosmic structure consisting of a complex of 
structures and features, moulded into ever larger entities, ultimately 
grouping into an encompassing foamlike Megaparsec structure. The 
individual substructures will have reached different evolutionary stages 
and hence display a diverse range of morphological geometries. Compact and 
dense clumps have emerged at highly advanced stages. More moderate 
quasi-linear phases have been reached by the features displaying the 
most pronounced anisotropic -- elongated or flattened -- geometries.  
On the very largest scales only some mild density depressions or 
enhancements can be discerned, being fluctuations still residing in the 
very early linear regime of growth. 

\subsubsection{{\it Hierarchical Assembly: Clump formation}}
In hierarchical formation scenarios small-scale fluctuations have a higher 
primordial amplitudes as the larger scale ones in which they are embedded. These 
small sized perturbations will therefore have passed through the full itinerary of 
evolutionary phases at a substantially faster rate than the ones on a 
larger scale. We will therefore observe the small scale clumps to be the 
first ones to emerge as genuine recognizable cosmic objects. The larger 
scale perturbation will not yet have matured as fully yet. Instead, we will  
observe the gradual development of the encompassing perturbation through 
merging and accretion of the smaller scale clumps, a process which may 
be readily appreciated from the computer simulation shown in Fig. 28. 
Aptly described by the concept of {\it merger tree} (see e.g. Kauffman \& 
White 1993), the precise path that an encompassing perturbation follows towards 
final collapse and virialization may be highly diverse. Both accretion onto a 
small-scale core as well as merging of more comparable substructures are 
viable evolutionary tracks. The precise historic track that will be 
followed will be determined by scale and specific physical circumstances.

At a typical galaxy scale ($\sim 0.5h^{-1} Mpc$), perturbations 
underwent gravitational collapse, virialization and -- as at these scale 
gas and radiative processes do play a dominating role -- further dissipative 
contraction of its baryonic matter content within the dark matter potential well. 
Ultimately, this yields the gaseous and luminous basic constituents of our Universe, the 
galaxies. 

Encompassing a larger primordial region, rich galaxy clusters are generically considered to 
represent the most massive and most recent (nearly) fully collapsed and 
virialized structures in our Universe. Their inner and most compact 
regions have certainly fully matured. Often they form the core of 
a wider vigorously evolving complex, while substantial amounts of 
surrounding material continuously accrete while the encompassing 
perturbation is approaching a comparable stage of development. 

\subsubsection{{\it Hierarchical Assembly:  Linear Power Law Excursions}}
Turning attention to the quasi-linear structures within these hierarchical 
scenarios -- at a present-day scale in the order of $\sim 10h^{-1}\hbox{Mpc}$ 
-- we still find the development of a foamlike geometry. However, its assembly 
will not have been the result of a simple monologous anisotropic contraction. Instead, 
it will have emerged through a preferential gravitational clumping of substructures 
along the principal directions of the corresponding large-scale anisotropic collapse. 
Hence, walls and filaments will not be uniform and coherent structures, but instead 
resemble a grouping of beads on a rod. 

In fact, nearly all viable formation scenarios may be regarded specific versions 
of the same hierarchical theme. The major differences in coherence of structures 
and other physical characteristics between the various hierarchical scenarios can 
ultimately (mostly) be traced back to differences in {\it power spectrum slope}, 
\begin{equation}
n\ = \ {\displaystyle d\ln P(k) \over \displaystyle d\ln k}\ . 
\end{equation}
\noindent The latter determines the relative dynamical state of substructures by the time 
they combine into a larger encompassing structure. If 
they are fully virialized objects a ``grainy'' configuration will emerge, while a 
roughly similar dynamical timescale over a wide range of scales -- i.e. a 
spectral slope $n \sim -3$ -- will yield a rather coherent structure. This can be 
readily appreciated by considering the relative timescales in which structures 
on various mass scales $M$ will form within hierarchical formation scenarios. To 
retain lucidity, let us restrict to a scenario with a pure power spectrum, 
\begin{equation}
P(k)\ = \ A\ k^{n}\ . 
\end{equation}
We follow a description on the basis the excursion set formalism (Bond et al. 1991) which 
extended the Press-Schechter formalism (Press \& Schechter 1974), putting it on a more robust physical footing. This analytical formalism has proven to yield remarkably good predictions on 
the sample average characteristics of an emerging population of nonlinear objects evolving 
from a linear field of density fluctuations in the primordial cosmos. 

\medskip
The pure Press-Schechter treatment makes the assumption of nonlinear objects evolving 
from initial fluctuation peaks whose shape is perfectly spherically symmetric. Then, the 
fully analytically tractable spherical model (Gunn \& Gott 1972) can be invoked to find 
relatively simple expressions for the dynamical timescales of the evolving density perturbation. 
As the analytical arguments are algebraically most tractable within the context of the 
Einstein-de Sitter Universe ($\Omega_0=1.0$), we will here restrict ourselves accordingly. 

For any spherical initial (linear) overdensity $\Delta_i$ we may then identify the 
epoch $a_{turn}$ at which its initial expanding motion turns around into contraction, the 
epoch $a_{vir}$ at which it will have contracted towards its virial radius, and in the 
academically interesting yet physically less relevant collapse time $a_{coll}$ at which it 
will have collapsed onto a singular point. Relating the dynamics of the spherically 
contracting cloud to the (hypothetical) linear evolution of the initial density field, 
the observation can be made that these dynamically characteristic (nonlinear) epochs 
can be simply related to the corresponding linearly extrapolated density excess, 
$\Delta_{lin}(t)$ of the same perturbation,
\begin{equation}
\Delta_{lin}(t)\ = \ {\displaystyle D(t) \over \displaystyle D(t_i)}\ \Delta_i\ > \ \delta_c , 
\end{equation} 
\noindent in which $D(t)$ is the density perturbation growth factor according to linear 
gravitational instability theory (Peebles 1980). Hence, once the linearly extrapolated 
$\Delta_{lin}(t)$ exceeds the proper critical (excursion) value, $\delta_c$, we will 
observe the nonlinearly evolving spherical density perturbation itself to reach the  
the corresponding characteristic phase. The value of the (excursion) values $\delta_c$ 
appear to be universal for the specific spherical case, and for an Einstein-de Sitter 
Universe are given by the values:
\begin{equation}
\delta_{c,turn}\ = \ 1.08\ ;\hskip 0.5truecm 
\delta_{c,vir}\ = \ 1.59\ ;\hskip 0.5truecm
\delta_{c,coll}\ = \ 1.69\ ,
\end{equation} 
\nonumber where we presumed the fluctuation to start its evolution on the basis of the 
growing mode solution of perturbation theory (Peebles 1980). These values will be 
slightly modified if e.g. the perturbation sets off without an initial corresponding 
velocity perturbation. 

The {\it excursion set} description thus readily identifies the characteristic stages of 
its full nonlinear evolution  -- turnaround time, virialization time and ultimate collapse 
time -- with that of reaching a critical linearly extrapolated density value, we may simply 
turn to the far simple Gaussian nature of the initial density fluctuation field to predict 
the full history of the emergence of objects from those initial conditions. As the 
formalism treats fluctuations over all scales on an equal footing, it deals with 
the collapse of density excesses over a large range of scales. This translates into  
an impressively insightful and surprisingly accurate description. The ongoing 
process of fully collapsed objects merging is the process contributing to the 
growth of the larger-scale density fluctuations. Gradually, the encompassing 
large scale entity will then reach a collapse phase where its linear (extrapolated) 
density excess passes through the critical {\it excursion} value. By working out 
the history of emerging objects according to this formulation in terms of the 
Press-Schechter philosophy, we are able to construct for any given realization of 
the primordial density field the resulting {\it merger tree} history of the 
hierarchical process (Bower 1991, Kauffman \& White 1993).

By reducing the nonlinear formation history to one in which we can 
focus on the initial fluctuation field, the problem has been simplified to 
one in which we deal with simple Gaussian statistics. This then leads to a situation 
in which we can readily compute at each cosmic epoch $t$ the fraction $F(M,t)$ of 
matter fluctuations on a given mass scale $M$ that will have fully collapsed. Its 
time dependence is fully incorporated through the evolving value of the density field 
dispersion value $\sigma(M,t)$, 
\begin{equation}
\sigma(M,t)\ = \ {\displaystyle D(t) \over \displaystyle D(t_i)}\ \sigma(M,t_i)\ . 
\end{equation}
\noindent Then, the fraction of linearly extrapolated density fluctuations in excess of the 
critical value $\delta_c$,  
\begin{eqnarray}
F(M,t)&\ = \ &{\displaystyle 1 \over \sqrt{2 \pi}\ \sigma(M,t)}\ \int_{\delta_c}^{\infty}\ 
\exp\left\{ - {\displaystyle \delta^2 \over \displaystyle 2 \sigma^2(M,t)} \right\}\ d\delta 
\ \nonumber\\ 
\nonumber\\
\nonumber\\
&\ = \ &{1 \over 2}\ \biggl[1\ - \hbox{\rm erf}\ \biggl\{{\displaystyle \delta_c \over 
\displaystyle \sqrt{2}\sigma(M,t)}\biggr\}\ \biggr]\ ,
\end{eqnarray}
\noindent in which $\hbox{\rm erf(x)}$ is the conventional error function. Notice that by this 
definition of $F(M)$ we deal with the full range of density values, thereby delimiting a mere 
$50\%$ to overdense fluctuations. Instead of computing the fraction $F(M,t)$ for a given value 
of the $\sigma(M,t)$, we may invert this by computing the cosmic epoch $t$ at which a 
particular fixed fraction $F_c$ of the density fluctuations on mass scale $M$ have collapsed 
(or, equivalently, have turned around or collapsed to the virial radius), and hence the 
time at which $\sigma(M,t)$ has reached a corresponding critical value $\sigma_c$.  
\begin{equation}
\sigma(M,t)\ = \ \sigma_c \ \equiv \ {\displaystyle \delta_c \over \displaystyle \sqrt{2}}\ \hbox{\rm erfinv}\left\{1 - 2 F_c\right\}\ ,
\end{equation}
\noindent with $\hbox{\rm erfinv}(x)$ the inverse error function. By comparing the corresponding 
$\sigma(M,t)$ on two mass scales $M_1$ and $M_2$ we find the ratio between the density 
growth factors $D(t(M_1))\equiv D_t(M_1)$ and $D(t(M_2)\equiv D_t(M_2)$ at which the density 
field at both scales has reached the same level of ``fruition''. For linearly evolving 
fluctuations, $\sigma_c=(D_t(M)/D_{\circ}) \sigma_{\circ}$, and thus  
\begin{equation}
{\displaystyle D_t(M_1) \over \displaystyle D_t(M_2)}\ = \ {\sigma_{\circ}(M_2) \over 
\sigma_{\circ}(M_1)}\ .  
\end{equation}
\noindent where the linear density fluctuations on the scale $M$ are normalized to their 
present day values $\sigma_{\circ}$. These are directly related to the power 
spectrum $P(k) (\propto k^n)$ of the density field via the corresponding length scale,
\begin{equation}
\sigma^2(M)\ = \ \biggl\langle\ \left({\displaystyle \delta \rho \over 
\displaystyle \rho}\right)^2\biggr\rangle\ \ \propto \ \ M^{-(n+3)}\ .
\end{equation}
\noindent We therefore see that the characteristic dynamical times on different mass 
scales $M$, specified in terms of the linear growth factor $D_t$ of matter fluctuations, can be 
straightforwardly related to the spectrum of primordial density fluctuations, 
\begin{equation}
{\displaystyle D_t(M_1) \over \displaystyle D_t(M_2)}\ = \ \left(\displaystyle 
M_2 \over \displaystyle M_1\right)^{(n+3)/2}\ .
\end{equation}
\noindent On the basis of the above we find major differences in the qualitative progression of 
the hierarchical evolution process for fluctuation spectra with different spectral 
slopes $n$. Proceeding from a pure white noise spectrum with $n=0$ to spectra with 
ever more negative values of $n$ ($0> n > -3$), we notice a diminishing ratio of 
characteristic timescales. Hence, in a scenario with $n=0$ we will see the full 
collapse and virialization of small-scale $M_1$ clumps before fluctuations on an order 
of magnitude larger scale $M_2$ start to approach a similar stage. By contrast, such 
$M_1$ fluctuations will hardly get the time to fully virialize before the 
encompassing larger scale $M_2$ fluctuations reach a similar dynamic state. 

\subsubsection{{\it Hierarchical Assembly: Granulated Filaments}}
Taking the specific example of an emerging filament, its formation will consist of the 
gradual assembly of earlier virialized small-scale clumps. A rather grainy feature 
will be the result. In an $n=-2$ scenario, on the other hand, the contracting filament 
will be collapsing while its contents  
in smaller scale clumps have not yet had the opportunity to fully settle. Often 
these have not yet even reached a stage of full virialization, and perhaps still reside in 
a stage with a pronounced anisotropic geometry (i.e. only one or two of its principal 
axes have passed on to full collapse). Such scenarios will produce coherent large-scale 
filaments in which the internal small-scale structure is only moderately 
visible, if at all once they get fully merged with the encompassing superstructure. 
Most dramatic will be the $n=-3$ scenario, which is the asymptotic situation in 
which fluctuations over the full range of scales undergo contraction and collapse 
at the same time. It marks the fundamental limit for genuine hierarchical clustering. 
In other words, we have identified a sliding scale of morphological constituency and 
appearance within the context of possible hierarchical structure formation scenarios. 

\subsubsection{{\it Hierarchical Assembly: Shaping Up}}
Unlike the basic Press-Schechter formulation followed above, based on spherical 
configurations, a realistic formalism should take into account the intrinsically 
anisotropic shape of (primordial) density peaks. From the statistics of linear 
Gaussian random fields (e.g. Bardeen et al. 1986) 
we know that spherical density peaks do not exist, all peaks have at least some 
degree of primordial flattening, be it never extreme. Referring to the growing 
anisotropy of homogeneous ellipsoidal overdensities, it is evident that this 
will involve considerable repercussions for the evolution of each individual 
peak. Indeed, merely correcting for this by means of an approximate idealized (isolated) 
homogenous ellipsoidal model has already proven to yield a significant 
improvement on quantitative aspects and details such as the predicted mass function 
of collapsed clumps (Sheth, Mo \& Tormen 2001). 

For the purpose of our focus on the morphology of filaments, a major consequence  
is the corresponding diffusion in collapse time scales. Collpase times will no longer 
only be set by the value of the initial overdensity, its primordial shape will modify this 
to a considerabel extent. It may lead to situations in which clumps with a higher amplitude 
initial overdensity may reach ultimate collapse later than a more moderate overdensity, 
in the case of its primordail shape being sufficiently more elongated. 

For our considerations concerning flattened and filamentary features these modifications 
will be even more relevant. We are interested in exactly those scales 
at which the large-scale object is still residing in one of its intermediate 
anisotropic phases. Unlike issues of final halo mass spectrum, here collapse has not 
yet progressed through the full itinerary of the ultimate 3-D collapse unto a compact clump.  
One property of anisotropic collapse is the differentiation in collapse times along 
each of the principal directions of the object, inducing a diffusion in relative 
collapse timescales (see the axis evolution of a homogeneous ellipsoid, 
Fig. 20). The issue of comparative {\it dynamical timescales} at different spatial scales 
of hierarchically embedded structures will therefore be even more contrived than in the 
simple approximate scheme presented above. 

For instance, in a strongly elongated case the longest axis may not yet have reached the 
contraction phase while along the shortest axis the object is well on the way towards 
full collapse, reaching this stage on a much shorter timescale than it would have 
done in the equivalent spherical case. At some given scale a structure may therefore 
not yet have evolved to an evolutionary state as advanced as its peer would have done 
in the purely spherical case (because its final collapse direction, along the longest axis, 
has evolved more slowly). On the other hand, the encompassing overdensity on a larger scale 
may have advanced up to a stage of contraction along the shortest axis at a considerably 
faster rate than it would have done in the equivalent spherical situation. With respect 
to a purely spherical formalism, it would involve a convergence of evolutionary 
timescales of structure at different spatial scales. 

Nonetheless, we may consider expression (65) as a useful guiding principle. It  
represents a robust lower limit to an effect that may get aggravated by the 
diffusion of timescales resulting from the modifications on the basis of peak 
anisotropy. These would amplify the effect and therefore, if anything, 
underline the presented analytical arguments even more strongly.

In summary, in a scenario with a higher value of spectral slope $n \uparrow 0$ 
(towards white noise) the interscale difference in dynamical timescales is 
sufficient large that we will still be able to distinctly identify between 
the subsequent evolutionary phases at various scales. In such a scenario, the 
typical ``granular'' character will be retained. On the other hand, 
as the slope $n$ of power law scenarios tends more and more towards lower 
values, $n \downarrow -3$, the emerging coherent superstructures will assume 
an increasingly dominating stature over their substructure. The contrast of 
the latter will diminish as $n$ decreases, while the filamentary and/or 
wall-like nature of the superstructure will stand out stronger and stronger. 

Even further complications are awaiting us when focussing on the detailed 
nonlinear gravitational effects occurring in the interior of the web 
structures. ``Interscale'' nonlinear gravitational interactions 
will be linking up features at different scales, producing an 
additional amplification of their alignments.

\subsubsection{{\it Hierarchical Assembly:  Nonlinear Aggravation}}
The ``Press-Schechter'' type linear arguments presented above in essence 
involve a simple ``isolated nonlinear'' extrapolation of features 
identified in the primordial density field. It does not take into 
account the nonlinear interactions between the features forming at 
various scales. In fact, the hierarchical congregation of filamentary 
structures is considerably complicated by a variety of such nonlinear 
effects.

A systematic alignment of flattened peaks with respect to their 
surroundings is a generic property in any Gaussian linear initial density 
field (e.g. Bardeen et al., Van de Weygaert \& Bertschinger 1996). These primordial 
alignments get significantly amplified by the subsequent infall of clumps 
from the surroundings, in which several extra nonlinear effects may be 
discerned. 

Subclumps rain in along directions set out the large-scale configurations. The 
hierarchical buildup of a filament makes the impression of a process of a 
gradual and continuous inflow of small clumps alon the direction of the 
filament towards the highly dense connecting clusters. It's as if filaments act 
like migration channels of the emerging cosmic web. 

Earlier, we have seen that the location of these channels can be traced back to the 
primordial tidal field pattern. The morphology and nature of filaments -- strong, dominating, 
large and coherent or bearing the character of short, weak and erratic hairlike extensions 
connected to nearby peaks -- will be of decisive influence over aspects like the 
angular distribution of clumps infalling towards a cluster. An interesting nonlinear 
alignment amplification is involved with this process. 
\begin{figure}[b]
\centering\mbox{\hskip -0.8truecm\psfig{figure=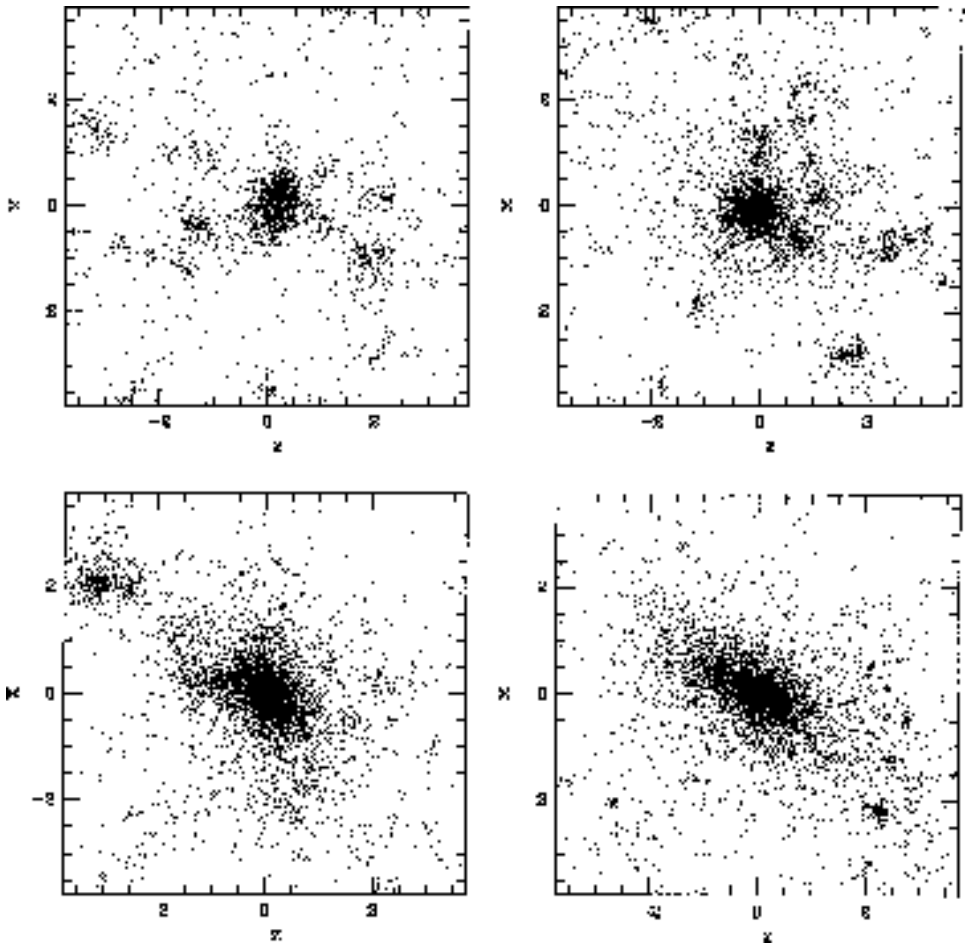,width=12.7cm}}
\vskip -0.3truecm
\caption{Infall onto forming cluster: infall from arbitrary directions, typical 
for a $P(k) \propto k^0$ white noise structure formation scenario (Einstein-de Sitter 
Universe, $\Omega_0=1$. From Van Haarlem \& Van de Weygaert 1993. Reproduced by permission of the 
AAS.}
\vskip -0.5truecm
\end{figure}
\begin{figure}[t]
\centering\mbox{\hskip -0.8truecm\psfig{figure=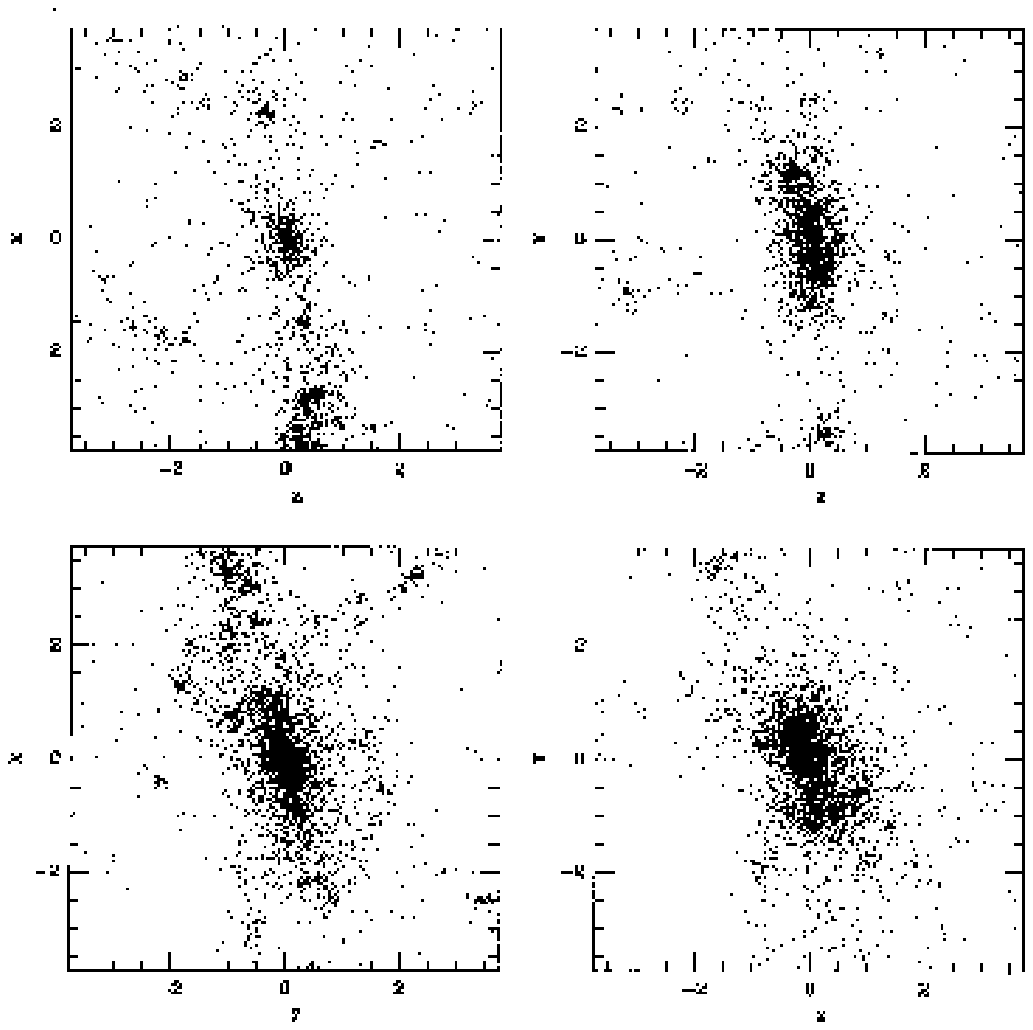,width=12.7cm}}
\vskip -0.3truecm
\caption{Infall onto forming cluster: channelling of infalling clumps via a 
filament connecting to the cluster, typical in a CDM structure formation 
scenario in a Einstein-de Sitter Universe ($\Omega_0=1$). From Van Haarlem \& 
Van de Weygaert 1993. Reproduced by permission of the AAS.}
\vskip -0.0truecm
\end{figure}

Van Haarlem \& Van de Weygaert (1993, HW) focussed on the infall pattern of clumps 
as they are channelled through the filament towards heavy clusters (see Fig. 29 and 
Fig. 30, from HW). As expected, the infall pattern is heavily influenced by the strength,  
contrast and multitude of the filamentary connections of the cluster towards the 
cosmic surroundings. In Fig. 29 and Fig. 30 we see the difference between the 
infall towards a cluster arising in a SCDM formation scenario and that towards a 
cluster in a pure white noise ($P(k) \propto k^0$) scenario, lacking any substantial 
large-scale power. In the case of the SCDM scenario (Fig. 30, with a slope $n \approx -1$ 
on the relevant cluster scale) we observe a pronounced and dominating filament. It induces 
a pattern of continuously infalling subclumps, all entering along the one outstanding 
direction defined by the filament. In the other set of frames (Fig. 29), we see the effects 
of a typical ``white noise'' ($P(k)\propto k^0$) scenario, with small scale clumps 
having fully settled before one can even start to notice the presence of features on 
a larger spatial scale. Upon the cluster-like core finally having arrived at a stage 
of contraction, it is accompanied by an isotropic pattern of small clumps continuously raining in 
from all over the ``sky'' 

Interestingly, a cluster appears to orient itself towards the direction along which the 
last substantial subclump came falling in (HW). This 
is a pure nonlinear gravitational effect, involving the equipartition and virialization of 
the energy and momentum contained in the ``particles'' of the infalling clump. The 
preferential direction defined by the infall direction of the clump, and hence of the 
major share of the linear momenta of its constituent ``particles'', then leads to 
an anisotropic redistribution in the phase space of the resulting merger. 

As a consequence of the cluster orienting itself towards the last infalling clump 
the angular distributions of the infalling objects assume an even more significant 
influence. In the $P(k)\propto k^0$ scenario, the orientation of clusters will 
hardly have any systematic correlation with the surrounding matter distribution. This 
stands in marked contrast to the situation in the presence of a pronounced filament. 
The exclusive and continuous infall of clumps along the spine of the dominating 
filament (see Fig. 30, from HW) indueces a strong aligment of cluster orientation, its 
substructure and the cosmic surroundings. This was most manifestly depicted in the 
sky distribution of infall directions of clumps  onto the evolving cluster complexes. 
In the white noise $n=0$ situation, the ``sky'' pattern did not reveal any 
preferred direction. On the other hand, an outstanding and stable infall angle could 
easily be identified in the case of the CDM cluster (see HW, Fig. 30). 

In summary, in hierarchical scenarios with a relatively high level of 
large scale power we can discern a variety of mutually 
amplifying factors contributing to the development of pronounced morphologies. 
Partially, this had already be predestined by their primordial shape. Due to 
the spatial correlations in primordial density fields, matter fluctuations are 
intrinsically aligned. Adding to to such linear primordial circumstances, and 
occasionally dominating over them, are the various nonlinear couplings 
between surrounding and embedding structures within the matter distribution. 

The implied alignment of clusters with surrounding large scale structure has been 
adressed in a variety of observational studies. Conclusive evidence is hard to unveil due to a 
plethora of disturbing physical influences and processes. A few studies tried to find indications 
through the presence of a significant cluster-cluster alignment (Binggeli 1982, Rhee \& Katgert 
1987). Other analyses seek to investigate possible vestiges of the cluster infall process on 
the remaining substructures. One related interesting effect may be that  
the inflow rate of subclumps becomes a significantly more efficient process through 
the presence of filaments. A strong indication for the reality of such an effect is the 
recent work by Plionis \& Basilakos (2001), who disclosed a tight link between 
alignment of clusters with respect to their surroundings and the presence of substructure. 
Other tantalizing consequences may be a possible trace left in the morphology of 
infalling galaxies, which will certainly be effected by the influences to which they 
get subjected upon their arrival in the clusters. Indications for such morphological 
tendencies have been found by Thomas \& Katgert (personal communication, see Thomas 2002).
\medskip
\subsection{Voids holding Sway}
It is with some justification that most observational attention is directed to regions where most 
matter in the Universe has accumulated. Almost by definition they are the sites of most 
observational studies, and the ones that are most outstanding in appearance. 
However, inspired by early computer calculations, Icke (1984) pointed out that for the 
understanding of the formation of the large coherent patterns pervading the 
Universe it may be more worthwhile to direct attention to the complementary  
evolution of underdense regions, the progenitors of the observed voids. 

\subsubsection{{\it The Bubble Theorem}}
Icke (1984) made the interesting observation that the arguments presented 
for the anisotropic collapse of overdensities, when approximated by 
that of homogenous ellipsoids, are equally valid when considering the 
evolution of {\it low\/}-density regions. These low-density regions are the 
progenitors of the observed voids. Note that although uniform ellipsoids 
at first appear to be a rather artificial configuration, they do 
represent proper second-order approximations to the density field in the 
immediate vicinity around a peak or dip, a fact that may be easily appreciated 
from the fact that the smallest closed contours in any topographical map are ellipses. 
While for underdensities the same equations are used for this approximation, 
the {\it quintessential} observation is that the sense of the final effect is 
reversed. Because a void is effectively a region of negative 
density in a uniform background: 
\begin{itemize}
\item{} {\it Expansion}\\
Voids expand as overdense regions collapse
\item{} {\it Spherical shape}\\
slight asphericities decrease as the voids become larger. 
\item{} {\it Velocity field}\\
The (peculiar) velocity field has a Hubble-type character, linear in position: 
super-Hubble expansion.  
\end{itemize}
The second point can be simply deduced from the observation that with respect to an equally 
deep spherical underdensity, an ellipsoidal void has a decreased rate of expansion along 
the longest axis of the ellipsoid and an increased rate of expansion along 
the shortest axis. Moreover, one may readily appreciate that the uniform density 
of homogeneous ellipsoids corresponds to a velocity field that will be a linear function of 
position, so that in the interior of such a void we will observe a Hubble-type 
velocity field. In summary, voids will behave like low-density `super-Hubble' expanding 
patches in the Universe. To describe this behaviour the term 
``Bubble Theorem'' (Icke 1984) was coined. 

Evidently, we have to be aware of the serious limitations of the ellipsoidal 
model. It disregards important aspects like the presence of substructure. 
More serious is the neglect of any external influence, whether secondary infall, 
``collision'' with surrounding matter, or the role of nonlocal tidal fields.  
Yet, comparison with the evolution of voids in realistic clustering scenarios 
shows that in the case of voids, it tends to become a better description as 
time proceeds, in particular for the very inner regions. N-body simulations 
clearly bear out that the density fields in the central region of the (proto)void 
will flatten out while the voids expand and get drained (Fig. 31, from Van de Weygaert 
\& Van Kampen 1993). Hence, voids develop into regions of a nearly 
uniform density and the region of validity of the approximation grows accordingly. 

\subsubsection{{\it Soapsud of Expanding Voids}}
By contrast to the overdense features, the low-density regions start to take 
up a larger and larger part of the volume of the Universe. Upon their discovery, the 
independent dynamical role of underdense regions was not immediately appreciated, many 
considering them mere byproducts in the form of space evacuated by contracting high 
density clumps. Once it got realized that also dips and valleys in the pristine density 
field may develop a distinct dynamical evolution of their own (e.g. Hoffman \& Shaham 1982), 
it was straightforward to see that such underdense regions must play an essential 
and independent dynamical role in the formation of cosmic structure. 
\begin{figure}[h]
\centering\mbox{\hskip -0.8truecm\psfig{figure=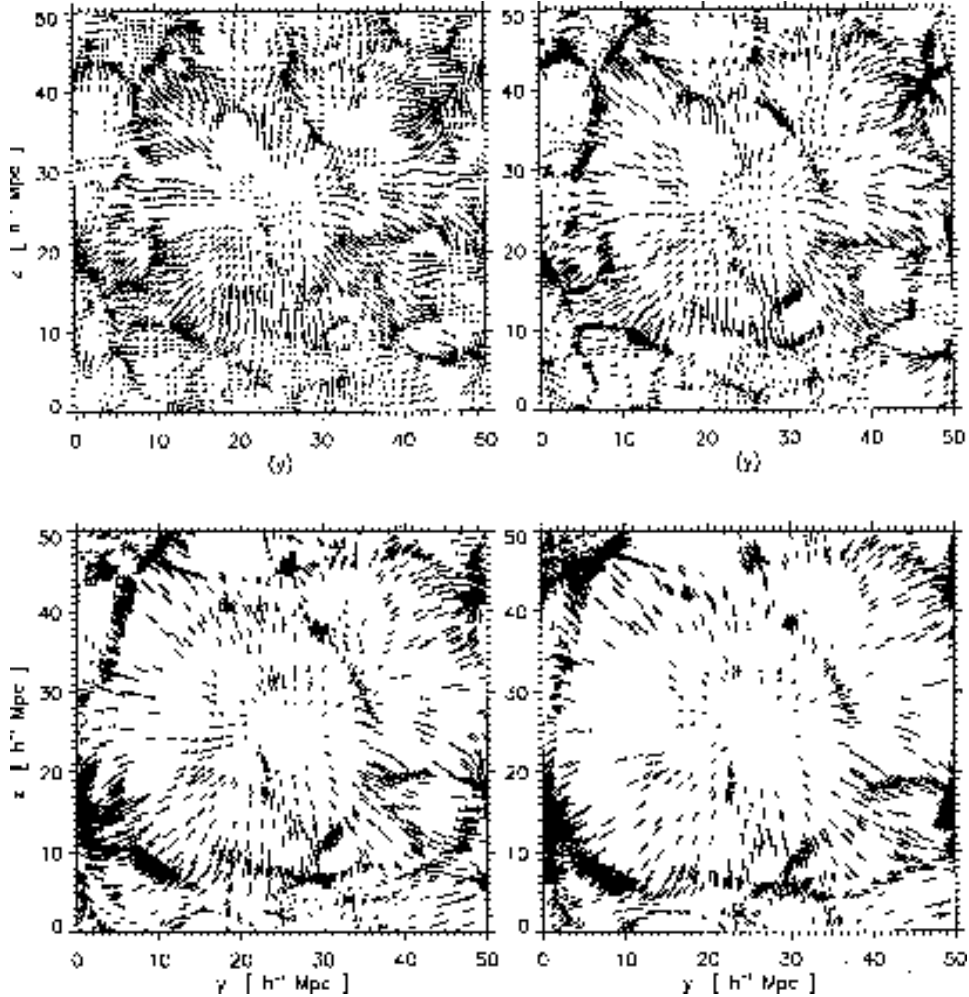,width=12.7cm}}
\vskip -0.3truecm
\caption{The evolution of a void in a constrained SCDM N-body simulation of 
a $3\sigma_0(4h^{-1}\hbox{Mpc})$ void. Shown are particle distributions 
at expansion factors $a=0.2, 0.4, 0.7$ and $1.0$. The particle position is 
indicated by a dot, forming the base of the corresponding velocity vector 
within the slice. From van de Weygaert \& van Kampen 1993. Reproduced by 
permission of the Royal Astronomical Society.}
\vskip -0.1truecm
\end{figure}
Even though the value of their underdensity cannot surpass the natural 
value of $-1.0$ -- nothing can be emptier than empty -- their growing size 
may compensate to achieve a dynamical influence akin to that of 
a considerably compacter high-density clump, as long as their coherence scale 
is such that their effective mass is comparable. While they grow to occupy a larger 
and larger fraction of the Universe, it will be as if matter in the intervening 
high-density domains will gradually be swept up in the wall-like and filamentary 
interstices, 
yielding a natural explanation for the resulting coherence of the cosmic foam. 
\begin{figure}[t]
\vskip -0.0cm
\centering\mbox{\hskip -0.5truecm\psfig{figure=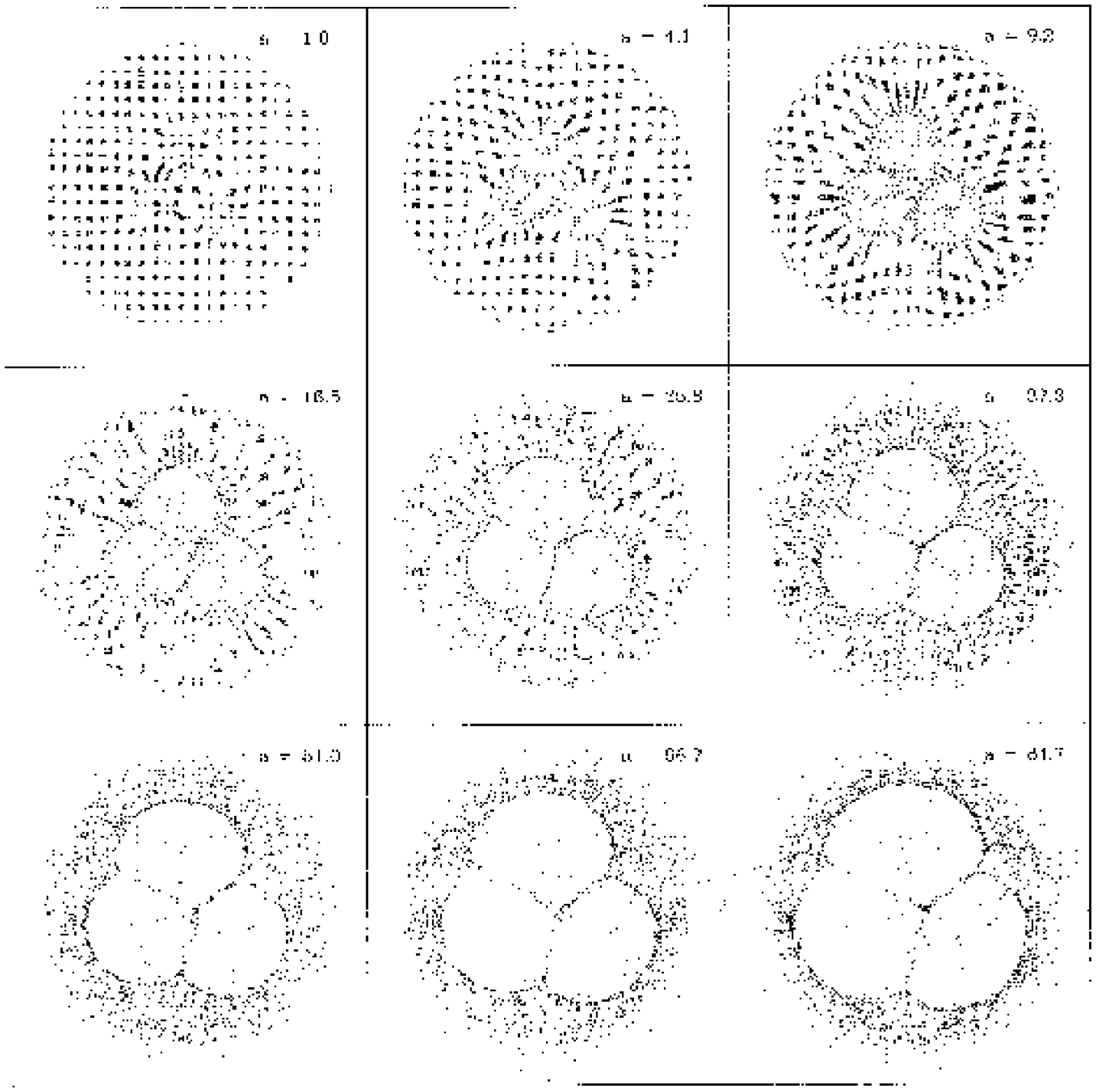,width=12.7cm}}
\vskip -0.25cm
\caption{Evolving void hierarchy. Illustration from Dubinski etal. 1993. A nested 
set of a large spherical tophat void filled with 3 smaller yet deeper ones, each of 
them in turn filled with another set of 3 even deeper and smaller voids. Illustrated 
are the comoving poisition for one slice through the centre of the sphere at 
different cosmic epochs. At different epochs, different void scales emerge based 
on their initial depths. As the evolution proceeds, substructure freezes in a 
network of walls. Courtesy: John Dubinski. Reproduced by permission of the 
AAS.}
\vskip -0.25truecm
\end{figure}
In realistic circumstances, expanding voids will sooner or later encounter 
their peers or run into dense surroundings. The volume of space available to 
a void for expansion is therefore restricted. Voids will also be influenced by the 
external cosmic mass distribution, and substructure may represent an additional 
non-negligible factor within the void's history. In general, we deal with a 
complex situation of a field of expanding voids and collapsing peaks, of voids and 
peaks over a whole range of sizes and masses, expanding at different rates and 
at various stages of dynamical development. For the purpose of our geometric 
viewpoint, the crucial question is whether it is possible to identify some 
characteristic and simplifying elements within such a complex. Indeed, simulations 
of void evolution (e.g. Dubinski et al. 1993) represent a suggestive illustration of 
a hierarchical process akin to the {\it void hierarchy} seen in realistic 
simulations (e.g. Van de Weygaert 1991b). It shows the maturing of small-scale voids 
until their boundaries would reach a shell-crossing catastrophe, after which they merge 
and dissolve into a larger embedding void. This process gets continuously repeated 
as the larger parent voids in turn dissolve into yet larger voids. At any one cosmic 
epoch there appears to be a characteristic void size, the one corresponding to the 
typical tophat void shell-crossing scale (see Fig. 32, from Dubinski et al. 1993). 
Interestingly, a crude estimate shows that for a large range of primordial spectra voids 
of such size would approximately constitute a volume-filling network. 

\subsubsection{{\it Void Hierarchy}}
Indeed, a detailed assessment of the void hierarchy as it evolves from a primordial Gaussian 
density field (Sheth \& Van de Weygaert 2002) suggests the gradual disappearance of 
small voids as they merge and get absorbed into the encompassing underdensities, while 
colossal and large voids would be rare by virtue of the fluctuation field statistics, 
the mainstay of voids would have sizes within a rather restricted range. 
Corresponding calculations yield a void size distribution (broadly) peaked 
around a characteristic void size.  

\subsubsection{{\it Voids: Fragmenting the Universe}}
A bold leap then brings us to a geometrically interesting situation. Taking the 
voids as the dominant structure-shaping component of the Universe, and following 
the ``Bubble Theorem'', we may think of the large scale structure as a close 
packing of spherically expanding regions. Then, approximating a peaked 
void distribution by one of a single scale, we end up with a situation 
in which the matter distribution in the large scale Universe is set up 
by matter being swept up in the bisecting interstices between spheres 
of equal expansion rate. This {\it ASYMPTOTIC} description of the 
cosmic clustering process leads to a geometrical configuration that 
is one of the main concepts in the field of stochastic geometry: 
{\cal{VORONOI TESSELLATIONS.}}
\medskip

\subsection{The Cosmic Foam:}
\begin{flushright}
{\bf{\large Pulling together the Strings}}
\vskip -0.2truecm
\end{flushright}
The preceding sections compel us to conclude that one of the most 
{\it prominent manifestations} of structure formation driven by 
the {\it force of gravity} is a strong and persistent tendency of matter 
to aggregate into weblike networks of filaments and walls. 

\subsubsection{{\it The Cosmic Foam: Dynamic Essence}}
The basic mechanism behind this tendency can be most straightforwardly 
illuminated on the basis of the dynamical evolution of the simplified asymptotic 
configuration of isolated homogenous ellipsoidal overdensities. 
A subsequent elaboration towards the generic context of a general 
field of stochastic density fluctuations is worked out by the 
first-order Lagrangian formalism of the Zel'dovich approximation, 
comprising quasi-linear displacements which have proven to retain 
such surprising validity over a long cosmic time. From these two 
idealizations focussing on the relevant core issues, we have come 
to learn that the continuously increasing tendency towards 
matter migration flows into ever more flattened and, ultimately, elongated 
structures is ultimately stemming from the accompanying generic anisotropies 
in the gravitational force field. In first instance, these anisotropies are 
a natural complement of the spatially stochastic, random, nature of the 
primordial density field. However, once the emerging matter features have 
developed pronounced anisotropic shapes, the tendency gets strongly 
reinforced, which leads to a cosmic pattern with pronounced features 
of high contrast, connecting into the cosmic foam. 

\subsubsection{{\it The Cosmic Foam:  Dynamic Elaborations}}
Subsequent elaborations of more detailed and careful considerations of 
the processes involved with the gravitational clustering allow us to 
appreciate important correlated issues. Failing to deal with the growing 
selfgravity of emerging structures, and therefore not being able to 
explain the apparent solidity, cohesion and temporal persistency of 
e.g. filamentary structures, the Zel'dovich approximation needs to 
be supplemented by more elaborate schemes. Indeed, a variety of 
analytical approximation schemes and descriptions have attempted to 
assimilate the selfgravity of structures, usually enhancing the solidity 
and cohesion of massive structures by design. Analytical 
non-linear approximation schemes which sought to extend the Zel'dovich scheme -- 
like the adhesion model, the frozen flow approximation and the truncated Zel'dovich 
approximation (see Sahni \& Coles 1995, for an extensive and balanced review) -- 
without exception produce pronounced and compact weblike structures. 

Less through the insight of an analytical approximation, based on 
some well-chosen and balanced assumptions, than through their capacity to reproduce 
the real world as good as possible in as much detail as technically feasible, 
elaborate and sophisticated full-scale gravitational N-body computer 
simulations have presented the most convincing evidence for the 
overall prominence of foamlike patterns. These N-body simulation, 
handling ever more complex and sophisticated situations, have made clear that 
overdensities -- on any scale and in any scenario -- indeed tend to collapse such 
that they become increasingly anisotropic. At first they turn into a flattened `pancake', 
possibly followed by contraction into an elongated filament. Note that 
such structures may still expand along one direction, even while having 
collapsed along any of the other ! 

\subsubsection{{\it The Cosmic Foam:  Relaxation and Cosmic Amnesia}}
Ultimately, the evolutionary phase marked by the pronounced geometrical 
pattern of the cosmic web will give way to yet more advanced stages wherein virialization 
starts to assume a dominant role. The object finally settles down into a 
quasi-equilibrium virialized state of its internal structure and kinematics. Galaxies 
and clusters are evident examples of objects that have reached this stage. Even 
though highly nonlinear objects will retain some memory of past flattened and 
elongated geometries, in the virialization process a substantial fraction gets 
evened out. Hence, we encounter the most pronounced anisotropies in stages of 
moderate quasi-linear dynamical evolution, that in which the object has contracted 
along one or two dimensions, but not yet reached full nonlinear collapse. 

\subsubsection{{\it The Cosmic Foam:  Fossils in Space}}
Regardless of their internal morphology, we have therefore arrived at the 
point at which we can fully appreciate the unique position of the cosmic 
web within the overall scheme of cosmic organization. 

The cosmic web and its constituting structural elements form marked and 
characteristic features within the spatial cosmic distribution of matter, 
structures with dimensions in the 10-500 Mpc regime that still reside at 
a unique stage of their dynamical development. On these intermediate Megaparsec 
scales features have as yet only evolved mildly since the recombination epoch. 
While sufficiently pronounced to analyze and scrutinize their structure and 
dynamics, they have not yet passed through the more complex nonlinear phases 
wherein orbit mixing of the accompanying migration 
flows and virialization of the matter content have upset causal relationships 
and rendered orbit inversion a nontrivial and cumbersome procedure. 

Partially related is the fact that on these scales it suffices to use a 
simple ``dust'' equation of state. In most formation scenarios nonbaryonic 
dark matter is the dominant gravitational component, essentially setting the 
gravititional potential wells, while for all possible scenarios dissipative gas 
and radiative processes may be conveniently set aside at these large scales. Also 
beneficial is the fact that on these scales we can also circumvent the complexities 
of a full General Relativistic description of the gravitational forces involved, 
simple Newtonian gravitational instability provides a more than appropriate 
accurate description. 

Hence, the Megaparsec structures joining into the cosmic web may be justifiably 
portrayed as genuine ``cosmological fossils''. It is them who contain, more 
directly tangible than any other object in our Cosmos, the keys for unlocking 
the enigma to the emergence of the Universe's infrastructure !!! 

\subsubsection{{\it The Cosmic Foam:  Unravelling the Cosmic Pattern}}
All in all, we conclude that the cosmos has been supplied with a dominant 
force of gravity which not only determines its global development and fate, 
but also takes care of a truely enticing internal matter distribution. 
It is the  {\it generic anisotropic nature of gravitational contraction and 
collapse} that acts as the {\it principal cause responsible} for the 
characteristic {\it foamlike appearance} of the cosmic matter distribution. 

Direction closer attention to its ensuing dynamical evolution, we have 
come to realize that the scale of the presently observed cosmic foam is exactly 
the one corresponding to a stage of mild nonlinearity, the stage at which 
structures tend to acquire their most pronounced stage. Therefore, more so 
even then through the sheer intrinsic beauty of the complex geometric patterns 
themselves, we have come to appreciate why in the study of structure 
formation it is the {\it the cosmic foam} which should be branded as its 
most {\it fundamental manifestation} ! 

Also, we have come to appreciate the existence of major technical obstacles towards 
unravelling the cosmic secrets contained within the cosmic foam. The absence of any 
informative instrument for analyzing and exploiting the characteristic, intrinsically 
geometric, properties of the cosmos' interior matter arrangement implies us 
to accept a disregard for and squandering of highly relevant information. 
Hence, following the alternative and complementary track of adressing the stochastic 
geometric nature of the cosmic web, we seek to define a path towards a more fundamental 
understanding of its geometric aspects. Such insight will pave the way towards a 
better and more meaningful exploitation of the treasure trove of information on the 
process of structure formation contained in the salient frothy patterns we have 
found to permeate our Universe. 
\medskip
\section{\rm{\Large CELESTIAL POLYHEDRA:...}}
\begin{flushright}
{\rm{\Large  Tessellating the Universe}}
\end{flushright}
Following the philosophy delineated above, and continuing the arguments leading to the 
concept of Voronoi tessellations, we proceed by construct the ``skeleton'' of  the mass 
distribution by considering the locus of points towards which the matter streams out of  
voids. The premise is that some primordial 
cosmic process generated a density fluctuation field. In this random density field we 
can identify a collection of regions where the density is slightly less 
than average or, rather, the peaks in the primordial gravitational potential perturbation 
field. As we have seen, these regions are the seeds of the voids. These underdense patches 
become ``expansion centres'' from which matter flows away until it runs into its 
surroundings and encounters similar material flowing out of adjacent void, as 
indeed is observed with the CDM void in Fig. 31. Notice also that the dependence 
on the specific structure formation scenario at hand is entering via the spatial 
distribution of the sites of the density dips in the 
primordial density field, whose statistical properites are fully determined
by the spectrum of primordial density fluctuations.

Matter will collect at the interstices between the expanding voids. In the 
asymptotic limit of the corresponding excess Hubble parameter being the 
same in all voids, these interstices are the bisecting planes, perpendiculary 
bisecting the axes connecting the expansion centres. 
For any given set of expansion centres, or {\it nuclei}, the arrangement of these 
planes define a unique process for the partitioning of space, a {\it Voronoi 
tessellation\/} (Voronoi 1908). A particular realisation of this process (i.e. a 
specific subdivision of $N$-space according to the Voronoi tessellation) may be 
called a {\it Voronoi foam\/} (Icke \& Van de Weygaert 1987). 

\subsection{Voronoi Tessellations:}
\begin{flushright}
{\bf{\large the Geometric Concept}}
\vskip -0.2truecm
\end{flushright}
\noindent A Voronoi tessellation of a set of spatially distributed 
nuclei is a space-filling network of polyhedral cells (see Fig. 34), each of which delimits that 
part of spacethat is closer to its nucleus than to any of the other nuclei. Hence, each 
Voronoi region $\Pi_i$ is the set of points which is nearer to nucleus $i$ 
than to any of the other nuclei $j$ in a set $\Phi$ of nuclei $\{x_i\}$ in
$d$-dimensional space $\Re^d$, or a finite region thereof,
\begin{equation}
\Pi_i = \{{\vec x} \vert d({\vec x},{\vec x}_i) < d({\vec x},{\vec x}_j)\ ,
\ \forall\ j \not= i \}\ ,
\end{equation}
where ${\vec x}_j$ are the position vectors of the nuclei in $\Phi$, and $d({\vec x},{\vec y})$ 
the Euclidian distance between ${\vec x}$ and ${\vec y}$ (evidently, one can extend the 
concept to any arbitrary distance measure). From this basic definition, we can 
directly infer that each Voronoi region $\Pi_i$ is the intersection of the open 
half-spaces bounded by the perpendicular bisectors (bisecting planes in 3-D) of 
the line segments joining the nucleus $i$ and any of the the other nuclei. This  
implies a Voronoi region $\Pi_i$ to be a convex polyhedron (or polygon when in 2-D), 
a {\it Voronoi polyhedron}. 

\begin{figure}
\centering\mbox{\psfig{figure=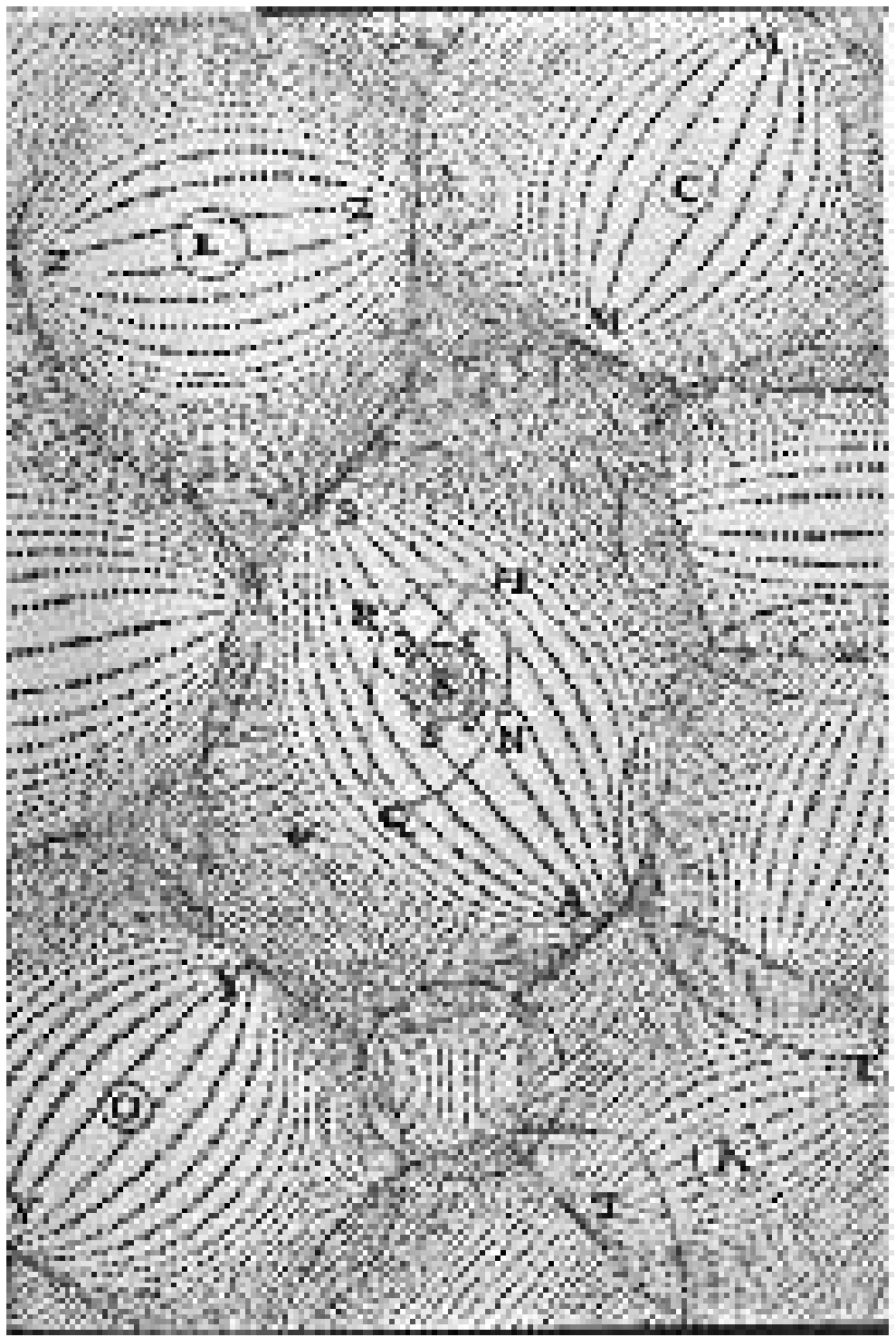,width=13.cm}}
\caption{R. Descartes; Le Monde, Ou Trait\'e de la Lumi\`ere, \&c., Paris, MDCLXIV}
\end{figure}
\begin{figure}
\centering\mbox{\hskip -0.5truecm\psfig{figure=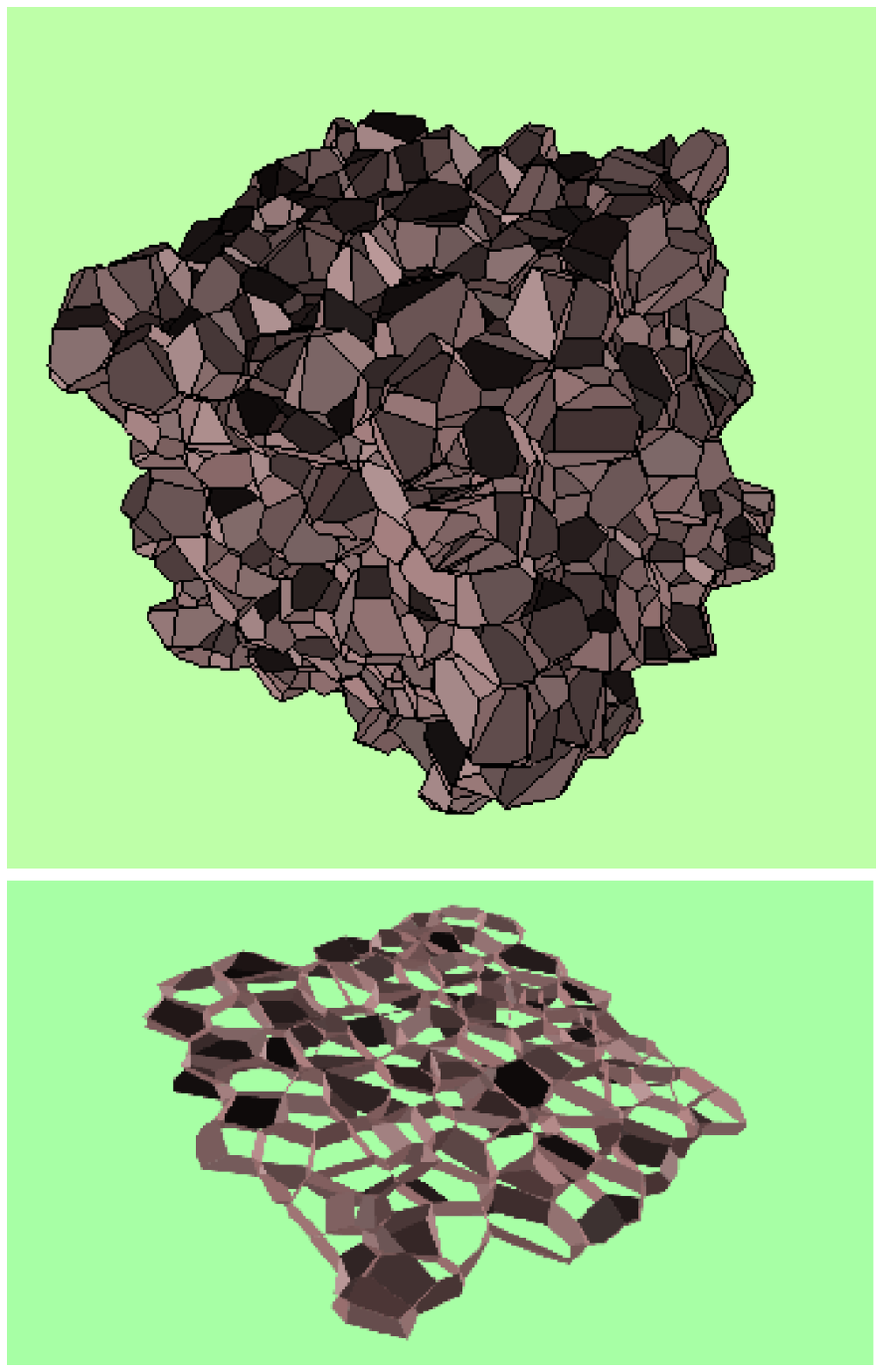,width=12.6cm}}
\caption{A full 3-D tessellation comprising 1000 Voronoi cells/polyhedra 
generated by 1000 Poissonian distributed nuclei. Courtesy: Jacco Dankers}
\end{figure} 
The complete set of Voronoi polyhedra constitute a space-filling tessellation of
mutually disjunct cells, the {\it Voronoi tessellation}. A good impression of the 
morphology of a complete Voronoi tessellation can be seen in figure 22, a tessellation 
of 1000 cells generated by a Poisson distribution of 1000 nuclei in a cubic box.

Figure 34 shows how in three dimensions a Voronoi foam forms a packing of Voronoi cells,
each cell being a convex polyhedron enclosed by the bisecting planes between
the nuclei and their neighbours. A Voronoi foam consists of four 
geometrically distinct elements: the polyhedral cells ({\it voids\/}), their 
walls ({\it pancakes\/}), edges ({\it filaments\/}) where three walls intersect, and 
nodes ({\it clusters\/}) where four filaments come together.

Taking the three-dimensional tessellation as the archetypical representation
of structures in the physical world, we can identify four constituent {\it elements} 
in the tessellation, intimately related aspects of the full Voronoi tessellation. 
In addition to (1) the polyhedral {\it Voronoi cells} $\Pi_i$ these are (2) the polygonal
{\it Voronoi walls} outlining the surface of the Voronoi cells,
(3) the one-dimensional {\it Voronoi edges} defining the rim of
both the Voronoi walls and the Voronoi cells, and finally (4) the
{\it Voronoi vertices} which mark the limits of edges, walls and cells. 
To appreciate the interrelation between these different geometric aspects, 
figure 23 lifts out one particular Voronoi cell from a clump of a dozen Voronoi cells. 
The central cell is the one with its polygonal Voronoi walls surface-shaded,
while the wire-frame representation of the surrounding Voronoi cells reveals the Voronoi 
edges defining their outline and the corresponding vertices as red dots.

While each Voronoi cell is defined by one individual nucleus in the
complete set of nuclei $\Phi$, each of the polygonal Voronoi walls 
$\Sigma_{ij}$ is defined by two nuclei $i$ and $j$, consisting of points ${\vec x}$ having
equal distance to $i$ and $j$. Evidently, the Voronoi wall $\Sigma_{ij}$ is a
subregion of the full bisecting plane of $i$ and $j$, the subregion
consisting of all points ${\vec x}$ closer to both $i$ and $j$ than
other nuclei in $\Phi$. The number of walls constituting the surface
of each cell $\Pi_i$ is a stochastic quantity. In fact, in three
dimensions even the average number of walls per cell in a specific
finite Voronoi tessellation is a stochastic quantity, with an expectation value
of $7.768$ for a Poisson-Voronoi tessellation (ie. a tessellation generated
by Poisson distributed nuclei). In analogy to the definition of a
Voronoi wall, a Voronoi edge $\Lambda_{ijk}$ is a subregion of the 
equidistant line defined by three nuclei $i$, $j$ and $k$, the subregion consisting of
all points ${\vec x}$ closer to $i$, $j$ and $k$ than to any of the other
nuclei in $\Phi$. Moreover, in analogy to the definition of Voronoi walls
the Voronoi edge $\Lambda_{ijk}$ is a part of the -- surface of -- three
Voronoi cells, $\Pi_i$, $\Pi_j$ and $\Pi_k$. Evidently, it is part of
the perimeter of three walls as well, $\Sigma_{ij}$, $\Sigma_{ik}$ and
$\Sigma_{jk}$, the first of which is a segment of the surface of $\Pi_i$
and $\Pi_j$, the second one of $\Pi_i$ and $\Pi_k$ and the third one of
$\Pi_j$ and $\Pi_k$. Pursuing this enumeration, Voronoi vertices $V_{ijkl}$
are defined by four nuclei, $i$, $j$, $k$ and $l$, being the one point
equidistant to them and closer to them than to any of the other nuclei
belonging to $\Pi_i$. In other words, the vertex is the circumsphere of 
the tetrahedron defined by the four nuclei. In other words, each set of 
nuclei $i$, $j$, $k$ and $l$ corresponding to a Voronoi vertex defines a 
unique tetrahedron, which is known as Delaunay tetrahedron (Delone 1934), with the 
defining characteristisc that no other nucleus can be inside their circumsphere. 
It also implies that from the set of Voronoi vertices we can define an 
additional ``dual'' space-filling tessellation, the {\it DELAUNAY TESSELLATION}. 

\begin{figure}[t]
\vskip 0.5truecm
\centering\mbox{\psfig{figure=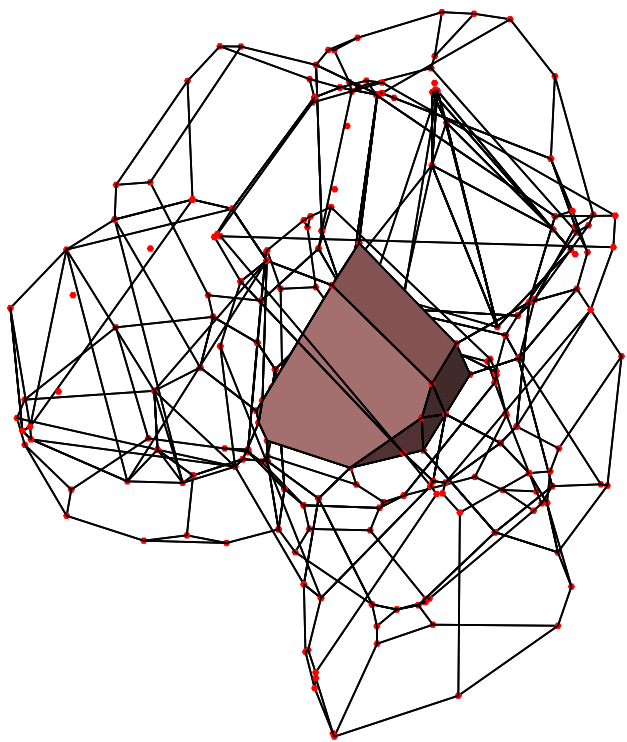,height=10.0cm}}
\vskip -0.3truecm
\caption{Wireframe illustration of interrelation between various Voronoi 
tessellation elements. The central ``Voronoi cell'' is surrounded by its wire-frame 
depicted ``contiguous'' Voronoi neighbours. The boundaries of the cells 
are the polygonal ``Voronoi walls''. The wire edges represent the 
Voronoi edges. The ``Voronoi vertices'', indicated by red dots, are 
located at each of the 2 tips of a Voronoi edge, each of them located at 
the centre of the circumsphere of a corresponding set of four nuclei. 
Courtesy: Jacco Dankers.}
\vskip -0.5truecm
\end{figure}
To appreciate the geometric definitions and relationships it is instructive to turn to 
Fig. 35, showing a wire-frame network of all Voronoi edges 
belonging to the contiguous (neighbouring) Voronoi cells touching one particular 
central Voronoi cell (solid). It indicates the sites, by means heavy red dots, 
of the corresponding Voronoi vertices (and Delaunay circumcentres). Notice then, 
that the stochastic point process of nuclei brings forth a new and 
uniquely defined, that of the {\it vertices} !!!

\vfill\eject
\subsection{Voronoi Tessellations:}
\begin{flushright}
{\bf{\large the Cosmological Context}}
\vskip -0.2truecm
\end{flushright}
In the cosmological context {\it Voronoi Tessellations} represent the {\it 
Asymptotic Frame} for the ultimate matter distribution distribution in any 
cosmic structure formation scenario, the 
skeleton delineating the destination of the matter migration streams involved 
in the gradual buildup of cosmic structures. Within such a cellular framework 
the interior of each ``{\it VORONOI CELL}'' is considered to be 
a void region. The planes forming the surfaces of the cells 
are identified with the ``{\it WALLS}'' in the galaxy distribution (see e.g. Geller 
\& Huchra 1989). The ``{\it EDGES}'' delineating the rim of each wall are to be  
identified with the filaments in the galaxy distribution. In general, what is 
usually denoted as a flattened ``supercluster'' or cosmic ``wall'' will comprise an assembly 
of various connecting walls in the Voronoi foam, as the elongated ``superclusters'' 
or ``filaments'' will usually consist of a few coupled edges (Fig. 41 and 42 clearly 
illustrate this for the Voronoi kinematic model). Finally, the 
most outstanding structural elements are the ``{\it VERTICES}'', tracing the surface of 
each wall, outlining the polygonal structure of each wall and limiting the 
ends of each edge. They correspond to the very dense compact nodes within the cosmic 
network, amongst which the rich virialised Abell clusters form the most 
massive representatives. In a way, the Voronoi foam outlines the ``skeleton'' of the 
cosmic matter distribution. It identifies the structural frame around which matter 
will gradually assemble in the course of the development of cosmic structure. In this 
view the process of cosmic structure formation is one in which we see a 
gradually unfolding of the cellular pattern in the matter distribution as 
matter is set to migrate away from the primordial location towards the high-density 
features in the cosmic foam.  

Although the idea of tessellations in an astronomical context dates back 
centuries (see Fig. 33), the first actual application of Voronoi tessellations to 
astrophysics is of a more recent date. Kiang (1966) invoked them to obtain a 
mass spectrum for the fragmentation of interstellar molecular clouds, be it without 
success in reproducing the Initial Mass Function of stars. It were Matsuda \& Shima 
(1984) who noticed the similarity between 2-D Voronoi tessellations and the outcome 
of the first computer experiments of cosmic structure formation (Melott 1983), 
a similarity which found a solid foundation when Icke \& van de Weygaert 
(1987) independently stuck upon the concept of Voronoi tessellations 
pursuing the physical argument that expanding density depressions 
play a dominating and regulating role in the formation of cosmic 
structure (Icke 1984). 
\begin{figure}[h]
\vskip 0.5truecm
\centering\mbox{\hskip -0.15truecm\psfig{figure=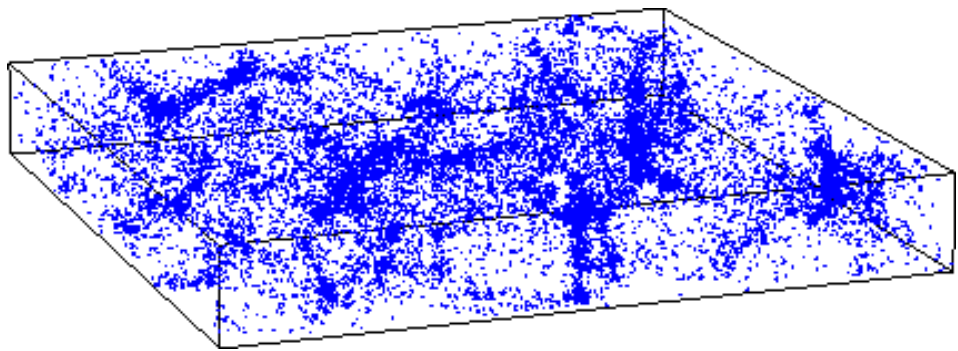,width=12.6cm}}
\vskip -0.truecm
\caption{Voronoi Galaxy Distribution: \hfill an example of a galaxy distribution whose 
geometrical pattern is defined through a Voronoi network. Also see Fig. 41 \& 42}
\vskip -0.truecm
\end{figure}
\subsection{Voronoi Galaxy Distributions}
Cosmologically, the great virtue of the Voronoi foam is that it provides a
conceptually simple model for a cellular or foamlike distribution of
galaxies, whose ease and versatility of construction makes it an ideal
tool for statistical studies. The stochastic, non-Poissonian and geometric 
nature of the spatial distribution of walls, filaments and clusters framing 
the cosmic web is responsible for large-scale spatial clustering 
in the matter distribution, and the related galaxy populations. 
The Voronoi model hands us a flexible template for studying 
galaxy distributions around geometrical features that themselves 
have some distinct and well-defined stochastic spatial 
distribution, represented by the corresponding components 
in the Voronoi tessellations. 

To study the specific properties of such weblike galaxy 
distributions, geometrically constructed models offer a variety of 
advantages. Its great virtue is its realistic rendering and representation of 
the spatial distribution of walls and filaments defining the overall distribution.
In its focus on these geometric components, it provides a 
laboratory for studying a variety of different cellular 
distributions. It may therefore fulfil a key role in 
dissecting the fundamental spatial characteristics of such 
geometries, and potentially is a very useful instrument 
for understanding and interpreting the observed galaxy 
distribution. An additional virtue is that the model distributions will be 
far less restricted in resolution and number of particles than 
conventional N-body experiments, as cellular structure can be generated over 
a part of space beyond the reach of any N-body experiment. 
The Voronoi model will therefore also be particularly suited for studying the 
properties of galaxy clustering in cellular structures on very large 
scales, for example in very deep pencil beam surveys, as well as for studying 
the clustering of clusters in these models. 

A mere qualitative assessment of such three-dimensional geometries already yields 
the interesting and important observation that the non-Poissonian distribution 
of the Voronoi walls, edges and vertices is a stochastic process characterized by 
strong spatial correlations. This is readily apparent from e.g. Fig 35, and 
even more obvious from the lower frame slice in Fig. 34. The important  
repercussion is that the geometric Voronoi components themselves 
are grouping into coherent ``super''structures, inducing intrinsic spatial 
correlations over scales substantially superseding the basic cell scale 
(see Van de Weygaert 2002a,b). Also note that the nontrivial morphology of spatially 
clustered geometrical elements not only determines the overall clustering 
properties of its galaxy population but that it also forms a stark contrast 
to less physically motivated and less realistic stochastic toy models as e.g. 
the double Poisson process. 

The obvious shortcoming of the model is the fact that it does not 
and cannot addres the galaxy distribution on small scales, i.e. the distribution 
within the various components of the cosmic skeleton. This will involve the 
complicated details of highly nonlinear small-scale interactions of the 
gravitating matter. N-body simulations are by far the most reliable for treating 
that problem in the highly nonlinear clustering stages. 

\begin{figure}[t]
\vskip 0.5truecm
\centering\mbox{\hskip -0.5truecm\psfig{figure=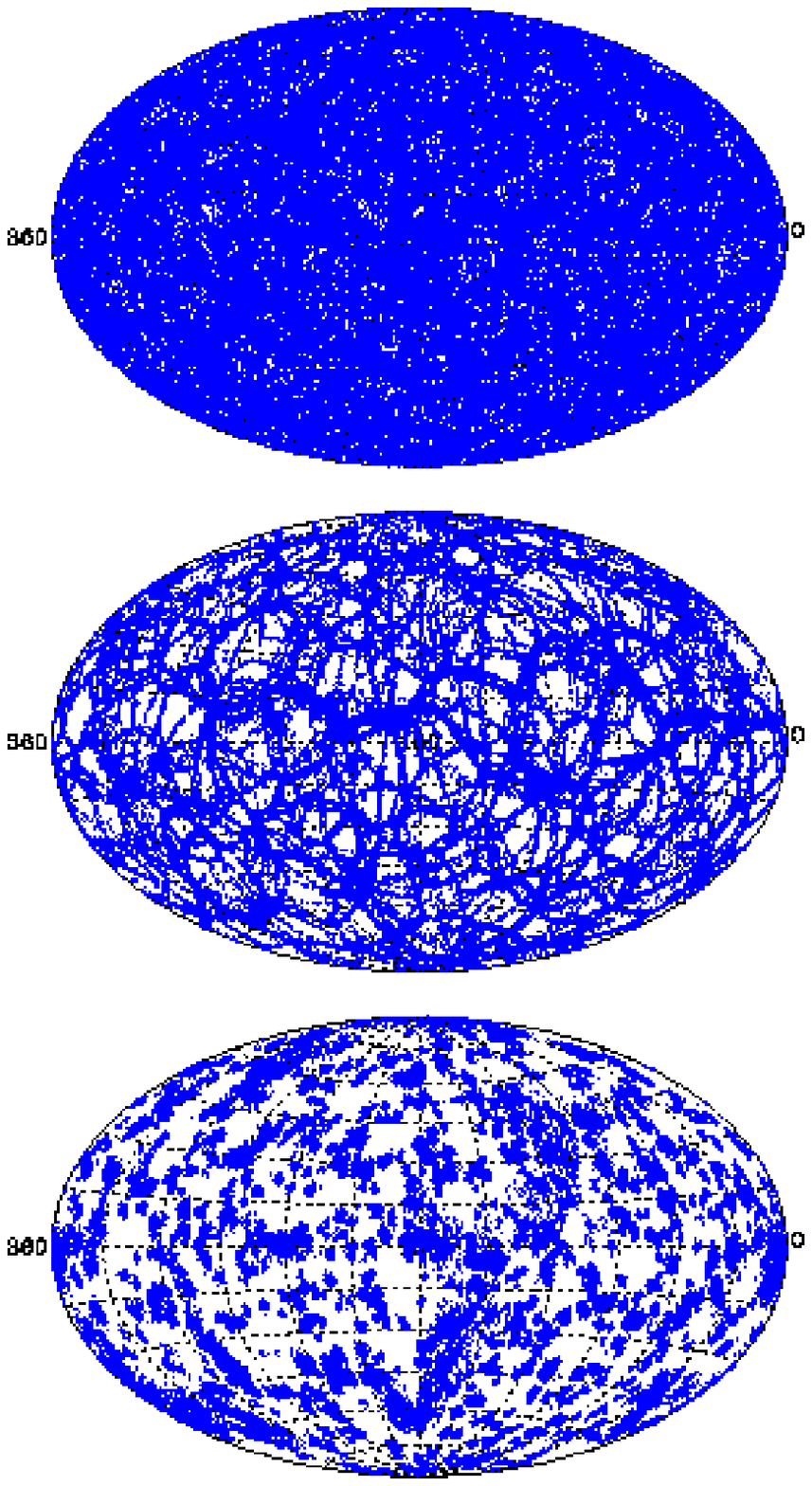,width=9.5cm}}
\end{figure}

\begin{figure}
\caption{Sky distributions for three different patterns of Voronoi galaxy 
distributions, depicted by means of an Aitoff projection. The depicted skies 
correspond to a wall-dominated Voronoi Universe (top), a filamentary 
Voronoi Universe (centre) and a cluster-dominated Voronoi Universe 
(bottom). The observer has been mapping all galaxies in a magnitude-limited 
survey ($m_{lim}=15.5$) comprising a surrounding $250h^{-1}\hbox{Mpc}$ spherical 
region lifted out of a world in which the mean size of the void (cell) regions is 
$\approx 25h^{-1}\hbox{Mpc}$. The number of galaxies corresponds to the 
number density set the Schechter luminosity function of Efstathiou, Ellis 
\& Peterson, for galaxies brighter than $M_{gal}=-17$. From: Van de Weygaert 2002b.}
\vskip -0.5truecm
\end{figure}
For our purposes, we take the route of complementing the large-scale cellular 
distribution induced by Voronoi patterns by a user-specified small-scale 
distribution of galaxies. On the one hand, it would be ideal to use well-defined 
and elaborate physical models to fill in this aspect. On the other hand, 
it would remove the essence of the charm and flexibility of the Voronoi 
concept. Far more beneficial is to set up tailor-made and user-defined 
spatial model distributions. In this, we distinguish two different yet 
complementary approaches. One is the fully heuristic approach of {\it ``Voronoi 
element models''}, genuine tools for the systematic investigation of very 
specific individual details of the full cellular structure. The second, 
supplementary, approach is that of the ``Voronoi kinematic distributions'', which 
attempt to ``simulate'' foamlike galaxy distributions in the true meaning of 
the word.

\subsubsection{{\it Voronoi galaxy distributions: Voronoi Element Models}}
A more practical alternative approch involves the generation of  
tailor-made purely heuristic ``galaxy'' distributions in and around the 
various elements of a Voronoi tessellation, ``Voronoi Element Models''. Such models are 
particularly apt for fathoming profound systematic properties of spatial 
galaxy distributions confined to one or more structural elements of 
nontrivial geometric spatial patterns. 

Telling examples are the ones represented by means of the Aitoff projected sky 
distributions depicted in Fig. 37. These yield a impression of what the observed galaxy 
distribution on the sky would be for a fictitious observer within such a model 
Universe, assuming ideal and uniform 
observational conditions. The resulting processed model sky distributions are 
a lucid means of conveying the impression one would obtain if one were living  
in such a model world and observe the surrounding world. Figure 37 shows the full 
sky distribution in Aitoff projection of all galaxies brighter than $m_{lim}=16.5$ out 
to a maximum depth of $100h^{-1}\hbox{Mpc}$, if the observer were to reside 
in a Universe with an interior foamlike pattern consisting of only walls (top), 
filaments (centre) or cluster clumps (bottom). 

\subsubsection{{\it Voronoi galaxy distributions: Web Pattern Dynamics}}
\begin{figure}[b]
\vskip -0.2cm
\centering\mbox{\hskip -1.1truecm\psfig{figure=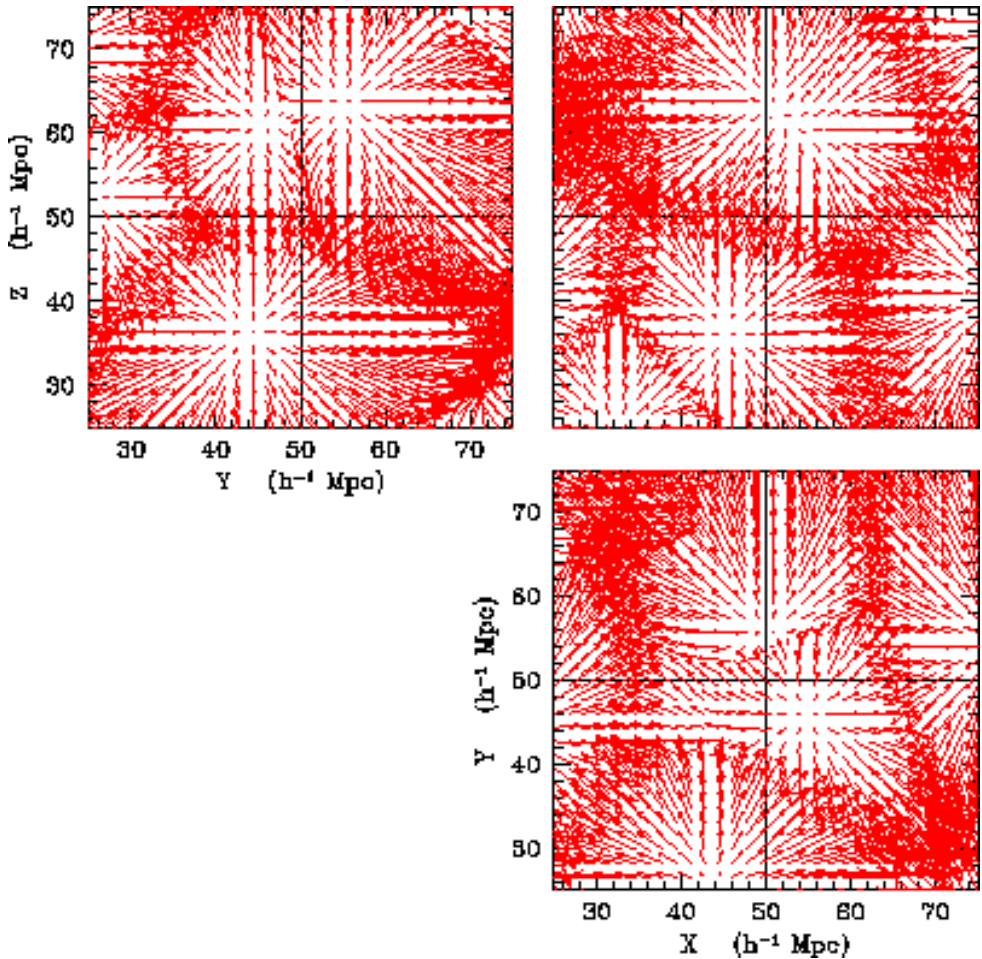,width=12.9cm}}
\vskip -0.1cm
\caption{The gravity field around one particular location in a wall-dominated 
(Voronoi) matter distribution, illustrated by grid vector maps of the gravity component 
in three mutually perpendicular planes passing through the central location. The 
length of each vector is proportional to the strength of the gravity and its 
direction pointing in the direction of the gravitational acceleration at the 
grid location represented by the base of the vector.}
\end{figure}
\begin{figure}[b]
\centering\mbox{\hskip -0.4truecm\psfig{figure=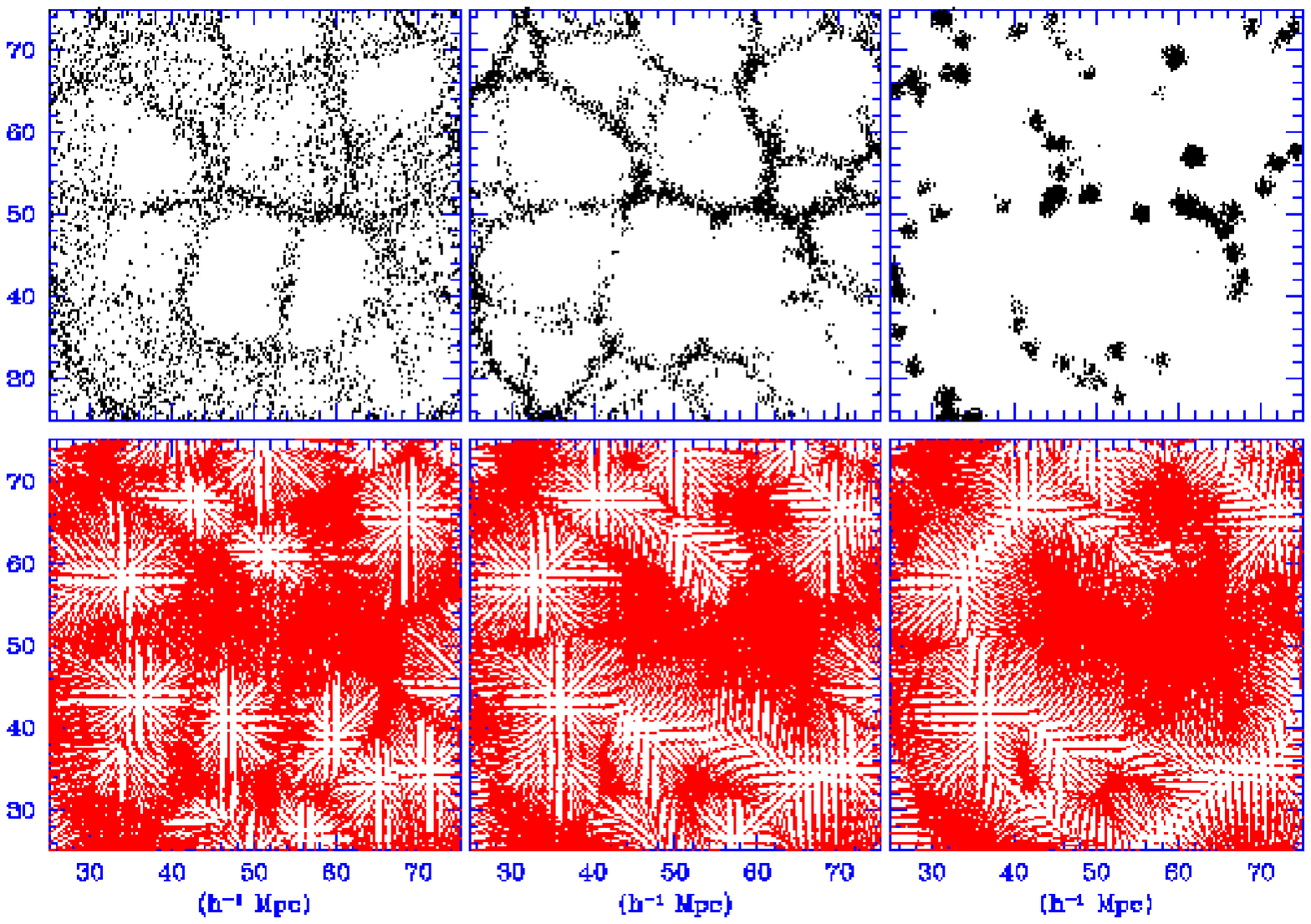,width=12.5cm}}
\vskip -0.1cm
\caption{Gravity fields of 3 specific cellular (Voronoi) matter distributions. The top row 
of 3 panels depicts the particle distribution in a central slice through the box of 
the Voronoi model realizations, while the bottom row involves the corresponding 
gravity vector fields in the same central slice of the box. In the bottom row, 
the arrows indicate the direction and strength of the gravity force component 
in the shown slice. All particle distributions correspond to a uniform density 
in each individual element, wall, filament or cluster. Left: a Voronoi wall 
distribution. Centre: a filamentary Voronoi distribution. Right: a Voronoi 
cluster distribution. From: Van de Weygaert 2002b.}
\vskip -0.3truecm
\end{figure}
\noindent An illustrative example of potential applications of such heuristic Voronoi 
models is the study of dynamics of matter distributions confined 
to one or more structural elements. For instance, the gravitational 
force field corresponding to a wall-dominated matter distribution would be 
resembling the gravity vector field shown in Fig. 38, being 3 
perpendicular planes centered on one specific location. It shows that the 
gravitational influence of the matter content in the walls is particularly 
strong in the direct environment of the walls. The interior of the 
voids are remarkably less pronounced, partly due to the evening out of 
the conflicting gravitational attractions exerted by the various 
individual walls. Extending such considerations, a comparison between the 
gravity field configurations effected by a wall-dominated, filamentary 
or cluster-dominated matter distribution (Fig. 39) reveals a rising 
contrast in gravitational strength between empty void regions and 
ever more compact and denser high-density regions (notice that in all 
cases, the matter density within each individaul wall is uniform, be it with  
a surface density different for each wall, determined as it is by its  
immediate environment). In the wall-dominated 
world (Fig. 39, lefthand frames), gravitatitional forces are particularly 
strong near the densest vertices (clusters), yet also have a noticeable 
strength immediately in and around nearly every wall. The walls can be 
readily identified from the spatial pattern of the gravity field itself, 
delineating dynamical boundaries between low-density voids. The topology 
of the gravity field changes drastically as the matter distribution assumes 
a more filamentary character (Fig. 39, central frames). The walls get more tenuous 
and command less and less dynamical weight, gradually dissolving into the background. 
The low-density voids, on the other hand, appear to merge into large 
regions characterized by a low and divergent gravity field. Dynamically 
more pronounced are the high-density cluster regions near the interstices of the 
densest filaments. They represent regions of considerably stronger gravity 
than in the corresponding case of a wall-dominated matter distribution. Also notice 
that the occasional ``isolated'' filament -- located at the boundary between 
large voids -- is characterized by a pronounced and concentrated gravity field,  
rapidly falling off into the void region. This trend of strongly concentrated 
gravity fields is continued towards configurations with compact cluster 
clumps. Very strong gravitational forces are felt near the complexes of 
such (clustered) clumps, with a weak gravity pertaining in the remaining 
low density regions.  

Armed with the insight provided by the nature, patterns and behaviour of such  
artificial gravity fields -- induced by specific asymptotic cellular matter 
distributions -- we get equipped with a necessary toolbox for a far more 
systematic and meaningful assessment of the dynamics of the more complex 
and realistic matter distributions usually encountered in N-body computer 
simulations. It will pave the way for a far more systematic study of the 
typical characteristics of the dynamics involved with cellular matter 
concentrations. 

\subsubsection{{\it Voronoi galaxy distributions: the Kinematic Model}}
Alternatively, we can generate distributions that more closely resemble the 
outcome of dynamical simulations, and represent an idealized and asymptotic 
description thereof. Such a model is the {\it kinematic model} defined 
by Van de Weygaert \& Icke (1989). 

The kinematic Voronoi model is based on the notion that when matter streams out 
of the voids towards the Voronoi skeleton, cell walls form when material from one void 
encounters that from an adjacent one. In the original ``pancake picture'' of Zel'dovich and 
collaborators, it was gaseous dissipation fixating the pancakes (walls), 
automatically leading to a cellular galaxy distribution. But also when the 
matter is collisionless, the walls may be hold together by their own self-gravity. 
Accordingly, the structure formation scenario of the kinematic model proceeds as 
follows. 
\begin{figure}[h]
\vskip 0.2truecm
\centering\mbox{\hskip 0.0truecm\psfig{figure=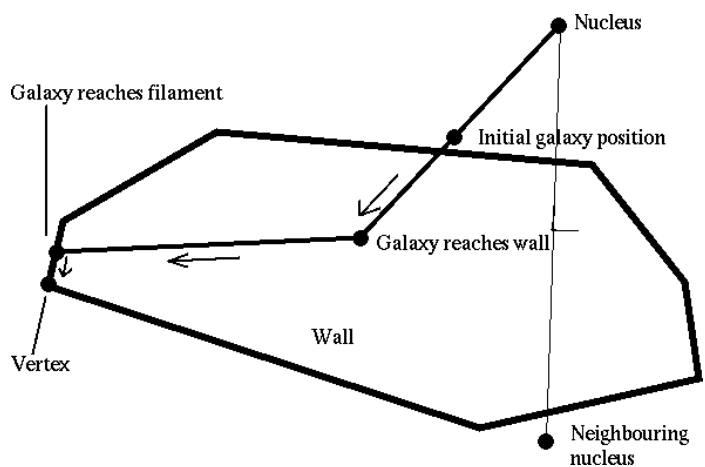,height=6.0cm}}
\caption{Schematic illustration of the Voronoi kinematic model. Courtesy: Jacco Dankers.}
\vskip -0.25truecm
\end{figure}
Within a void, the mean distance between galaxies increases uniformly in the course 
of time. When a galaxy tries to enter an adjacent cell, the gravity of the wall, aided 
and abetted by dissipational processes, will slow down its motion. On the average, 
this amounts to the disappearance of its velocity component perpendicular to the 
cell wall. Thereafter, the galaxy continues to move within the wall, until it tries 
to enter the next cell; it then loses its velocity component towards that cell, so
that the galaxy continues along a filament. Finally, it comes to rest
in a node, as soon as it tries to enter a fourth neighbouring void. In
a Voronoi foam, there are exactly four cells adjoining each node, and
the above process is unique. An immediate consequence of this
kinematic behaviour is that the density in the walls quickly becomes
smaller than in the filaments which, in turn, remain less dense than
the nodes, where all matter eventually congregates. This is the main reason
why we identify the nodes with the rich Abell clusters. 
\begin{figure}[t]
\vskip 0.4truecm
\centering\mbox{\hskip -0.55truecm\psfig{figure=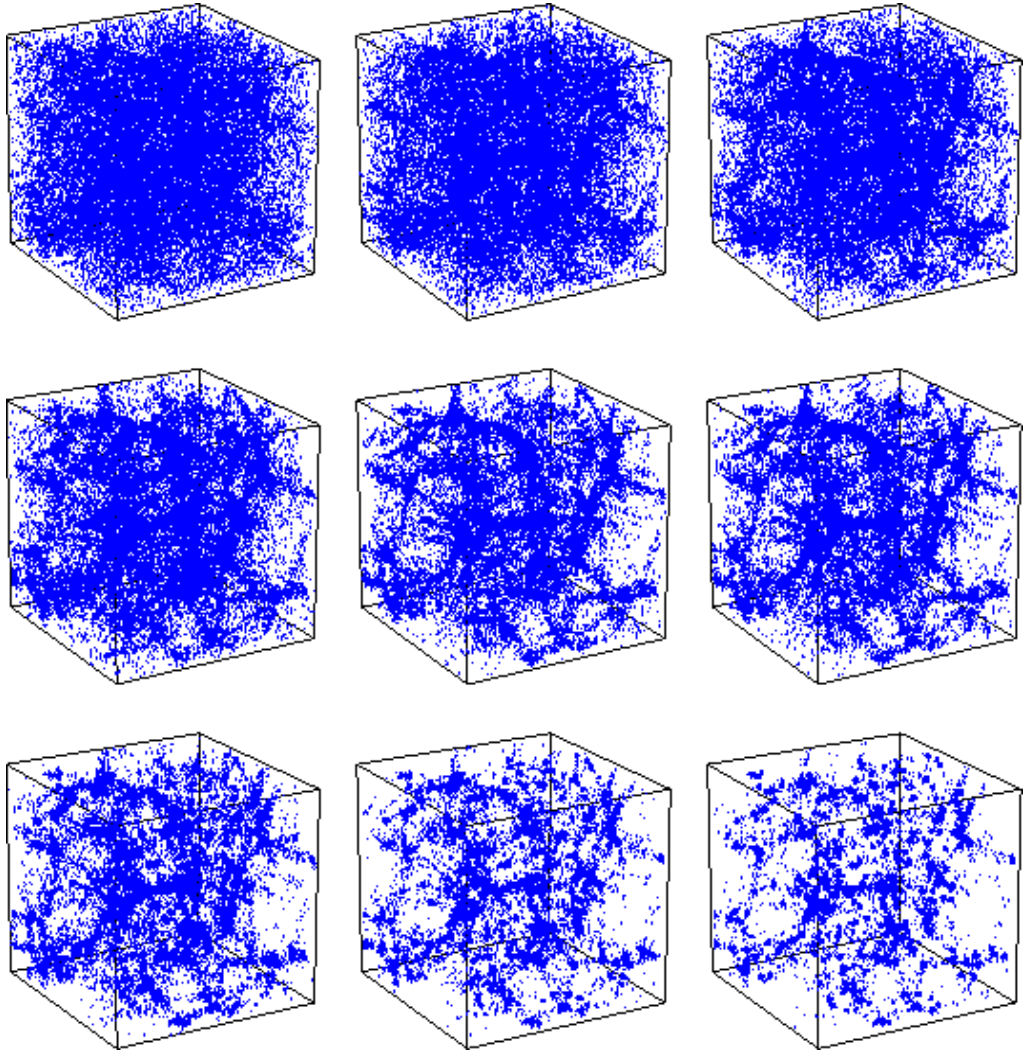,width=13.5cm}}
\vskip -0.1truecm
\caption{Evolution of galaxy distribution in the Voronoi kinematic model. A sequel 
of 9 consecutive timesteps within the kinematic Voronoi cell formation process,  
proceeding from left to right, and from top to bottom. The depicted boxes  
have a size of $100h^{-1} \hbox{Mpc}$. Within these cubic volumes some $64$ Voronoi 
cells with a typical size of $25h^{-1}\hbox{Mpc}$ delineate the cosmic framework around 
which some 32000 galaxies have aggregated. Taken from a total (periodic) 
cubic ``simulation'' volume of $200h^{-1}\hbox{Mpc}$ containing 268,235 
``galaxies''.}
\end{figure}
The evolutionary progression of an almost featureless random distribution, 
via a wall-like and filamentary morphology towards a distribution in which matter 
ultimately aggregates into conspicuous compact cluster-like clumps can be immediately  
appreciated from the sequence of 9 cubic 3-D particle distributions in Figure 41. 
Proceeding from the top left, left to right, and from top to bottom, it depicts a 
sequel of consecutive timesteps within the kinematic Voronoi cell formation process. 
The depicted boxes have a size of $100h^{-1} \hbox{Mpc}$. Within these cubic volumes 
some $64$ Voronoi cells with a typical size of $25h^{-1}\hbox{Mpc}$ delineate the 
cosmic framework around 
which some 32000 galaxies have aggregated\footnote{corresponding roughly  
to the number density of galaxies yielded by a Schechter luminosity 
function with parameters according to Efstathiou, Ellis \& Peterson (1988): 
$\phi_* = 1.56\,10^{-1}\,(h^{-1} \hbox{Mpc})^{-3}$, $\alpha=-1.07$ and $M_* = -19.68 + 
5logh$, where we restricted ourselves to galaxies brighter than 
$M_{gal} = -17.0$. In the full ``simulation box'' of $200h^{-1}\hbox{Mpc}$, this 
amounts to 268,235 galaxies.}. 

The steadily increasing contrast of the various structural features is accompanied by 
a gradual shift in topological nature of the distribution. The virtually uniform particle 
distribution at the beginning (upper lefthand frame) ultimately unfolds into the 
highly clumped distribution in the lower righthand frame. 

At first only a faint imprint of density enhancements and depressions can be discerned. 
In the subsequent 
first stage of nonlinear evolution we see a development of the matter distribution 
towards a wall-dominated foam. The contrast of the walls with respect to the general 
field population is rather moderate (see e.g. second frame), and most obviously 
discernable by tracing the sites where the walls intersect and the galaxy density is 
slightly enhanced. The ensuing frames depict the gradual progression via a 
wall-like through a filamentary towards an ultimate cluster-dominated matter 
distribution. By then nearly all matter has streamed into the nodal sites of 
the cellular network. The initially almost hesitant rise of the clusters quickly 
turns into a strong and incessant growth towards their appearance as dense 
and compact features which ultimately stand out as the sole dominating element 
in the cosmic matter distribution (bottom righthand frame). 

A particularly transparent view of the way in which the various morphological 
elements connect into the cosmic foam is shown by a set of cross-sections through 
the simulation boxes. A progressive evolutionary sequence of 6 such slices is 
shown in Figure 42, corresponding to the same realization as the cubic distributions 
in Fig. 41. On purpose the slicewidth of $20h^{-1}\hbox{Mpc}$ was chosen 
to be slightly smaller than the characteristic cellsize of $25h^{-1}\hbox{Mpc}$. 
The cross-sections give a good impression of the way in which the various filaments, 
each of varying orientation, link up with their peers into a seemingly undulating 
and space pervading network, sprinkled with a population of dense and compact 
clusters delineating the interstices of the network. The semblance of 
of the slice distributions to that of galaxies in the slice redshift surveys released 
since the first one by De~Lapparent, Geller, \& Huchra (1986) is indeed striking 

\begin{figure}[h]
\centering\mbox{\hskip -0.0truecm\psfig{figure=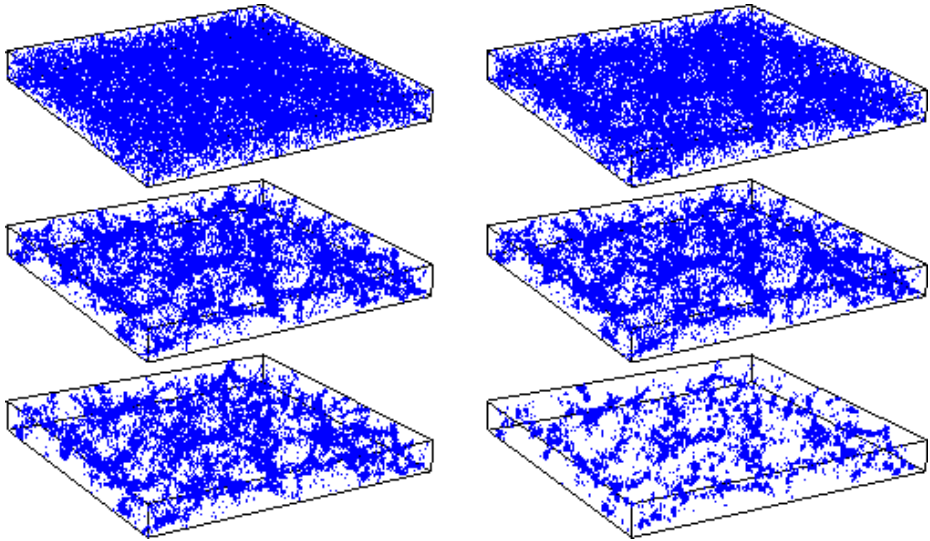,width=12.25cm}}
\vskip -0.1truecm
\caption{A sequel of six slices from a kinematic Voronoi model realization. 
Frames represent successive cosmic snapshots, taken the same ``simulation'' 
volume as in Fig. 41. Each slice has a slicewidth 
of $20h^{-1}\hbox{Mpc}$ and full boxsize of $200h^{-1}\hbox{Mpc}$.}
\vskip -0.25truecm
\end{figure}

\subsubsection{{\it Voronoi galaxy distributions: Observing the World of Voronoi}}
Most often observational samples represent a non-uniform sampling of the 
full spatial galaxy distribution. The resulting selection function makes it 
almost impossible to obtain an objective and complete reconstruction of the 
true spatial distribution. This is in particular true when highly nonlinear 
and anisotropic features, so characteristic for foamlike patterns, are 
involved. Comparison of observational samples with model distributions 
processed through a typical observational setup are therefore often a 
good way for evaluating the model's validity. Besides the telling examples in Fig 36 
of three additonal sky galaxy distributions observed by a fictituous observer, we present 
two other typical observational strategies. 

\begin{figure}[b]
\vskip -0.2truecm
\centering\mbox{\hskip -0.0truecm\psfig{figure=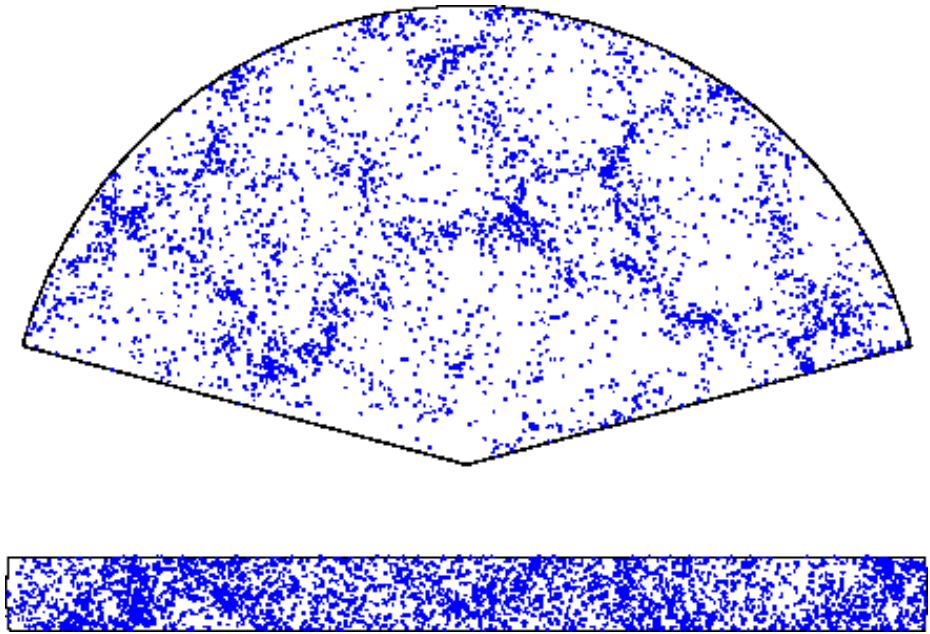,width=12.25cm}}
\caption{Top: A simulated redshift slice through a kinematic Voronoi distribution of 
galaxies. Galaxies distributed within the Voronoi foam, in which walls and 
filaments were having a proper width. The galaxies were given a luminosity, randomly 
drawn from a proper (Schechter) galaxy luminosity function (parameters: Efstathiou, 
Ellis \& Peterson 1988. The apparent magnitude limit of the survey is 
$m_{lim}=16.5$, confined to a slice with angular width and length of 
$12^{\circ}$ by $150^{\circ}$, going out to a depth of $100h^{-1}\hbox{Mpc}$. Bottom: 
the sky galaxy distribution of the galaxies depicted in the slice.}
\vskip -0.5truecm
\end{figure}
Slice redshift surveys through a kinematic Voronoi distribution were simulated. A typical 
example is shown in Figure 43. It concerns a ``redshift'' survey with 
a limiting magnitude of $m_{lim}=16.5$, out to a depth of $100h^{-1}\hbox{Mpc}$ within 
a $12^{\circ}$ wide slice of $150^{\circ}$ length. The resemblance to published 
survey results is indeed striking. Notice for instance the ``Great Wall'' running 
laterally along the slice, consisting of a series of connected individual 
Voronoi walls. In addition, we can discern various outstanding high-density 
``clusters''. The corresponding sky projection of the slice galaxies 
in the lower frame reinforces the realistic impression. 

The Voronoi model has been in particularly succesfull in explaining and interpreting 
the spiky pencil-beam redshift surveys (as shown in Fig. 44). It is rather 
easy to understand that the typical spiky pattern in these narrow redshift 
beams is due to the passing of the lines of sight through walls, and occasionally 
filaments, of the frothy galaxy distribution. Van de Weygaert (1991a,b) and 
Subbarao \& Szalay (1992) invoked the Voronoi model to show that such a 
spiky pattern in the redshift distribution can be understood quite naturally 
with a cellular pattern extending out to high redshifts, a fact that had 
been pointed out on the basis of an analytical evaluation of Voronoi 
statistics by Coles (1991). To this end, Van de Weygaert (1991a,b) simulated 
and analyzed pencil-beam patterns in Voronoi `universes', some suggestively 
similar to the observational results (see Fig. 44). In particular, it was pointed out 
while the wall distribution is not intrinsically regular a certain fraction 
(some 15$\%$) of the beams would yield a `quasi-periodic' galaxy distribution, 
offering a non-contrived explanation for the puzzling regularity in the 
Broadhurst et al. (1990) results.
\begin{figure}[h]
\vskip -0.1truecm
\centering\mbox{\hskip -.truecm\psfig{figure=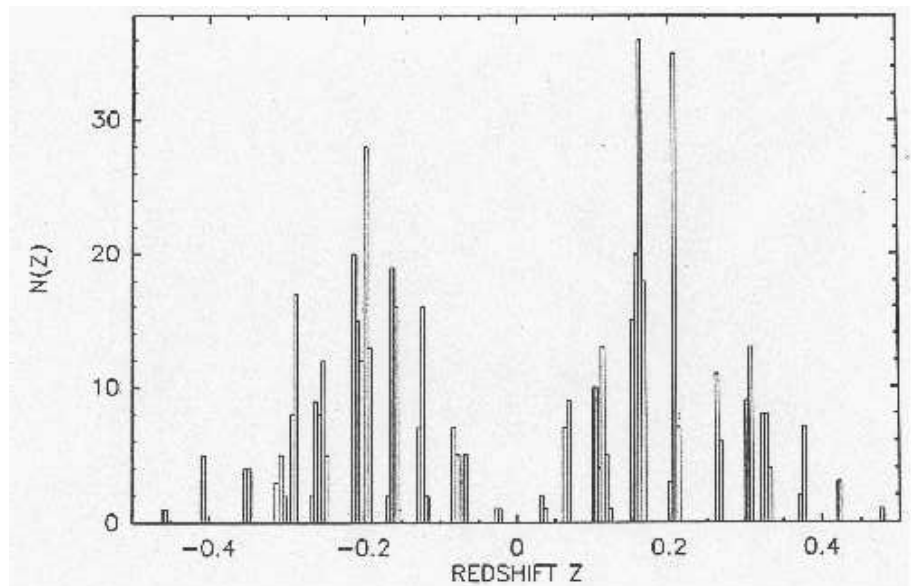,angle=270.,width=12.0cm}}
\vskip -0.25truecm
\caption{A simulated pencil beam survey through a deep Voronoi web. From Van de Weygaert 
1991a,b. Reproduced by permission of the Royal Astronomical Society.}
\end{figure}

\subsubsection{{\it Voronoi galaxy distributions: Quantification of Galaxy Clustering}}
To be of more than illustrative use, Voronoi tessellations should supersede 
the status of a pure qualitative sketch and be able to reproduce 
quantitative aspects of the observed galaxy and mass distribution. 

A full quantitative description of any point distribution in principle 
consists of the full hierarchy of M-point correlation functions $\xi_M$. 
They are defined along the lines set by that of the 2-point function 
$\xi(r)$, quantifiying the excess probability of finding a pair of points in 
volume elements $d V_1$ and $d V_2$ separated by a distance $r$ in a point 
sample of average number density ${\bar n}$, the two-point function $\xi(r)$ 
is defined by 
\begin{equation}
dP(r)\ =\ {\bar n}^2\ (1 + \xi(r))\ dV_1 dV_2\ .
\end{equation}
\noindent Conventional cosmological terminology expresses the amplitude of $\xi(r)$ 
in terms of the scale $r_{\rm o}$, 
\begin{equation}
\xi(r_{\rm o})\ =\ 1\ .
\end{equation}
\noindent Conventionally denoted by the name ``correlation length'', we prefer the 
more correct name of ``clustering length''. Rather than a characteristic geometric scale, 
$r_{\rm o}$ is a measure for 
the ``compactness'' of the spatial clustering. A more significant scale within the 
context of the geometry of the spatial patterns in the density distribution is 
the scale at which 
\begin{equation}
\xi(r_{\rm a})\ =\ 0\ .
\end{equation} 
\noindent As a genuine scale of coherence 
it is a highly informative measure for the morphology of nontrivial 
spatial structures, so that we reserve the name ``correlation length'' for 
this scale. 

In  many cosmological studies the two-point correlation function $\xi(r)$ 
figures predominantly and often exclusively. This is partially based 
on historical development, $\xi(r)$ being the obvious first step  
in characterizing deviations from uniform point distributions. Other, 
substantial, cosmological considerations are of a more profound nature. 
The Gaussianity of the matter field perturbations in early linear phases 
of evolution implies a full description by $\xi(r)$. In more advanced nonlinear 
phases, $\xi$ figures in the close link between the kinematics of associated matter 
flows and the matter distribution and forms the basic element in a 
a hierarchical correlation function series. And, more straightforward, is 
the practical consideration of the limited measurability of higher order 
functions as noisy samples of objects imply an incessant error increase 
with order $M$.

As for the real world, the most solid estimate of the spatial two-point correlation 
function of galaxies is inferred on the basis of the millions of objects in sky catalogues, 
through deprojection of the angular two-point correlation $\omega(\theta)$. 
On scales $\leq 5h^{-1}\hbox{Mpc}$ this is very well approximated by 
a power law, which implies a power-law spatial correlation function $\xi(r)$ 
(see e.g. Efstathiou 1996), 
\begin{equation}
\xi_{gg}(r)\ =\ \left({\displaystyle r_{\rm o} \over \displaystyle r}\right)
^\gamma\,;\hskip 2.0truecm \gamma \approx 1.8,\hskip 0.5truecm 
r_{\rm o} \approx 5h^{-1}\hbox{Mpc}\,.
\end{equation}
\noindent Although direct estimates from 3-D redshift survey samples are 
complicated by discreteness noise, sampling and selection effects and 
redshift distortions, overall they tend to corroborate this power-law behaviour, also 
wrt. the parameter values (e.g. Davis \& Peebles 1983). 

\subsubsection{{\it Voronoi galaxy distributions: Point correlation analysis}}
The two-point correlation for a representative ``Voronoi kinematic'' galaxy distributions 
is depicted in Fig. 45. It shows a kinematic galaxy 
distribution for a distribution optimally resembling the observed 
galaxy distibution. The two bottom frames depict the value of $\xi(r)$ as 
a function of the distance $r$ between the ``galaxies'', expressed in units 
of the typical cell size $\lambda_{c}$. The figure contains both a 
log-log plot (lefthand), highlighting the small-scale clustering behaviour, and 
a lin-lin plot (righthand). The lin-lin plot is particularly apt in disclosing 
large-scale correlations in the spatial galaxy distribution, in particular also 
over distances considerably extending beyond the first zero-crossing of $\xi$, 
which typically is of the order of one cellsize. To be able to trace $\xi$ over 
such a large spatial range, including ones where the amplitude of correlations is 
miniscule, a specially designed algorithm was implemented (Van de Weygaert 2002a,b). 
Note that the levelling off of $xi(r)$ at the radii $r \lessapprox 0.5\lambda_c$ is 
an artefact. It is a consequence of the prescription for setting up the cellular 
galaxy samples, which slightly smears out the galaxy distribution in and around 
the walls, filaments and vertices. 
\begin{figure}[t]
\vskip 0.2truecm
\centering\mbox{\hskip -1.1truecm\psfig{figure=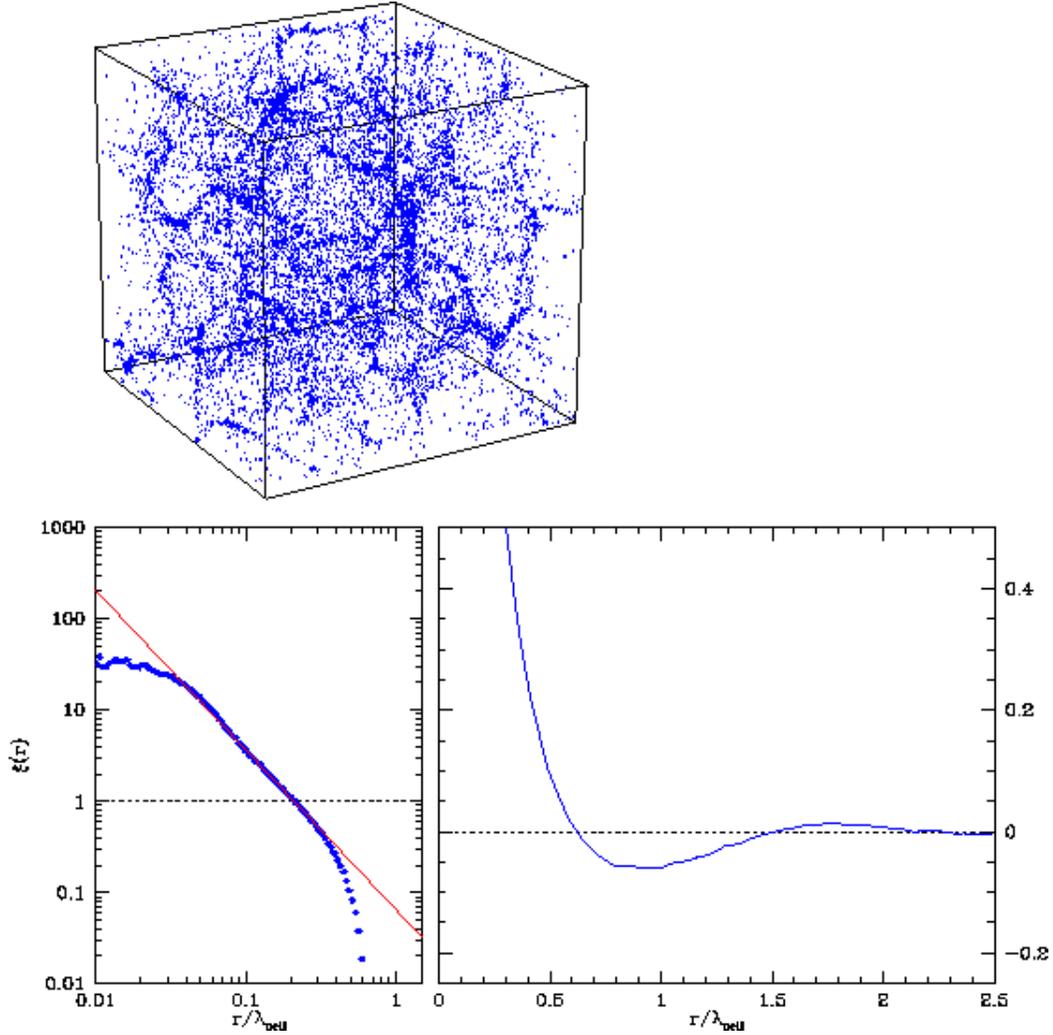,width=13.8cm}}
\vskip -0.25truecm
\caption{Two-point correlation function analysis of a selection of galaxies in 
a Voronoi kinematic model realization. Top frame: a spatial 3-D depiction of a full 
galaxy sample in a box of size $150h^{-1}\hbox{\rm Mpc}$, at a stage corresponding 
to the present cosmic epoch $\sigma(8h^{-1}\hbox{\rm Mpc}\approx 1$. The cellular 
morphology with walls and filaments forms a marked pattern throughout the box, with 
sites of a few conspicuously dense cluster ``nodes'' standing out. 
Bottom left: a log-log plot of the $\xi(r)$, with distance $r$ in units 
of the basic cellsize $\lambda_{\rm cell}$. The power-law character of $\xi$ 
up to $r \sim 0.5 \lambda_c$ is evident. Bottom right: a lin-lin plot of 
$\xi$. The beautiful ringing behaviour out to scales $r \sim 2 \lambda_{\rm cell}$ 
has been amply recovered. From: Van de Weygaert 2002b.}
\vskip -0.1truecm
\end{figure}
A striking aspect of the lin-lin $\xi$ plot is the beating pattern (righthand 
frame). Such a pattern of oscillating function values, alternating between positive 
and negative values, together with a gradually dimishing oscillation amplitude is 
very characteristic for cellular geometries. At sub-cellular scales $\xi(r)$ 
has very high positive values, seemingly diverging for $r \rightarrow 0$. 
Falling off like a power-law at these small distances (lefthand frame), 
it reaches a zero value at a correlation distance $r_{\rm a} \approx 0.6\lambda_c$. 
Subsequently, it becomes anticorrelated before reaching again a 
spatial range with positive correlations as $r > 1.65\lambda_c$, be 
it with very small correlation values. Our code was able to follow non-zero 
correlation values over distances $r > 2-2.5 \lambda_c$, beyond which 
the progressive damping of the correlation amplitude renders any 
non-zero correlation practically indetectable. 

Turning to the log-log plot, only evaluated for $r < r_{\rm a} \lessapprox 
\lambda_c$, we encounter some interesting behaviour. Most enticing is the power-law 
behaviour of $\xi(r)$ over almost the complete subcellular range. For the 
distribution shown the power-law slope is $\gamma \approx 1.85$ (!!!), while 
$r_{\rm o} \approx 0.23 \lambda_c$. When considering the development of $\xi(r)$ 
over the gradual progress of the kinematic model  
we find a correlation function whose shape is similar to the one shown in 
Fig. 45, though with a constantly increasing amplitude. The ``correlation'' 
scale $r_{\rm a}$ does not shift (the cellular pattern in the Voronoi model 
is static) as the coherence scale of the galaxy distribution does not 
evolve. On the other hand, the increasing level of clustering finds 
its expression in a steadily growing ``clustering length'' $r_{\rm o}$ as well 
as a continuously increasing power-law slope. For instance, the first 
box in Fig. 41 corresponds to a $(r_{\rm o}/\lambda_c,\gamma) \approx (0.06,1.3)$ 
versus the values of $(0.23,1.85)$ that we quoted for the final stage. 

Intriguing is the finding that a value of $r_{\rm o} \approx 5h^{-1}\hbox{Mpc}$, the 
current value for the observed galaxy distribution, would suggest a cellular 
scale $\lambda_c \approx 20-25h^{-1}\hbox{Mpc}$ when we take the timesteps 
with $r_{\rm o}/\lambda_c \approx 0.2-2.25$ as best match to the observed galaxy 
distribution. Such a size $\lambda_c \approx 25h^{-1}\hbox{Mpc}$ is teasingly 
close to the quoted values for the size of typical voids. In fact, the suggested 
intimate relation between cellsize and clustering length $r_{\rm o}$ had already been 
pointed out by Heavens (1985) for the simple -- and highly artificial -- 
configuration of an infinite network of cubic cells. If interesting, an even more 
intriguing thought may be that this is not contradictory to the conventional explanation 
within the context of nonlinear gravitational clustering starting from a field 
of Gaussian random density perturbations, but should rather be seen as complementary 
manifestations. Both the clustering length $r_{\rm o}$ and the cellular 
pattern are then intimately related, both being a product of the underlying process 
of gravitational clustering. 

The discussed kinematic Voronoi distribution represents a teasingly good agreement with  
that in the observed galaxy distribution. Naturally, the versatility of the 
Voronoi model allows it to be used as a template for a range of significantly 
different distributions. For example, we tested the correlation behaviour for pure wall-like, 
pure filamentary, and pure cluster galaxy distributions. Restricting the galaxy locations 
to uniform distributions within these structural features, we found that all three yield 
a power-law $\xi$ at sub-cellular scales, with a filamentary distribution corresponding to 
a substantially higher clustering amplitude $r_{\rm o} \approx 0.23 \lambda_c$ and 
steeper slope of $\gamma \approx 1.9$, while a wall-like distribution 
has a more moderate $r_{\rm o} \approx 0.14 \lambda_c$ and a shallow slope $\gamma \approx 1.4$ 
(Van de Weygaert 1991b, 2002b). 

\subsubsection{{\it Voronoi galaxy distributions: Words of Prudence}}
Of course the detailed and full physical picture underlying the cosmic galaxy distribution is 
expected to differ from that encapsulated in the Voronoi model, considerably so in the 
very dense, highly nonlinear regions of the network, around 
the filaments and clusters. Nonetheless, the success of the Voronoi kinematic model in 
reproducing and describing the  structural morphology and relevant characteristics 
of the cosmic foam, both the one seen in large redshift surveys as well as the 
one found in the many computer model N-body simulations, indicates its significance 
for the goal of defining a proper geometric model which may hope to succeed in modelling 
its essentials. 

\subsection{Superclustering}
Within the context of the identification of the Voronoi framework with the 
large-scale matter distribution, a special role is assumed by the 
{\it Voronoi vertices}. They are the tentative sites of the most 
pronounced components in the large scale galaxy distribution, the 
clusters of galaxies, located at the interstices in the 
cosmic framework. This can be clearly discerned from the evolving 
structure in Figure 41.

It is with respect to the identification of Voronoi vertices with 
the clusters of galaxies that the most telling and intriguing successes 
of the Voronoi model have been registered. Of instrumental 
significance in this context is the fact that the identification of vertices 
with clusters is straightforward, fully and exclusively defined by the 
geometry of the Voronoi tessellation realization. A primary assessment of 
the clustering of these vertices is fully set by the geometry of the tessellation 
and can therefore be done without further assumptions. When doing this, we basically 
use the fact that {\it the Voronoi node distribution is a topological 
invariant\/} in co-moving coordinates, and does not depend on the way in which 
the walls, filaments, and nodes are populated with galaxies. The statistics 
of the nodes should therefore provide a robust measure of the Voronoi 
properties. By contrast, for the modelling of related galaxy distributions 
additional specification for the fine small-scale details is very necessary. 

\subsubsection{{\it Superclustering: Cluster Clustering}}
As borne out by Fig. 7/8, clusters display a significant degree of 
clustering. An important issue is whether their clustering is merely 
a randomly sampled and diluted reflection of the underlying mass distribution 
or whether there are some clearly distinguishing characteristics to it.  
A comparison with the galaxy distribution have revealed three 
distinct aspects in the clustering of clusters. 

\begin{itemize}
\item{} The first aspect is the finding that the clustering of clusters is 
considerably more pronounced than that of galaxies. The two-point correlation 
function $\xi_{cc}(r)$ of clusters appears to be a scaled version of 
the power-law galaxy-galaxy correlation function, $\xi(r) = (r_o/r)^{\gamma}$. 
Most studies agree on the same slope $\gamma \approx 1.8$ while all yield a 
significantly higher amplitude. The estimates of the latter differ considerably 
from a factor $\simeq 10-25$. The original value found for the ``clustering 
length'' $r_{\rm o}$ for rich $R\geq 1$ Abell clusters was 
$r_{\rm o} \approx 25h^{-1}\hbox{Mpc}$ (Bahcall \& Soneira 1983),
\begin{equation}
\xi_{cc}(r)\ =\ \left({\displaystyle r_{\rm o} \over \displaystyle r}\right)
^\gamma\ ;\hskip 2.0truecm \gamma=1.8 \pm 0.2;\hskip 0.5truecm 
r_{\rm o}=26 \pm 4\ h^{-1} \hbox{Mpc}\ ,
\end{equation}
\noindent up to a scale of $100 h^{-1}$ Mpc (Bahcall 1988). Later work favoured 
more moderate values in the order of $15-20h^{-1}\hbox{Mpc}$ (e.g. Sutherland 1988, 
Dalton et al. 1992, Peacock \& West 1992).  
In terms of statistical significance, the recent clustering analysis of the 
cleanly defined REFLEX cluster sample has produced the currently most significant 
and elucidating determination of cluster-cluster correlation function (see Fig. 46, 
from Borgani \& Guzzo 2001) and its corresponding power spectrum (Borgani \& Guzzo 2001, 
Collins et al. 2001, Schuecker et al. 2001). As can be clearly discerned 
from Fig. 46, it strongly endorses the amplified cluster clustering wrt. 
the galaxy distribution (from the LCRS survey, Tucker et al. 1997). 
\item{} A related second property of cluster clustering is that the differences in 
estimates of $r_{\rm o}$ are at least partly related to the specific selection of 
clusters. There appears to be a trend of an increasing clustering strength as the 
clusters in the sample become more rich ($\approx$ massive). On the 
basis of the first related studies, Szalay \& Schramm (1985) even 
put forward the (daring) suggestion that samples of clusters selected 
on richness would display a `fractal' clustering behaviour, in which the 
clustering scale $r_{\rm o}$ would scale linearly with the typical scale $L$ 
of the cluster catalogue, 
\begin{equation}
\xi_{cc}(r)\ =\ \beta\,\left({\displaystyle L(r) \over \displaystyle r}\right)^\gamma\ ; \hskip 1.0truecm L(R)\ =\ n^{-1/3}\ . 
\end{equation}
\noindent The typical scale $L(R)$ is then the mean separation between the clusters of richness 
higher than $R$. 
Although the exact scaling of $L(r)$ with mean number density $n$ is questionable, 
observations seem to follow the qualitative trend of a monotonously 
increasing $L(R)$. It also appears to be reflected to some extent in a 
similar increase in clustering strength encountered in selections of 
model clusters in large-scope N-body simulations (e.g. Colberg 1998).
\item{} A final and third aspect of cluster clustering, is the issue of 
the spatial range over 
which clusters show positive correlations, the ``coherence'' scale of cluster 
clustering. Usually it is an aspect that escapes proper attention, yet may be 
of crucial significance. There is ample evidence that $\xi_{cc}(r)$ extends 
out considerably further than the galaxy-galaxy correlation $\xi_{gg}$, possibly out to 
$50h^{-1}-100h^{-1}\hbox{Mpc}$. This is not in line with 
conventional presumption that the stronger level of cluster clustering 
is due to the more clustered locations of the (proto)cluster peaks in the 
primordial density field with respect to those of (proto)galaxy peaks. According 
to this conventional ``peak bias'' scheme we should not find significant 
non-zero cluster-cluster correlations on scales where the galaxies no 
longer show any significant clustering. If indeed $\xi_{gg}$ is negligible  
on these large scales, explaining the large scale cluster-cluster clustering 
may be posing more complications than a simple interpretation would suggest.
\end{itemize}
\begin{figure}[t]
\vskip -0.5truecm
\caption{The two-point correlation functions $\xi$ of galaxies (squares) and X-ray 
clusters of galaxies (circles), plotted as a function of (redshift space) separation 
$r_s$, computed from the Las Campanas galaxy redshift survey (Tucker et al. 1997) and 
the REFLEX X-ray cluster survey (Collins et al. 2001). The two curves are the 
predictions for 2 CDM models, both in spatially flat universes 
($\Omega_m+\Omega_{\lambda}=1$), one with $\Omega_m=0.3, h=0.7$ (solid line), the 
other with $\Omega_m=0.5, h=0.6$ (dashed line). Courtesy: Borgani \& Guzzo 2001. Reproduced by permission of Nature.}
\vskip -0.25truecm
\centering\mbox{\hskip 2.75truecm\psfig{figure=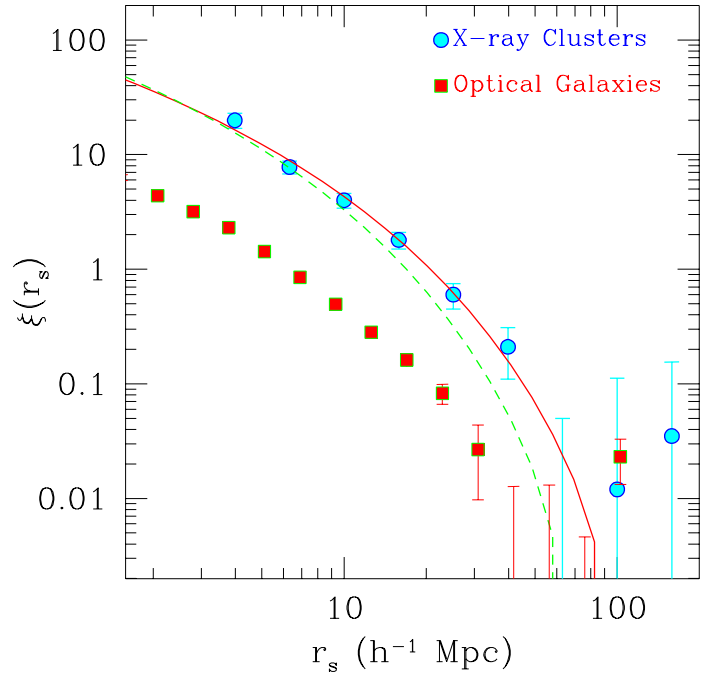,height=10.0cm}}
\vskip -0.5truecm
\end{figure}

\subsubsection{{\it Superclustering: Voronoi Vertices}}
An inspection of the spatial distribution of Voronoi vertices (Fig. 47, 
righthand frame) immediately reveals that it is not a simple random Poisson 
distribution. The full spatial distribution of Voronoi vertices in the 
$250h^{-1}\hbox{Mpc}$ cubic volume of figure 30 involves a substantial 
degree of clustering, a clustering which is even more strongly borne out 
by the distribution of vertices in a thin slice through the box (bottom 
lefthand frame) and equally well reflected in the sky distribution (bottom 
righthand frame). 
\begin{figure}[b]
\centering\mbox{\hskip -0.truecm\psfig{figure=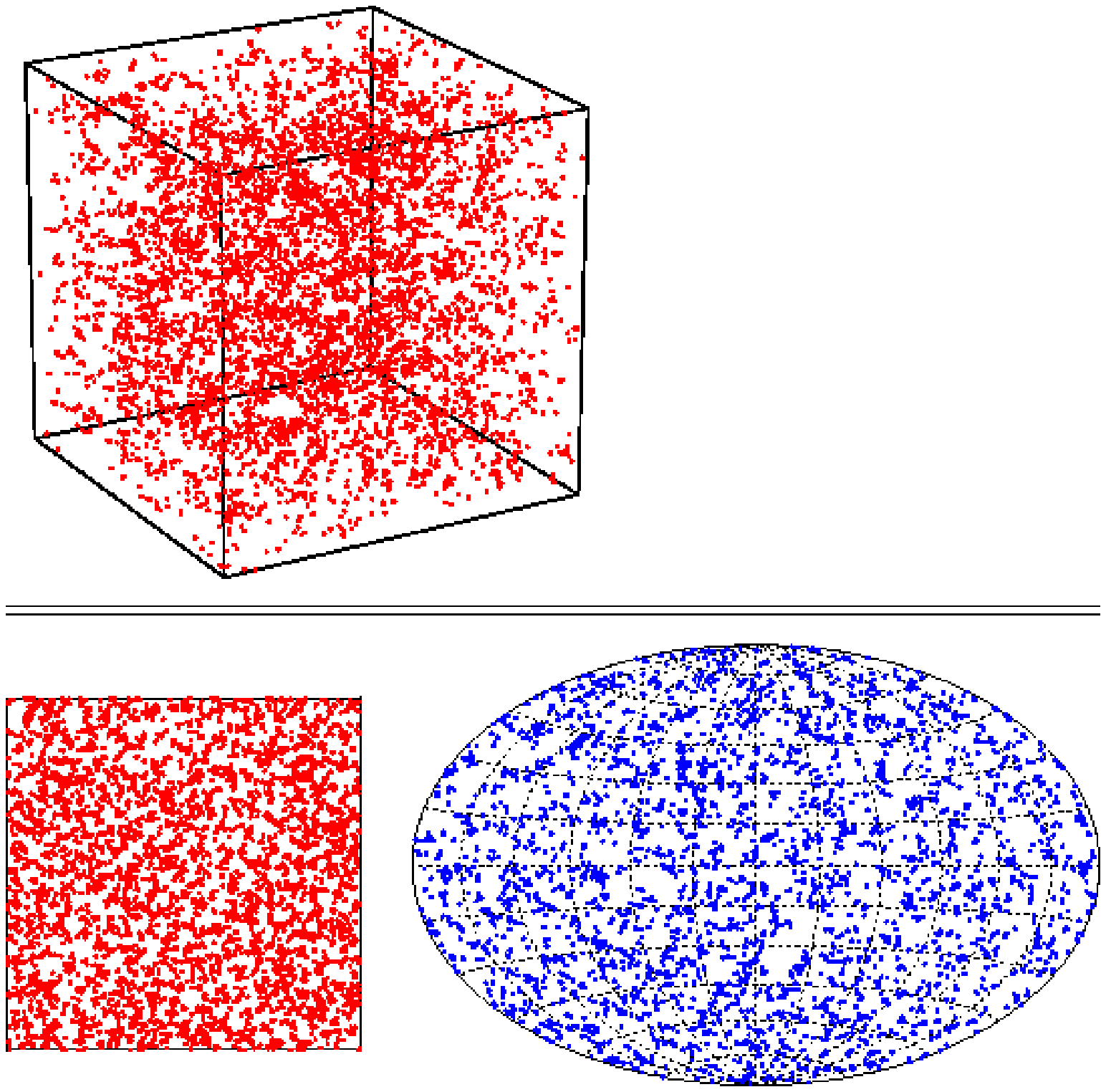,width=12.cm}}
\vskip -0.truecm
\caption{The spatial distribution of a full sample of Voronoi vertices. Top 
frame; the 3-D distribution in a $250h^{-1}\hbox{Mpc}$ box containing 1000 
Voronoi cells ($\sim 6725$ vertices). Notice the hint for vertices grouping 
in superstructures. Bottom left: the vertex distribution in 
a $25h^{-1}\hbox{Mpc}$ wide slice through box. Bottom right: an (Aitoff) sky 
projection of vertices out to a distance of $125h^{-1}\hbox{Mpc}$ from the 
box centre.}
\vskip -0.25truecm
\end{figure}
The impression of strong clustering, on scales smaller than or of the order of 
the cellsize $\lambda_{\rm C}$, is most evidently expressed by the corresponding 
two-point correlation function $\xi(r)$ (Fig. 48, left: log-log, right: 
lin-lin). Not only can we discern a clear positive signal, 
but out to a distance of at least $r \approx 1/4\,\lambda_{\rm c}$ the vertex-vertex 
correlation function is indeed an almost perfect power-law, 
\begin{equation}
\xi_{vv}(r)=\left({\displaystyle r_{\rm o} \over \displaystyle r}\right)
^\gamma\,;\hskip 2.0truecm \gamma = 1.95;\hskip 0.5truecm 
r_{\rm o} \approx 0.3\,\lambda_{\rm c}\,.
\end{equation}
\noindent with a slope $\gamma \approx 1.95$ and ``clustering length'' $r_{\rm o} 
\approx 0.3\,\lambda_{\rm c}$. Beyond this range, the power-law behaviour breaks down 
and following a gradual decline the correlation function rapidly falls off to a zero 
value once distances are of the order of (half) the cellsize. A value of 
$r_{\rm a}\approx 0.5\lambda_c$ for the zeropoint ``correlation length'' may be 
established most clearly from a linear-linear diagram of $\xi(r)$, while 
beyond $r_{\rm a}$ the distribution of Voronoi vertices appears to be practically 
uniform. Its only noteworthy behaviour is the gradually declining and alternating 
quasi-periodic ringing between positive and negative values similar to that we also 
recognized in the ``galaxy'' distribution, a vague echo of the cellular patterns which 
the vertices trace out. Ultimately, beyond $r \approx 2 \lambda_c$ any noticeable trace 
of clustering seems to be absent. 
\begin{figure}[t]
\centering\mbox{\hskip -1.10truecm\psfig{figure=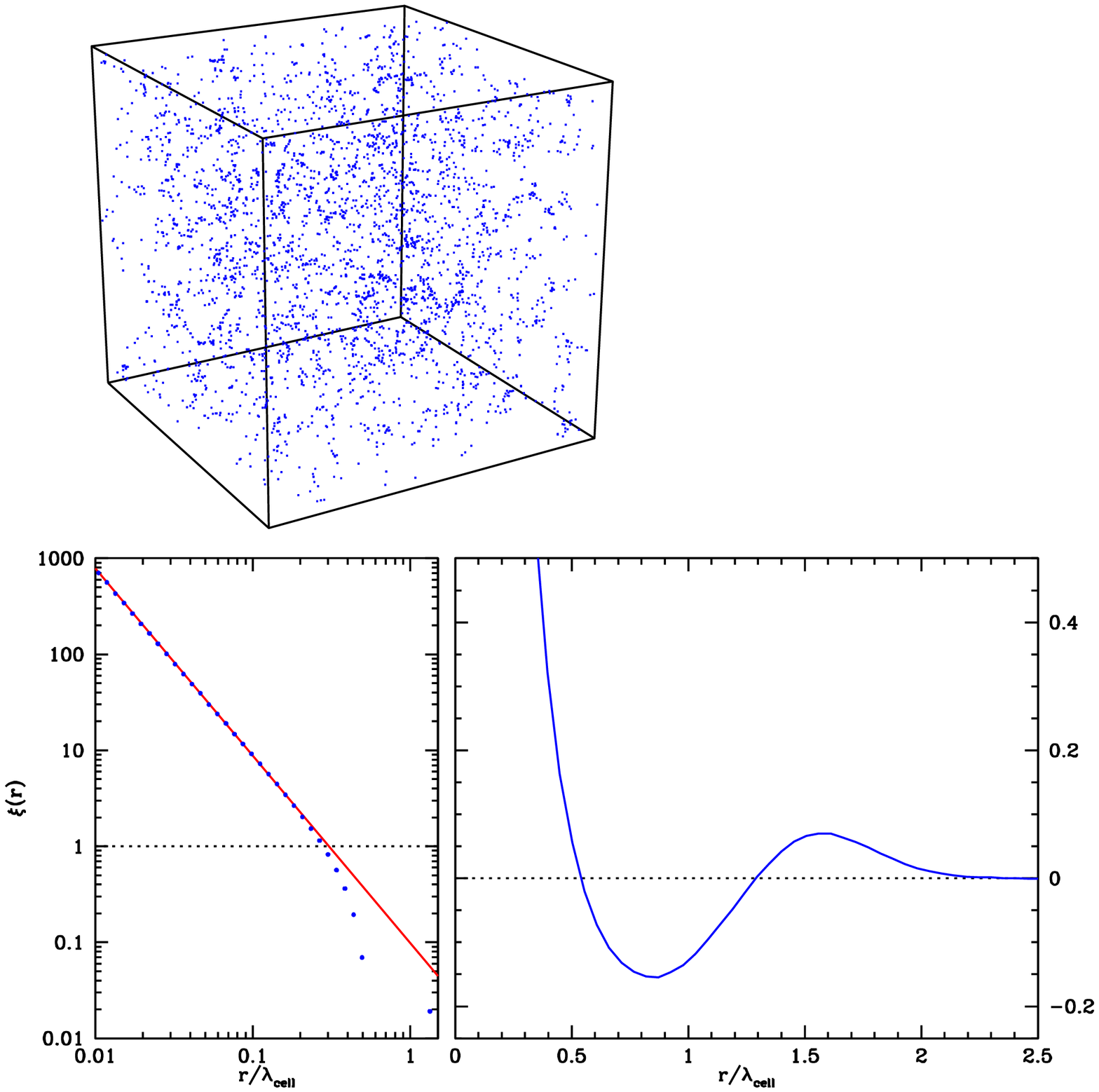,width=14.0cm}}
\vskip -0.4cm
\caption{Two-point correlation function analysis of a (full, non-selected) set of Voronoi 
vertices. Top frame: a spatial 3-D depiction of Voronoi vertex distribution. 
Upon close attention, the underlying cellular geometry may be discerned. 
Bottom left: a log-log plot of the $\xi(r)$, with distance $r$ in units 
of the basic cellsize $\lambda_{\rm cell}$. The power-law character of $\xi$ 
up to $r \sim 0.3 \lambda_c$ is evident. Bottom right: a lin-lin plot of 
$\xi$. The beautiful ringing behaviour out to scales $r \sim 2 \lambda_{\rm cell}$ 
has been amply recovered. From: Van de Weygaert 2002a.}
\vskip -0.3truecm
\end{figure}
The power-law behaviour of $\xi_{vv}$ is in remarkable agreement with that 
of the cluster distribution. It may hint at a geometrical origin for 
the power law slope $\gamma \approx 2$ of the cluster distribution. Also, 
its amplitude is in accordance with the observed cluster clustering length 
$r_{\rm o} \approx 20h^{-1}\,\hbox{Mpc}$, i.e. if we assume a basic cosmic foam 
cellsize of $\lambda_{\rm c} \approx 70h^{-1}\,\hbox{Mpc}$. The latter 
might actually be a complication, be it for the most simplistic interpretation 
assuming that every vertex would indeed represent a cluster.
 
\subsubsection{{\it ``Geometric Biasing'': Cluster Selections}}
 The vertex correlation function in eqn.~(18) does not take into account 
possible selection effects for the vertices. In reality, not every 
vertex will represent sufficient mass, or a sufficiently deep potential 
well, to be identified with a true compact galaxy cluster. If we take 
the Voronoi model as an asymptotic approximation to the true galaxy 
distribution, its vertices will comprise a range of ``masses''. Upon 
closer attention, the time sequence of evolving galaxy distributions 
in Fig. 41 indicates a continuously widening difference in the 
concentration of particles in and near vertices. Dependent on the 
specific geometrical setting of each vertex -- the size of the 
corresponding cells, walls and edges, the proximity of nearby 
vertices, etc. -- the total mass acquired by a vertex will span 
a wide range of values. 

Brushing crudely over the details of the temporal evolution, we may assign each 
Voronoi vertex a ``mass'' estimate by equating that to the total amount of matter 
ultimately will flow towards that vertex. Invoking the ``Voronoi streaming model'' 
as a reasonable description of the 
clustering process, it is reasonably straightforward if cumbersome to compute 
the ``mass'' or ``richness'' ${\cal M}_{\rm V}$ of each Voronoi vertex by pure 
geometric means (Van de Weygaert 2002a). The geometric computation far more 
efficient than Monte Carlo ``particle-based'' evaluations, yet also 
challenging and cumbersome in its implementation. In essence, the computation of 
the final mass consists of the evaluation of the Lagrangian volume of the mass 
content of the vertex. This Lagrangian volume is a non-convex polyhedron centered 
on the Voronoi vertex. The connected Voronoi nuclei, in the ``streaming model'' 
supplying the Voronoi vertex with inflowing matter, define the polyhedral 
vertices. 

\begin{figure}[t]
\centering\mbox{\hskip 0.truecm\psfig{figure=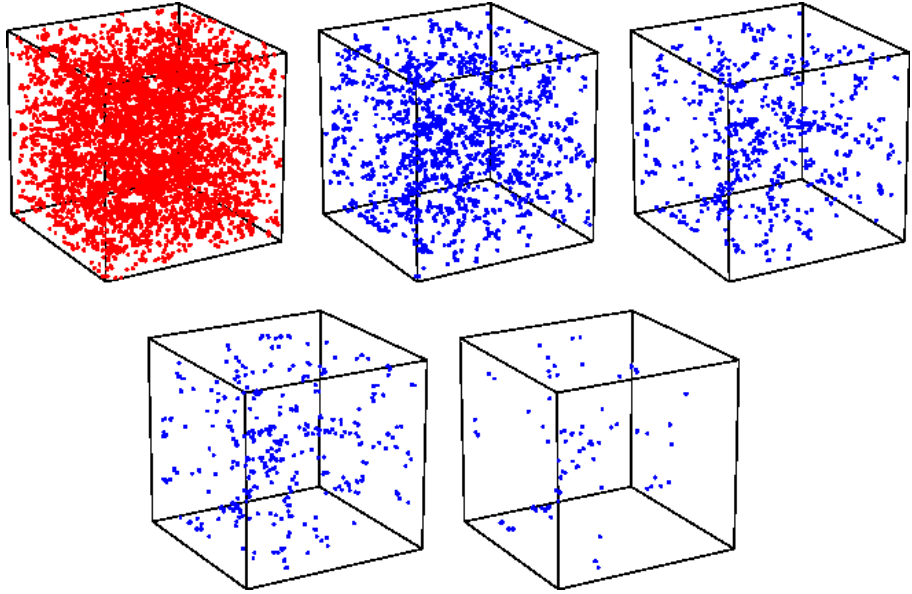,angle=270.,width=12.cm}}
\vskip -0.2truecm
\caption{Selections of vertices from a full sample of vertices. Depicted are the 
(100$\%$) full sample (top left), and subsamples of the 25$\%$, 10$\%$, 5$\%$ and 
$1\%$ most massive vertices (top centre, top right, bottom left, bottom right). 
Note how the richer vertices appear to highlight ever more pronounced a 
filamentary superstructure running from the left box wall to the box centre. 
From: Van de Weygaert 2002a.}
\vskip -0.5truecm
\end{figure}
To get an impression of the resulting selected vertex sets, Figure 49 shows 5 times 
the same box of $250h^{-1}\hbox{Mpc}$ size, each with a specific subset of the 
full vertex distribution (top lefthand cube). In the box we set up a realization of 
a Voronoi foam comprising 1000 cels with an average size of $25h^{-1}\hbox{Mpc}$. 
From the full vertex distribution we selected the ones whose ``richness'' 
${\cal M}_{\rm V}$ exceeds some specified lower limit. The depicted vertex subsets 
correspond to progressively higher lower mass limits, such that 100$\%$, 25$\%$, 
10$\%$, 5$\%$ and $1\%$ most massive vertices are included (from top lefthand to 
bottom righthand). The impression is not the one we would get if the subsamples 
would be mere random diluted subsamples from the full vertex sample. On the 
contrary, we get the definite impression of a growing coherence scale !!!  
For instance, it is as if the $1\%$ subsample subtends a single huge filament 
running the extent of the full box, even though this would be suggesting a single 
feature of $200-250h^{-1}\hbox{Mpc}$ size, an order of magnitude larger than 
the basic Voronoi cellsize. 

\begin{figure}[b]  
\centering\mbox{\hskip -0.75truecm\psfig{figure=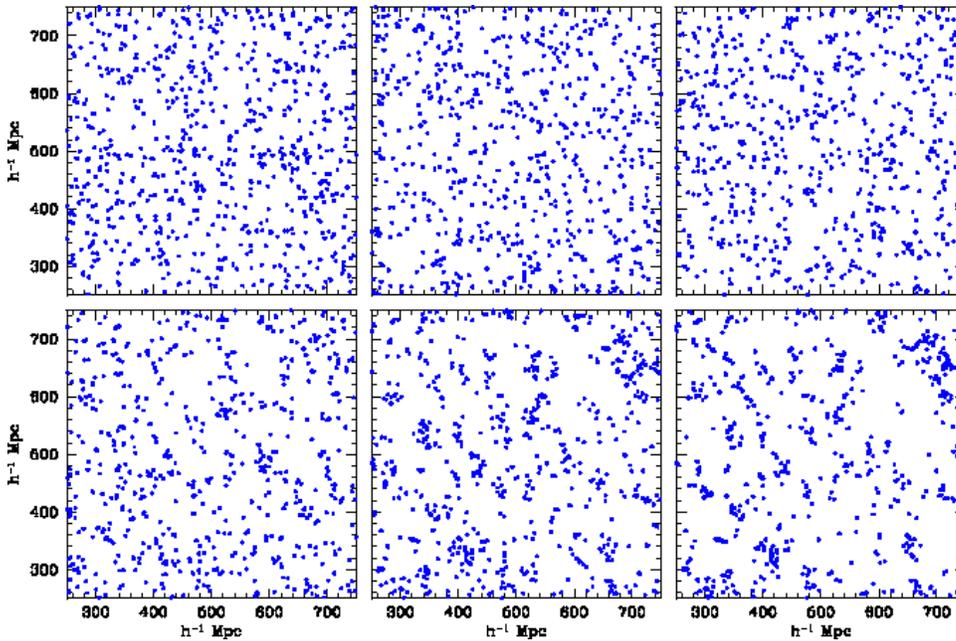,angle=270.,width=12.75cm}}
\vskip -0.4truecm
\caption{Selections of Voronoi vertices. Each subsample consists of the same 
number of vertices, randomly selected from samples of ever richer vertices 
from top left to bottom right. Top left: random selection from complete 
sample of vertices. Bottom right: $0.25\%$ richest vertices. Notice the 
continuous increase in clustering strength, and the stark contrast between 
the mild clustering of the full sample and that amongst the 
richest vertices. From: Van de Weygaert 2002a.}
\vskip -0.3truecm
\end{figure}
\subsubsection{{\it ``Geometric Biasing'': Transforming Clustering Patterns}}
The observed tendency of more massive vertex subsamples to display a 
stronger level of clustering which extends out to large distances 
has been scrutinized. After all, the human eye has a great talent 
for picking up patterns, thereby regularly exaggerating their 
reality or even imagining them while they do not even exist. 
To correct for possible diluted sampling effects provoking an exaggerated 
impression of an intrinsically moderate or even non-existent clustered 
distribution, we calibrated all point samples to the same number density, 
thereby assuring that their spatial statistics would be retained.
This is accomplished by pure random sampling of the same number of 
points from each subsample. 

For a huge cubic volume of $800h^{-1}\hbox{Mpc}$, containing $64^3$ cells of 
$25h^{-1}\hbox{Mpc}$ size, Figure 50 shows the enticing result. 
Beyond any doubt it confirms the impression of a intrinsic significantly 
stronger clustering for the more massive vertices. There is a salient contrast 
between the rather moderate level of clustering in the top lefthand frame 
($100\%$ level) and the striking point patterns in the sample of bottom righthand 
frame ($0.25\%$ level) is remarkable.  

At least three aspects concerning the more pronounced clustering of the 
more massive cluster samples may be discerned:
\begin{itemize}
\item{} {\it Stronger clustering}\\
The clustering itself is stronger, expressing itself in 
tighter and more compact point concentrations.
\item{} {\it Increased clustering scale}\\
The clustering extends over a substantially larger spatial 
range. Structures, clumps and huge voids, subtending several elementary 
cell scales are clearly visible (see in particular centre and right bottom 
frames Fig. 50). 
\item{} {\it Anisotropic extensions}\\
The subtended large scale features appear to become more 
distinctly anisotropic, wall-like or filamentary, for more massive samples  
(note the huge filamentary complexes in lower righthand frame Fig. 50).
\end{itemize}

\begin{figure}[b]
\vskip -0.2truecm
\centering\mbox{\hskip -0.8truecm\psfig{figure=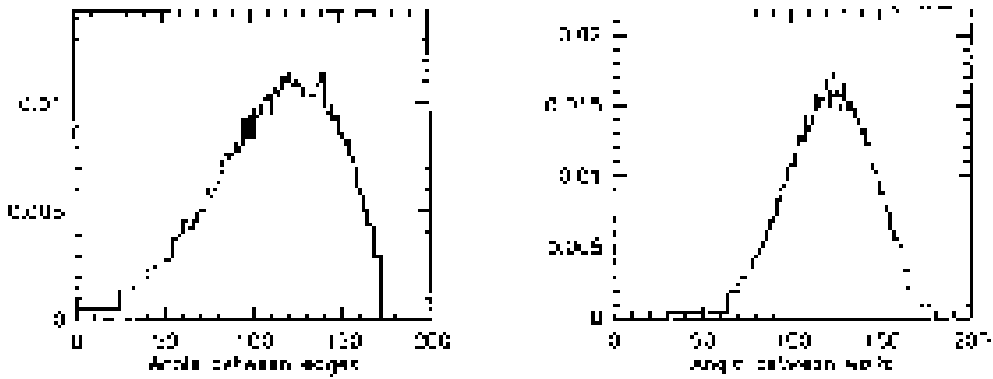,width=13.0cm}}
\caption{Angles between neighbouring Voronoi walls (left) and Voronoi edges (right) in 
a Poisson Voronoi tessellation (tessellation resulting from Poisson distributed 
nuclei). Notice that both distribution functions are peaked around the obtuse 
angle of $\sim 120^{\circ}$. From Van de Weygaert 1991b, 1994.}
\vskip -1.25truecm
\end{figure}
\noindent These visual impressions seem to reveal a striking 
``superclustering'' tendency hidden within the basic cosmic foam pattern 
-- modelled by the basic Voronoi foam -- and disclosing itself only 
through the distribution of its most prominent elements, the most massive 
clusters. The supercluster complexes -- huge filaments and walls -- form 
by linking several (Voronoi) edges and walls. 

Note that linking a set of randomly oriented filaments or walls  
generically would not subtend such stretched superstructures. Instrumental in 
understanding the presence of such features is their embedding within the underlying 
``cellular'' geometry of the matter distribution, the significance 
of which has usually escaped proper appreciation, if any at all. A key 
aspect of cellular geometries is the rigid embedding of walls and edges into 
a distinct connected network. As is easily inferred from geometrical 
modelling, walls and edges will not be oriented randomly and 
isotropically with respect to each other. On the contrary, their mutual 
orientation is typically centering around values of obtuse angles. The Voronoi 
geometry presents us with a telling illustration of this fact (see Fig. 51, and 
Van de Weygaert 1991b, 1994): walls and edges do indeed connect to their 
neighbouring peers with angles whose statistical distribution peaks around 
obtuse angles $\sim 120^{\circ}$. The distribution function for both walls 
and edges is indeed a peaked distribution, with the minor part of angles 
below $\sim 100^{\circ}$, and the majority subtending obtuse angles 
larger than $\sim 120^{\circ}$.  

\begin{figure}[t]
\vskip -0.5truecm
\centering\mbox{\hskip -1.3truecm\psfig{figure=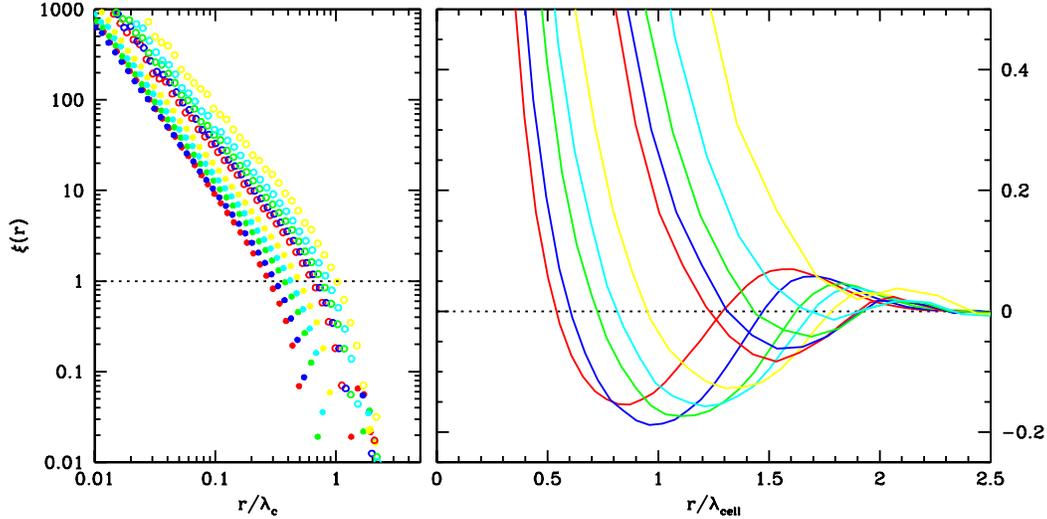,width=14.2cm}}
\vskip -0.5truecm
\caption{Scaling of the two-point correlation function of Voronoi vertices, for a 
variety of subsamples selected on the basis of ``richness'', ranging from samples 
with the complete population of vertices down to subsamples containing the 
$2.5\%$ most massive vertices. Left: log-log plot of $\xi(r)$ against 
$r/\lambda_c$, with $\lambda_c$ the basic tessellation cellsize 
($\equiv$ intranucleus distance). Notice the upward shift of $\xi(r)$ for  
subsamples with more massive vertices. Right: lin-lin plot of $\xi(r)$ against 
$r/\lambda_c$. Notice the striking rightward shift of the ``beating'' pattern as 
richness of the sample increases. From: Van de Weygaert 2002a.}
\vskip -0.3truecm
\end{figure}
Thus, large coherent filaments are a direct consequence of an underlying 
cellular geometry. They get assembled by virtue of the implicit obtuse 
intra-wall and intra-filament angles. 

We have therefore found that richer objects not only cluster 
more strongly, but also out to a larger range. Hence, our audacious 
claim that it is the geometry of foamlike networks which is responsible 
for observed supercluster patterns in the distribution of the rich clusters 
and rare cosmic powerhouses of the AGNs. Also note that such 
geometries would induce a distinct flattening in the distribution 
of clusters and AGNs at scales where we would not be able to trace 
any such anisotropy in the galaxy distribution itself. The seeming contrast 
between these large scale concentrations, out to beyond $100h^{-1}\hbox{Mpc}$, 
and the smaller scales on which detectable galaxy clustering is encountered, 
is likely to find its origin in the very nontrivial geometry of the galaxy 
distribution itself ! 

\subsubsection{{\it ``Geometric Biasing'': Correlation Scaling}}
The qualitative impression of a gradually stronger, more pronounced and 
richer pattern of clustering becomes even more striking upon quantitatively 
analyzing correlation function systematics (Fig. 52, Van de Weygaert 2002a). 
A thorough numerical study of vertex clustering patterns disclosed 
an unexpected and surprising ``self-similarity''. 

\begin{figure}[b]
\vskip -1.0truecm
\centering\mbox{\hskip -0.9truecm\psfig{figure=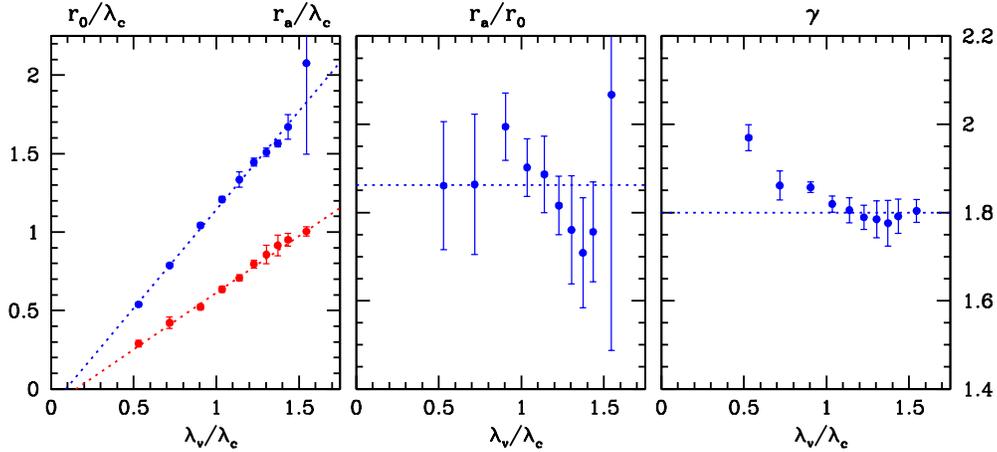,width=14.0cm}}
\vskip -7.4truecm
\caption{Scaling of Voronoi vertex two-point correlation function parameters for 
vertex subsamples over a range of ``richness''/``mass''. Left: 
the clustering length $r_0$ (red, $\xi(r_0) \equiv 1.0)$ and the correlation 
(coherence) length $r_a$ (blue, $\xi(r_a) \equiv 0$) as a function of average spatial 
separation between vertices in (mass) selected subsample, $\lambda_v/\lambda_c$. 
Centre: the ratio between clustering length $r_0$ and coherence length $r_a$ as 
function of subsample intravertex distance $\lambda_v/\lambda_c$. Right: the 
power-law slope $\gamma$ as function of $\lambda_v/\lambda_c$.}
\end{figure}
The impression of stronger clustering is indeed confirmed through a 
systematic, linear, increase in the value of the ``clustering length'' 
$r_o$. Possibly more surprising is the equally systematic increase 
of the ``correlation length'' $r_a$, the quantitative expression for 
the observed impression of point clustering noticeably extending 
over larger regions of space. Especially noteworthy are the 
following aspects of clustering scaling (see Fig. 52 \& 53):
\begin{figure}[t]
\vskip -0.3cm
\centering\mbox{\hskip -1.00truecm\psfig{figure=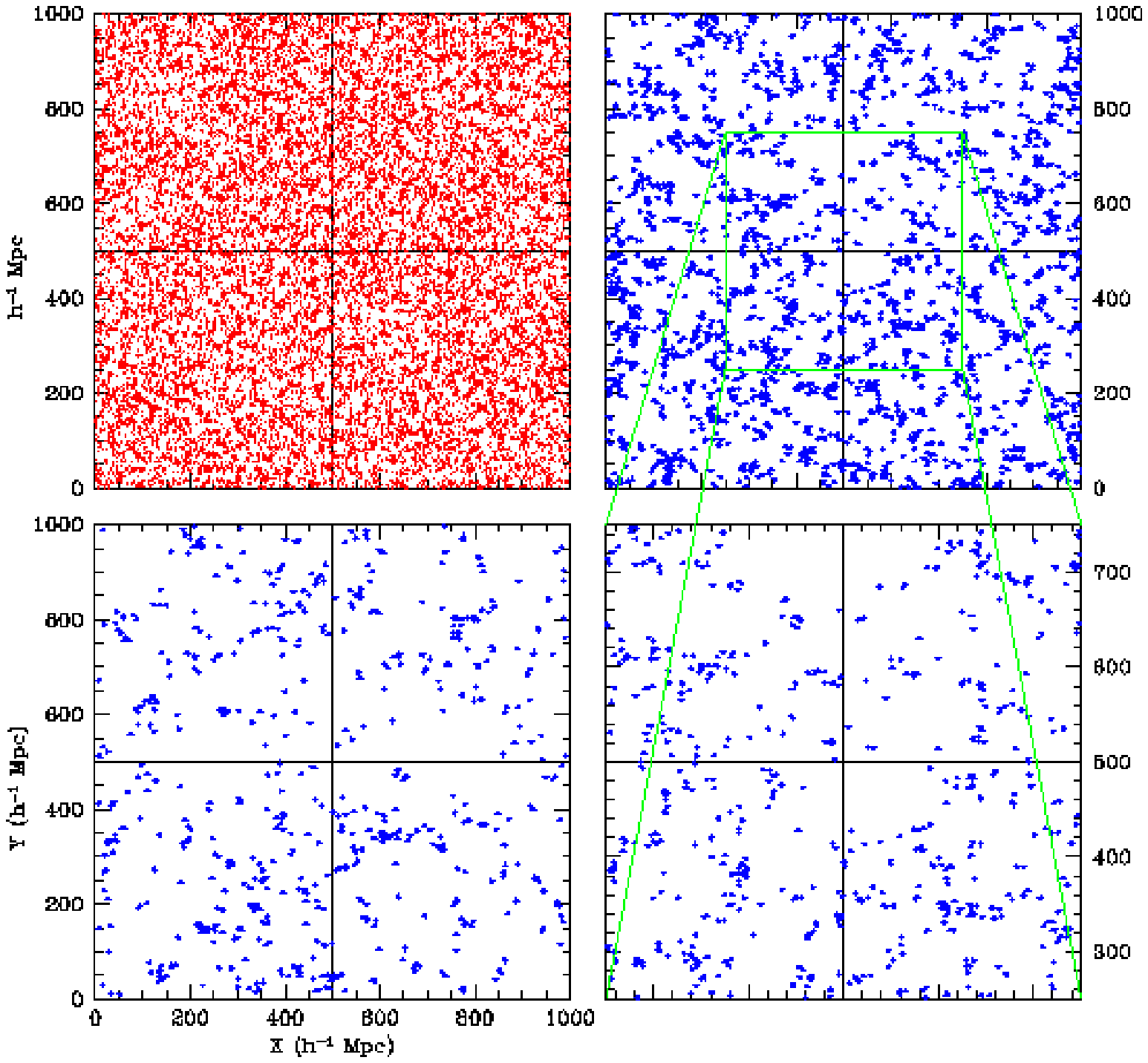,width=13.5cm}}
\vskip -0.5truecm
\caption{A depiction of the idea of `self-similarity' in the vertex distribution. 
Out of a full sample of vertices (top left) in a central slice, (top right) the 
20.0$\%$ richest vertices. Similarly, (bottom left) the 2.5$\%$ richest vertices. 
When lifting the central 1/8$^{th}$ region out of the $20\%$ vertex subsample in 
the (top righthand) frame and sizing it up to the same scale as the full box, 
we observe the similarity in point process between the resulting (bottom 
righthand) distribution and that of the $2.5\%$ subsample (bottom lefthand). 
Self-similarity in pure form !}
\vskip -0.5truecm
\end{figure}
\begin{itemize}
\item{} {\it Two-point correlation function}\\
The two-point correlation functions of selected massive cluster 
samples display a behaviour similar to that found for unbiased samples 
(Fig. 52): an almost perfect power-law at short range which beyond its 
coherence scale changes gradually into a oscillating behaviour between 
positive and negative correlations, swiftly decaying within a few 
``ringings'' to zero level. 
\item{} {\it Parameters $\xi(r)$}\\
The parameters characterizing the generic behaviour of $\xi$ -- 
amplitude, coherence scale and power-law slope -- are subject to 
systematic scaling behaviour.
\item{} {\it Correlation amplitude}\\
The amplitude of the correlation functions increases with rising 
vertex sample richness. The ``clustering length'' $r_{\rm o}$ increases 
almost perfectly linear as a function of the characteristic intra-vertex 
distance $\lambda_{\rm v}$ of the particular richness selected vertex sample.
\item{} {\it Correlation extent}\\
The large-scale (lin-lin) behaviour of $\xi_{vv}$ extends out to 
larger and larger distances with increasing sample richness. As in the 
case of $r_{\rm o}$ the ``correlation (coherence) scale'' $r_{\rm a}$ 
possesses an almost perfectly linear relation as function of the 
average sample vertex distance $\lambda_v$.
\item{} {\it Clustering and coherence scaling}\\
Therefore, combining the behaviour of $r_{\rm o}$ and $r_{\rm a}$ 
a striking ``self-similar'' scaling behaviour is revealed: the 
ratio of correlation versus clustering length is virtually constant for 
all vertex samples, $r_{\rm a}/r_{\rm o} \approx 1.86$ (for Poisson Voronoi 
tessellations). 
\item{} {\it Correlation function slope}\\
At the short power-law range, the correlation functions have 
rather similar slopes. Nonetheless, a slight and significant trend in the 
power-law slope has been found, involving an gradually increasing tilt.
Interestingly, we see a gradual change from a slope $\gamma \approx 1.95$ 
for the full sample to a robust (and suggestive) $\gamma \approx 1.8$ for 
the selected samples.
\end{itemize}
\noindent All in all, these intrinsically geometrical properties hint at a scaling 
behaviour which may befittedly be called ``geometrical biasing''. It is 
be qualitatively different from the more conventional ``peak biasing'' 
picture (Kaiser 1984) in that it involves an effect of spatial extending 
clustering, yet equivalent in its ramifications for offering an 
explanatio for the more pronounced level of clustering displayed by galaxy 
clusters.

\subsubsection{{\it ``Geometric Biasing'': Self-Similarity and 
the Cosmic Foam}}
Arguably most enticing in the scaling behaviour of the correlation functions 
has been the finding that the vertex clustering patterns display an 
intriguing intrinsic {\it self-similarity.} The correlation functions 
of the various richness selected samples all appear to constitute, 
within reasonable limits, a scaled version of  correlation 
function. The full correlation function $\xi_s(r)$ of the various 
richness selected subsamples $s$, not just the part in the power-law range, 
consistute self-similar mappings of an elementary function $\xi_{\rm el}$, 
scaled by means of a characteristic lengthscale parameter $L_s$.,
\begin{equation}
\xi_s(r)\,=\,{1 \over A_s}\,\,\xi_{\rm el}\Bigl({r/L_s}\Bigr)\,.
\end{equation}
\noindent In other words, in terms of point statistical behaviour, 
each selected vertex sample behaves like a spatially scaled version 
of a basic point distribution. This is tellingly illustrated in Fig. 54 
through a realization of such a vertex distribution, in comparison with 
a few selected subsets. A central slice through the full sample of 
Voronoi vertices in a box of a $1000h^{-1}\hbox{Mpc}$ is shown in 
the top lefthand frame (all vertices in red). From these the 
$20\%$ richest are selected and shown in the top righthand frame, while 
the $2.5\%$ richest are shown in the lower lefthand frame. Notice the 
impression of vast coherent linear structures !!! Then, sizing up 
the central half-size part of the ``$20\%$'' sample and comparing the 
resulting point process to the full sample of the ``$2.5\%$'' sample 
we indeed do find point distributions whose spatial statistics is practically 
equivalent. In other words, a pure illustration of a genuine self-similar 
point process ! 

\vfill\eject
%\bigskip
\begin{figure}[t]
\centering\mbox{\hskip -0.1truecm \psfig{figure=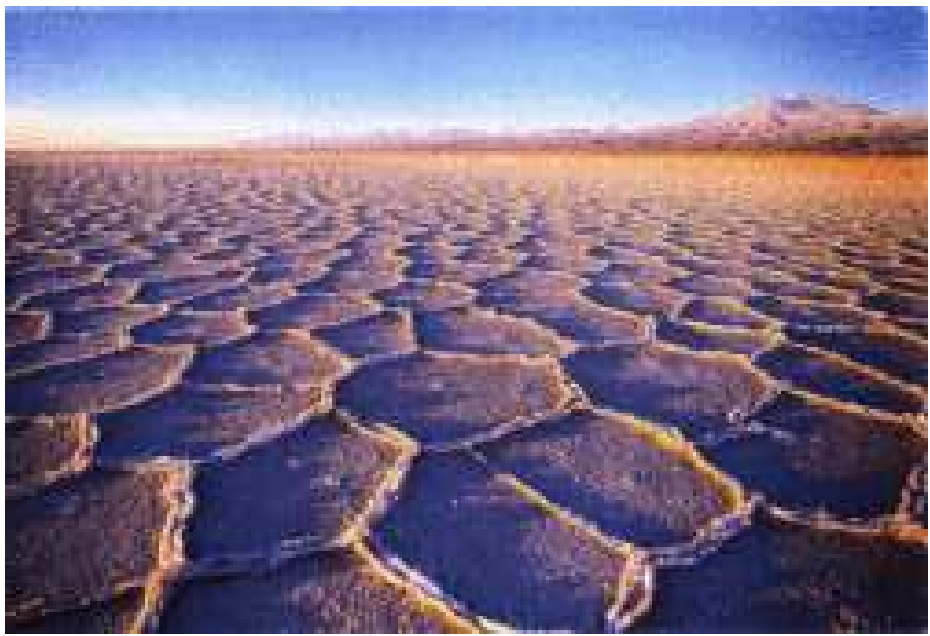,angle=270.,width=12.25cm}}
\vskip 0.25truecm
\caption{Observing in the Atacama Desert: a world filling tessellation. Courtesy: 
Dave Watson (CSIRO, Div. Exploring and Mining, Perth, Australia).}
\end{figure}
\bigskip
\vfill\eject
\section{\rm {\Large HORA EST:...}}
\begin{flushright}
{\rm{\Large   The World, A Foam}}
\end{flushright}
\smallskip\noindent
We have explored the background behind the uncovered foamlike geometry 
of the cosmic galaxy distribution. Over the past decades gradually a 
new paradigm for the cosmic matter distribution has been established, 
that of a Megaparsec scale cosmic foam, in which walls, filaments -- 
with scales up to over $100h^{-1}\hbox{Mpc}$ -- and galaxy clusters link 
up into a vast cosmos pervading network, meandering in between huge 
empty void regions with sizes up to tens of Megaparsec. 

We have indicated that such foamlike geometries are manifestations of the 
process of structure formation under the influence of gravity. 
Acting in a Universe evolving from a primordial distribution which 
is nearly uniform but for a sea of tiny density ripples, it is the 
cosmic force field which is ultimately responsible for the emergence 
of structure in the Universe. Not only the final fate of the 
forming individual matter concentrations is a consequence of its 
workings. Its sway extends sofar that it as well controls the intricate 
morphology and geometry in which the cosmos' matter content settles 
itself. It is the intrinsically anisotropic character of the cosmic force 
field which transforms gravity into the sculptor of the salient 
patterns in the cosmic matter distribution. We extensively 
discussed the way in which gravity from the onset onward moulds 
the matter distribution into the salient filamentary and wall-like 
structures that form such distinctive elements of the observed 
and mapped galaxy and matter distribution. In particular, by means 
of the simple asymptotic {\it ellipsoidal model} we indicated why on 
scales where gravity has started to decouple these perturbations from 
the cosmic background, while not yet having proceeded to a stage full collapse 
an intrinsic anisotropic shape is a natural configuration. The discussion 
continued with the indication of how such features assemble into 
the nontrivial and astonishing complexity of the {\it cosmic foam}. 

Isolating within this cosmic foam assembly the void, underdense 
regions, we argued how we can understand the formation of the 
cosmic web in an equivalent, seemingly alternative yet 
complementary, approximation. Starting from the premise of the 
intrinsically simpler dynamics of underdense regions we find a 
Universe in which we will observe the rise of perpetually  
expanding ``void sectors'', whose continuous and proliferating 
drainage and unceasing tendency towards a spherical shape must 
be seen as one of the major symptoms of the gravitational 
structure growth process. On the basis of such a view, 
we are led to an asymptotic description ultimately yielding a 
geometrical model for the cellular distribution of matter. 
{\it Voronoi Tessellations} represent a central concept in the 
mathematical branch of {\it Stochastic Geometry}. 

Voronoi tessellations represent a versatile and flexible mathematical 
model for foamlike patterns. Based on a seemingly simple definition, 
Voronoi tessellations define a wealthy stochastic network of interconnected 
anisotropic components, each of which can be identified with the various 
structural elements of the cosmic galaxy distribution. On the basis of 
this concept we have been able to investigate the ramifications of 
nontrivial foamlike patterns for a variety of characteristics 
of the spatial organization of matter in our Universe. 

On the basis of the geometry of Voronoi tessellations we 
have seemingly uncovered a surprising yet fundamental kinship between 
the cosmic foam and the large scale cluster distribution. 
Plainly tentalizing is the finding -- within the context of cellular 
geometries so characacteristic for the cosmic matter distribution --  
that a geometry as that of the cosmic foam holds the implication of 
an effect which may be denoted as a ``geometrical biasing''. In this 
contribution, we describe the existence of 
the underlying self-similar clustering behaviour in such 
cellular or foamlike geometries. It suggests a tantalizing and 
intimate relationship between the cosmic foamlike geometry 
and a variety of aspects of the spatial distribution of galaxies and 
clusters. It would explain why the level of clustering amongst 
massive galaxy clusters is so much stronger than that between 
their more moderate brethren. As significant is the finding 
of positive spatial cluster-cluster correlations over scales 
substantially exceeding the ``elementary'' scale of voids and 
other cosmic foam elements. 

\bigskip
\bigskip
Indeed, it appears that our Universe in more than one way resembles 
Plato's Academia, that venerable institution devoted to learning, seeking to uncover  
the secrets of the world and our existence, over whose door it was said to be written: 

\bigskip
\bigskip
{\centerline{\Large {\it ``Let no one unversed in geometry enter here. ''}}}
\bigskip

\vfill\eject
\begin{acknowledgments}
In preparing this manuscript, I wish to thank John Peacock for his permission to use the 2dF 
image in Fig. 1 and Fig. 5, Michael Strauss for providing me with the Sloan redshift survey 
impression in Fig. 1, Luiz da Costa for providing the CfA2/SSRS survey map in Fig. 1, and to 
the LCRS team for the Las Campanas redshift map in Fig. 1, and S. Maddox for Fig. 2. Also grateful I 
am to Martha Haynes for supplying me with Fig. 5, to Stefano Borgani for his permission to use Fig. 7, 8, and 
46, to Alex Szalay for issueing Fig. 11, to V. de Lapparent for the permission to use Fig. 3, to B. Kirshner for 
using Fig. 9, to H. Quintana for Fig. 10, to G. Bothun for Fig. 13 and to John Dubinski for Fig. 32. 
Dave Watson provided the beautiful Atacama desert impression. In particular 
I am grateful to Jacco Dankers for helping to prepare the Geomview package graphs of Fig. 34, 35 \& 40.  
In particular acknowledged are W. Schaap and E. Romano-D\'{\i}az for useful and inspiring discussions, 
and for preparing Figs. 14 \& 15 and Figs. 17 \& 18, and allowing me to use these prior to publication. 

\vfill
%\begin{flushright}
%{\rm cont'd next page}
%\end{flushright}
\eject
\bigskip
\ \ \ \ \ \ 
\bigskip
\vfill
Fond memories and gratitude characterize these days in the cradle of 
western civilization.

First and foremost, the author wishes to thank Manolis Plionis for the kind invitation 
for this wonderful workshop, for his caring hospitality and even more for the almost 
infinite and greatly appreciated patience and co\"operation 
in preparing this contribution, beset as it was by an unacceptable delay in 
submission. Yet, ultimately, the contribution is not nearly as large as the gratitude 
for the joys of Archaion Gefseis and the late night sounds and dances of rebetiko ... 
it was like Zorba spelling out the essentials of true living ! 

Moving these days was the heartwarming hospitality of 
the Papadopoulos family; Padelis, his parents and brother made us feel 
like family by indulging us to the intoxicating delights of traditional Greek 
gastronomy, feasting and sirtaki !

Then, in the quest for the origin and workings of our world, it was no more than proper 
to pay due respect to the two Athenians who were so crucial in moulding the human 
mind into its prime instrument of inquiry ... inciting a pilgrimage to the cell 
where Socrates' showed the ultimate resolve for principle and moral, and a 
visit to the foundations of western knowledge and learning, there where Plato's 
Academia assembled ... 

Profoundly awe-inspiring it was to witness Zeus himself descending, amidst his 
thunderclouds over the Acropolis he came to pay tribute to a special Greek friend, 
namesake and descendant of the legendary queen of Greece, the woman who 
launched a thousand ships, Helena ...
\end{acknowledgments}
\bigskip
\vfill
{\it ``... Each time I climbed the Acropolis again, the Parthenon seemed to be 
swaying slightly, as in a motionless dance -- swaying and breathing. ...''}
\begin{flushright}
{\rm cont'd beyond next page}
\end{flushright}

\vfill\eject
\begin{figure}[h]
\vskip 1.0cm
\centering\mbox{\psfig{figure=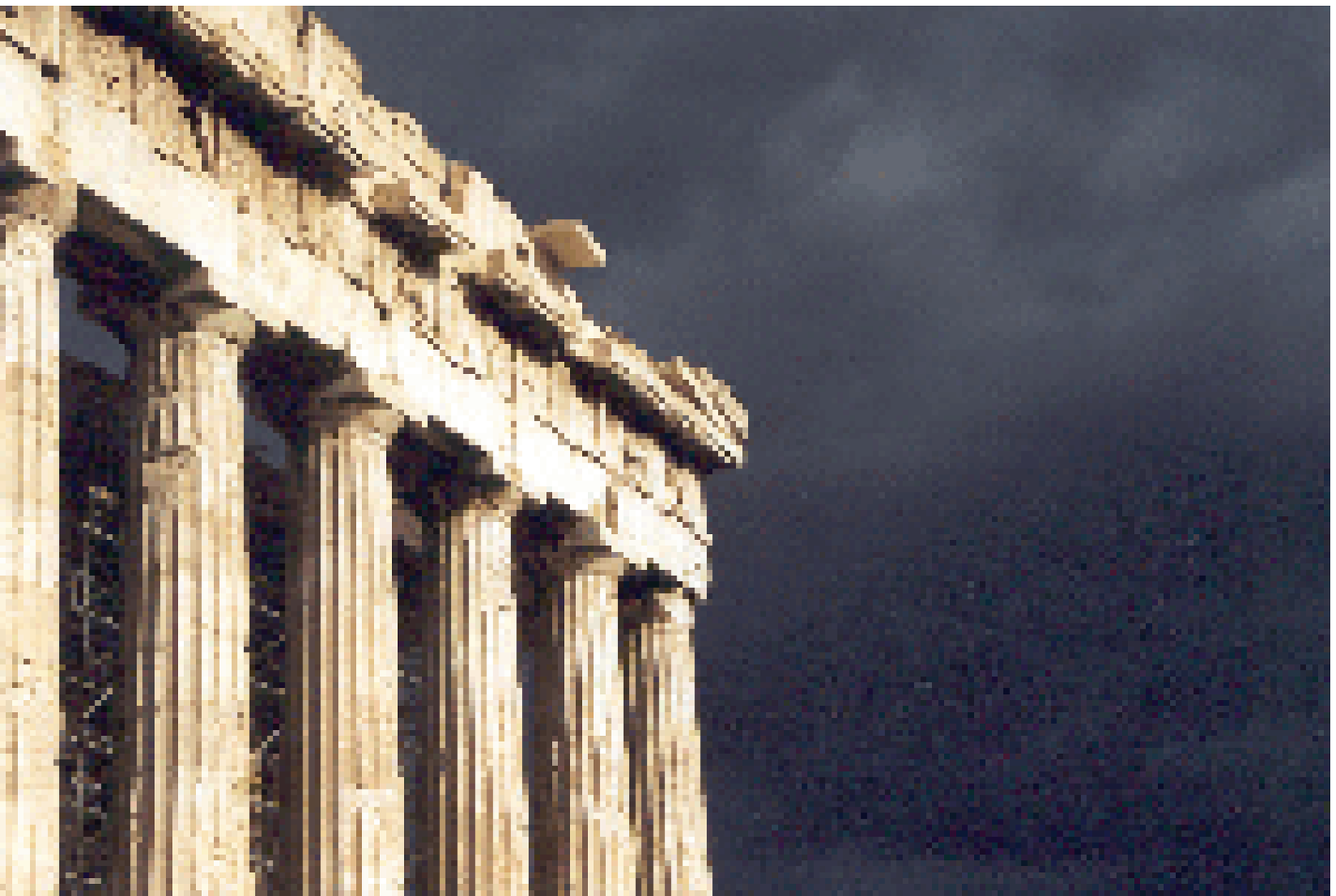,width=12.0cm}}
\vskip 1.0cm
\centering\mbox{\psfig{figure=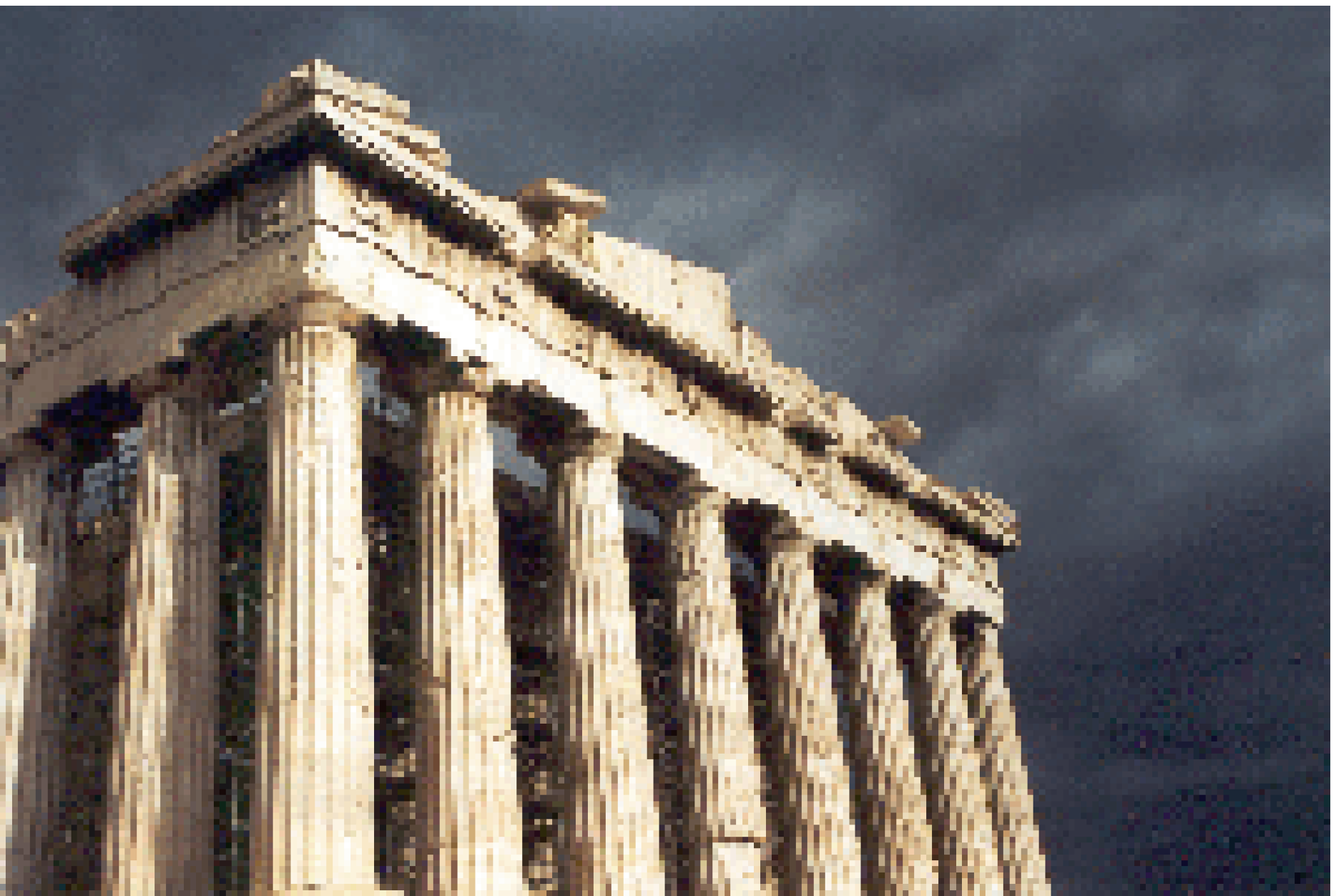,width=12.0cm}}
\end{figure}

\vfill\eject
{\it ``[..] This temple that towered before me, what a trophy it was,
what a collaboration between mind and heart, what a supreme fruit of human effort ! 
Space had been conquered; distinctions between large and small had largely vanished. Infinity 
entered this narrow, magical parallelogram carved out by man, entered leisurely and took 
its repose there. Time had been conquered as well; the lofty moment had been 
transformed into eternity.''}
\smallskip
\begin{flushright}
{\rm Nikos Kazantzakis, Report to Greco, 1961}
\end{flushright}

\end{document}